\documentclass[twoside,11pt]{article}

\usepackage{preprint}
\usepackage{xcolor} 
\usepackage{amsmath,amssymb}
\usepackage{dirtytalk} 
\usepackage{tikz}
\usetikzlibrary{shapes, arrows, positioning, fit, backgrounds, calc, matrix}
\usepackage{algorithm} 
\usepackage{bbm} 
\usepackage[noend]{algpseudocode} 
\usepackage{longtable,booktabs,array, makecell}
\newcommand\numberthis{\addtocounter{equation}{1}\tag{\theequation}}

\newtheorem{condition}{Condition}

\usepackage{lastpage}
\jmlrheading{}{}{1-\pageref{LastPage}}{9/26; 
}{
}{21-0000}{Jongmin Mun, Paromita Dubey and Yingying Fan}

\ShortHeadings{Thompson Sampling SDP Clustering}{Mun, Dubey and  Fan}
\firstpageno{1}

\begin{document}

\title{Iterative Exploration-Driven Sparse SDP Clustering via Thompson Sampling}

\author{\name Jongmin Mun \email jongmin.mun@marshall.usc.edu \\
\name Paromita Dubey \email paromita@marshall.usc.edu \\
\name Yingying Fan \email fanyingy@marshall.usc.edu \\
       \addr Department of Department of Data Sciences and Operations, Marshall School of Business \\
       University of Southern California\\
       Los Angeles, CA 90089, USA}

\editor{}

\maketitle

\begin{abstract}
High-dimensional sparse clustering is a challenging combinatorial problem due to the tight coupling between cluster assignment and variable selection. We demonstrate that semidefinite programming (SDP) relaxation of K-means is robust to variable over-selection by establishing minimax separation bounds. Leveraging this robustness,
we offer a twist on the recent trend of using exogenous randomness to evaluate feature importance and control false discovery by partially shifting the objective toward active feature exploration. We propose a randomized alternating block-coordinate framework that alternates between SDP clustering and  feature selection. Here, feature selection is formulated as a multi-armed bandit problem, solved via Thompson Sampling with stochastic rewards generated by    clustering and  robust permutation tests.
This approach introduces adaptive memory by aggregating historical variable-selection outcomes into posterior distributions, and selects features via posterior sampling, enabling stochastic exploration that promotes the inclusion of underexplored features and facilitates escape from local maxima. We establish uniform exact-recovery guarantees for adaptively selected feature sets.
We extend the method to settings with unknown covariance through a scalable estimation procedure. Synthetic experiments and a real-data application in document clustering show that the proposed memory-driven randomized approach outperforms state-of-the-art sparse clustering methods across the studied settings.
\end{abstract}


\section{Introduction}\label{section:introduction} 
We study high-dimensional clustering when the underlying cluster structure is determined by a small, unknown subset of features. For example, in customer segmentation, businesses may measure thousands of customer attributes, yet meaningful groups often depend on only a few.
Identifying these features yields interpretable segments that inform personalized products and advertisements \citep{wangSparseClusteringCustomer2024}. Exploiting sparsity to improve clustering and support downstream applications is also common in genomics \citep{golub_molecular_1999, luEnhancedInterpretableClustering2017}, neuroimaging \citep{mishra_av1451_2017, namgungSexDifferencesAutism2024}, and finance \citep{nystrup_feature_2021}.

This paper introduce structured exogenous randomness to guide active feature exploration, using successive clustering outcomes to refine feature selection.  
Structured exogenous randomness has been widely used in supervised variable selection to assess feature importance and control false discoveries, such as stability selection \citep{meinshausenStabilitySelection2010}, model-X knockoffs \citep{candesPanningGoldModelX2018}, and  Gaussian mirror method \citep{xing_controlling_2023}. We build on this idea of leveraging structured exogenous randomness to promote active feature exploration in sparse clustering, where feature selection and cluster assignment are closely coupled: informative features help reveal cluster structure, while cluster assignments provide evidence of feature importance. We show that accurate clustering can tolerate moderate feature over-selection under suitable conditions. This motivates using randomness to explore potentially informative features, with accurate clustering as the primary objective.

Building on this motivation, we propose an iterative algorithm that alternates between clustering and feature selection, accumulating evidence from clustering and permutation tests to guide subsequent feature exploration. Feature selection is formulated as a multi-armed bandit problem, with Thompson sampling balancing exploration and exploitation and permutation tests generating stochastic rewards. This mechanism adaptively searches for a feature set sufficient for accurate clustering without requiring exact feature recovery.
 
\subsection{Clustering Model and Target Parameter}
With accurate clustering as our primary goal, we treat variable inclusion indicators as discrete nuisance parameters. This paper formally shows that consistent and efficient estimation of cluster assignments does not require precise identification or optimal convergence of these nuisance components.

We use the planted structure model of \cite{giraudPartialRecoveryBounds2019} to formalize the sparsity structure. Assume that the number of clusters $K$ is  known. Let $G^\ast := (G_1^\ast, \dots, G_K^\ast )$ denote non-random partition of observation indices $\{1, \ldots, n\}$. Let
$\mathbf{c}^\ast_1, \ldots, \mathbf{c}^\ast_K \in \mathbb{R}^p$ and $\boldsymbol{\Sigma}^\ast \in \mathbb{R}^{p \times p}$ denote the unknown cluster centers and the common covariance matrix, respectively. We observe a data matrix $\mathbf{X} \in \mathbb{R}^{p \times n}$, with  columns  $\mathbf{X}_1, \ldots, \mathbf{X}_n$ independently generated as 
\begin{equation}\label{def:gaussian_mixture_model}
    \mathbf{X}_i 
    =
    \mathbf{c}^\ast_k + \mathbf{W}_i~\text{if}~i \in G_k^\ast,
    \quad
    \mathbb{E}[\mathbf{W}_i] = 0,
    \quad
    \mathrm{Cov}(\mathbf{W}_i) = \boldsymbol{\Sigma}^\ast.
\end{equation}  
The goal is to recover $G^\ast$ from the observations $\mathbf{X}$ alone, treating the $\mathbf{c}^\ast_k$ and $\boldsymbol{\Sigma}^\ast$ as nuisance parameters.

Let $
\boldsymbol{\Omega}^\ast :=
(\boldsymbol{\Sigma}^{\ast})^{-1}$. For two-class Gaussian mixtures, the Bayes-optimal classification boundary that maximizes the signal-to-noise ratio is proportional to $\boldsymbol{\Omega}^\ast(\mathbf{c}_1^\ast - \mathbf{c}_2^\ast)$, rather than $(\boldsymbol{\Omega}^\ast)^{1/2}(\mathbf{c}_1^\ast - \mathbf{c}_2^\ast)$ or $(\mathbf{c}_1^\ast - \mathbf{c}_2^\ast)$ \citep{fan_optimal_2013}.
Motivated by this, the high-dimensional clustering literature \citep{cai_chime_2019,azizyan_efficient_2015} defines an unknown set of signal variables:
\begin{equation}\label{def:S0}
 S^\ast :=
 \hskip -6.4mm
 \bigcup_{1 \leq k_1 \neq k_2 \leq K} 
  \hskip -5.4mm
  \mathrm{supp} \bigl( \boldsymbol{\Omega}^\ast
(\mathbf{c}^\ast_{k_1} - \mathbf{c}^\ast_{k_2}) \bigr) \subset \{1, \ldots, p\},  
\end{equation}
which we hereafter refer to as the \emph{support}.
We emphasize that while $S^\ast$ is critical for clustering performance, we ultimately treat it as a nuisance parameter; our primary target of inference is the partition $G^\ast$.

\subsection{Our Key Contributions}
The interdependence between variable selection and clustering naturally motivates an iterative approach that alternates between selecting variables based on current cluster assignments and refining those assignments using the selected variables. Prominent sparse clustering methods \citep{witten_framework_2010, cai_chime_2019} adopt this approach, but their greedy optimization can lead to convergence at poor local optima. Our method combines \emph{partial relaxation} to make optimization more tractable with \emph{bandit-driven random search} to evaluate feature importance and facilitate escape from poor local solutions through exploration and memory.
 
 For the partial relaxation, we build on recent theoretical advances in semidefinite programming (SDP) relaxations for $K$-means. 
 In an iterative algorithm, given a feature subset $S$  selected in the previous step, a natural approach to identifying clusters is to apply a clustering algorithm using exclusively the variables in $S$.
 Under a sub-Gaussian noise assumption, we extend existing  theory \citep{chen_cutoff_2021} to quantify the sensitivity of SDP $K$-means to  (possibly random) choice of variable set $S$. 
 In particular, we establish a uniform minimax separation bound that guarantees exact cluster recovery for SDP $K$-means confined to $S$, whenever the cluster center separation confined to $S$ satisfies
\begin{equation*}
    \min_{1 \leq k_1 \neq k_2 \leq K} 
\| \big(\boldsymbol{\Omega}^\ast
(\mathbf{c}_{k_1}^\ast - \mathbf{c}_{k_2}^\ast )\big)_{S \cap S^\ast}
 \|_2^2 \gtrsim \log n + \frac{|S| \log p}{n} + \sqrt{ \frac{|S| \log p}{n} }.
\end{equation*}
The dependence on the set size $\vert{}S\vert{}$ shows that SDP-relaxed $K$-means is robust to moderate variable over-selection, as the penalty for including a few additional noise variables scales only with the size of the selected set. Thus, accurate clustering requires only that a moderately sized variable set captures sufficient   signal, rather than exactly recovering the true signal variable set.

Motivated by  this result, we develop a randomized block coordinate optimization algorithm that alternates between SDP-relaxed $K$-means and stochastic variable selection. We formulate variable selection as a multi-armed bandit problem and solve it via Thompson sampling, modeling each variable's clustering utility as a latent Bernoulli parameter  and estimating it through an iteratively updated Beta posterior.
This framework introduces two key innovations for sparse clustering: 
\begin{enumerate}
    \item \emph{Active exploration}: Variable inclusion indicators are sampled from their posterior distributions, encouraging the exploration of underrepresented variables and helping the algorithm escape poor local optima.
    \item  \emph{Memory}: Through iteration, each variable is evaluated multiple times under exogenous randomness. Accumulated evidence from past successes and failures in variable selection stabilizes learning and mitigates overfitting to transient under- or over-selection in any single iteration.
\end{enumerate}
For bandit formulation, we first cast variable selection as a  two-sample mean testing problem, assessing whether the clusters differ in distribution mean along each individual variable. To obtain stochastic rewards at each iteration, we employ permutation tests. We frame these as Maximum Mean Discrepancy (MMD) permutation tests, as they are known to be theoretically optimal for two-sample testing under various forms of data corruption \citep{schrabRobustKernelHypothesis2025, munMinimaxOptimalTwoSample2025, kimDifferentiallyPrivatePermutation2026}.
Specifically, we use the adversarial-corruption correction of \cite{schrabRobustKernelHypothesis2025} to make each reward test conservative under a prespecified budget of erroneous cluster assignments.

Finally, for unknown covariance structure, we avoid full covariance estimation by employing the \say{innovated transformation} $\boldsymbol{\Omega}^\ast \mathbf{X}$ \citep{fan_innovated_2016}, which estimates only the necessary precision matrix components, ensuring computational efficiency.

We summarize the major contributions of this paper and contrast them with the previous studies as follows:
\begin{itemize}
  \item \emph{Uniform minimax separation of subset SDP $K$-means:} We establish a uniform separation bound guaranteeing exact cluster recovery for SDP $K$-means confined to any feature subset $S$. Unlike prior work \citep{chen_cutoff_2021}, our bound holds uniformly over all possible subsets, quantifying the algorithm's sensitivity to the chosen variable set and showing its robustness against moderate over-selection. This result   justifies active variable exploration over strict false discovery control in clustering task.
  \item \emph{Iterative algorithm via Thompson sampling and corruption-robust permutation testing:} We propose an iterative algorithm in which active exploration and memory are formally governed by Thompson sampling. Unlike prior work in the supervised setting \citep{liuVariableSelectionThompson2023}, defining a stochastic reward in clustering setting is challenging because intermediate rewards   depend on imperfect intermediate clustering assignments. We employ a corruption-robust permutation test to cast variable selection in clustering as a stochastic reward. Furthermore, unlike previous iterative sparse clustering methods \citep{witten_framework_2010, cai_chime_2019}, our approach incorporates a principled mechanism to escape local optima.
\end{itemize}

\subsection{Existing Literature}
Unsupervised dimension reduction methods such as PCA  may emphasize variation that is unrelated to the sparse clustering signal and, as a result, hinder rather than facilitate downstream clustering \citep{chang_using_1983}. The sparse clustering literature therefore differs primarily in how the clustering signal is defined and how it is extracted. Existing approaches can be broadly classified into two categories: two-step procedures and iterative methods.
We focus our review on iterative methods, as they are more directly related to our work. For comprehensive discussions of two-step procedures, we refer readers to \cite{jinInfluentialFeaturesPca2016} and \cite{witten_framework_2010}. In particular, \citet{jinPhaseTransitionsHigh2017a} show that within the two-step framework, accurate clustering can still be achieved without perfect variable selection under certain sparsity and signal regimes, conveying a message similar to ours. However, in their proposed method, the variable selection relies on higher criticism \citep{donoho_higher_2008}, which is effective for detecting rare and weak signals. This contrasts with our perspective that some weak signals can be safely ignored.

Compared to two-step methods, iterative approaches jointly update variable selection and cluster assignments, typically alternating clustering with regularized estimation of model parameters such as variable weights \citep{    witten_framework_2010, huo_integrative_2017},
cluster centers \citep{
sun_regularized_2012}, and variable rankings \citep{arias-castroSimpleApproachSparse2017, zhang_simple_2020}.  
Regularized EM methods follow the same principle, updating labels in the E-step and sparse model parameters in the M-step \citep{dong_tuningfree_2024, pan_penalized_2007,     zhou_penalized_2009}.  
Notably, \citet{cai_chime_2019} derive a minimax rate for excess clustering risk on new observations via $\ell_1$-regularized EM estimation of transformed cluster center differences. 
 
Casting variable selection as a reinforcement learning problem is intractable, as the state space grows exponentially with the number of features. To address this, bandit-based methods have been proposed, including Upper Confidence Trees for classification \citep{gaudelFeatureSelectionOneplayer2010} and Thompson sampling for regression \citep{liuVariableSelectionThompson2023}. We extend these ideas to unsupervised clustering.
Our bandit-based approach  is also related to recent learning-to-optimize literature \citep{scavuzzo_machine_2024} for  mixed-integer programs. In this literature, reinforcement learning is integrated into branch-and-bound to learn variable selection policies or primal heuristics from signals generated during the optimization process.

\subsection{Notations}
Vectors and matrices are denoted in bold, with entries in corresponding non-bold.  
For a matrix \( \mathbf{A} \), a single subscript denotes a row, column, or corresponding subset, as determined by context, while double subscripts specify row and column indices, respectively; subscripts may be indices or index sets. For vectors, subscripts denote elements or sub-vectors, and parentheses are used to avoid ambiguity, for example, \( (\widetilde{\mathbf{c}}_1^\ast)_S \).
We write $\mathbf{A} \geq 0$, $\mathbf{A} > 0$, and $\mathbf{A} \succeq 0$ for elementwise nonnegativity, positivity, and positive semidefiniteness, respectively.  For a set $S$, let $|S|$ denotes its cardinality.
For an  integer $a>0$, let $\mathbf{1}_a \in \mathbb{R}^a$ denote the \textit{all-one vector},
and $[a]$ denote $\{1,\cdots, a\}$.
For a set $G_k \subset [n]$, let $\mathbf{1}_{G_k} \in \{0,1\}^n$ denote the indicator vector that takes the value 1 at indices in $G_k$ and 0 elsewhere.  
 
\section{SDP K-Means with  Selected Variables}
\label{section:theoretical_motivation} 
We first review the SDP relaxation of $K$-means and its high-dimensional limitations via a minimax analysis. We then propose a new version of the SDP relaxation for $K$-means, termed \emph{innovated subset SDP}, which incorporates sparsity through variable subsetting with an innovated transform.   Finally, we establish the robustness of the innovated subset SDP to variable over-selection.
\subsection{SDP K-Means}
For exposition, we assume that the  precision matrix $\boldsymbol{\Omega}^\ast$ is known.
We introduce the cluster assignment matrix $\mathbf{H} \in \{0,1\}^{n \times K}$ as an alternative representation of the partitions $G_1, \ldots, G_K$, with entries defined by $H_{i,k} := \mathbbm{1}(i \in G_k)$ for $i \in [n]$ and $k \in [K]$.
Additionally, let $\mathbf{D}_G:=\mathrm{diag}(1/\vert{}G_1\vert{}, \ldots, 1/\vert{}G_K\vert{})$, and let $\langle \cdot, \cdot \rangle$ denote the Frobenius inner product.
Maximizing the $K$-means objective over all possible partitions $G$ arises from the maximum likelihood problem for a Gaussian mixture model with general covariance matrices \citep{zhuang_likelihood_2023}, which can be formulated as the following integer program:
\begin{equation}\label{objective_from_chen}
 \max_{
 \mathbf{H} \in \{0,1\}^{n \times K}
 } \;
 \langle
 \mathbf{X}^\top \boldsymbol{\Omega}^\ast \mathbf{X},
 \mathbf{H} \mathbf{D}_G \mathbf{H}^\top \rangle,~\text{s.t.}~\mathbf{H} \mathbf{1}_K = \mathbf{1}_n,
\end{equation}
  Under the change of variables $\mathbf{Z} = \mathbf{H} \mathbf{D}_G \mathbf{H}^\top$, we observe that the $n \times n$ symmetric matrix $\mathbf{Z}$ satisfies $\mathbf{Z}^\top = \mathbf{Z}, \mathbf{Z} \succeq 0, \mathrm{tr}(\mathbf{Z}) = K, \mathbf{Z} \boldsymbol{1}_n = \boldsymbol{1}_n$, and $\mathbf{Z} \geq 0$.
Replacing the constraint $\mathbf{H}\mathbf{1}_K = \mathbf{1}_n$ with these constraints and expanding the search space to the set of continuous matrices, we obtain the following SDP relaxation of \eqref{objective_from_chen}:
\begin{equation}\label{kmeans_SDP_form}
\max_{ \mathbf{Z} \in \mathbb{R}^{n \times n} }~\langle \mathbf{X}^\top \boldsymbol{\Omega}^\ast \mathbf{X}, \mathbf{Z} \rangle
\quad\text{s.t.}\quad\mathbf{Z}^\top = \mathbf{Z},~
	\mathbf{Z} \succeq 0,~
	\mathrm{tr}(\mathbf{Z}) = K,~ 
	\mathbf{Z} \boldsymbol{1}_n = \boldsymbol{1}_n,~
    \mathbf{Z} \geq 0.
\end{equation}
Since the solution to \eqref{kmeans_SDP_form} is a continuous-valued matrix,   the SDP $K$-means algorithm recovers cluster labels by applying spectral clustering to the solution of \eqref{kmeans_SDP_form}.
Under fixed dimensionality $p$, when $\boldsymbol{\Sigma}^\ast = \mathbf{I}_p$ and the $\mathbf{W}_i$ are Gaussian, the SDP $K$-means algorithm achieves the minimax rate for exact cluster recovery. Specifically, minimax rate is characterized by a minimum $\ell_2$ separation between cluster centers of order
$\log n + \sqrt{ (Kp\log n)/n }$
\citep{chen_cutoff_2021}.
Since      
$\log n + \sqrt{ (Kp\log n)/n }$  grows with $p$ at a square-root rate,  methods that ignore sparsity (including SDP $K$-means) may fail when $p$ is comparable to or larger than $n$, unless separation scales accordingly.
In other words, although SDP $K$-means achieves the optimality in the low-dimensional setting, it does not adapt to sparsity. This motivates incorporating sparsity into SDP-based clustering to develop a sparse clustering algorithm that can potentially achieve the optimality.

\subsection{Incorporating Sparsity via Subset SDP \texorpdfstring{$K$}{K}-Means with Innovated Transformation}\label{section:best_subset_sdp}
We extend the SDP relaxation in \eqref{kmeans_SDP_form} to incorporate variable selection while aligning with the Bayes-optimal classification boundary.  
For ease of exposition, we assume that the precision matrix $\boldsymbol{\Omega}^\ast$ is known.
Let us define 
the innovated transformation
\begin{equation*}
  \widetilde{\mathbf X}:=\boldsymbol\Omega^\ast\mathbf X.
\end{equation*}
Such a transform maximizes the signal-to-noise ratio for Gaussian classification, outperforming whitening transformation $(\boldsymbol{\Omega}^\ast)^{1/2}\mathbf X$ \citep{fan_optimal_2013}, and, by extension, improves clustering performance \citep{cai_chime_2019}.
Using the innovated transformation, we can express
$
\mathbf{X}^\top \boldsymbol{\Omega}^\ast \mathbf{X}
=
\widetilde{\mathbf{X}}^\top \boldsymbol{\Sigma}^\ast \widetilde{\mathbf{X}},
$
and $\widetilde{\mathbf{X}} = \widetilde{\mathbf{C}}^\ast + \widetilde{\mathbf{W}}$, 
where $ \widetilde{\mathbf{C}}^\ast$ is the transformed cluster center matrix and $\widetilde{\mathbf{W}}$ is the transformed   noise matrix.
Then, the SDP objective separates into a signal term $(\widetilde{\mathbf{C}}^\ast)^\top \boldsymbol{\Sigma}^\ast \widetilde{\mathbf{C}}^\ast$ and noise terms independent of $\widetilde{\mathbf{C}}^\ast$.
Under sparsity assumption \eqref{def:S0}, the signal concentrates on $S^\ast$:   
\begin{equation*}
(\widetilde{\mathbf{C}}^\ast)^\top \boldsymbol{\Sigma}^\ast \widetilde{\mathbf{C}}^\ast
=
(\widetilde{\mathbf{C}}^\ast_{S^\ast})^\top
\boldsymbol{\Sigma}^\ast_{S^\ast,S^\ast}
\widetilde{\mathbf{C}}^\ast_{S^\ast}
+ \text{terms involving features in } [p] \setminus S^\ast.
\end{equation*}
Motivated by this observation, we approximate the signal component
$(\widetilde{\mathbf{C}}^\ast_{S^\ast})^\top
\boldsymbol{\Sigma}^\ast_{S^\ast,S^\ast}
\widetilde{\mathbf{C}}^\ast_{S^\ast}$
by its empirical analogue
 $\widetilde{\mathbf{X}}_{S^\ast}^\top \boldsymbol{\Sigma}^\ast_{S^\ast,S^\ast} \widetilde{\mathbf{X}}_{S^\ast}$.
If a reliable feature set $S$ is available, it is therefore natural to solve
\begin{equation}\label{def:subset_sdp}
\max_{\mathbf{Z} \in \mathbb{R}^{n \times n}}
\;\;
\big\langle
\widetilde{\mathbf{X}}_{S}^\top
\boldsymbol{\Sigma}^\ast_{S,S}
\widetilde{\mathbf{X}}_{S},
\mathbf{Z}
\big\rangle
\quad
\text{s.t.}
\quad
\mathbf{Z}^\top = \mathbf{Z},\;
\mathbf{Z} \succeq 0,\;
\mathbf{Z} \ge 0,\;
\mathrm{tr}(\mathbf{Z}) = K,\;
\mathbf{Z}\boldsymbol{1}_n = \boldsymbol{1}_n.
\end{equation}
See Appendix~\ref{section:simul_known_noniso}   for empirical evidence that this objective indeed captures the clustering signal.
We refer to the clustering procedure, which maps $(\widetilde{\mathbf{X}}, S)$ to cluster labels, as \textit{innovated subset SDP} (Algorithm~\ref{alg:sdp_subroutine}).

If an adequate subset $S$ close to the true support $S^\ast$ were known a priori, executing Algorithm~\ref{alg:sdp_subroutine} would solve the sparse clustering problem. In practice, however, $S^\ast$ is unknown, and estimating it requires a reasonably accurate initial clustering. This interdependence motivates a natural and widely used iterative framework: starting with a pilot clustering, the procedure alternates between estimating the active variables $\widehat{S}$ given the current partition $\widehat{G}$, and updating the partition $\widehat{G}$ based on $\widehat{S}$. Consequently, Algorithm~\ref{alg:sdp_subroutine} serves as the core clustering subroutine within this alternating framework. 
When designing a procedure to estimate $\widehat{S}$ from $\widehat{G}$, one must consider how sensitive the subsequent clustering step is to the quality of the selected variables. The following subsection addresses this challenge.

\begin{algorithm}[b!]
\caption{Innovated Subset SDP $K$-means  for $K=2$ : \texttt{SDPclust}}
\label{alg:sdp_subroutine}
\begin{algorithmic}[1]
\Require Transformed and truncated data matrix  $\widetilde{\mathbf{X}}_{S, \cdot}$, covariance  sub-matrix $\boldsymbol{\Sigma}_{S, S}$
\State
$\hat{\mathbf{Z}} 
\gets
\arg \max_{\mathbf{Z} \in \mathbb{R}^{n \times n}}
\langle 
	\widetilde{\mathbf{X}}_{S, \cdot}^\top 
    \boldsymbol{\Sigma}_{S, S}
    \widetilde{\mathbf{X}}_{S, \cdot}, \mathbf{Z} \rangle
    \newline
    \hspace*{2.7em}
    \text{s.t.}~\mathbf{Z}^\top =~\mathbf{Z},~\mathbf{Z} \succeq~0,~ 
	\mathrm{tr}(\mathbf{Z}) =~2, 
	~\mathbf{Z} \boldsymbol{1}_n =~\boldsymbol{1}_n,~
	\mathbf{Z} \geq 0$
\State Output $\hat{G} = (\hat{G}_1, \hat{G}_2)\gets$ spectral clustering result  on  $\hat{\mathbf{Z}}$ 
\end{algorithmic}
\end{algorithm}

\subsection{Robustness of Subset SDP to Variable Over-Selection} 
\label{section:minimax}
When $\boldsymbol{\Sigma}^\ast = \sigma^2 \mathbf{I}_p$,
 SDP $K$-means restricted to  an arbitrary $S$ solves: 
\begin{equation}\label{SDP_objective_submatrix}
	\max_{\mathbf{Z} \in \mathbb{R}^{n \times n}}~ \langle 
	\mathbf{X}_{S, \cdot}^\top \mathbf{X}_{S, \cdot}, \mathbf{Z} \rangle
    \quad\text{s.t.}\quad
    \mathbf{Z}^\top = \mathbf{Z},~
    \mathbf{Z} \succeq 0,~
	\mathrm{tr}(\mathbf{Z}) = K,~
	\mathbf{Z} \boldsymbol{1}_n = \boldsymbol{1}_n,~
	\mathbf{Z} \geq 0.
\end{equation}
Under the assumption that  
$\mathbf{W}_i$ are  Gaussian,
 our   analysis proceeds in two steps:
 \begin{enumerate}
     \item Show that if the cluster center separation confined in $S \cap S^\ast$  exceeds a threshold depending on $n$, $p$, and subset size $|S|$, exact recovery holds uniformly over all $S$ with high probability.
     \item Characterize the corresponding minimax optimal regime.
 \end{enumerate}
 For  $\boldsymbol{\psi} = (\mathbf{c}_1, \ldots, \mathbf{c}_K, \mathbf{H})$, with $\mathbf{c}_k \in \mathbb{R}^p$ cluster centers and $\mathbf{H} \in \{0,1\}^{n \times K}$ cluster assignment matrix,
let $\mathbb{P}_{\boldsymbol{\psi}}$ denote the joint distribution of $n$ independent Gaussian vectors with means $\mathbf{c}_1, \ldots, \mathbf{c}_K$ and covariance $\sigma^2 \mathbf{I}_p$.
The parameter space 
$$\Psi(s,\Delta,K) := \mathcal{C}(s,\Delta,K)\times\mathcal{H}(K)$$ 
is defined as the product of the following sets, which characterize the cluster centers and partitions:
\begin{align*}
	\mathcal{C}(s,  \Delta, K) &:= \bigl\{ 	
	( 
	\mathbf{c}_1, \ldots, \mathbf{c}_K 
	):%
	\mathbf{c}_k \in \mathbb{R}^p,
	\| \mathbf{c}_{k_1}  - \mathbf{c}_{k_2}\|_2^2 \geq \Delta^2, ~\text{for all}~1 \leq k_1 \neq k_2 \leq K,
    	\\& \hskip 1.1cm~
\cup_{k_1, k_2}\mathrm{supp}(\mathbf{c}_{k_1} - \mathbf{c}_{k_2}) =  S^\ast \subset [p], |S^\ast| \leq s
	\bigr\},
\\\mathcal{H}(K) &:= \bigl\{ \mathbf{H} = (H_{i,k}) \in \{0,1\}^{n \times K}:
	\sum_{k=1}^K H_{i,k} =1 ~\text{for all } i=1, \ldots, n\bigr\}.
\end{align*} 
For a given clustering algorithm $\hat{\mathbf{H}}: \mathbb{R}^{p \times n} \to  \mathcal{H}(K)$, we assess its performance using the worst-case probability of exact cluster recovery over the distribution class \( \Psi(s, \Delta, K) \):  
\begingroup
\begin{equation*}
	\sup_{\boldsymbol{\psi} \in \Psi(s,  \Delta, K)}
	\min_{\pi \in \mathrm{perm}(K)}
	\mathbb{P}_{\boldsymbol{\psi}} \bigl(
    \hat{\mathbf{H}}(\mathbf{X}) \neq \pi(\mathbf{H})
    \bigr),
\end{equation*}
\endgroup  
where $\mathrm{perm}(K)$ is the set of label permutations of $[K]$ and $\pi(\mathbf{H})$ is the matrix after label permutation specified by $\pi\in \mathrm{perm}(K)$.
For brevity, we omit  \( \min_{\pi \in \mathrm{perm}(K)} \) henceforth.

Given a problem instance
$\boldsymbol{\psi}^\ast = (
\mathbf{c}_1^\ast, \ldots, \mathbf{c}_K^\ast, \mathbf{H}^\ast
)
\in \Psi(s,  \Delta, K)
$,
the true cluster  assignment $\mathbf{H}^\ast$ has one-to-one correspondence to the block-diagonal matrix
$
    \mathbf{Z}^\ast :=  
    \mathbf{H}^\ast \mathbf{D}_{G^\ast} (\mathbf{H}^\ast)^\top
    =
    \sum_{k=1}^K |G_k^\ast|^{-1} \mathbf{1}_{G_k^\ast} \mathbf{1}_{G_k^\ast}^\top,
$
up to label permutations.
Let us denote the squared $\ell_2$ cluster center separation  confined to a feature set $S$ as
\begin{equation*}
\Delta_{S \cap S^\ast}^2 :=
  \hskip -3mm
\min_{1 \leq k_1 \neq k_2 \leq K} 
\| \big(\boldsymbol{\Omega}^\ast
(\mathbf{c}_{k_1}^\ast - \mathbf{c}_{k_2}^\ast )\big)_{S \cap S^\ast}
 \|_2^2.
 \end{equation*}
 Fix a sufficiently large universal constant $C_{\mathcal S}>0$.  Let $\mathcal S$ denote the collection of variable subsets with sufficiently strong signal:
\begin{align*}
\mathcal S &:= 
\bigl\{
S: S\subset [p], 
    |S|\leq \sqrt p, 	{ \Delta_{S \cap S^\ast}^2 \geq C_{\mathcal S}\bar{\Delta}_S^2}
    \bigr\},   \numberthis \label{def:S-set}
\\
\bar{\Delta}_S^2
&:= {\sigma^{-2}}
		\biggl(\log n +
        \frac{|S| \log p}{m}
		+
		\sqrt{
        \frac{|S| \log p}{m}
        }
		\biggr),
        \quad
        m := 2\min_{1 \leq k_1 \neq k_2 \leq K}
\left\{
		\biggl(
		\frac{1}{ |G_{k_1}^\ast| }
		+
		\frac{1}{ |G_{k_2}^\ast| }
		\biggr)^{-1}
        \right\}.
\end{align*}
Because $\boldsymbol\Omega^\ast=\sigma^{-2}\mathbf I_p$ in
this subsection, the innovated-scale condition in \eqref{def:S-set} is
equivalent to requiring the raw squared center separation to exceed
$C_{\mathcal S}\sigma^2\{\log n+(|S|\log p)/m+
\sqrt{(|S|\log p)/m}\}$, which is the scale used in the proof.
Theorem \ref{theorem:separation_condition} gives the separation required for SDP $K$-means to achieve exact recovery:

\begin{theorem}\label{theorem:separation_condition}
Assume 
$m		
		\geq C_1
		n/\log n
$ with some universal constant $C_1>0$ and  $|G_k^\ast| \geq 2 $  for all $ k \in  [K]$.
    Let $\hat{\mathbf Z}(S)$ be the solution of \eqref{SDP_objective_submatrix} corresponding to the subset $S\in \mathcal S$. Then 
\begin{equation*}
    \mathbb P
    \bigl(
    \hat{\mathbf Z}(S) = \mathbf Z^\ast, \forall S\in \mathcal S
    \bigr)\geq 1-  {C_2}K/n,    
\end{equation*}
   where $C_2>0$ is some constant.  
   Furthermore, if $(s \log p)/n=o(1) $
	and \( n \geq 2 \), no clustering rule can guarantee exact recovery uniformly over the  class  
$
     \overline{\Psi}:= \Psi \bigl(s, 
         C_3 
\sigma^2 \log n
    , K \bigr)$ 
     with $C_3>0$ some small  constant.  
	In other words, there exists a constant $C_5>0$ such that
\begin{equation*}
    	\inf_{ 
\hat{\mathbf{H}}: \mathbb{R}^{p \times n} \to  \mathcal{H}(K)
}
		\sup_{\boldsymbol{\psi} \in 	\overline{\Psi}}
		\mathbb{P}_{\boldsymbol{\psi}} 
        \bigl(
        \hat{\mathbf{H}}(\mathbf{X}) \neq \mathbf{H}
        \bigr)
		\geq
		C_5.
\end{equation*}

\end{theorem}
The proof of Theorem~\ref{theorem:separation_condition} is given in Appendix~\ref{section:proof:theorem:separation_condition}. The stated probability in Theorem~\ref{theorem:separation_condition} holds uniformly over all variable subsets in \( \mathcal S \), justifying iterative procedures that solve a sequence of subset SDPs: once a ``good'' $S$ is reached, clustering is accurate.
The assumption 
$m \geq C_1 n / \log n$ prevents overly imbalanced clusters and is standard in clustering theory \citep{
cai_chime_2019,
jinInfluentialFeaturesPca2016,
lofflerOptimalitySpectralClustering2021}.
Since $m$ is a harmonic mean, this implies
$K \lesssim \log n$,
allowing slowly diverging known $K$, which is milder than $K$  bounded above by a constant assumption in prior work   \citep{verzelenDetectionFeatureSelection2017, jinInfluentialFeaturesPca2016}. The second claim of Theorem \ref{theorem:separation_condition} shows no clustering rule can guarantee exact recovery for all $S$ with signal strength of order $\sigma^2 \log n$.
When $(|S| \log p) / m \lesssim \log n$, the separation bound matches this lower bound up to constants, establishing minimax optimality.

Theorem \ref{theorem:separation_condition}  also indicates that in the sparse regime where $|S^\ast|\ll p$, variable selection can be critical for clustering accuracy. For example, when $K=2$, $|G_1^\ast|=|G_2^\ast|$ and $(\mathbf{c}_1^\ast-\mathbf{c}_2^\ast)_{S^\ast} = c_0 \mathbf{1}_{|S^\ast|}$,  
then for any set $S$ satisfying 
$|S|\lesssim \sqrt{p}$
and
$|S|\lesssim (n\log n)/ \log p$ 
, exact recovery holds if and only if 
\begingroup
\begin{equation*}
|S\cap S^\ast|c_0^2\gtrsim \log n + \frac{|S|\log p}{n} + \sqrt{\frac{|S|\log p}{n}} \asymp \log n.
\end{equation*}
\endgroup
This condition is stringent if  $S$ is a very noisy estimate of $S^\ast$ (i.e., missing too many signal variables), but a slight overestimate of $S^\ast$ may not be very damaging.  
An important message of Theorem~\ref{theorem:separation_condition} is that recovering the exact sparse set is not essential for accurate clustering. Rather, identifying any sparse feature set with sufficiently large cumulative signal strength is key. Such a ``good'' set can be any $S$ in $\mathcal S$, which motivates the iterative algorithm in the next subsection, where we aim to obtain some set $S$ in $\mathcal S$ rather than a specific one.

\section{Proposed Iterative Block Coordinate Optimization Algorithm}\label{sec:main}
We propose a block coordinate optimization framework that alternates between variable selection and clustering. For simplicity, we assume that $\boldsymbol{\Omega}^\ast$ is known throughout this section; the extension to the unknown-covariance setting is discussed in Section~\ref{section:baselinemethod}. We first provide an overview of the proposed algorithm, which consists of two blocks and two substeps, in Section~\ref{section:tvs_algo_overview}, followed by detailed descriptions of each block and substep in Section \ref{section:one_step}.

\subsection{Algorithm Overview}\label{section:tvs_algo_overview}
The algorithm is initialized with a pilot partition $\hat{G}^0$, obtained, for example, via spectral clustering using all available variables. Choose positive Beta prior parameters $a_j^0,b_j^0$ for each $j\in[p]$. For the regularization parameter $\epsilon>0$, put $\tau_\epsilon:=\log(1/\epsilon)/\log((\epsilon+1)/\epsilon)$, draw $\vartheta_j^0\sim\operatorname{Beta}(a_j^0,b_j^0)$ independently, and set $\hat S^0:=\{j\in[p]:\vartheta_j^0>\tau_\epsilon\}$. Given the initial partition $\hat{G}^0$, the feature selection block produces the first active variable set $\hat{S}^1$, which is then passed to the SDP clustering block to obtain the updated partition $\hat{G}^1$. This alternating procedure continues until a prespecified maximum number of iterations $T$ is reached. Figure~\ref{fig:alg_schematic} summarizes the resulting TVS--SDP workflow. We next provide a brief overview of the feature selection and clustering blocks at a generic iteration $t$; detailed descriptions are given in Section~\ref{section:one_step}.

\medskip
\noindent
\textbf{Feature Selection Block:}  
The feature selection block consists of two substeps, each of which employs permutation testing and Thompson sampling.
\begin{enumerate}
    \item  \textbf{Reward Step (Section \ref{section:robust_testing}):} 
Given the current cluster partition $\hat{G}^{t-1}$, each variable $j \in \hat{S}^{t-1}$   receives a stochastic binary reward
\begin{equation}\label{bernoulli_first_appearance}
Y_j^t (\hat{G}^{t-1}) = \Lambda_j(\hat{G}^{t-1}, \widetilde{\mathbf{X}}; U^t) \in \{0,1\},  
\end{equation}
generated by a robust permutation test (Algorithm \ref{alg:robust_dc} in Section \ref{section:robust_testing}), where $U^t$ denotes the exogenous randomness associated with the permutation procedure. The test is based on the absolute difference in means between clusters (under innovated transform):
\begin{equation}\label{eq:def-Tj}
\Lambda(G, \widetilde{\mathbf{X}}_j) :=
\biggl|
\frac{1}{|G_1|}
\sum_{i_1 \in G_1} \widetilde{X}_{i_1, j}
-
\frac{1}{|G_2|}
\sum_{i_2 \in G_2} \widetilde{X}_{i_2, j}
\biggr|.
\end{equation}
The permutation test is conducted at level $\tau_\epsilon-\alpha$, where
\[
\tau_{\epsilon}
:=
\frac{\log(1/\epsilon)}
{\log\!\left(\frac{\epsilon+1}{\epsilon}\right)}
\]
is the feature-selection threshold used in subsequent step, induced by the regularization hyperparameter $\epsilon>0$. We introduce a tuning margin $\alpha\in(0,\tau_\epsilon)$ so that the testing level $\tau_\epsilon-\alpha$ is strictly smaller than the selection threshold $\tau_\epsilon$, which is essential for the $\log(T)$ order regret analysis of the Thompson-sampling procedure. The corresponding regret analysis is omitted here and was provided in an earlier version of the manuscript.
  \item \textbf{Choose Step (Section \ref{section:bernoulli_TVS}):} 
Using Beta--Binomial conjugacy, we save the historical binary rewards for each variable in the parameters of a Beta distribution. Specifically, the rewards from the Reward Step are used to update these parameters as follows:
\[
\{(a_j^t, b_j^t)\}_{j \in \hat{S}^{t}} = \{(a_j^{t-1} + Y_j(\hat{G}^{t-1}), b_j^{t-1} + 1 - Y_j(\hat{G}^{t-1}))\}_{j \in \hat{S}^{t}}.
\]
Rather than relying solely on the current binary reward, which may be noisy due to imperfect cluster assignments, we aggregate historical rewards through the updated parameters $\{(a_j^t,b_j^t)\}_{j=1}^p$. To account for the remaining uncertainty in this accumulated evidence, we introduce Thompson-sampling randomization by drawing a latent reward probability
\[
\vartheta_j^t \sim \operatorname{Beta}(a_j^t,b_j^t),
\qquad j=1,\ldots,p.
\]
The new variable set is then formed by selecting features whose sampled probabilities exceed the threshold $\tau_\epsilon$:
\[
\hat S^t=\{j:\vartheta_j^t>\tau_\epsilon\}.
\]
 
\end{enumerate}

\medskip
\noindent
\textbf{Clustering Block:} Given the active variable set $\hat{S}^t$, we compute the innovated transformation $\widetilde{\mathbf{X}} = \boldsymbol{\Omega}^\ast \mathbf{X}$ and solve the subset SDP (Algorithm~\ref{alg:sdp_subroutine}). This produces the updated partition, 
\[
\hat{G}^t = \texttt{SDPclust}(\widetilde{\mathbf{X}}_{\hat{S}^t, \cdot}, \boldsymbol{\Sigma}^\ast_{\hat{S}^t, \hat{S}^t}),
\]
which serves as the input for the reward block in the subsequent iteration.

Our algorithmic design addresses the   fragility of memoryless feature selection. While we guard against moderately inaccurate cluster assignments by framing the test statistic $\Lambda(\hat{G}^{t-1},\widetilde{\mathbf X}_j)$ in \eqref{eq:def-Tj} as an empirical maximum mean discrepancy (MMD) with a linear kernel 
 and applying its corruption-corrected variant (\citealp{schrabRobustKernelHypothesis2025}; see Section \ref{section:robust_testing}), this defense still has limitations. Under highly inaccurate initializations of $\hat G^{t-1}$, the permutation test struggles to separate weak signals from noise, causing informative variables to be discarded without
a mechanism for their systematic reconsideration.
In particular, \emph{nearly significant} features under noisy cluster assignments
do not accumulate evidence across iterations.
This lack of memory can yield a self-reinforcing loop of ``poor cluster labels $\Rightarrow$ low test power $\Rightarrow$ under-selection of $S^\ast$ (which is critically damaging as shown in Theorem~\ref{theorem:separation_condition}) $\Rightarrow$ poor cluster labels.''
Increasing the  level of permutation test mitigates this but inflates $|\hat S ^t|$, potentially violating the regime (e.g.\ $|\hat S ^t|\le \sqrt p$) where uniform SDP recovery holds. 
These issues suggest that the selection should (i) \emph{remember} past evidence on variable relevance and (ii) \emph{occasionally re-select} uncertain variables so that improved labels can later validate them. 
 
We bridge the permutation test and the multi-armed bandit framework through the following perspective. Conditional on ${\mathbf X}$ sampled from the model in \eqref{def:gaussian_mixture_model}, the outcome of a randomized permutation test for feature selection acts as a Bernoulli random variable. Its success probability depends on the feature index $j$ and the previous cluster estimate $\hat{G}^{t-1}$, and can be formally analyzed through the lens of two-sample testing with corrupted samples. Exploiting this Bernoulli structure allows us to cast the Choose Step of our block coordinate optimization as a multi-armed bandit problem.  
Casting the problem as a bandit model allows us to incorporate memory and controlled exploration. In particular, we model variable selection as a \emph{multi-play} bandit problem \citep{komiyama_optimal_2015, liuVariableSelectionThompson2023}: at each iteration, the agent selects a subset of arms (variables), observes Bernoulli rewards, and aggregates this feedback with past evidence to guide future actions. This formulation enables systematic use of historical information and occasional re-selection of uncertain variables. 

We present the details of  TVS--SDP under known covariance, as in Theorem~\ref{theorem:separation_condition}; the unknown-covariance extension appears in Section~\ref{section:method_unknown_cov}.
For clarity, we focus on the case $K=2$, though the methods naturally generalize to larger $K$.

\begin{figure}
\centering
\begin{tikzpicture}[
    node distance=0.40cm,
    tvsblock/.style={
        rectangle,
        draw=black,
        text=black,
        align=center,
        text width=3.05cm,
        minimum height=3.25cm,
        inner sep=4pt,
        font=\linespread{0.90}\small\rmfamily\selectfont
    },
    tvsline/.style={draw=black, -latex, thick},
    tvsblock_pilot/.style={
        rectangle,
        draw=black,
        text=black,
        align=center,
        text width=2.15cm,
        minimum height=3.25cm,
        inner sep=4pt,
        font=\linespread{0.90}\small\rmfamily\selectfont
    },
    tvsblock_reward/.style={
        rectangle,
        draw=black,
        text=black,
        align=center,
        text width=4cm,
        minimum height=3.25cm,
        inner sep=4pt,
        font=\linespread{0.90}\small\rmfamily\selectfont
    },
        tvsblock_selection/.style={
        rectangle,
        draw=black,
        text=black,
        align=center,
        text width=4.1cm,
        minimum height=3.25cm,
        inner sep=4pt,
        font=\linespread{0.90}\small\rmfamily\selectfont
    }
]
\node[tvsblock_pilot] (pilot) {
    \textbf{Pilot}\\[1mm]
    Obtain $\widehat G^0$ and initialize $\{(a_j^0,b_j^0)\}_{j=1}^p$
};
\node[tvsblock_reward, right=of pilot] (reward) {
    \textbf{Reward}\\[1mm]
    1. Transform the data
    \\
    2. Compute \textup{rMMD} rewards   $\{ Y_j(\hat{G}^{t-1})\}_{j \in \hat{S}^{t-1}} $

};
\node[ tvsblock_selection, right=of reward] (select) {
    \textbf{Choose}\\[1mm]
    1. Update $\{(a_j^t,b_j^t)\}_{j \in \hat{S}^{t-1}}$
    2. Draw $\vartheta_j^t\sim\operatorname{Beta}(a_j^t,b_j^t)$ 
    \\
    3. Set
    $\widehat S^t=\{j:\vartheta_j^t>\tau_\epsilon\}$
};
\node[tvsblock_pilot, right=of select] (cluster) {
    \textbf{Subset SDP}\\[1mm]
    Cluster on $\widehat S^t$ to obtain $\widehat G^t$
};
\draw[tvsline] (pilot) -- (reward);
\draw[tvsline] (reward) -- (select);
\draw[tvsline] (select) -- (cluster);
\coordinate (loopright) at ($(reward.south -| cluster.south)+(0,-0.65cm)$);
\coordinate (loopleft) at ($(reward.south)+(0,-0.65cm)$);
\draw[tvsline] (cluster.south) -- (loopright) -- node[below=1pt, text=black, font=\small\rmfamily] {next iteration} (loopleft) -- (reward.south);
\end{tikzpicture}
\caption{ {TVS--SDP workflow.  The transformed data are $\widetilde{\mathbf X}=\boldsymbol\Omega^\ast\mathbf X$ when the covariance is known; the ISEE extension introduced in Section \ref{section:method_unknown_cov} and Appendix \ref{section:isee_detail} supplies an estimated transformation when it is unknown. We also provide  memoryless greedy benchmark  in Algorithm~\ref{alg:known_cov}.}}
\label{fig:alg_schematic}
\end{figure}

\subsection{Details on   Feature Selection Block}\label{section:one_step} 
We provide details of the feature selection block introduced in Section~\ref{section:tvs_algo_overview}, explaining the Reward Step in Section~\ref{section:robust_testing} and the Choose Step in Section~\ref{section:bernoulli_TVS}.

 \subsubsection{Reward Step: Robust Two-Sample Mean Permutation   Test}\label{section:robust_testing}
 This section gives details on how binary (Bernoulli) reward \eqref{bernoulli_first_appearance} is generated in Reward Step introduced in Section~\ref{section:tvs_algo_overview}. 
 We start by formally defining the Bernoulli rewards using an arbitrary randomized test that depends on the data and the cluster partition.
\begin{definition}[Bernoulli reward]
\label{def:test_rule_reward}
For $j=1,\ldots,p$, let $\Lambda_j$ be an arbitrary randomized test such that, for any cluster partition $\hat{G}^{t-1}$ and  random seed $U^t$, 
$\Lambda_j(\hat{G}^{t-1}, \widetilde{\mathbf{X}};U^t)\in\{0,1\}$ is measurable with respect to the data $\widetilde{\mathbf{X}}$ and $U^t$.  
Given $\hat{G}^{t-1}$, we view the test outcome as a Bernoulli reward
$Y_j(\hat{G}^{t-1})
\,:=\, 
\Lambda_j(\hat{G}^{t-1}, \widetilde{\mathbf{X}};U^t)$
with mean reward
 $\theta_j(\hat{G}^{t-1})
\,:=\,
\mathbb{P}\bigl( Y_j(\hat{G}^{t-1})=1 \,\big|\, \widetilde{\mathbf{X}},\hat{G}^{t-1} \bigr)$,
where the probability is taken over the randomness of $U^t$.
\end{definition}
We use independent randomizations $U^t$ across iterations $t$ and features $j$, where the dependence of $U^t$ on $j$ is suppressed whenever it is apparent from context.

The specific test $\Lambda_j$ used in the TVS algorithm, detailed in Algorithm~\ref{alg:robust_dc}, is a robust two-sample mean permutation test based on the adversarial corruption model of \cite{schrabRobustKernelHypothesis2025}. 
The adversarial corruption allows samples to be swapped between $G_1^\ast$ and $G_2^\ast$, resulting in misclustered samples in $\hat{G}_1^t$ and $\hat{G}_2^t$.
While Algorithm~\ref{alg:robust_dc} fundamentally operates as a robust two-sample mean test, we formally frame the statistic as a linear-kernel MMD to leverage theoretical guarantees of robust MMD testing. 
Because its theory (Appendix~\ref{section:MMD_theory}) requires bounded input, we apply   a bounded transformation (Appendix \ref{section:bounded_transformation}) to our observations.
At iteration $t$,    Let
$n_g=|\widehat G_g^{t-1}|$ and $n_{\min}:=\min(n_1,n_2)$.  Let $r$ bound the
\emph{total} number of corrupted   observations.  Boundedness gives
global sensitivity at most $1/n_{\min}$. As detailed in Algorithm~\ref{alg:robust_dc}, the robust MMD test is implemented simply by adding a margin of $2r/n_{\min}$ to the standard permutation test cutoff.  We use
$0\leq r<n_{\min}/2$; larger values make rejection impossible because
$0\leq\Lambda\leq1$.  Appendix~\ref{section:details_MMD}
contains the transformation, sensitivity, power, and implementation details.

\begin{algorithm}[t!]
\caption{Robust MMD Permutation Test for Feature $j$: \texttt{rMMD}}
\label{alg:robust_dc}
\begin{algorithmic}[1]
\Require (1) Current partition $\hat{G}^{t-1}$ with sizes $n_1, n_2$, 
(2) $j$th feature vector $\widetilde{\mathbf X}_j$, (3) total-corruption budget $0\leq r<n_{\min}/2$ with $n_{\min}:=\min(n_1,n_2)$, (4) significance level $\tau_{\epsilon} - \alpha$, (5) permutation count $B$.
\State $\widetilde{\mathbf{X}}_j \mapsto \texttt{BoundedTransform}(\widetilde{\mathbf{X}}_j)$\Comment{Appendix \ref{section:bounded_transformation}}
    \State $\pi_0, \pi_1, \dots, \pi_B\gets \text{Identity permutation}$ and   i.i.d. random permutations   of $[n]$.
    \For{$b = 0, 1, \ldots, B$}
        \State Construct permuted partition $\hat{G}^{t-1, (b)}$ using $\pi_b$.
        \State Compute $\Lambda_j^{(b)} \gets \Lambda (\hat{G}^{t-1, (b)}, \widetilde{\mathbf X}_j)$ \Comment{$\Lambda$ \text{defined in} \eqref{eq:def-Tj};  $\Lambda _j^{(0)}$ is the observed statistic}
    \EndFor
    \State Compute the $(1-\tau_{\epsilon} + \alpha)$-quantile $q_{\tau_{\epsilon} - \alpha}$ of the empirical distribution $\{\Lambda_j^{(b)}\}_{b=0}^B$.
    \State \Return 1 if $\Lambda_j^{(0)} > q_{\tau_{\epsilon} - \alpha} + 2r
    \frac{1}{ {n_{\min}}}
    $
    \Comment{$ {1/n_{\min}}=\text{global sensitivity } \Gamma_{j}(\hat{G}^{t-1})$}
\end{algorithmic}
\end{algorithm}
 We motivate Algorithm~\ref{alg:robust_dc} using
Theorem~\ref{theorem:mmd} in Appendix \ref{section:MMD_theory}, which gives a marginal power guarantee under sufficient cluster separation.   At the rates specified in Theorem~\ref{theorem:mmd}, its marginal level and power bounds control the oracle-target terms in Theorem~\ref{theorem:posterior_surplus}. They do not by themselves control the remaining history-dependent posterior error, because TVS conditions on the reused data $\widetilde{\mathbf X}$. 



\subsubsection{Choose Step:    Empirical Counterpart of Oracle Target Set}\label{section:bernoulli_TVS}
 This section gives details on the Choose Step introduced in Section~\ref{section:tvs_algo_overview}.
To provide an intuitive explanation of the Choose Step,
we first consider the oracle setting where the clustering block consistently outputs the true partition $G^\ast$. 
Then conditional on 
$(\widetilde{\mathbf{X}}, G^\ast)$,
the rewards $\{Y_j^t(G^\ast)\}_{t\ge 1}$ are i.i.d.\
Bernoulli 
$\bigl(
\theta_j(G^\ast)
\bigr)$ for each fixed $j$. 
Denote the   \emph{oracle}   mean reward as
\begin{equation}\label{theta_j_orc}    \theta_j^{\mathrm{orc}}:=\theta_j(G^\ast) 
    =
    \mathbb{P}\bigl( Y_j(G^\ast)=1 \,\big|\, \widetilde{\mathbf{X}},G^\ast \bigr).  
\end{equation}
For a given regularization parameter $\epsilon > 0$,
motivated by \citet{liuVariableSelectionThompson2023}, we define the \emph{oracle} utility of a variable subset $S$  as:
\begingroup
\begin{equation*}
r_\epsilon^{\mathrm{orc}}(S)
:= \sum_{j\in S}\left\{
\theta_j^{\mathrm{orc}}
\log\!\left(\frac{\epsilon+1}{\epsilon}\right)+\log \epsilon\right\}.
\end{equation*}
\endgroup
This formulation gives a direct criterion for beneficial inclusion. Adding $j\notin S$ strictly improves the utility if and only if its oracle mean reward exceeds an $\epsilon$-dependent cutoff:
\begin{equation}\label{def:tau_epsilon}
r_\epsilon^{\mathrm{orc}}(S \cup \{j\}) > r_\epsilon^{\mathrm{orc}}(S)
\quad \Longleftrightarrow \quad
\theta_j^{\mathrm{orc}} > \tau_{\epsilon},
\qquad
\tau_{\epsilon} := \frac{\log(1/\epsilon)}{\log\!\left(\frac{\epsilon+1}{\epsilon}\right)}.
\end{equation}
 We therefore define the (random) canonical oracle target as the utility maximizer:
\begin{equation}\label{eq:SCstar}
S_\epsilon^\ast
:=\{j\in[p]:\theta_j^{\mathrm{orc}}>\tau_\epsilon\},
\qquad
S_\epsilon^\ast\in
\arg\max_{S\subseteq[p]}r_\epsilon^{\mathrm{orc}}(S),
\end{equation}
where the randomness is with respect to the data $\mathbf{X}$ (see \eqref{theta_j_orc}).

We use this utility-defined set $S_\epsilon^\ast$ as the oracle benchmark, representing the optimal feature set attainable by the TVS algorithm. Notably, $S_\epsilon^\ast$ need not equal the true signal support $S^\ast$. 
This aligns with our theoretical framework, which prioritizes finding a valid set within $\mathcal S$ over the perfect identification of $S^\ast$. Specifically, Theorem~\ref{theorem:posterior_surplus} guarantees high clustering quality under appropriately calibrated level and power controls of the reward tests, so long as $S_\epsilon^\ast$ captures a subset of $S^\ast$ carrying sufficient cumulative signal. 
The result permits both false inclusions in $S_\epsilon^\ast$ and arbitrarily weak signals in $S^\ast\setminus S_\epsilon^\ast$.
 
The set $\hat{S}^t$ produced by the Choose Step   serves as a stochastic, empirical counterpart to the oracle benchmark $S_\epsilon^\ast$. Because the true partition $G^\ast$ and the oracle probabilities $\theta^{\mathrm{orc}}_j$ are unknown, we approximate them using estimates from the preceding iteration. At iteration $t$, recall from Section~\ref{section:robust_testing} that the rewards
$
Y_j^t
=
Y_j(\hat G^{t-1})
=
\Lambda_j(\hat G^{t-1},\widetilde{\mathbf X};U^t)
$
are generated using the current partition $\hat G^{t-1}$ and the randomized testing rule $\Lambda_j$, with exogenous randomness $U^t$. We denote the corresponding expected reward by
\[
\theta_j(\hat G^{t-1})
:=
\mathbb E\!\left[Y_{j}^t\mid \widetilde{\mathbf{X}},\hat G^{t-1}\right]
.
\]
 If $\hat G^{t-1}=G^\ast$ (up to label permutation) for some $t-1\geq 0$, then
$\theta_j(\hat G^{t-1})=\theta_j(G^\ast)=\theta_j^{\mathrm{orc}}$.
Thus, whenever the current partition is correct, the subsequent reward means coincide with their oracle counterparts.
The iterative algorithm maintains feature-wise Beta posteriors $\mathrm{Beta}(a_j^t,b_j^t)$, where $a_j^t$ and $b_j^t$ record the cumulative number of significant ($Y_j^t=1$) and non-significant ($Y_j^t=0$) test outcomes up to iteration $t$. Because the intermediate cluster estimates $\hat G^{t-1}$ are imperfect, relying solely on the current parameter $\theta_j(\hat G^{t-1})$ yields a noisy approximation of the oracle target $\theta_j^{\text{orc}}$. Instead, the posterior parameters $(a_j^t, b_j^t)$ aggregate the historical evidence across all previous iterations. Guided by this accumulated memory and exploration principle, we sample $\vartheta_j^t \sim \mathrm{Beta}(a_j^t,b_j^t)$ and construct the active feature set as $\hat S^t = \{j : \vartheta_j^t > \tau_\epsilon\}$.
The resulting set $\hat S^t$ is passed to the clustering block to obtain $\hat G^t$.  

As $t$ increases, historical evidence accumulates, causing the variance of the Beta posteriors to shrink. Consequently, the stochasticity of the selection process diminishes, and the active feature set $\hat{S}^t$ converges to a stable subset.
The combination of reward and Choose Step appears in Algorithm~\ref{alg:tvs_step}. 
The full iterative algorithm effectively alternates between Algorithm~\ref{alg:sdp_subroutine} and Algorithm~\ref{alg:tvs_step}.

The use of $\alpha$ is to encourage exploration for reducing under-selection, as motivated by Theorem \ref{theorem:separation_condition}.  
   The cutoff $\tau_{\epsilon}$ is crucial to the success of the TVS algorithm, as illustrated in Figure~\ref{fig:combined_performance}. The simulation shows that $\epsilon$ strongly affects clustering accuracy and that recovering more true positives can improve performance even at the cost of additional false positives. This supports Theorem~\ref{theorem:separation_condition} and contrasts with the regression setting in \cite{liuVariableSelectionThompson2023}, which fixes $\tau_{\epsilon}=0.5$ and emphasizes false-positive control over true-positive recovery.
Motivated by these findings, we select $\epsilon$ by grid search to balance signal retention and over-selection. Since the true clustering accuracy is unavailable, we choose $\epsilon$ by maximizing the silhouette index \citep{rousseeuw1987silhouettes}, a standard measure of clustering quality computed on the union of variables selected for each candidate $\epsilon$.

\begin{algorithm}[t]
\caption{Feature Selection Block at the $t$-th Iteration }
\label{alg:tvs_step}
\begin{algorithmic}[1]
\Require Transformed data $\widetilde{\mathbf{X}}$, partition $\hat{G}^{t-1}$, variable set $\hat{S}^{t-1}$, priors $
\bigl\{ 
(a^{t-1}_j, b^{t-1}_j)
\bigr\}_{j=1}^p
$, cutoff $\tau_{\epsilon}$, type I error rate $\tau_{\epsilon} - \alpha$ 

\State $a_j^t \gets a_j^{t-1}, \quad b_j^t \gets b_j^{t-1}$ for all $j \notin  \hat{S}^{t-1}$
\For{$j \in \hat{S}^{t-1}$}
        \State $Y_j^t \gets 
        \texttt{rMMD}(\hat{G}^{t-1},  {\widetilde{\mathbf X}_j}, r, \tau_{\epsilon} - \alpha, B)$ 
\Comment{\textbf{Reward   Step}  (Algorithm \ref{alg:robust_dc}; only for $j\in \hat{S}^{t-1}$)}
        \State $a_j^t \gets a_j^{t-1} + Y_j^t, \quad b_j^t \gets b_j^{t-1} + (1 - Y_j^t)$
\Comment{Update   (only for $j\in \hat{S}^{t-1}$)}
\EndFor
\State Sample $\vartheta_j^t \sim \operatorname{Beta}(a_j^t, b_j^t)$ for all $j \in [p]$
\Comment{\textbf{Choose Step}}
\State
\Return $\hat{S}^t \gets \{ j : \vartheta_j^t > \tau_{\epsilon} \}$, Updated Priors $\bigl\{ 
(a^{t}_j, b^{t}_j)
\bigr\}_{j=1}^p$
\end{algorithmic}
\end{algorithm}

\subsection{Cluster Recovery Guarantees for the TVS algorithm} \label{sec: theory_loop}
We establish a clustering guarantee for the TVS algorithm, which uses Algorithm~\ref{alg:tvs_step} for variable selection and Algorithm~\ref{alg:sdp_subroutine} for clustering. This result applies to the known isotropic covariance setting introduced in Theorem~\ref{theorem:separation_condition}. We outline the required assumptions in Section \ref{section:assumptions} before detailing the main theoretical results in Section \ref{section:theory_present}.

\subsubsection{Assumptions}\label{section:assumptions}
Consistent with Theorem \ref{theorem:separation_condition}, all conditions are imposed on set sizes and cumulative signals; no assumptions are made about individual feature entries. We begin with Condition~\ref{condition:core_margin}, which specifies the rates for the set size and cumulative signal required for exact cluster recovery. Furthermore, these rates dictate the permissible bounds on cumulative testing errors, ensuring that the TVS oracle targets an appropriate feature set (Condition \ref{condition:oracle_screening}).
Finally, Condition~\ref{condition:posterior_learning} controls the fluctuation around this target set, bounding the additional error introduced by the adaptive TVS draws.

For technical simplicity, we state the result for $K=2$. We consider a sequence of problem instances indexed by the sample size $n$, where 
$n \to \infty$ and $p = p_n \to \infty$.  This data-generating asymptotic regime is distinct from the algorithmic limit of the Thompson sampling steps ($T \to \infty$).
Let us define
$
  L_p := \lfloor\sqrt{p}\rfloor.   
$
and denote the transformed signal contribution of coordinate $j$ as
\begin{equation}\label{feature_wise_clustering_signal}
w_j :=
\left\{\bigl[\boldsymbol\Omega^\ast
(\mathbf c_1^\ast-\mathbf c_2^\ast)\bigr]_j\right\}^2.
\end{equation}
As a proof device,
let $D\subseteq S^\ast$  denote a \emph{candidate signal
core}, and define the cumulative signal carried by the core $D$ as
\begingroup
\begin{equation*}
\Delta_D^2:=\sum_{j\in D}w_j.
\end{equation*}
\endgroup
We parameterize the SDP signal threshold defined in \eqref{def:S-set} for a selected set of size (denoted by integer $L>0$) as
\begin{equation}\label{T_l}
T_L
:=
C_{\mathcal S}\sigma^{-2}
\left\{
\log n+\frac{L\log p}{m}
+\sqrt{\frac{L\log p}{m}}
\right\}.
\end{equation}
We emphasize that the core $D$ is primarily a proof device. It needs not contain every signal coordinate. It only needs to carry enough cumulative signal to guarantee clustering. We formalize this into the following condition.
\begin{condition}
\label{condition:core_margin}
The data is generated according to the model \eqref{def:gaussian_mixture_model} and
$m		
		\geq C_1
		n/\log n
$ with some universal constant $C_1>0$ and  $|G_1^\ast| \geq 2 $  and   $|G_2^\ast| \geq 2 $.
There exist 
cutoff $\tau_\epsilon$ and
the nonempty core $D\subseteq S^\ast$, which are deterministic  and may
depend on $n$, such that  for some fixed $\tau_0\in(0,1/2)$,
$\kappa\in(0,1)$, and $\gamma>0$, it holds that
$\tau_\epsilon\in[\tau_0,1-\tau_0]$ and
\begin{equation}
\label{eq:clean_core_margin}
|S^\ast|\leq(1-\kappa)L_p,
\qquad
\Delta_D^2\geq(1+\gamma)T_{L_p}.
\end{equation}
\end{condition}
Recall from \eqref{theta_j_orc} that we defined   
$\theta_j^{\mathrm{orc}} = \mathbb{P}\bigl( Y_j(G^\ast)=1 \,\big|\, \widetilde{\mathbf{X}},G^\ast \bigr)$. 
Because this quantity depends explicitly on the realization of the data, we marginalize over $\mathbf{X}$ to define its population-level counterpart: 
\begingroup
\begin{equation*}
\bar\theta_j^{\mathrm{orc}}
:=
\mathbb E_{\mathbf X}(\theta_j^{\mathrm{orc}})
=
\mathbb P_{\mathbf X,U}\{Y_j(G^\ast)=1\},
\end{equation*}
\endgroup
where $\mathbb E_{\mathbf X}$ denotes expectation with respect to the randomness in the observed data, while $\mathbb P_{\mathbf X,U}$ denotes probability with respect to both the data randomness and the exogenous randomness introduced by the permutation test. Note that $\theta_j^{\mathrm{orc}}$ is defined with respect to the data randomness only and is therefore independent of the permutation-testing randomness.

For any feature index $j$, because $0\leq\theta_j^{\mathrm{orc}}\leq1$,  Markov's inequality implies that
\begin{equation}
\label{eq:marginal_target_bridge}
\mathbb P_{\mathbf X}(j\in S_\epsilon^\ast)
=
\mathbb P_{\mathbf X}
(\theta_j^{\mathrm{orc}}
>
\tau_\epsilon
)
\leq\frac{\bar\theta_j^{\mathrm{orc}}}{\tau_\epsilon},
\quad
\mathbb P_{\mathbf X}(j\notin S_\epsilon^\ast)
=
\mathbb P_{\mathbf X}
(
1-\theta_j^{\mathrm{orc}}
\ge
1-\tau_\epsilon
)
\leq
\frac{1-\bar\theta_j^{\mathrm{orc}}}{1-\tau_\epsilon}.
\end{equation}
Recall from \eqref{eq:SCstar} that $S_\epsilon^\ast$ is a random set depending on $\mathbf{X}$. The set difference $S_\epsilon^\ast \setminus S^\ast$ consists of all noise coordinates (features not in the true set $S^\ast$) that are incorrectly included in $S_\epsilon^\ast$. We can sum the first inequality in \eqref{eq:marginal_target_bridge} over the noise coordinates, and the second over the core coordinates in $D$ weighted by $w_j$, to obtain
\begin{equation}\label{descrp_upper_bounds}
\begin{aligned}
\mathbb{E}_{\mathbf{X}}|S_\epsilon^\ast\setminus S^\ast| 
    &= \sum_{j \notin S^\ast} \mathbb{P}_{\mathbf{X}}(j \in S_\epsilon^\ast) 
    \leq \frac{1}{\tau_\epsilon} \sum_{j\notin S^\ast}\bar\theta_j^{\mathrm{orc}}, \\
    \mathbb{E}_{\mathbf{X}}\!\left(\sum_{j\in D\setminus S_\epsilon^\ast}w_j\right) 
    &= \sum_{j \in D} w_j \, \mathbb{P}_{\mathbf{X}}(j \notin S_\epsilon^\ast) 
    \leq \frac{1}{1-\tau_\epsilon} \sum_{j\in D}w_j(1-\bar\theta_j^{\mathrm{orc}}).
\end{aligned}
\end{equation}
In expectation, the first inequality bounds the discrepancy between $S_\epsilon^\ast$ and $S^\ast$ in terms of the number of features, whereas the second bounds the discrepancy between $D$ and $S_\epsilon^\ast$ in terms of cumulative signal.
Our next assumption specifies the rates required for these discrepancies to remain asymptotically negligible:
\begin{condition}
\label{condition:oracle_screening}
The marginal oracle rewards satisfy
\begingroup
\begin{equation*}
\sum_{j\notin S^\ast}\bar\theta_j^{\mathrm{orc}}=o(L_p),
\qquad
\sum_{j\in D}w_j(1-\bar\theta_j^{\mathrm{orc}})
=o(\Delta_D^2).
\end{equation*}
\endgroup
\end{condition}
Together, Conditions~\ref{condition:core_margin} and~\ref{condition:oracle_screening} function as a margin-and-error pair. Condition~\ref{condition:core_margin} leaves $\kappa L_p$ empty spaces for $S^\ast$ below the maximum size limit and secures a signal surplus of $\gamma$ for $D$ above the required threshold. Condition~\ref{condition:oracle_screening} then guarantees that the discrepancies between $S^\ast$ and $S_\epsilon^\ast$, and between $D$ and $S_\epsilon^\ast$, are asymptotically negligible compared to these margins. Because the errors are entirely absorbed by the safety margins, the target set $S_\epsilon^\ast$ can safely deviate from $S^\ast$ while still satisfying the constraints of $\mathcal{S}$ with high probability.

Having established the assumptions on the oracle target $S_\epsilon^\ast$, we now turn to the assumptions governing the actual estimates produced by TVS. We begin by defining the relevant $\sigma$-field.
\begin{definition}\label{def:sigma_field}
At iteration $t$ of the algorithm, within the selection block of Algorithm~\ref{alg:tvs_step}, let $\mathcal{H}_t$ denote the $\sigma$-field generated by the data and the algorithmic history up to and including the Reward-Step, but strictly before the draws made in the Choose-Step.
\end{definition}
The Beta random numbers sampled in the Choose Step are generated conditionally on $\mathcal{H}_t$ and independently of earlier randomizations. Because $S_\epsilon^\ast$ is determined by the data, it is   $\mathcal{H}_t$-measurable. We define the two resulting conditional losses relevant to clustering as:
\begin{equation}
\label{eq:posterior_target_errors}
\operatorname{Out}_t
:=\mathbb E\!\left(
|\widehat S^t\setminus(S_\epsilon^\ast\cup S^\ast)|\mid\mathcal H_t
\right),
\qquad
\operatorname{Miss}_t
:=\mathbb E\!\left(
\sum_{j\in D\cap S_\epsilon^\ast}
w_j \mathbbm{1}\{j\notin\widehat S^t\}
\,\middle|\,\mathcal H_t
\right).
\end{equation}
In this formulation, $\operatorname{Out}_t$ measures the expected number of noise coordinates (i.e., features outside both the oracle target and the true signal set) erroneously selected by the empirical TVS draw. Conversely, $\operatorname{Miss}_t$ captures the expected cumulative signal present in both the oracle target and the core $D$, but missed by the empirical TVS draw. These two quantities connect the empirical TVS-drawn set $\widehat{S}^t$ to the size and retained-signal requirements of $\mathcal{S}$ in Theorem \ref{theorem:separation_condition}. Because we want $\widehat{S}^t$ to be as compact as possible while capturing maximal signal, both the excess noise $\operatorname{Out}_t$ and the dropped signal $\operatorname{Miss}_t$ must be small. 
Our next assumption specifies the rates required for these
discrepancies to remain asymptotically negligible:
\begin{condition}
\label{condition:posterior_learning}
For some deterministic sequence of iteration indices $t_n\in\mathbb N$, the following convergences in probabilities hold:
\begingroup
\begin{equation*}
\frac{\operatorname{Out}_{t_n}}{L_p}\xrightarrow{\mathbb P}0,
\qquad
\frac{\operatorname{Miss}_{t_n}}{\Delta_D^2}\xrightarrow{\mathbb P}0.
\end{equation*}
\endgroup
Here, convergence in probability is under the joint distribution of the data and all algorithmic randomness contained in $\mathcal H_{t_n}$.
\end{condition}
We deliberately do not track  $S^\ast \setminus S_\epsilon^\ast$, as they do not affect the condition in $\mathcal{S}$.

\subsubsection{Clustering Guarantee}\label{section:theory_present}
Theorem~\ref{theorem:separation_condition} dictates that we do not need to recover the exact set $S^\ast$ to guarantee exact clustering; it is sufficient to show that the selected feature set $\widehat{S} \in \mathcal{S}$ (detailed in Appendix \ref{section:target_for_feature_recovery}). Building on the conditions from Section \ref{section:assumptions}, the following theorem establishes that both the oracle target $S_\epsilon^\ast$ and the TVS-selected feature set $S^{t_n}$ converge to a valid set in $\mathcal{S}$. Consequently, the resulting cluster assignments $G^{t_n}$ converge to $G^\ast$. The proof is provided in Appendix~\ref{section:proof:theorem:posterior_surplus}.
\begin{theorem}[Oracle-to-adaptive clustering]
\label{theorem:posterior_surplus}
Under Conditions~\ref{condition:core_margin}
and~\ref{condition:oracle_screening},
the oracle reward target
$S_\epsilon^\ast$:
\begingroup
\begin{equation*}
\mathbb P_{\mathbf X}( S_\epsilon^\ast \in\mathcal S)\longrightarrow1.
\end{equation*}
\endgroup
If Condition~\ref{condition:posterior_learning} also holds,
\begin{equation}
\label{eq:clean_exact_recovery}
\mathbb P\!\left(
\widehat S^{t_n}\in\mathcal S,
\ \widehat G^{t_n}=G^\ast
\right)\longrightarrow1.
\end{equation}
Here and below, equality of partitions is understood up to a permutation of
the cluster labels.
\end{theorem}

We provide examples and sufficient conditions that can verify Conditions \ref{condition:oracle_screening} and \ref{condition:posterior_learning}. To verify Condition \ref{condition:oracle_screening}, we consider the rMMD test in Algorithm \ref{alg:robust_dc}. Let $\Phi$ denote the standard normal cumulative distribution function. Define the transformed cluster mean separation as
\begin{equation*}
d_j^{\mathrm R}
=2\Phi\!\left(
\frac{|\widetilde c_{1j}^\ast-\widetilde c_{2j}^\ast|}
{2\sqrt2\,s_j}
\right)-1, \qquad j=1,\cdots, p,
\end{equation*}
where we denote \(
\widetilde c_{kj}^\ast
:=(\boldsymbol\Omega^\ast\mathbf c_k^\ast)_j
\) and $s_j:=\{(\boldsymbol\Omega^\ast)_{jj}\}^{1/2}$.

The following corollary (proof   in  
Appendix~\ref{section:proof:corollary:rmmd_oracle_screening}) provides an example where Condition~\ref{condition:oracle_screening} holds:
\begin{corollary}[Verifying Condition~\ref{condition:oracle_screening} via \textup{\texttt{rMMD}}]
\label{corollary:rmmd_oracle_screening}
Suppose Condition~\ref{condition:core_margin} holds. Define the minimum true cluster size as $n_{\min}:=\min\{|G_1^\ast|,|G_2^\ast|\}$. Let $a_n=o(p^{-1/2})$ and $b_n\rightarrow0$ be deterministic positive sequences . For all sufficiently large $n$, suppose we implement the \textup{\texttt{rMMD}} test (Algorithm~\ref{alg:robust_dc}) at level $a_n$ by setting $\alpha_n=\tau_\epsilon-a_n$, using a corruption budget $0\leq r_n<n_{\min}/2$. 
Let $C_{\mathrm R}>0$ be the constant from Theorem~\ref{theorem:mmd}. Assume further that the cluster sizes are balanced ($|G_1^\ast| \asymp |G_2^\ast|$), the number of permutations $B$ used in \textup{\texttt{rMMD}} satisfies $B>3a_n^{-2}\{\log(8/b_n)+a_n(1-a_n)\}$, and every core coordinate $j\in D$ satisfies
\begin{equation}
\label{eq:detectable_core}
d_j^{\mathrm R}
\geq
C_{\mathrm R}\max\!\left\{
\sqrt{\frac{\max\{\log(e/a_n),\log(e/b_n)\}}{n_{\min}}},
\frac{r_n}{n_{\min}}
\right\}.
\end{equation}
 Then Condition~\ref{condition:oracle_screening} is satisfied.
\end{corollary}

In Corollary~\ref{corollary:rmmd_oracle_screening}, the
rMMD significance level $a_n$ controls the first sum in
Condition~\ref{condition:oracle_screening}, while the power
$1-b_n$ controls the second. These yield the respective requirements
$p a_n=o(L_p)$ and $b_n\Delta_D^2=o(\Delta_D^2)$.
Importantly, the transformed mean separation is required only for
features in $D$; no such separation condition is imposed on
$S^\ast\setminus D$. Finally, the ratio $r_n/n_{\min}$ quantifies the
cost of robustness and need not vanish, provided that it is dominated
by $d_j^{\mathrm R}$.

Next, Corollary~\ref{corollary:posterior_certificate} (proof  in 
Appendix~\ref{section:proof:corollary:posterior_certificate}) converts
Condition~\ref{condition:posterior_learning} into posterior separation and
concentration.
\begin{corollary}[Verifying Condition~\ref{condition:posterior_learning} by posterior separation]
\label{corollary:posterior_certificate}
Let $t_n\in\mathbb N$ be deterministic, and define
\begin{equation*}
 \mu_{j,t}:=\frac{a_j^t}{a_j^t+b_j^t},
 \quad
 \nu_{j,t}:=a_j^t+b_j^t,
\quad
\mathcal O_n:=[p]\setminus(S_\epsilon^\ast\cup S^\ast),
\quad\text{and}\quad
\mathcal C_n:=D\cap S_\epsilon^\ast.
\end{equation*}
Under Conditions~\ref{condition:core_margin}
and~\ref{condition:oracle_screening}, suppose that for some constants
$\delta\in(0,\tau_0)$, a sufficiently large $c_{\rm out}>0$ depending only
on $\delta$, and a deterministic
sequence $q_n\to\infty$, with probability tending to one,
\[
\begin{array}{lll}
\mu_{j,t_n}\leq\tau_\epsilon-\delta
&\text{and}&
\nu_{j,t_n}\geq c_{\rm out}\log p,
\quad j\in\mathcal O_n,\\[2pt]
\mu_{j,t_n}\geq\tau_\epsilon+\delta
&\text{and}&
\nu_{j,t_n}\geq q_n,
\quad j\in\mathcal C_n.
\end{array}
\]
 Then
Condition~\ref{condition:posterior_learning} holds and, consequently,
\(
\mathbb P\!\left(
\widehat S^{t_n}\in\mathcal S,
\ \widehat G^{t_n}=G^\ast
\right)\longrightarrow1.
\)
\end{corollary}
In Corollary \ref{corollary:posterior_certificate}, posterior means for extraneous noise coordinates lie below the
cutoff, posterior means for retained core coordinates lie above it, and the sizes of their accumulated histories
  make fresh Beta draws unlikely to cross the cutoff in
the wrong direction.  Extraneous noise coordinates require order $\log p$
evidence because there are up to $p$ opportunities for false inclusion;
retained core coordinates require only a diverging amount because clustering
depends on their aggregate signal rather than coordinatewise recovery.  Figure~\ref{fig:combined_performance}
illustrates the resulting separation.

\section{Numerical Studies}\label{section:simulation}
We evaluate our algorithms on both synthetic and real datasets. The SDP is solved using a custom first-order ADMM solver. 
For all experiments, we report clustering accuracy averaged over 200 independent replications.   Details on  solvers and algorithm hyperparameters are provided in Appendices \ref{section:sdp_solver_detail} and \ref{section:hyperparameter}.

\subsection{Benchmarks and Baselines}\label{section:method_unknown_cov}
We benchmark our approach against three categories of baselines: (i) existing sparse clustering algorithms, (ii) a simplified greedy version of our algorithm, and (iii) an extension of this greedy method and TVS algorithm that estimates the covariance matrix instead of assuming it is known.

For existing sparse clustering algorithms, we compare our approach to several widely used iterative methods: sparse $K$-means (SKM; \citealp{witten_framework_2010}), CHIME \citep{cai_chime_2019}, and sparse alternate sum clustering (SAS; \citealp{arias-castroSimpleApproachSparse2017}). We also benchmark against a two-step method (IFPCA; \citealp{jinInfluentialFeaturesPca2016}) and a complete convex relaxation approach, sparse convex clustering (SCVX; \citealp{wangSparseConvexClustering2018}). Details on these competing methods are provided in Appendix \ref{section:simulation_details}.

In addition to existing methods, 
we introduce a simplified, memoryless, and greedy version of our proposed algorithm. This benchmark removes memory and stochasticity from the Choose step (see Figure \ref{fig:alg_schematic}) by discarding the algorithm history $\{a_j^t, b_j^t\}$ and eliminating beta sampling. Instead, the selected set simply consists of the variables that received a reward of 1, defined as $\widehat S_{\mathrm{greedy}}^t=\{j:Y_j(\hat{G}^{t-1})=1\}$. 
Algorithm~\ref{alg:known_cov} gives the complete selection block (reward + choose).
The full permutation construction and pseudocode for this benchmark are provided in Appendix~\ref{section:method_known_cov}.

Finally, we introduce a modified version of our greedy algorithm which address the unknown covariance case by adopting a strategy that avoids estimating the full precision matrix $\boldsymbol{\Omega}^\ast$. At iteration $t$, we directly estimate the innovated data matrix $\widetilde{\mathbf{X}}$ using the current cluster labels $\hat{G}^{t-1}$. This estimate is then used in the reward step ({Algorithm~\ref{alg:known_cov}} or Algorithm~\ref{alg:tvs_step}) and the clustering block (Algorithm~\ref{alg:sdp_subroutine}). The direct estimation of $\widetilde{\mathbf X}$ is done via the Innovated Scalable Efficient Estimation (ISEE) procedure \citep{fan_innovated_2016}, 
 a nodewise regression approach originally developed for high-dimensional sparse precision matrix estimation. We modify it to selectively estimate only the required quantities for $\texttt{SDPcluster}$ procedure (Algorithm \ref{alg:sdp_subroutine}), avoiding full precision matrix recovery.

The ISEE procedure exploits two observations. First, the innovated vector \( \widetilde{\mathbf{X}}_i = \boldsymbol{\Omega}^\ast \mathbf{X}_i\in \mathbb{R}^p \) is Gaussian with  
$
\operatorname{Cov}(\widetilde{\mathbf{X}}_i)  = \boldsymbol{\Omega}^\ast \boldsymbol{\Sigma}^\ast \boldsymbol{\Omega}^\ast = \boldsymbol{\Omega}^\ast.
$
Second, for any subset $
A \subset [p]
$ of size 2,
and \( i \in G_1^\ast \), we obtain
$ 
    \widetilde{\mathbf{X}}_{A, i} = (\widetilde{\mathbf{c}}_1^\ast)_{A} + \widetilde{\mathcal{E}}_{A, i}
$ by using
the conditional distribution property of multivariate Gaussian and Shur complement:
\begin{equation}\label{eq:regression}
\hspace{-2.7em}
\underbrace{
\mathbf{X}_{A,i}
}_{
\hspace{1.1em}
\text{response} 
\in \mathbb{R}^{|A|}} 
\hspace{-1.6em}
=
\hspace{0.7em}
\underbrace{(\mathbf{c}_1^\ast)_{A} + (\boldsymbol{\Omega}^\ast_{A,A})^{-1} \boldsymbol{\Omega}^\ast_{A,A^c} (\mathbf{c}_1^\ast)_{A^c}}_{:=(\boldsymbol{\alpha}_1)_A \in \mathbb{R}^{|A|}~\text{(intercept)}}  
~-
~
\underbrace{
(\boldsymbol{\Omega}^\ast_{A,A})^{-1} \boldsymbol{\Omega}^\ast_{A,A^c}}_{
\hspace{-2.6em}
\text{slope} \in \mathbb{R}^{|A| \times (p - |A|)}
} 
\hspace{-1.6em}
\underbrace{
\mathbf{X}_{A^c, i}
}_{
\hspace{1.1em}
\text{predictor} \in \mathbb{R}^{p - |A|}
} 
\hspace{-1.3em}
+ 
\hspace{-1.53em}
\underbrace{ \mathbf{E}_{A,i}}_{
\hspace{2.1em}
\text{residual} \in \mathbb{R}^{|A|}
}\hspace{-2.1em},
\end{equation}
where
$
\mathbf{E}_{A,i} 
\sim 
\mathcal{N}
\bigl(
0, (\boldsymbol{\Omega}^\ast_{A,A})^{-1}
\bigr)
$. 
This implies \citep{fan_innovated_2016}:
\begin{equation}\label{ISEE_estimation_key_identity}
	(\widetilde{\mathbf{c}}_1^\ast)_{A}
	= \boldsymbol{\Omega}^\ast_{A,A} (\boldsymbol{\alpha}_1)_{A}
    \quad 
    \text{and}
    \quad
    		\widetilde{\mathcal{E}}_{A, i}
=		\boldsymbol{\Omega}^\ast_{A,A } \mathbf{E}_{A,i}.
\end{equation}
Appendix \ref{section:isee_detail}  provides detailed derivations for \eqref{eq:regression} and \eqref{ISEE_estimation_key_identity}.  Since \( p - |A|\) can be comparable or larger than \( n \), we fit \eqref{eq:regression} via high-dimensional regression (e.g., lasso) to estimate $\boldsymbol{\Omega}^\ast_{A,A }$, $ \mathbf{E}_{A,i}$ and $(\boldsymbol{\alpha}_1)_{A}$, which further yields plug-in estimate of $\widetilde{\mathbf{X}}_{A, i}$. Repeating over $A$ in a partition $ \mathcal A$ (blocks of size 2 or 3) and stacking gives an estimate of the full matrix $\widetilde{\mathbf X}_{G_1^\ast} \in \mathbb R^{|G_1^\ast|\times p}$.   A similar expression holds for \( i \in G_2^\ast \), with \( \mathbf{c}_1^\ast \) and $\boldsymbol{\alpha}_1$ replaced by \( \mathbf{c}_2^\ast \) and $\boldsymbol{\alpha}_2$, respectively. Following the same steps as described above, we can obtain the estimate of the submatrix $\widetilde{\mathbf X}_{G_2^\ast} \in \mathbb R^{|G_2^\ast|\times p}$. In practice, true labels are unknown; at iteration $t$, we use \( \hat{G}^{t-1} \) as a proxy, perform the above regressions within each estimated cluster, and output $\hat{\widetilde{\mathbf{X}}}^t=\texttt{ISEE}(\mathbf{X}, \hat{G}^{t-1})$.  The full algorithm  is detailed in Algorithm~\ref{alg:ISEE_subroutine}. Appendix \ref{section:ISEE_variants}
    numerically demonstrates the benefits of ISEE bypassing the full covariance structure estimation.

 \begin{algorithm}[t]
\caption{ISEE Transformation Subroutine (\texttt{ISEE})}
\label{alg:ISEE_subroutine}
\begin{algorithmic}[1]
\Require Data $\mathbf{X} \in \mathbb{R}^{p \times n}$, cluster assignments $\hat{G} = (\hat{G}_1^{t-1}, \hat{G}_2^{t-1})$
\State Partition $[p]$ into disjoint subsets $A_1, \ldots, A_L$ of size 2 or 3
\For{$\ell = 1, \ldots, L$}
    \For{$k \in \{1, 2\}$} \Comment{Estimate parameters for each cluster}
        \State Regress $\mathbf{X}_{A_\ell}$ on $\mathbf{X}_{A_\ell^c}$ using samples $i \in \hat{G}_k^{t-1}$ (e.g., via Lasso)
        \State \textbf{Obtain:} Intercept $\hat{\boldsymbol{\alpha}}_{(k)}$ and residuals $\{\hat{\mathbf{E}}_{A_\ell, i} : i \in \hat{G}_k^{t-1}\}$
    \EndFor
    \State $\hat{\mathbf{E}}_{A_\ell} \gets [\hat{\mathbf{E}}_{A_\ell, i}]_{i=1}^n$ \Comment{Pool residuals across clusters}
    \State $\hat{\boldsymbol{\Omega}}_{A_\ell A_\ell} \gets \bigl( \frac{1}{n} \hat{\mathbf{E}}_{A_\ell} \hat{\mathbf{E}}_{A_\ell}^\top \bigr)^{-1}$ \Comment{Estimate block precision}
    \For{$i = 1, \ldots, n$} \Comment{Reconstruct innovated data}
        \State Let $k$ be the cluster index of sample $i$
        \State $\hat{\widetilde{\mathbf{X}}}_{A_\ell, i} \gets \hat{\boldsymbol{\Omega}}_{A_\ell A_\ell} \bigl( \hat{\boldsymbol{\alpha}}_{(k)} + \hat{\mathbf{E}}_{A_\ell, i} \bigr)$
    \EndFor
\EndFor
\State \Return $\hat{\widetilde{\mathbf{X}}} \gets [\hat{\widetilde{\mathbf{X}}}_{A_1}^\top, \ldots, \hat{\widetilde{\mathbf{X}}}_{A_L}^\top]^\top$
\end{algorithmic}
\end{algorithm}
\subsection{Simulation Studies}\label{section:simulations}

Synthetic datasets  are generated from  model \eqref{def:gaussian_mixture_model} with two symmetric clusters:
\begin{equation}\label{simulation_setting_symmetric_gaussian}
     \mathbf{X}_i = 
       \mathbf{W}_i + \mathbf{c}^\ast,~ i = 1, \dots, n/2,
\quad \mathbf{X}_i = 
 \mathbf{W}_i
 -\mathbf{c}^\ast,~i = n/2 + 1, \dots, n.
\end{equation}
For the known-covariance setting, we generate $\mathbf{W}_i$ with entrywise i.i.d. Gaussian and Laplace distributions, each standardized to have unit variance. In the unknown-covariance setting, $\mathbf{W}_i$ is obtained by premultiplying these unit-variance Gaussian and Laplace vectors by $(\boldsymbol{\Sigma}^\ast)^{1/2}$. We fix the support of $\boldsymbol{\Omega}^\ast \mathbf{c}^\ast$   as $S^\ast = [10]$ and vary the ambient dimension $p$. We consider two forms of $\boldsymbol{\Sigma}^\ast$, specified via its inverse: a chain graph and AR(1) (see Table~\ref{tab:synthetic_summary}).
 All entries of $(\boldsymbol{\Omega}^\ast
 \mathbf{c}^\ast)_{S^\ast}$ have equal magnitude, chosen so that the   signal strength
$
\Delta_{S^\ast}
$ defined in \eqref{def:S0}
remains constant as $p$ increases. Details on parameter settings are provided in Table \ref{tab:synthetic_summary}.

\begingroup
\begin{table}[t]
\caption{Summary of parameters for \eqref{simulation_setting_symmetric_gaussian} for synthetic data with known (Figure \ref{fig:varknown_coviso}) and unknown (Figure \ref{fig:varunknown}) covariance  settings.  For all settings, we have $S^\ast = \{1, \ldots, 10\}$.
}
\label{tab:synthetic_summary}
\vspace{1em}
\centering
\begin{tabular}{llllllll}
\toprule
Figure
&
Noise 
& $p$
&  $\boldsymbol{\Omega}^\ast$
& $n$ 
& $\Delta_{S^\ast} $
\\
\midrule 
\midrule 
 \ref{fig:varknown_coviso}-(a)
&
Gaussian  
&
50-5000 
&
$\mathbf{I}_p$ (known) 
& 200
& 4
\\
 \ref{fig:varknown_coviso}-(b), (c)
&
(a) Gaussian  (b) Laplace 
&
1000-30000 
&
$\mathbf{I}_p$ (known) 
& 200
& 4
\\
 \ref{fig:varunknown}-(a), (b)
&
(a) Gaussian  (b) Laplace
&
100-1000
&
Chain graph
& 200
& 6
\\
 \ref{fig:varunknown}-(c), (d)
&
(c) Gaussian  (d) Laplace 
&
100-1000
&
AR(1)
& 200
& 4
\\
 \ref{fig:varunknown}-(e), (f)
&
(a) Gaussian  (b) Laplace
&
100-400  
&
Chain graph
& 500
& 3
\\
 \ref{fig:varunknown}-(g), (h)
&
(c) Gaussian  (d) Laplace
&
100-400 
&
AR(1)
& 500
& 3
\\
\bottomrule
\end{tabular}
\end{table}
\par
\endgroup

\subsubsection{Sensitivity of   TVS Algorithm to   Beta Sample Cutoff}
We study the sensitivity of the TVS algorithm to the Beta sample cutoff $\tau_\epsilon$. This cutoff is parameterized by the regularization parameter $\epsilon$ in  \eqref{def:tau_epsilon}. The analysis assumes model \eqref{simulation_setting_symmetric_gaussian} and relies on the configuration from line 2 of Table \ref{tab:synthetic_summary}. For this setup, we fix the cluster separation defined in  \eqref{def:S0} as $\Delta_{S^\ast}=4$   and set the ambient dimension to $p=8000$.
Although \cite{liuVariableSelectionThompson2023} recommends $\epsilon \approx 0.618$ (corresponding to $\tau_\epsilon = 0.5$), we consider $\epsilon \in {0.2,0.4,0.6}$ to examine the exploration--exploitation trade-off. A larger value of $\epsilon$ retains more variables in $\hat{S}^t$.

\begin{figure}[t]
    \centering
    
    \includegraphics[width=0.8\linewidth]{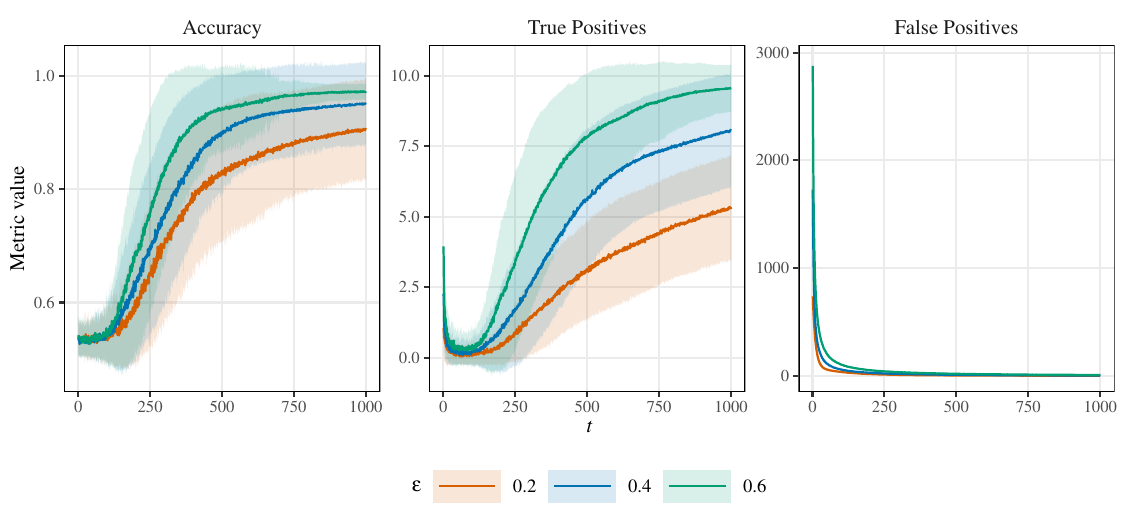}
    
    {\small (a) Clustering accuracy and variable selection over $t = 1, \dots, 1000$.}
    \label{fig:trajectory_C} 
    
    \vspace{0.6cm} 
    
    \includegraphics[width=0.8\linewidth]{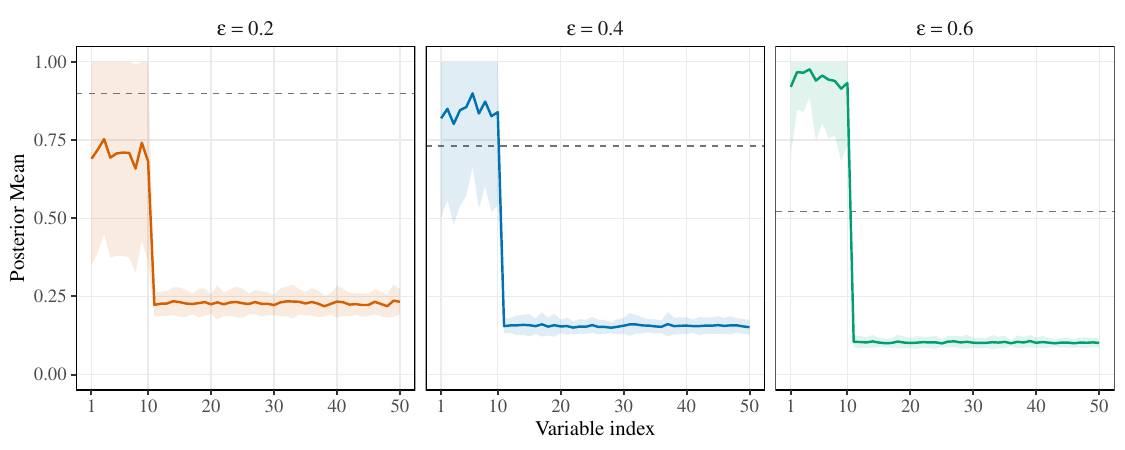}
    
    {\small (b) Beta posterior means at $T=1000$ for the first 50 variables.}
    \label{fig:theta_bar} 
    
    \vspace{0.2cm}
    
    \caption{Performance trajectories and posterior means across varying exploration rates $\epsilon \in \{0.2, 0.4, 0.6\}$ with fixed ambient dimension $p = 8000$. \textbf{(a)} Average trajectories of clustering accuracy ($\hat{G}^t$) and true/false positives ($\hat{S}^t$). \textbf{(b)} Beta posterior means $a_j^T / (a_j^T + b_j^T)$. Horizontal dashed lines indicate the inclusion thresholds $\tau_{\epsilon}
=
\frac{\log(1/\epsilon)}
{\log\!\left(\frac{\epsilon+1}{\epsilon}\right)}$ specific to each rate. In both panels, solid lines denote the mean and shaded bands denote $\pm 1$ standard deviation across 100 independent replications.}\label{fig:combined_performance}
\end{figure}

Figure~\ref{fig:trajectory_C} shows the average clustering accuracy of $\hat{G}^t$ and the numbers of true and false positives for $\hat{S}^t$ over $t=1,\dots,1000$, averaged across 100 replications. The more permissive choice $\epsilon=0.6$ achieves the best overall performance in this high-dimensional setting: although it yields more false positives, it also identifies substantially more true positives, resulting in higher clustering accuracy.

Figure~\ref{fig:combined_performance} qualitatively illustrates the two-error mechanism in Theorem~\ref{theorem:posterior_surplus}. Figure~\ref{fig:theta_bar} shows posterior mass increasing on displayed signal variables, which reduces missed core signal $\operatorname{Miss}_t$, while Figure~\ref{fig:trajectory_C} shows false-positive selections falling rapidly, consistent with decreasing posterior leakage $\operatorname{Out}_t$. Theorem~\ref{theorem:posterior_surplus} permits both errors when they are negligible relative to $\sqrt p$ and the core signal, respectively.

\subsubsection{Comparison with Baseline Methods under Known Covariance}
\begin{figure}[t]
\centering
\includegraphics[width=0.99\textwidth]{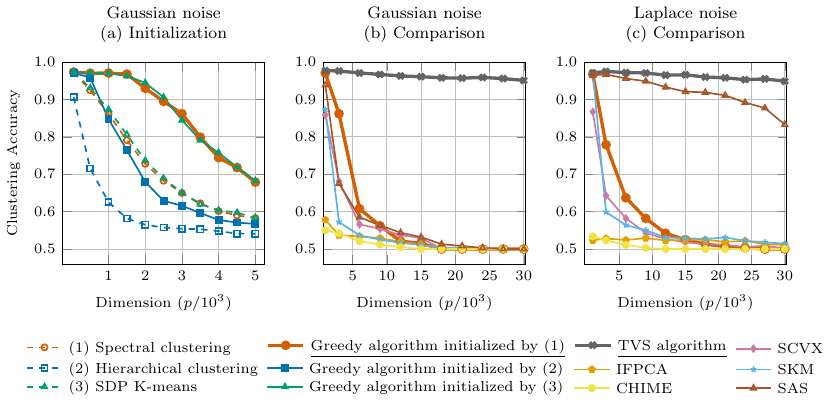}
\caption{ \small Clustering accuracy under known covariance settings. 
In (a), we compare 
the non-sparse baselines (dotted curves)
with the greedy algorithms initialized by each of them (solid curves).
 In (b) and (c), we compare the greedy and TVS algorithms (thick lines in the plots and underlined in the legends) against sparsity-aware baselines. Full parameter settings are provided in Table~\ref{tab:synthetic_summary}.}
\label{fig:varknown_coviso}
\end{figure}

We first run experiments under the synthetic data with known covariance setting (line 1-3 of Table \ref{tab:synthetic_summary}). 
To isolate the benefits of memory and random exploration, 
we first compare standard non-sparse methods (spectral clustering, hierarchical clustering, and full-feature SDP $K$-means) against our greedy algorithm initialized with each method's output ($\hat{G}^0$).
 Introduced in Section \ref{section:method_unknown_cov}, this greedy algorithm is a simplified version of TVS that omits memory and Beta sampling,   replacing the Choose Step with Algorithm~\ref{alg:known_cov}. The results in Figure~\ref{fig:varknown_coviso}(a) demonstrate that while non-sparse methods fail to adapt to high-dimensional sparsity, the greedy approach consistently improves upon them. This confirms that sparsity-aware iterations alone yield substantial improvements, even without active randomization and memory.

To evaluate the benefits of random exploration and memory, we compare our approach against  our greedy algorithm and existing sparse clustering methods. Figures~\ref{fig:varknown_coviso}(b) and \ref{fig:varknown_coviso}(c) demonstrate that the TVS algorithm consistently outperforms its competitors across all dimensions under both Gaussian and Laplace noise. Unlike IFPCA and SAS, which exhibit inconsistent performance depending on the noise setting, TVS remains highly stable. This strong performance highlights how memory and active exploration make the algorithm robust to local optima  and noise assumption violation.

\subsubsection{Comparison with Baseline Methods under Unknown Covariance}
Integrating covariance estimation to our TVS algorithm creates a tension: it requires large sample sizes, yet clustering difficulty scales with $\log n$ (Theorem \ref{theorem:separation_condition}). 
From a reinforcement learning perspective, the cluster partition $\hat{G}$ acts as a ``state'' determining reward probabilities. Introducing covariance estimation expands this state space from $\hat{G}$ to $\hat{G} \times \hat{\boldsymbol{\Omega}}$. Even when the estimated precision matrix $\hat{\boldsymbol{\Omega}}$ is relatively accurate, inherent statistical fluctuations unnecessarily enlarge the state space, making the learning problem overly complex. 

To address this, we propose a regime-specific strategy rather than directly incorporating ISEE into the TVS algorithm. When the precision matrix can be reliably estimated (typically moderate to large $n$ and small to moderate $p$), where well-established theory of covariance or precision matrix estimation ensures consistency, we apply a greedy algorithm combined with ISEE. Conversely, in the small-$n$, large-$p$ regime where reliable covariance estimation is infeasible, we apply our known-covariance method using the identity matrix as a plug-in estimate. This \say{naive} approach is supported by observations in linear classification, where bypassing covariance estimation entirely often yields better performance \citep{bickel_theory_2004a}.

\begin{figure}[t]
\centering
\includegraphics[width=0.99\textwidth]{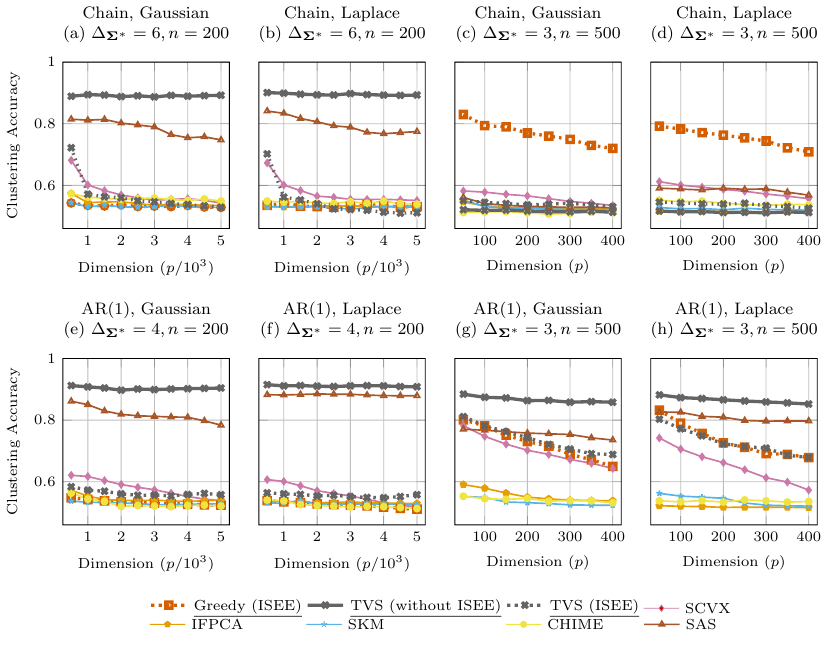}
\caption{ \small Clustering accuracy under unknown covariance settings. Our proposed algorithms are highlighted with thick lines in the plots and underlined in the legends.
In (a), we compare the greedy algorithms with non-sparse baselines; the dotted curves represent the non-sparse methods, corresponding to the initializations prior to the first variable selection block. In (b) and (c), we compare the greedy and TVS algorithms against sparsity-aware baselines. Full parameter settings are provided in Table~\ref{tab:synthetic_summary}.}
\label{fig:varunknown}
\end{figure}

We evaluate these strategies under unknown covariance setting (lines 3--6 of Table \ref{tab:synthetic_summary}) using variants of our greedy and TVS algorithms with and without ISEE. In most cases (6 out of 8 settings), TVS without ISEE outperforms other methods, indicating that bypassing covariance estimation is beneficial when the sample size is small, the dimension is large, or the separation is weak.
However, in a small-$p$, large-$n$, low-separation regime presented in Figures \ref{fig:varunknown}-(c) and \ref{fig:varunknown}-(d), TVS without ISEE performs poorly, while the greedy algorithm with ISEE performs well. This suggests that when the sample size is sufficiently large relative to the dimension, making covariance estimation reliable, the greedy approach with ISEE is effective. In contrast, TVS with ISEE underperforms in most settings.

We include additional simulations in the supplementary material, to numerically demonstrate that:
   the subset SDP \eqref{def:subset_sdp}  captures the clustering signal (Appendix~\ref{section:simul_known_noniso}),
ISEE bypassing the full covariance structure estimation improves the clustering performance (Appendix \ref{section:ISEE_variants}),
robustness of our algorithms to mild violations  of  covariance homogeneity (Appendix \ref{section:assumption_robustness:same_cov}) and exact sparsity
(Appendix \ref{section:assumption_robustness:sparsity}), 
alternative SDP solver for scalability to large sample sizes (Appendix \ref{section:solver}),
and comparison to subspae clustering approaches (Appendix \ref{section:JL}).

\subsection{Real data analysis}
 
We evaluate the proposed methods on a subset of the 20 Newsgroups corpus from the UCI Machine Learning Repository, consisting of $n=400$ documents sampled from the semantically distinct discussion groups \texttt{sci.space} and \texttt{rec.autos}. This pair of groups provides a challenging setting for sparse variable selection: although the two domains share substantial mechanical terminology, their underlying topics remain fundamentally different.  Standard preprocessing was applied, including lowercasing, punctuation and number removal, stop-word filtering, and whitespace normalization (see Appendix \ref{section:autoencoder} for details). Documents were represented using term frequency--inverse document frequency (TF--IDF) features, and terms appearing in fewer than $1\%$ of documents were removed. The resulting TF--IDF matrix has dimension $1331 \times 400$ with approximately $97.3\%$ zero entries and a median of 24 nonzero terms per document.
Since TF--IDF vectors are nonnegative and document lengths vary considerably, each document vector was $\ell_2$-normalized prior to clustering. The true newsgroup labels were used as the ground-truth partition. 

For our algorithms, we do not estimate the covariance structure because $n \ll p$ ($n=400$, $p=1331$). As observed Figure \ref{fig:varunknown}, bypassing covariance estimation is a well-justified approach in this high-dimensional setting.
Table~\ref{tab:real} reports clustering accuracy across competing methods. Both proposed approaches substantially outperform existing sparse clustering baselines.  

A key advantage of the TVS algorithm over greedy algorithm is that it provides a posterior-mean measure of variable importance. Under TVS, the ten variables with the highest posterior means are \textit{auto}, \textit{automatic}, \textit{best}, \textit{bought}, \textit{brake}, \textit{car}, \textit{cars}, \textit{chevy}, \textit{clutch}, and \textit{dealer}. Most of them are specific to automotive terminology. While the highest-ranked words based on TF-IDF values often contain semantically uninformative terms such as ``like'' and ``new,'' the words selected by TVS successfully exclude them. Consequently, the TVS selections are substantially more discriminative and interpretable, maintaining a clear focus on the automotive category.
\begin{table}[t]
\centering
\caption{Clustering accuracy on the 20 Newsgroups dataset.}
\label{tab:real}
\begin{tabular}{lccccccc}
\toprule
Method & SKM & SAS & IF-PCA & CHIME & SCVX  & \textbf{Greedy} & \textbf{TVS}\\
\midrule
Accuracy & 0.505 & 0.508 & 0.523 & 0.504 & 0.520 & \textbf{0.753} & \textbf{0.735}  \\
\bottomrule
\end{tabular}
\end{table}

\section{Discussion}\label{section:discussion} 
We addressed the challenge of coupling between two discrete structures: variable selection and clustering. By treating variable inclusion as a nuisance parameter, we established uniform minimax separation bounds for exact recovery in SDP $K$-means. Our theory demonstrates that while variable selection is necessary in sparse regimes, SDP $K$-means remains highly robust to moderate over-selection, provided that sufficient cumulative signal is retained.

Exploiting this robustness, we developed a randomized block-coordinate framework alternating between SDP clustering and variable selection.
To prevent stagnation in local optima (a common failure of existing greedy iterative  methods) we formulated variable selection as a multi-armed bandit problem. 
Solved via Thompson Sampling, this introduces \emph{adaptive memory} and \emph{active exploration} to prevent the premature discarding of weak signals.
 We use a robust permutation MMD test designed to make the bandit rewards less sensitive to imperfect intermediate clusterings. For unknown covariance settings, we adapted the ISEE procedure \citep{fan_innovated_2016} to bypass full precision matrix estimation.

We provide a theoretical guarantee for exact recovery that bridges the true support $S^\ast$, the SDP exact recovery collection $\mathcal{S}$ defined in Theorem \ref{theorem:separation_condition}, and the TVS oracle target $S^\ast_\epsilon$. All conditions are imposed on set sizes and cumulative signals; no assumptions are made about individual feature entries. The signal requirements need only hold in aggregate over a "signal core" carrying sufficient cumulative signal, meaning weak signals outside this core can be safely missed. Our theory requires the posterior to converge in two cumulative aspects: the number of noise selections ($\text{Out}_t$) and the missed core cumulative signal ($\text{Miss}_t$), rather than imposing  strict coordinatewise recovery requirement.
 
 Several future directions remain.
 First, deriving iteration-dependent rates for the $\text{Out}_t$ and $\text{Miss}_t$ terms would yield a formal finite-time convergence theorem.
Second, theoretically justifying the robust MMD test under the full iterative framework, where conditioning on the data used to construct the pilot cluster estimate complicates the derivation of two-sample testing error bounds, is an important next step.  Finally, extending the minimax lower bounds to accommodate general unknown covariance matrices would complete the information-theoretic characterization of sparse clustering.


\acks{The authors acknowledge the Center
for Advanced Research Computing (CARC) at the University of Southern California for
providing computing resources that have contributed to the research results reported within
this publication. URL: \url{https://carc.usc.edu.}}


\newpage

\appendix

\section{Overview of Appendix}\label{appendix_overview}
This appendix contains the technical lemmas and proofs for Theorems~\ref{theorem:separation_condition} (Appendix~\ref{section:proof:theorem:separation_condition}) and \ref{theorem:posterior_surplus} (Appendix~\ref{section:proof:TVS_theory}). It also provides details on the robust MMD test (Appendix~\ref{section:details_MMD}), the numerical study (Appendix~\ref{section:simulation_details}), and additional numerical results (Appendix~\ref{section:additional_simulation}).

\section{Proof of Theorem \ref{theorem:separation_condition}}\label{section:proof:theorem:separation_condition}
For simplicity, this proof   state the separation condition as 
\begin{equation*}
\Delta_{S \cap S^\ast}^2 :=
  \hskip -3mm
\min_{1 \leq k_1 \neq k_2 \leq K} 
\| 
(\mathbf{c}_{k_1}^\ast - \mathbf{c}_{k_2}^\ast )_{S \cap S^\ast}
 \|_2^2
 \ge
  {\sigma^{2}}
		\biggl(\log n +
        \frac{|S| \log p}{m}
		+
		\sqrt{
        \frac{|S| \log p}{m}
        }
		\biggr).
 \end{equation*}
Because the proof requires extensive notation, we provide a complete glossary in Appendix~\ref{section:list_of_notations} for the reader's convenience.
The first three sections (Appendices \ref{section:reform_to_stoc_dom}-\ref{section:proof:theorem:separation_condition:conclusion})
prove the first claim of theorem, 
\begin{equation*}
    \mathbb P
    \bigl(
    \hat{\mathbf Z}(S) = \mathbf Z^\ast, \forall S\in \mathcal S
    \bigr)\geq 1-  {C_2}K/n.
\end{equation*}
Then  Appendix \ref{section:proof:theorem:minimax_lower_bound} proves the second claim of theorem:
\begin{equation*}
    	\inf_{ 
\hat{\mathbf{H}}: \mathbb{R}^{p \times n} \to  \mathcal{H}(K)
}
		\sup_{\boldsymbol{\psi} \in 	\overline{\Psi}}
		\mathbb{P}_{\boldsymbol{\psi}} 
        \bigl(
        \hat{\mathbf{H}}(\mathbf{X}) \neq \mathbf{H}
        \bigr)
		\geq
		C_5.
\end{equation*}
The proof of the first claim follows the strategy of Theorem II.1 of \citet{chen_cutoff_2021}, with modifications to account for the sparsity assumption.
Within this proof only, $\Delta_{S\cap S^\ast}^2$ denotes the raw squared
center separation, which equals $\sigma^4$ times the innovated-scale
quantity in the main paper; see the explanation following
\eqref{def:S-set}.
The proof proceeds in three key steps.
\begin{enumerate}
    \item \textbf{Reformulation into high-probability bound problem (Appendix \ref{section:reform_to_stoc_dom}):}
For each $S$, we construct  three statistics that bound a dual certificate $\lambda^S$, which verifies exact cluster recovery by the SDP problem \eqref{SDP_objective_submatrix} corresponding to $S$. A nontrivial gap between these bounds guarantees the existence of such a certificate (Corollary \ref{corollary:problem_reformulation}).

\item \textbf{First bounds and construction of the dual variable (Appendix \ref{section:proof:set_lambda}).}
Under the separation and set size conditions imposed by $\mathcal{S}$, we derive uniform high-probability bounds on the two statistics defined in the first step and determine the value of $\lambda^S$ accordingly (Lemma \ref{lemma:lambda_between}).

\item \textbf{Second  bounds and conclusion (Appendix \ref{section:proof:theorem:separation_condition:conclusion}).} 
We establish a uniform high-probability bound on the third statistic and verify that the previously chosen value of $\lambda^S$ exceeds this bound, thereby confirming the existence of a dual certificate and completing the proof.
\end{enumerate}

Throughout this appendix, write $n_k:=|G_k^\ast|$ for the size of true cluster $k$.
For notational simplicity, we adopt the following matrix-based notations of our model $\mathbf{X}_i = \mathbf{c}_k^\ast + \boldsymbol{\varepsilon}_i$ for each $i \in G_k^\ast$, where $\boldsymbol{\varepsilon}_i \overset{\text{i.i.d.}}{\sim} \mathcal{N}(0, \sigma^2 \mathbf{I}_p)$:
\begin{definition}[Matrix-based notations]\label{def:theorem_1_notation}
We view the data as a matrix $\mathbf{X} \in \mathbb{R}^{p \times n}$ and write
$
\mathbf{X} = \mathbf{M}^\ast + \mathcal{E}$,
where $\mathbf{M}^\ast$ contains the cluster means column-wise, and $\mathcal{E}$ collects the noise vectors. For any support set $S \subset [p]$, we denote the $i$th observation and its noise restricted to $S$ by $$\mathbf{X}_{S, i} \quad \text{and} \quad \mathcal{E}_{S, i},$$ respectively. Given true clusters $G_1^\ast, \ldots, G_K^\ast$, we define the empirical mean and average noise of cluster $k$ on support $S$ as
$$
\overline{\mathbf{X}}^\ast_{S, k}
:=
\frac{1}{|G_k^\ast|} \sum_{i \in G_k^\ast} \mathbf{X}_{S,i},\quad
\overline{\mathcal{E}}^\ast_{S, k}
:=
\frac{1}{|G_k^\ast|} \sum_{i \in G_k^\ast} \mathcal{E}_{S,i}.
$$
\end{definition}

\subsection{Reformulation into  high-probability bound problem}\label{section:reform_to_stoc_dom}	
The primary goal of this section is to derive Corollary \ref{corollary:problem_reformulation}, which converts the optimality condition into high-probability bounds.
Let $S \subset [p]$ be a support set,
and consider 
the primal SDP problem \eqref{optim:SDP_submatrix_S} corresponding to $S$, recalled below:
\begin{equation}\label{optim:SDP_submatrix_S}
	\max_{\mathbf{Z} \in \mathbb{R}^{n \times n}}  \langle 
	\mathbf{X}_{S, \cdot}^\top \mathbf{X}_{S, \cdot}, \mathbf{Z} \rangle
    \quad
    \text{s.t.}
    \quad
    \mathbf{Z}^\top = \mathbf{Z},~
    \mathbf{Z} \succeq 0,~
	\mathrm{tr}(\mathbf{Z}) = K,~
	\mathbf{Z} \boldsymbol{1}_n = \boldsymbol{1}_n,~
	\mathbf{Z} \geq 0.
\end{equation}
Let
	$
	\mathcal{L}(\mathbf{Z}, \mathbf{Q}^S, \lambda^S,  \boldsymbol{\alpha}^S, \mathbf{B}^S)
	$
	denote the associated Lagrangian function.
	Here, the Lagrange multipliers
	$$
	\mathbf{Q}^S \in \mathbb{R}^{n \times n}, \quad \lambda^S \in \mathbb{R}, \quad \boldsymbol{\alpha}^S \in \mathbb{R}^n, \quad \text{and} \quad \mathbf{B}^S \in \mathbb{R}^{n \times n}
	$$
	correspond to the primal constraints
	$$
	\mathbf{Z} \succeq 0, \quad \mathrm{tr}(\mathbf{Z}) = K, \quad \mathbf{Z}^\top = \mathbf{Z}, \quad \text{and} \quad \mathbf{Z} \geq 0,
	$$
	respectively.
	Note that $\mathbf{Q}^S \succeq 0$ and $\mathbf{B}^S \geq 0$.
Based on this formulation, this section proceeds in three main steps to ultimately establish Corollary \ref{corollary:problem_reformulation}:
    \begin{enumerate}
        \item 
        \textbf{Necessary conditions for dual certificates (Appendix \ref{section:dual_cond_necessary})}:
We formalize sufficient conditions for the existence of dual certificates (Lemma \ref{lemma:opt_cond}) and leverage them to derive necessary conditions (Lemma \ref{lemma:opt_necessary}) that guide the construction of dual variables.

        \item \textbf{Reduction to a single dual variable construction (Appendix \ref{section:dual_construction_proposal}):} 
 Leveraging the necessary conditions, we parameterize the  dual variables in terms of $\lambda^S$ , reducing the construction to a one-dimensional problem (Lemma \ref{lemma:dual_construction}).

        \item \textbf{Reformulation into high-probability bound problem (Appendix \ref{section:lambda_bounds}):} 
We derive statistics that bound the dual certificate $\lambda^S$ from above (Lemma \ref{lemma:lambda_upper_bound}) and below (Lemma \ref{lemma:lambda_s_bound}). This shows that a nontrivial gap between them guarantees the existence of dual variables satisfying the conditions of Lemma \ref{lemma:opt_cond} (Corollary \ref{corollary:problem_reformulation}).

    \end{enumerate}

    \subsubsection{Necessary conditions for dual certificates}\label{section:dual_cond_necessary}
Let $S \subset [p]$ be a support set, and consider the corresponding primal SDP problem \eqref{optim:SDP_submatrix_S}. Lemma \ref{lemma:opt_cond} provides sufficient conditions under which the dual variables $\lambda^S$, $\boldsymbol{\alpha}^S$, and $\mathbf{B}^S$ form a valid dual certificate (with $\mathbf{Q}^S$ implicitly captured by the conditions). In particular, the lemma guarantees that the duality gap vanishes at the true cluster membership matrix
$$
\mathbf{Z}^\ast = \sum_{k=1}^K \frac{1}{|G_k^\ast|} \mathbf{1}_{G_k^\ast} \mathbf{1}_{G_k^\ast}^\top.
$$

	\begin{lemma}[Optimality conditions]\label{lemma:opt_cond}
		Let $S \subset [p]$ be a support set, and
		consider the corresponding primal SDP problem \eqref{optim:SDP_submatrix_S}. 
		The matrix $\mathbf{Z}^\ast$ is the unique solution to this problem if there exist dual variables $\lambda^S \in \mathbb{R}$, $\boldsymbol{\alpha}^S \in \mathbb{R}^n$, and $\mathbf{B}^S \in \mathbb{R}^{n \times n}$ such that the following conditions are satisfied:
		\begin{align*}
			&(\mathrm{C}1) \quad \mathbf{B}^S \geq 0; \\
			&(\mathrm{C}2)
			\quad
			\mathbf{W}^S
			:=
			\lambda^S \mathbf{I}_n
             - \mathbf{B}^S
			+
			\frac{1}{2} 
			\mathbf{1}_n (\boldsymbol{\alpha}^S)^\top
            + 
            \frac{1}{2} 
            \boldsymbol{\alpha}^S \mathbf{1}_n^\top
			- \mathbf{X}_{S, \cdot}^\top \mathbf{X}_{S, \cdot} \succeq 0; \\
			&(\mathrm{C}3) \quad \langle \mathbf{W}^S, \mathbf{Z}^\ast \rangle  = 0; \\
			&(\mathrm{C}4) \quad \langle \mathbf{B}^S, \mathbf{Z}^\ast  \rangle = 0;\\
			&(\mathrm{C}5) \quad \mathbf{B}^S_{\mathbf{G}_k^\ast, \mathbf{G}_l^* } > 0 \quad \text{for all} \; 1 \leq k \neq l \leq K.
		\end{align*}
	\end{lemma}
	In Lemma \ref{lemma:opt_cond}, conditions (C1) and (C2) ensure dual feasibility, (C3) and (C4) guarantee a zero duality gap, and (C5) ensures the uniqueness of the solution.
	The proof is provided in Appendix~\ref{section:proof:lemma:opt_cond}.
Building on these conditions, we now establish necessary conditions for dual certificates.
Let 
$\bigl(
\mathbf{B}^S_{G_k^\ast G_l^\ast} \mathbf{1}_{|G_l^\ast|}
\bigr)_i$ denote the $i$th  entry of the 
	row-wise sum of
    $\mathbf{B}^S_{G_k^\ast G_l^\ast} \in \mathbb{R}^{|G_k^\ast| \times |G_l^\ast|}$, which is the
    $(k,l)$th off-diagonal block of $\mathbf{B}^S$.
    Let $\overline{\mathbf{X}}^\ast_{S,k}$ denote the average of the observations belonging to the true cluster $k$, restricted to the entries corresponding to the variable set $S$.
Leveraging these notations, we state the necessary conditions in the following lemma.	
	\begin{lemma}[Necessary conditions for optimality]\label{lemma:opt_necessary}
		Let $S \subset [p]$ be a support set, and
		consider the corresponding primal SDP problem \eqref{optim:SDP_submatrix_S}.
		Suppose dual variables
		$\lambda^S \in \mathbb{R}$,
		$\boldsymbol{\alpha}^S \in \mathbb{R}^n$,
		and
		$\mathbf{B}^S \in \mathbb{R}^{n \times n}$
		satisfy conditions \textnormal{(C2)}, \textnormal{(C3)}, and \textnormal{(C4)} of Lemma \ref{lemma:opt_cond}.
		Then, for all $1 \leq k \neq l \leq K$ and $i \in G_k^\ast$, the following statements  hold:
		\begin{align*}
			\boldsymbol{\alpha}^S_i
			&= 
			2
			\mathbf{X}_{S, i}^\top
			\overline{\mathbf{X}}^\ast_{S,k}
			- 
			\frac{\lambda^S}{|G_k^\ast|}
			- 
			\|
			\overline{\mathbf{X}}^\ast_{S,k}
			\|_2^2,~\text{and}~\numberthis \label{alpha_condition_final}
		\\
					\bigl(
		\mathbf{B}^S_{G_k^\ast G_l^\ast} \mathbf{1}_{|G_l^\ast|}
		\bigr)_i
		&=-\frac{|G_k^\ast| + |G_l^\ast|}{2 |G_k^\ast|} \lambda^S
		+ \frac{ |G_l^\ast| }{2} 
		\bigl(
		\| \overline{\mathbf{X}}^\ast_{S,l}- \mathbf{X}_{S, i} \|_2^2 
		- 
		\| \overline{\mathbf{X}}^\ast_{S,k}- \mathbf{X}_{S, i} \|_2^2 
		\bigr),
		~\numberthis	\label{B_condition_rowsum}
		\end{align*}		
	\end{lemma}
	The proof of Lemma \ref{lemma:opt_necessary} is provided in Appendix \ref{section:proof:lemma:opt_necessary}.

\subsubsection{Reduction to a single dual variable construction }\label{section:dual_construction_proposal}
For any given $\lambda^S \in \mathbb{R}$, condition \eqref{alpha_condition_final} uniquely determines the corresponding value of $\boldsymbol{\alpha}^S$, which we denote by $\dot{\boldsymbol{\alpha}}^S(\lambda^S)$. We then turn to the dual variable  $\mathbf{B}^S$ and construct $\dot{\mathbf{B}}^S(\lambda^S)$, designed to satisfy condition \eqref{B_condition_rowsum}. The matrix $\dot{\mathbf{B}}^S (\lambda^S)$ is structured as a block matrix, with symmetric rank-one matrices as its off-diagonal blocks and zero matrices as its diagonal blocks. These constructions follow the proof of Theorem II.1 in \cite{chen_cutoff_2021}, with modifications to account for varying support sets.

We now describe the construction of $\dot{\mathbf{B}}^S(\lambda^S)$. Fix any value of $\lambda^S \in \mathbb{R}$, and for notational simplicity, we suppress the dependence on $\lambda^S$ in what follows.
For each $k = 1, \ldots, K$, define the diagonal block $\dot{\mathbf{B}}^S_{G_k^\ast, G_k^\ast} \in \mathbb{R}^{|G_k^\ast| \times |G_k^\ast|}$ to be the zero matrix.
Next, consider the off-diagonal blocks. Given $\lambda^S$, the data, and the true cluster structure, condition \eqref{B_condition_rowsum} uniquely determines the row-wise sum of the $(k, l)$ off-diagonal block of the dual certificate. Denote this unique vector by $\mathbf{r}^{(k,l,S)} \in \mathbb{R}^{|G_k^\ast|}$.
Then, for each $1 \leq k \neq l \leq K$, we define the off-diagonal block $\dot{\mathbf{B}}^S_{G_k^\ast, G_l^\ast} \in \mathbb{R}^{|G_k^\ast| \times |G_l^\ast|}$ as the symmetric rank-one matrix:
	\begingroup
\begin{equation*}
		\dot{\mathbf{B}}^S_{G_k^\ast, G_l^\ast}
		:= 
        \frac{1}{			\mathbf{1}_{ |G_l^\ast| }^\top
			\mathbf{r}^{(l,k, S)}}
			\mathbf{r}^{(k,l, S)}
			(\mathbf{r}^{(l,k, S)})^\top,
	\end{equation*}
\endgroup
This construction ensures that the row-sum condition in \eqref{B_condition_rowsum} is  satisfied:
	\begin{align*}
		\dot{\mathbf{B}}^S_{G_k^\ast, G_l^\ast} \mathbf{1}_{ |G_l^\ast| }
		&=
        \frac{1}{			
        \mathbf{1}_{ |G_l^\ast| }^\top
			\mathbf{r}^{(l,k, S)}
            }
			\bigl\{
            \mathbf{r}^{(k,l, S)}
			(\mathbf{r}^{(l,k, S)})^\top
            \bigr\}
		\mathbf{1}_{ |G_l^\ast| }
		\\&=
        \frac{1}{			
        \mathbf{1}_{ |G_l^\ast| }^\top
			\mathbf{r}^{(l,k, S)}
            }
			\mathbf{r}^{(k,l, S)}
                    \bigr\{
			(\mathbf{r}^{(l,k, S)})^\top
		\mathbf{1}_{ |G_l^\ast| }
        \bigr\}
		\\&=
		\mathbf{r}^{(k,l, S)}.
	\end{align*}
We summarize the construction of   $\dot{\boldsymbol{\alpha}}^S(\lambda^S)$ and $\dot{\mathbf{B}}^S (\lambda^S)$ in the following lemma.
\begin{lemma}[Parametrized dual certificate candidates]\label{lemma:dual_construction}
Fix the data matrix $\mathbf{X}$ and the true cluster structure $G_1^\ast, \ldots, G_K^\ast$. 
Let $S \subset [p]$ be a support set, and consider the corresponding primal SDP problem \eqref{optim:SDP_submatrix_S}. Fix a value of the dual variable $\lambda^S \in \mathbb{R}$.
Define $\dot{\boldsymbol{\alpha}}^S(\lambda^S) \in \mathbb{R}^n$ by setting, for each $k = 1, \ldots, K$,
        \begin{equation*}
			\dot{\boldsymbol{\alpha}}^S_{G_k^\ast}(\lambda^S)
			:= 
			2
			\mathbf{X}_{S, G_k^\ast}^\top
			\overline{\mathbf{X}}^\ast_{S,k}
			- 
			\frac{\lambda^S}{|G_k^\ast|}
            \mathbf{1}_{|G_k^\ast|}
			- 
			\|
			\overline{\mathbf{X}}^\ast_{S,k}
			\|_2^2
            \mathbf{1}_{|G_k^\ast|}
            ,
            \end{equation*}
Next, define $\dot{\mathbf{B}}^S(\lambda^S) \in \mathbb{R}^{n \times n}$ as the block matrix:
$$
\dot{\mathbf{B}}^S (\lambda^S) :=
\begin{bmatrix}
\mathbf{0} 
& 
\dot{\mathbf{B}}^S_{G_1^\ast, G_2^\ast} (\lambda^S)
& 
\cdots & \dot{\mathbf{B}}^S_{G_1^\ast, G_K^\ast}  (\lambda^S)
\\
\dot{\mathbf{B}}^S_{G_2^\ast, G_1^\ast}  (\lambda^S)
& \mathbf{0} & \cdots & 
\dot{\mathbf{B}}^S_{G_2^\ast, G_K^\ast} (\lambda^S)
\\
\vdots & \vdots & \ddots & \vdots \\
\dot{\mathbf{B}}^S_{G_K^\ast, G_1^\ast}(\lambda^S)
& \dot{\mathbf{B}}^S_{G_K^\ast, G_2^\ast} (\lambda^S)
& \cdots & \mathbf{0}
\end{bmatrix},
$$
where for each $1 \leq k \neq l \leq K$, the off-diagonal block $\dot{\mathbf{B}}^S_{G_k^\ast, G_l^\ast}  (\lambda^S) \in \mathbb{R}^{|G_k^\ast| \times |G_l^\ast|}$ is defined as
\begin{equation}\label{construction:B_off_diagonal}
\dot{\mathbf{B}}^S_{G_k^\ast, G_l^\ast} (\lambda^S) :=
\frac{1}{\mathbf{1}_{ |G_l^\ast| }^\top \mathbf{r}^{(l,k,S)}}
\, \mathbf{r}^{(k,l,S)} \, (\mathbf{r}^{(l,k,S)})^\top,
\end{equation}
with $\mathbf{r}^{(k,l,S)} \in \mathbb{R}^{|G_k^\ast|}$ defined componentwise by
\begingroup
\begin{equation*}
r_i^{(k,l,S)}
		:=-\frac{|G_k^\ast| + |G_l^\ast|}{2 |G_k^\ast|} \lambda^S
		+ \frac{ |G_l^\ast| }{2} 
		\bigl(
		\| \overline{\mathbf{X}}^\ast_{S,l}- \mathbf{X}_{S, i} \|_2^2 
		- 
		\| \overline{\mathbf{X}}^\ast_{S,k}- \mathbf{X}_{S, i} \|_2^2 
		\bigr).
\end{equation*}
\endgroup
Finally, we construct
\begin{equation*}
    \dot{\mathbf{W}}^S(\lambda^S)
			:=
			\lambda^S \mathbf{I}_n
             - \dot{\mathbf{B}}^S(\lambda^S)
			+
			\frac{1}{2} 
			\mathbf{1}_n 
            \bigl(
            \dot{\boldsymbol{\alpha}}^S(\lambda^S)
            \bigr)^\top
            + 
            \frac{1}{2} 
            \dot{\boldsymbol{\alpha}}^S
            (\lambda^S)
            \mathbf{1}_n^\top
			- \mathbf{X}_{S, \cdot}^\top \mathbf{X}_{S, \cdot}.
\end{equation*}
Constructed $\dot{\boldsymbol{\alpha}}^S(\lambda^S)$ and $\dot{\mathbf{B}}^S(\lambda^S)$ satisfy the conditions in Lemma \ref{lemma:opt_necessary}.
\end{lemma}
\subsubsection{Reformulation into high-probability bound problem}\label{section:lambda_bounds}
Next, we identify conditions of \( \lambda^S \), that ensures \( \dot{\mathbf{B}}^S (\lambda^S) \) and \( \dot{ \boldsymbol{\alpha} }^S (\lambda^S) \) satisfy the conditions in Lemma \ref{lemma:opt_cond}.  We begin with conditions (C1), (C3), (C4) and (C5). To this end, we introduce the following data-dependent quantities:
\begin{equation*}
    D^S_{kli}(\mathbf{X}) :=
		\| \overline{\mathbf{X}}^\ast_{S,l}- \mathbf{X}_{S, i} \|_2^2 
		- 
		\| \overline{\mathbf{X}}^\ast_{S,k}- \mathbf{X}_{S, i} \|_2^2,
\end{equation*}
and
\begin{equation}\label{U_s_X}
		U^S(\mathbf{X})
		:=
		\min_{1 \leq k \neq l \leq K}
		\biggl\{
		\biggl(
		\frac{1}{ |G_k^\ast| }
		+
		\frac{1}{ |G_l^\ast| }
		\biggr)^{-1}
		\min_{i \in G_k^\ast}
		D^S_{kli}(\mathbf{X}) 
		\biggr\}.
\end{equation}
Based on these definitions, we now state an upper bound  on $\lambda^S$:
\begin{lemma}[Upper bound on $\lambda^S$]\label{lemma:lambda_upper_bound}
Let $S \subset [p]$ be a support set, and consider the corresponding primal SDP problem \eqref{optim:SDP_submatrix_S}.
For any value of $\lambda^S$,  corresponding $\dot{ \boldsymbol{\alpha} }^S (\lambda^S)$ , $\dot{\mathbf{B}}^S(\lambda^S)$, and $\dot{\mathbf{W}}^S (\lambda^S)$  satisfy conditions (C3) and (C4) of Lemma~\ref{lemma:opt_cond}.
Moreover, if $$\lambda^S < U^S(\mathbf{X}),$$ then $\dot{ \boldsymbol{\alpha} }^S (\lambda^S)$ and $\dot{\mathbf{B}}^S(\lambda^S)$ also satisfy conditions (C1) and (C5) of Lemma \ref{lemma:opt_cond}.
\end{lemma}
Proof of Lemma \ref{lemma:lambda_upper_bound} is provided in Appendix \ref{section:proof:lemma:lambda_upper_bound}.

Next we move onto condition (C2) of Lemma \ref{lemma:opt_cond}. To this end, we define some notation.
We first define a linear subspace of $\mathbb{R}^n$ as:
$$
\Gamma_K := \operatorname{span} \left( \mathbf{1}_{G_k^*} : k \in [K] \right)^\perp,
$$
where $\perp$ means orthogonal complement. 
For each $\mathbf{v} \in \Gamma_K$ and support set $S \subset [p]$,
 we define a collection of data-dependent random variables.
We begin with  the following supremum  of noise quadratic form process:
\begin{equation}\label{def:L_1}
	L_1^S(\mathbf{X}) 
    :=  
 \sup_{\mathbf{v} \in \Gamma_K, \|\mathbf{v}\|_2 = 1}
  \hspace{-1.3em}
        \|		
		\mathcal{E}_{S, \cdot} \mathbf{v} 
        \|_2^2
    ,
\end{equation} 
Next, we define two fundamental components representing signal–noise and noise–noise interactions:
\begin{align*}
I^S_{kl}(\mathbf{v}, \mathbf{X}) &:= 
|G_l^\ast|
	\sum_{i \in G_k^\ast} v_i 
	\langle
	(\mathbf{c}_k^\ast
	-
	\mathbf{c}_l^\ast)_{S \cap S^\ast},
\;
	\mathcal{E}_{S \cap S^\ast,i}
	\rangle,~\text{and}
    \numberthis \label{def:IklsiS}
\\
N^S_{kl}(\mathbf{v}, \mathbf{X}) &:=
|G_l^\ast|
	\sum_{i \in G_k^*} v_i 
	\bigl \langle
	\overline{\mathcal{E}}_{S,k}^\ast -
	\overline{\mathcal{E}}_{S,l}^\ast,
    \;
	\mathcal{E}_{S,i}
	\bigr \rangle.
    \numberthis \label{def:NklsiS}
\end{align*}
Note that both definitions are asymmetric with respect to the first two indices.
Using these definitions, we introduce three composite random quantities:
\begingroup
\begin{align*}
	T^S_{1, k l}(\mathbf{v})
	~:=~&
	I^S_{kl}(\mathbf{v}, \mathbf{X})
  \cdot
    I^S_{lk}(\mathbf{v}, \mathbf{X}),
    \numberthis \label{def:T1}
	\\
	T^S_{2,kl}(\mathbf{v}) ~:=~& 
	N^S_{kl}(\mathbf{v}, \mathbf{X})
	\cdot
    N^S_{lk}(\mathbf{v}, \mathbf{X})
    ,~\text{and}
    \numberthis \label{def:T2}
	\\
	T^S_{3,kl}(\mathbf{v}) ~:= ~&
	N^S_{kl}(\mathbf{v}, \mathbf{X})
   \cdot
    I^S_{lk}(\mathbf{v}, \mathbf{X})
	+
	I^S_{kl}(\mathbf{v}, \mathbf{X})
    \cdot
    N^S_{lk}(\mathbf{v}, \mathbf{X}).
\end{align*}
\endgroup 
Next, the following quantity represents the sum of all elements of 
$\dot{\mathbf{B}}^S_{G_l^\ast, G_k^\ast}$, 
$(l,k)$th off-diagonal block of $\dot{\mathbf{B}}^S (\lambda^S)$: 
\begin{equation}\label{def:tkls}
	t_{lk}^S(\lambda^S):=
    \sum_{j \in G_l^\ast}
r_j^{(l,k,S)}
    =
	\sum_{j \in G_l^\ast}
	\biggl\{
	\frac{ |G_k^\ast| }{2} 
	D^S_{lkj}(\mathbf{X})
	-\frac{|G_k^\ast|+ |G_l^\ast|}{2|G_l^\ast|} \lambda^S
	\biggr\}.
\end{equation}

Leveraging all these notations, we finally define:
\begin{equation}\label{def:L2}
		L_2^S(\mathbf{X}, \lambda^S) 
        :=
		\sup_{\mathbf{v} \in \Gamma_K, \| \mathbf{v} \|_2  =1}
		\biggl\{
		\sum_{1 \leq k \neq l \leq K} 
		\frac{1}{t^S_{l k}(\lambda^S)} 
		\biggl(
		T^S_{1,kl}(\mathbf{v})
		+
		T^S_{2,kl}(\mathbf{v})
		+
		T^S_{3,kl}(\mathbf{v}) \biggr)
        \biggr\}
        ,
	\end{equation}    
Using this quantity, Lemma \ref{lemma:lambda_s_bound} states the lower bound condition on $\lambda^S$ that guarantees condition (C2) of Lemma \ref{lemma:opt_cond} is satisfied.
\begin{lemma}[Lower bound on $\lambda^S$]
	\label{lemma:lambda_s_bound}
Let $S \subset [p]$ be a support set, and consider the corresponding primal SDP problem \eqref{optim:SDP_submatrix_S}.
If 
\begin{equation*}
    \lambda^S 
    > 
    L_1^S(\mathbf{X})+
    L_2^S(\mathbf{X}, \lambda^S),
\end{equation*}
then the constructed dual variables $\dot{\boldsymbol{\alpha}}^S(\lambda^S)$ and $\dot{\mathbf{B}}^S(\lambda^S)$ satisfy condition (C2) of Lemma \ref{lemma:opt_cond}.
\end{lemma}
Proof of Lemma \ref{lemma:lambda_s_bound} is provided in Appendix \ref{section:proof:lemma:lambda_s_bound}.
Combining Lemmas \ref{lemma:lambda_upper_bound} and \ref{lemma:lambda_s_bound}, we finally arrive at the following corollary:
\begin{corollary}[Problem reformulation]\label{corollary:problem_reformulation}
Let $S \subset [p]$ be a support set, and consider the corresponding primal SDP problem \eqref{optim:SDP_submatrix_S}. 
If 
there exists $\lambda^S$ such that 
\begin{equation*}
  L_1^S(\mathbf{X})+
    L_2^S(\mathbf{X}, \lambda^S) < \lambda^S < U^S(\mathbf{X}),
\end{equation*}
then   the dual variables constructed with this $\lambda^S$ via Lemma \ref{lemma:dual_construction} yield a valid certificate
that satisfies all conditions of Lemma \ref{lemma:opt_cond}, verifying that $\mathbf{Z}^\ast$ is the unique optimal point of the primal problem.
\end{corollary}

\subsection{First Bounds and Construction of the Dual Variable}\label{section:proof:set_lambda}
This section is devoted to the derivation of Lemma \ref{lemma:lambda_between}, which establishes that, under the separation condition, there exists a $\lambda^S$ that partially satisfies the condition stated in Corollary \ref{corollary:problem_reformulation}.
We proceed in
two main steps:
\begin{enumerate}
\item \textbf{High-probability bounds (Appendix \ref{section:proof:high_prob_bounds_1})}
:
We derive 
a lower bound for $U^S(\mathbf{X})$ (Lemma \ref{lemma:bound_of_upper_bound})
and 
an upper bound for $L_1^S(\mathbf{X})$ (Lemma \ref{lemma:L1}), under the separation and set size condition of $\mathcal S$.
\item \textbf{Specifying the value of the dual variable (Appendix \ref{section:proof:set_lambda_value})}: We set the value of $\lambda^S$ to lie between the bounds established in the previous step (Lemma \ref{lemma:lambda_between}).
\end{enumerate}

\subsubsection{High-probability Bounds}\label{section:proof:high_prob_bounds_1}
Recall from 
\eqref{U_s_X}
that $U^S(\mathbf{X})$ is defined as
\begin{equation*}
		U^S(\mathbf{X})
		=
		\min_{1 \leq k \neq l \leq K}
		\biggl\{
		\biggl(
		\frac{1}{ |G_k^\ast| }
		+
		\frac{1}{ |G_l^\ast| }
		\biggr)^{-1}
		\min_{i \in G_k^\ast}
		D^S_{kli}(\mathbf{X}) 
		\biggr\},
\end{equation*}
where
\begin{equation*}
    D^S_{kli}(\mathbf{X}) =
		\| \overline{\mathbf{X}}^\ast_{S,l}- \mathbf{X}_{S, i} \|_2^2 
		- 
		\| \overline{\mathbf{X}}^\ast_{S,k}- \mathbf{X}_{S, i} \|_2^2.
\end{equation*}
Recall from \eqref{def:S-set} that we define $\mathcal S $ as
\begin{equation*}
    \mathcal S := \left\{S: S\subset [p], 
    S \cap S^\ast \neq \emptyset,
    |S|\leq \sqrt p, 	\Delta_{S \cap S^\ast}^2 \gtrsim \sigma^2
		\left(\log n +
		\frac{|S| \log p}{m}
		+
		\sqrt{\frac{|S| \log p}{m}}
		\right)\right\},
\end{equation*}
where
\begin{equation*}
\Delta_{S \cap S^\ast}^2 = 
\min_{1 \leq k \neq l \leq K} \| (\mathbf{c}_l^\ast - \mathbf{c}_k^\ast )_{S \cap S^\ast} \|_2^2,
\end{equation*}
and
\begin{equation*}
		m = \min_{1 \leq k \neq l \leq K}
		\biggl(
		\frac{2 |G_l^\ast|  |G_k^\ast| }{|G_l^\ast|+ |G_k^\ast|}
		\biggr)
		\geq C_1
		\frac{n}{\log n}.
	\end{equation*}
The following lemma  establishes a high-probability lower bound for $D^S_{kli}(\mathbf{X})$.
\begin{lemma}\label{lemma:bound_of_upper_bound} 
	Uniformly across 
	all
	index  $(k, l, i)$  ranging over all  pairs  $1 \leq k \neq l \leq K$ and  $i \in G^*_k$, and
    all $S \in \mathcal S$,
    with probability at least $1-  	C_6/n$, it holds that
     \begin{equation*}
        D^S_{kli}(\mathbf{X})
		\gtrsim 
         \| (\mathbf{c}_k^\ast - \mathbf{c}_l^\ast)_{S \cap S^\ast} \|_2^2
        +
        \biggl( \frac{1}{|G_k^\ast|} + \frac{1}{|G_l^\ast|} \biggr)
        |S|  
		 - R^S_{kl}.
    \end{equation*}
     where  $C_6>0$ is a constant and
	 \begin{equation*}
         R^S_{kl} :=  \sqrt{ \frac{ {\zeta^S}}{|G_l^\ast|} } \| (\mathbf{c}_l^\ast - \mathbf{c}_k^\ast )_{S \cap S^\ast}\|_2 
          +
          \biggl( \frac{1}{|G_k^\ast|} + \frac{1}{|G_l^\ast|} \biggr)\sqrt{|S| {\zeta^S}}
          +
         \frac{1}{|G_k^\ast|}   \zeta^S,
    \end{equation*} 
    and $ \zeta^S := \log(n^2 |S|^2  {\binom{p}{|S|}})$.
 \end{lemma}
The proof of Lemma \ref{lemma:bound_of_upper_bound} is presented in Appendix \ref{section:chisq_tail_application}.

Next, Lemma \ref{lemma:L1} establishes a high-probability upper bound for \( L_1^S(\mathbf{X})\), where we recall from \eqref{def:L_1}:
\begin{align*}
	L_1^S(\mathbf{X}) 
    =  
 \sup_{\mathbf{v} \in \Gamma_K, \|\mathbf{v}\|_2 = 1}
  \hspace{-1.3em}
        \|		
		\mathcal{E}_{S, \cdot} \mathbf{v} 
        \|_2^2
    ,
\end{align*} 
and $\mathcal{E}_{S, \cdot}$ is a submatrix of $\mathcal{E} \in \mathbb{R}^{p \times n}$, whose entries are i.i.d.  Gaussian random variables.

\begin{lemma}\label{lemma:L1}
Uniformly across all $S \subset [p]$ such that $|S| \leq \sqrt{p}$,
with  probability at least $1 - C_7/n$, where $C_7>0$ is a constant, it holds that
	\begin{equation*}
    L_1^S(\mathbf{X})  
		\lesssim
		\sigma^2(n + \log n + |S| \log p).
\end{equation*}
	
\end{lemma}
The proof of Lemma \ref{lemma:L1} is provided in Appendix \ref{section:proof:lemma:L1}.

\subsubsection{Specifying the Value of the Dual Variable }\label{section:proof:set_lambda_value}
Since $L_2^S(\mathbf{X}, \lambda^S)$ in \eqref{def:L2} depends on $\lambda^S$, we first specify a value which lies between $L_1^S(\mathbf{X})$ and  $U^S(\mathbf{X})$ with high probability,
uniformly over $S \in \mathcal{S}$.
\begin{lemma}\label{lemma:lambda_between}
Let $\dot{\lambda}^S := \sigma^2 |S| + m \Delta_{S \cap S^\ast }^2/4$.
   Then  uniformly over $S \in \mathcal{S}$,  with probability at least $1 - C_{9}/n$, where $C_{9}>0$ is a constant, it holds that
  \begin{equation*}
  L_1^S(\mathbf{X}) 
\lesssim
\dot{\lambda}^S 
\lesssim
U^S(\mathbf{X}).
  \end{equation*}
\end{lemma}
Proof of Lemma \ref{lemma:lambda_between} is provided in Appendix \ref{section:proof:lemma:lambda_between}.  

\subsection{Second high-probability bounds and conclusion}\label{section:proof:theorem:separation_condition:conclusion}
This section completes the proof by showing that, with high probability,
$
    L_2^S(\mathbf{X}, \dot{\lambda}^S) \lesssim \dot{\lambda}^S
$.
As defined in equation \eqref{def:L2}, $L_2^S(\mathbf{X}, \dot{\lambda}^S)$ is the supremum of an empirical process involving statistics
$t^S_{lk}$ \eqref{def:tkls}, 
$I^S_{kl}(\mathbf{v} ,\mathbf{X})$ 
$I^S_{lk}(\mathbf{v}, \mathbf{X})$ \eqref{def:IklsiS},  
$N^S_{kl}(\mathbf{v}, 
\mathbf{X})$, 
and
$N^S_{lk}(\mathbf{v}, \mathbf{X})$ \eqref{def:NklsiS}.
Therefore, the bound on $L_2^S(\mathbf{X}, \dot{\lambda}^S)$ follows directly from  bounds on the $t^S_{lk}$,  $I^S_{kl}(\mathbf{v}, \mathbf{X})$ and $N^S_{kl}(\mathbf{v}, \mathbf{X})$. Accordingly, the remainder of this section proceeds as follows:
\begin{enumerate}
\item \textbf{Bounding the off-diagonal block sum of the dual variable matrix (Appendix \ref{section:proof:off_diagonal_block_sum})}: We derive a high-probability lower bound for $t^S_{lk}$ (Lemma \ref{lemma:boun(D^t)^Csl}).
\item \textbf{Bounding the signal–noise interaction term (Appendix \ref{section:proof:IkliS}):}
We derive a high-probability upper bound for $I^S_{kl}(\mathbf{v}, \mathbf{X})$
(Lemma \ref{lemma:T1Sv})
and  show that the corresponding component of $L_2^S(\mathbf{X}, \dot{\lambda}^S)$ is dominated by $\dot{\lambda}^S$ (Lemma \ref{lemma:T1_dominance}).
\item \textbf{Bounding the noise–noise interaction term (Appendix \ref{section:proof:NkliS}):}
We derive a high-probability upper bound for $N^S_{kl}(\mathbf{v}, \mathbf{X})$
(Lemmas \ref{lemma:upperbound_G1ksv} and \ref{lemma:ub_g2klsv}),
and show that 
the corresponding component of $L_2^S(\mathbf{X}, \dot{\lambda}^S)$ is dominated by $\dot{\lambda}^S$ (Lemma \ref{lemma:T2_dominance}).
\item \textbf{Conclusion (Appendix \ref{theorem_final_conclusion})}.
\end{enumerate}

\subsubsection{Bounding the off-diagonal block sum of the dual variable matrix}\label{section:proof:off_diagonal_block_sum}
With the value of $\dot{\lambda}^S$
set in Lemma \ref{lemma:lambda_between}, we provide a high-probability lower bound of $t^S_{lk}(\dot{\lambda}^S)$ defined in \eqref{def:tkls}.
\begin{lemma}\label{lemma:boun(D^t)^Csl}
For any $1 \leq l \neq k \leq K$,	if  $|G_l^\ast| \geq 2 $ and $|G_k^\ast| \geq 2$,
    uniformly over $S \in \mathcal{S}$, with probability at least $1 - C_6/n$, where ${C_6}$ is the constant used in Lemma \ref{lemma:bound_of_upper_bound}, it holds that
	\begin{equation*}
    t^S_{lk}(\dot{\lambda}^S)
    \gtrsim
        |G_l^\ast| \; |G_k^\ast|
        \bigl(
        \|
        (\mathbf{c}^\ast_k -\mathbf{c}^\ast_l)_{S \cap S^\ast}
        \|_2^2 + |S|
        \bigr)
        .
	\end{equation*}
\end{lemma}
\noindent
The proof of Lemma \ref{lemma:boun(D^t)^Csl} is provided in Appendix \ref{section:proof:lemma:boun(D^t)^Csl}. 
\subsubsection{Bounding the signal-noise interaction term}\label{section:proof:IkliS}
We first provide a high-probability upper bound of
\begin{equation*}
    I^S_{kl}(\mathbf{v}, \mathbf{X}) = 
|G_l^\ast|
	\sum_{i \in G_k^\ast} v_i 
	\langle
	(\mathbf{c}_k^\ast
	-
	\mathbf{c}_l^\ast)_{S \cap S^\ast},
\;
	\mathcal{E}_{S \cap S^\ast,i}
	\rangle, 
\end{equation*}
 defined in \eqref{def:IklsiS}.
 \begin{lemma}\label{lemma:T1Sv}
Uniformly across $S \in \mathcal S$ and $(k,l)$ such that $1 \leq k \neq l \leq K$, 
	with probability at least $1 - C_9/n$, where $C_9>0$ is a constant, it holds that
\begin{align*}
 &\sup_{\mathbf{v} \in \Gamma_K, \|\mathbf{v}\|_2=1}
I^S_{kl}(\mathbf{v}, \mathbf{X})   
		\\& \quad \quad \lesssim
        |G_l^\ast|
        \|
		(\mathbf{c}_k^\ast - \mathbf{c}_l^\ast)_{S \cap S^\ast}
		\|_2
                \biggl(
                |G_k^\ast|
                +
                \sqrt{
                |G_k^\ast| \log n
                }
                +
                \sqrt{
                |G_k^\ast| |S| \log p
                }+
                \log n + |S| \log p
        \biggr)^{1/2}.
\end{align*}
\end{lemma}
The proof of Lemma \ref{lemma:T1Sv} is provided in Appendix \ref{section:proof:lemma:T1Sv}. Building on this result, we establish the first partial upper bound for $L_2^S(\mathbf{X}, \dot{\lambda}^S)$, corresponding to the term
$T^S_{1,kl}(\mathbf{v}) = I^S_{kl}(\mathbf{v}, \mathbf{X}) \cdot I^S_{lk}(\mathbf{v}, \mathbf{X})$, defined in \eqref{def:T1}.
\begin{lemma}\label{lemma:T1_dominance}
Uniformly across $S \in \mathcal S$, 
	with probability at least $1 - C_{11}/n$, where $C_{11}>0$ is a constant, it holds that
    \begin{equation*}
  \sup_{\mathbf{v} \in \Gamma_K, \| \mathbf{v} \|_2  =1}
\sum_{1 \leq k \neq l \leq K} 
		\frac{T^S_{1,kl}(\mathbf{v})}{t^S_{lk}(\dot{\lambda}^S)}       
        \lesssim \dot{\lambda}^S  .
\end{equation*}
\end{lemma}
Proof of Lemma \ref{lemma:T1_dominance} is provided in Appendix \ref{section:proof:lemma:T1_dominance}.
\subsubsection{Bounding the noise-noise interaction term}\label{section:proof:NkliS}
Since $N^S_{kl}(\mathbf{v}, \mathbf{X})$, defined in  \eqref{def:NklsiS}, is an inner product of dependent Gaussian variables, we decompose it into a sum of inner products between independent Gaussian variables.
\begin{align*}
   N^S_{kl}(\mathbf{v}, \mathbf{X}) &=
|G_l^\ast|
	\sum_{i \in G_k^*} v_i 
	\bigl \langle
	\overline{\mathcal{E}}_{S,k}^\ast -
	\overline{\mathcal{E}}_{S,l}^\ast,
    \;
	\mathcal{E}_{S,i}
	\bigr \rangle
\\&= 
|G_l^\ast|
	\sum_{i \in G_k^*} v_i 
	\bigl \langle
	\overline{\mathcal{E}}_{S,k}^\ast ,
    \;
	\mathcal{E}_{S,i}
	\bigr \rangle
-
|G_l^\ast|
	\sum_{i \in G_k^*} v_i 
	\bigl \langle
	\overline{\mathcal{E}}_{S,l}^\ast,
    \;
	\mathcal{E}_{S,i}
	\bigr \rangle
\\&= 
        \underbrace{
\frac{ |G_l^\ast| }{|G_k^\ast|} 
		\sum_{
       \{(i, j) \in G_k^\ast\}
        }
        \hspace{-1.5em}
        v_i 
		\left\langle
		\mathcal{E}_{S,i},
		\mathcal{E}_{S,j}
		\right\rangle 
               }_{:=G^S_{1, kl}(\mathbf{v})}
                               \underbrace{
-
|G_l^\ast|
	\sum_{i \in G_k^*} v_i 
	\bigl \langle
	\overline{\mathcal{E}}_{S,l}^\ast,
    \;
	\mathcal{E}_{S,i}
	\bigr \rangle,
            }_{:=G^S_{2, kl}(\mathbf{v})}
	\end{align*}

We now present  high-probability upper bounds for \( G^S_{1, kl}(\mathbf{v}) \) and \( G^S_{2, kl}(\mathbf{v}) \) in order. Since all terms consist purely of noise random variables, no separation condition is needed in this case.
	\begin{lemma}\label{lemma:upperbound_G1ksv} Uniformly across $S \in [p]$ such that $|S| \leq \sqrt{p}$ and $(k,l)$ such that $1 \leq k \neq l \leq K$,
with probability at least $1 - C_{11}/n$, 
where $C_{11}>0$ is a constant, 
it holds that
\begin{equation*}
\sup_{ 
\mathbf{v} \in \Gamma_K, \|\mathbf{v} \|_2 = 1
} \hspace{-1em}
                  G^S_{1, kl}(\mathbf{v})
		\lesssim
|G_l^\ast|
		\sqrt{
		|S| (  \log n + |S| \log p
        )
	}.
	\end{equation*}
	\end{lemma}
	The proof of Lemma \ref{lemma:upperbound_G1ksv} is provided in Appendix \ref{section:proof:lemma:upperbound_G1ksv}.

	\begin{lemma}\label{lemma:ub_g2klsv}
  Uniformly across $S \in [p]$ such that $|S| \leq \sqrt{p}$ and $(k,l)$ such that $1 \leq k \neq l \leq K$,
with probability at least $1 - C_{12}/n$, where $C_{12}>0$ is a constant, it holds that
	\begin{equation*}
		\sup_{\mathbf{v} \in \Gamma_K, \| \mathbf{v} \|_2 = 1}
		|G_{2, kl}^S(\mathbf{v})|
		\lesssim
       \sqrt{|G_l^\ast|}
      \sqrt{ |S| + 2\sqrt{
		|S| 
		 \xi^S 
	} + 2 \xi^S }
     \sqrt{|G_k^\ast|  +\sqrt{\xi^S }
		},	
	\end{equation*}
    where $ \xi^S = \log(n  |S|^2  {\binom{p}{|S|}})$.
	\end{lemma}
	The proof of Lemma \ref{lemma:ub_g2klsv} is provided in Appendix \ref{section:proof:lemma:ub_g2klsv}.
  \begin{lemma}\label{lemma:T2_dominance}
 Uniformly across $S \in \mathcal S$ and $(k,l)$ such that $1 \leq k \neq l \leq K$,
with probability at least $1 - C_{13}/n$, where $C_{13}>0$ is a constant, it holds that
    \begin{equation*}
  \sup_{\mathbf{v} \in \Gamma_K, \| \mathbf{v} \|_2  =1}
\sum_{1 \leq k \neq l \leq K} 
		\frac{T^S_{2,kl}(\mathbf{v})}{t^S_{lk}(\dot{\lambda}^S)}       
        \lesssim \dot{\lambda}^S  .
\end{equation*}
\end{lemma}
Proof of Lemma \ref{lemma:T2_dominance} is provided in Appendix \ref{section:proof:lemma:dominance_T2}.

\subsubsection{Conclusion}\label{theorem_final_conclusion}
The inequality regarding $T^S_{3,kl}(\mathbf{v})$ is a direct consequence of Lemmas \ref{lemma:T1_dominance}, and \ref{lemma:T2_dominance}, by the Cauchy-Schwarz inequality.
Therefore, by combining Lemmas \ref{lemma:lambda_between}, \ref{lemma:T1_dominance}, and \ref{lemma:T2_dominance}, we conclude that, uniformly over $S \in \mathcal{S}$, with probability at least $1 - C_{14}/n$, where $C_{14} > 0$ is a constant, the quantity $\dot{\lambda}^S$ constructed in Lemma \ref{lemma:lambda_between} satisfies
$$
L_1^S(\mathbf{X}) + L_2^S(\mathbf{X}, \dot{\lambda}^S) \lesssim \dot{\lambda}^S \lesssim U^S(\mathbf{X}).
$$
Therefore, by Corollary \ref{corollary:problem_reformulation}, the dual variables constructed in Lemma \ref{lemma:dual_construction} serve as dual certificates verifying that $\mathbf{Z}^\ast$ is the unique optimal solution to the primal SDP problem corresponding to $S$, stated in \eqref{optim:SDP_submatrix_S}.
This completes the proof of Theorem \ref{theorem:separation_condition}.

\subsection{Proof of Supporting Lemmas for the Minimax Upper Bound Proof}
This section presents omitted proof of lemmas presented in the proof of Theorem \ref{theorem:separation_condition}. 
\subsubsection{Proof of Lemma \ref{lemma:opt_cond}}\label{section:proof:lemma:opt_cond}
The proof proceeds in three steps:
\begin{enumerate}
    \item Conditions (C1) and (C2) from dual feasibility,
    \item Conditions (C3) and (C4) from zero dual gap,
    \item Condition (C5) from the  block-diagonal structure of $\mathbf{Z}^\ast = \sum_{k=1}^K |G_k^\ast|^{-1} \mathbf{1}_{G_k^\ast} \mathbf{1}_{G_k^\ast}^\top$.
\end{enumerate}
\begin{proof}  We proceed by following the steps outlined above in sequence.

\medskip
\noindent
\textbf{1. Conditions (C1) and (C2) from dual feasibility.}
After simplification, the Lagrangian function becomes an affine function of $\mathbf{Z}$:
	\begin{align*}
		&\mathcal{L}(\mathbf{Z}, \mathbf{Q}^S, \lambda^S,  \boldsymbol{\alpha}^S, \mathbf{B}^S) 
		\\&\quad   := 
		\langle
            \mathbf{X}_{S, \cdot}^\top \mathbf{X}_{S, \cdot},
            \;
            \mathbf{Z}
        \rangle 
		+ 
		\langle \mathbf{Q}^S,\mathbf{Z} \rangle
		+ 
		\lambda^S(K - \operatorname{tr}(\mathbf{Z})) 
		+ (\boldsymbol{\alpha}^S)^\top 
        \biggl( \mathbf{1}_n 
        - \frac{1}{2}
        ( \mathbf{Z} + \mathbf{Z}^\top)
        \mathbf{1}_n 
        \biggr) 
        + 
        \langle \mathbf{B}^S, \mathbf{Z} \rangle
		\\&\quad   = 
		\lambda^S K
        +
        (\boldsymbol{\alpha}^S)^\top \mathbf{1}_n
		+ 
		\bigl\langle  
        \mathbf{Q}^S - \mathbf{W}^S
		,\;
		\mathbf{Z}
		\bigr\rangle,
	\end{align*}
    where 
    we use  $\mathrm{tr}(\mathbf{Z}) = \langle \mathbf{I}_n, \mathbf{Z} \rangle$ and
     recall from Lemma \ref{lemma:opt_cond} that
    \begin{equation*}
\mathbf{W}^S
			=
			\lambda^S \mathbf{I}_n
             - \mathbf{B}^S
			+
			\frac{1}{2} 
			\mathbf{1}_n (\boldsymbol{\alpha}^S)^\top
            + 
            \frac{1}{2} 
            \boldsymbol{\alpha}^S \mathbf{1}_n^\top
			- \mathbf{X}_{S, \cdot}^\top \mathbf{X}_{S, \cdot}.
    \end{equation*}
Therefore, the Lagrange dual function, shown below, can be derived analytically, as a linear function is bounded  only when the slope part is  zero.
	\begin{align*}
		g( \mathbf{Q}^S, \lambda^S,  \boldsymbol{\alpha}^S, \mathbf{B}^S)
		&  :=
		\sup_{\mathbf{Z} \in \mathbb{R}^{n \times n}}
        \mathcal{L}(\mathbf{Z}, \mathbf{Q}^S, \lambda^S,  \boldsymbol{\alpha}^S, \mathbf{B}^S) 
		\\&  =
		\lambda^S K + (\boldsymbol{\alpha}^S)^\top \mathbf{1}_n
		+ 
		\sup_{\mathbf{Z} \in \mathbb{R}^{n \times n}}
		\bigl\langle 
		\mathbf{Q}^S - \mathbf{W}^S, \;
		\mathbf{Z}
		\bigr\rangle,	
	\end{align*}
As a result, we have:
	\begin{align*}
		g( \mathbf{Q}^S, \lambda^S,  \boldsymbol{\alpha}^S, \mathbf{B}^S)
		=
		\begin{cases}
			\lambda^S K + (\boldsymbol{\alpha}^S)^\top \mathbf{1}_n,
			\quad
			&\mathbf{Q}^S 
			=
			\mathbf{W}^S~(\text{slope is }\mathbf{0})
			\\
			\infty, \quad &\text{otherwise}.
		\end{cases}
	\end{align*}
	Making the following domain of the dual function
	as an explicit linear equality constraint, and combining it  with 
	$\mathbf{Q}^S \succeq 0 $ and $\mathbf{B}^S \geq 0$, with some abuse of terminology,
	we refer to the following problem as the Lagrange dual problem associated with the primal SDP problem 
	\begin{align*}
		&\min_{ \lambda^S \in \mathbb{R},  \boldsymbol{\alpha}^S \in \mathbb{R}^{n}, \mathbf{B}^S \in \mathbb{R}^{n \times n}} \hskip -2em
		\lambda^S K + (\boldsymbol{\alpha}^S)^\top \mathbf{1}_n,
\quad \text{s.t.} \quad 
		\mathbf{B}^S \geq 0,
\mathbf{W}^S \succeq 0.
	\end{align*}
	The constraints in this optimization problem correspond to conditions (C1) and (C2).

\medskip
\noindent
    \textbf{2. Conditions (C3) and (C4) from zero dual gap.}
	The duality gap is simply the difference between the dual and primal objective function:
	\begin{align*}
		&\lambda^S K + (\boldsymbol{\alpha}^S)^\top \mathbf{1}_n - \langle 
		\mathbf{X}_{S, \cdot}^\top \mathbf{X}_{S, \cdot} ,\; \mathbf{Z}  \rangle
		\\& \quad \quad
		\stackrel{(i)}{=}
		\lambda^S \mathrm{tr}(\mathbf{Z}) + \frac{1}{2}(\boldsymbol{\alpha}^S)^\top 
		(\mathbf{Z} + \mathbf{Z}^\top) \boldsymbol{1}_n - \langle 
		\mathbf{X}_{S, \cdot}^\top \mathbf{X}_{S, \cdot} ,\; \mathbf{Z}  \rangle
		\\& \quad \quad \stackrel{(ii)}{=}
		\bigl \langle 
		\lambda^S \mathbf{I}_n 
        + 
        \frac{1}{2} 
        \mathbf{1}_n (\boldsymbol{\alpha}^S)^\top
        + 
        \frac{1}{2} 
        \boldsymbol{\alpha}^S \mathbf{1}_n^\top
        - \mathbf{X}_{S, \cdot}^\top \mathbf{X}_{S, \cdot} - \mathbf{B}^S 
		, \; \mathbf{Z} \bigr \rangle 
		+ \langle \mathbf{B}^S, \mathbf{Z} \rangle
        		\\& \quad \quad =
                \langle \mathbf{W}^S, \mathbf{Z} \rangle + \langle \mathbf{B}^S, \mathbf{Z} \rangle
	\end{align*}
	where step $(i)$ uses primal constraints $\mathrm{tr}(\mathbf{Z}) = K$, $\mathbf{Z}^\top = \mathbf{Z}$ and $\mathbf{Z} \boldsymbol{1}_n = \boldsymbol{1}_n$,
	step $(ii)$ uses $\mathrm{tr}(\mathbf{Z}) = \langle \mathbf{I}_n, \mathbf{Z} \rangle$, and added and subtracted $\langle \mathbf{B}^S, \mathbf{Z} \rangle$. This duality gap is zero 
    if $ \langle \mathbf{W}^S, \mathbf{Z} \rangle = 0$ (C3) and $\langle \mathbf{B}^S, \mathbf{Z} \rangle = 0$ (C4).

    \medskip \noindent \textbf{3. Condition (C5) from the  block-diagonal structure.}
Finally, we derive condition (C5), which ensures that $\mathbf{Z}^\ast$ is the unique solution to the primal SDP. To this end, we first show that it is the only primal feasible point with the following block-diagonal structure:
	where \( \mathbf{Z}_{G_k^\ast, G_l^\ast} = 0 \) for all \( 1 \leq  k \neq l \leq  K \):
	\[
	\mathbf{Z} =
	\begin{bmatrix}
		\mathbf{Z}^{(1)} & \mathbf{0} & \cdots & \mathbf{0} \\
		\mathbf{0} & \mathbf{Z}^{(2)} & \cdots & \mathbf{0} \\
		\vdots & \vdots & \ddots & \vdots \\
		\mathbf{0} & \mathbf{0} & \cdots & \mathbf{Z}^{(K)}
	\end{bmatrix}.
	\]

	We use proof by contradiction.
    Suppose there exists a primal feasible \( \mathbf{Z}  \neq \mathbf{Z}^\ast\) with the block diagonal structure described above. By the primal constraint \( \mathbf{Z} \mathbf{1}_n = \mathbf{1}_n \), each block \( \mathbf{Z}^{(k)} \) satisfies \( \mathbf{Z}^{(k)} \mathbf{1}_{|G_k^\ast|} = \mathbf{1}_{|G_k^\ast|} \), which implies that \( (1, n_k^{-1/2} \mathbf{1}_{|G_k^\ast|}) \) is an eigenvalue-eigenvector pair of \( \mathbf{Z}^{(k)} \). Since the trace of a matrix equals the sum of its eigenvalues, this implies \( \mathrm{tr}(\mathbf{Z}^{(k)}) \geq 1 \) for every \( k \).  
	Combining this with the overall trace constraint \( \mathrm{tr}(\mathbf{Z}) = \sum_{k=1}^K \mathrm{tr}(\mathbf{Z}^{(k)}) = K \), we deduce that \( \mathrm{tr}(\mathbf{Z}^{(k)}) = 1 \) for every \( k \).  
	Additionally, 
    each \( \mathbf{Z}^{(k)} \) is positive semidefinite,
    due to the primal constraint \( \mathbf{Z} \succeq 0 \), along with the Schur complement condition for the positive semidefiniteness of block matrices.
    Then \( \mathbf{Z}^{(k)} \) must take the form \( n_k^{-1} \mathbf{1}_{|G_k^\ast|} \mathbf{1}_{|G_k^\ast|}^\top \),
    since \( 1 \) is the only nonzero eigenvalue of \( \mathbf{Z}^{(k)} \), corresponding to the eigenvector \( n_k^{-1/2} \mathbf{1}_{|G_k^\ast|} \). Consequently, \( \mathbf{Z} = \mathbf{Z}^\ast \), a contradiction. 
	
	Given this block diagonal structure and the condition 
    \begin{equation*}
        \mathrm{tr}(\mathbf{B}^S \mathbf{Z}^\ast) = 0,
    \end{equation*}
    we conclude that \( \mathbf{Z}^\ast \) is the unique solution to the primal SDP problem 
    if (C1)-(C4) is satisfied and 
\begin{equation*}\mathbf{B}^S_{G_k^\ast,G_l^\ast}>0\text{ for all }1\leq k \neq l \leq K.\end{equation*}This corresponds to condition (C5) and completes the proof of Lemma~\ref{lemma:opt_cond}.
\end{proof}

\subsubsection{Proof of Lemma \ref{lemma:opt_necessary}}\label{section:proof:lemma:opt_necessary}
The proof proceeds in three steps:
\begin{enumerate}
    \item Preliminary computations,
    \item  Necessary condition for \( \boldsymbol{\alpha}^S \),
    \item Necessary condition for \( \mathbf{B}^S \).
\end{enumerate}

\begin{proof}
Throughout the proof, we assume that conditions (C1)–(C5) from Lemma \ref{lemma:opt_cond} are satisfied.
Recall that true cluster membership matrix is defined as $\mathbf{Z}^\ast =  \sum_{k=1}^K {|G_k^\ast|}^{-1} \mathbf{1}_{G_k^\ast} \mathbf{1}_{G_k^\ast}^\top$,
where $\mathbf{1}_{G_k^\ast}$, as defined in Section \ref{section:introduction}, is the indicator vector for membership in $G_k^\ast$.  

\medskip \noindent \textbf{Preliminary computations.}
We first summarize the results for reference. The derivations will follow.
\begin{align*}
    \bigl(
    (\mathbf{X}_{S, \cdot}^\top \mathbf{X}_{S, \cdot}) \mathbf{1}_{G_l^\ast}
    \bigr)_{G_k^\ast} 
    &=
    |G_l^\ast|
		\mathbf{X}_{S, G_k^\ast}^\top
		\overline{\mathbf{X}}^\ast_{S, l},~\text{for }(k,l) \in [K]^2~(\text{no condition required}),
        		\numberthis \label{empirical_cluster_mean_calculation}
   \\
   \mathbf{1}_{G_k^\ast}^\top
		\mathbf{W}^S 
		\mathbf{1}_{G_k^\ast} &=0~\text{for }k \in [K]~(\text{by (C2) and (C3)}), 
                \numberthis \label{Ws_quadform_equal_zero}
    \\
    		\mathbf{W}^S 
		\mathbf{1}_{G_k^\ast}  &=\mathbf{0}~\text{for }k \in [K]~(\text{by (C2) and (C3)}),
        \numberthis \label{w_one_equals_zero}
    \\
    \mathbf{B}^S_{G_k^\ast, G_k^\ast} &= \mathbf{0}~\text{for }k \in [K]~(\text{by (C4)})
   \numberthis     \label{B_diagonal_block_zero}
    \\
    ( \mathbf{B}^S \mathbf{1}_{G_k^\ast} )_{G_k^\ast} &=0~\text{for }k \in [K]~(\text{by (C4)}), \text{and}
   \numberthis \label{B_one_equals_zero}
   \\
   ( \mathbf{B}^S \mathbf{1}_{G_l^\ast} )_{G_k^\ast} &=
         \mathbf{B}^S_{G_k^\ast, G_l^\ast} \mathbf{1}_{|G_l^\ast|}~\text{for }1\leq k \neq l \leq K~(\text{no condition required}).
         \numberthis \label{B_one_off_diagonal}
\end{align*}
We start with \eqref{empirical_cluster_mean_calculation}.
  For $1 \leq k  \leq K $ and $1 \leq l \leq K$,
  using only the definition of $\mathbf{1}_{G_k^\ast}$ and $\mathbf{1}_{G_l^\ast}$ given in Section \ref{section:introduction}, 
  we have
	\begin{align*}
    \bigl(
    (\mathbf{X}_{S, \cdot}^\top \mathbf{X}_{S, \cdot}) \mathbf{1}_{G_l^\ast}
    \bigr)_{G_k^\ast}
    &=
    (\mathbf{X}_{S, \cdot}^\top \mathbf{X}_{S, \cdot})_{G_k^\ast, G_k^\ast}
    (\mathbf{1}_{G_l^\ast})_{G_k^\ast}
    +
    (\mathbf{X}_{S, \cdot}^\top \mathbf{X}_{S, \cdot})_{G_k^\ast, (G_k^\ast)^C}
    (\mathbf{1}_{G_l^\ast})_{(G_k^\ast)^C}
    \\&\overset{(i)}{=}
    (\mathbf{X}_{S, \cdot}^\top \mathbf{X}_{S, \cdot})_{G_k^\ast, (G_k^\ast)^C}
    (\mathbf{1}_{G_l^\ast})_{(G_k^\ast)^C}
    \\&=
		(\mathbf{X}_{S, \cdot}^\top \mathbf{X}_{S, \cdot})_{G_k^\ast, G_l^\ast} \mathbf{1}_{ |G_l^\ast| }
		\\&=
		(\mathbf{X}_{S, G_k^\ast})^\top
		\mathbf{X}_{S, G_l^\ast}
		\mathbf{1}_{ |G_l^\ast| }
		\\&=
		|G_l^\ast|
		\mathbf{X}_{S, G_k^\ast}^\top
		\overline{\mathbf{X}}^\ast_{S, l}.
	\end{align*}
    where step $(i)$ uses $(\mathbf{1}_{G_l^\ast})_{G_k^\ast} = 0$.
    
Using the definition of $\mathbf{Z}^\ast$ and condition (C3), which states that $\langle \mathbf{W}^S, \mathbf{Z}^\ast \rangle = 0$, we obtain:
	\begin{equation*}
		\langle \mathbf{W}^S, \mathbf{Z}^\ast \rangle
		=
        \sum_{k=1}^K
                \frac{1}{|G_k^\ast|}
        \langle \mathbf{W}^S,\mathbf{1}_{G_k^\ast} \mathbf{1}_{G_k^\ast}^\top \rangle
        =
		\sum_{k=1}^K 
        		\frac{1}{|G_k^\ast|}
		\mathbf{1}_{G_k^\ast}^\top
		\mathbf{W}^S 
		\mathbf{1}_{G_k^\ast} 
		=0.
	\end{equation*}
By the condition (C2),
which states that $\mathbf{W}^S \succeq 0$ , 
	 we have
	$\mathbf{1}_{G_k^\ast}^\top
	\mathbf{W}^S 
	\mathbf{1}_{G_k^\ast} \geq 0$ for all $k=1, \ldots, K$.
	This along with the equality above implies \eqref{Ws_quadform_equal_zero}:
	\begin{equation*}
		\mathbf{1}_{G_k^\ast}^\top
		\mathbf{W}^S 
		\mathbf{1}_{G_k^\ast} =0~\text{for all}~k =1, \ldots, K.
	\end{equation*}
	Given that  $\mathbf{W}^S \succeq 0$,
	the function $ \mathbf{v} \mapsto \mathbf{v}^\top \mathbf{W}^S  \mathbf{v}$ is convex and bounded below by 0.
	Therefore, $\mathbf{1}_{G_k^\ast}^\top
	\mathbf{W}^S 
	\mathbf{1}_{G_k^\ast} =0$
	implies that $\mathbf{1}_{G_k^\ast}$ minimizes $\mathbf{v}^\top \mathbf{W}^S  \mathbf{v}$. By the first order condition, we obtain \eqref{w_one_equals_zero}
	\begin{equation*}
		\mathbf{W}^S 
		\mathbf{1}_{G_k^\ast}  =\mathbf{0}~\text{for all}~k =1, \ldots, K.
	\end{equation*}
  
By condition (C4), which states that $\langle \mathbf{B}^S, \mathbf{Z}^\ast \rangle = 0$, and the block-diagonal structure of $\mathbf{Z}^\ast$,
we obtain \eqref{B_diagonal_block_zero}:
\begin{equation*}
    \mathbf{B}^S_{G_k^\ast, G_k^\ast} = \mathbf{0},
\end{equation*}
which implies \eqref{B_one_equals_zero}: 
    \begin{equation*}
        ( \mathbf{B}^S \mathbf{1}_{G_k^\ast} )_{G_k^\ast} 
        =
        \mathbf{B}^S_{G_k^\ast, G_k^\ast}
        (\mathbf{1}_{G_k^\ast})_{G_k^\ast}
        +
        \mathbf{B}^S_{G_k^\ast, {G_k^\ast}^C}
        (\mathbf{1}_{G_k^\ast})_{{G_k^\ast}^C}
        =
        0.
    \end{equation*}
On the other hand, since the off-diagonal block $\mathbf{B}^S_{G_k^\ast, G_l^\ast}$ may be nonzero for $k \neq l$, we can only conclude as \eqref{B_one_off_diagonal}:
    \begin{equation*}
        ( \mathbf{B}^S \mathbf{1}_{G_l^\ast} )_{G_k^\ast} =
         \mathbf{B}^S_{G_k^\ast, G_l^\ast} \mathbf{1}_{|G_l^\ast|}.
    \end{equation*}
    
    \medskip \noindent \textbf{Necessary condition for \( \boldsymbol{\alpha}^S \).}
Our goal is to derive condition \eqref{alpha_condition_final}.
Only using the definition of \( \mathbf{W}^S \) given in Lemma \ref{lemma:opt_cond}, we compute:
  \begin{align*}
        \mathbf{W}^S 
		\mathbf{1}_{G_k^\ast}
		&=
		\lambda^S
         \mathbf{1}_{G_k^\ast}
		+ 
		\frac{1}{2}
        \mathbf{1}_n (\boldsymbol{\alpha}^S)^\top  \mathbf{1}_{G_k^\ast}
        + 
        \frac{1}{2}
        \boldsymbol{\alpha}^S   
        \mathbf{1}_n^\top
        \mathbf{1}_{G_k^\ast}
		- 
		(\mathbf{X}_{S, \cdot}^\top \mathbf{X}_{S, \cdot}) \mathbf{1}_{G_k^\ast}
		- 
		\mathbf{B}^S   \mathbf{1}_{G_k^\ast}
		\\&=
		\lambda^S
         \mathbf{1}_{G_k^\ast}
		+ 
		\frac{1}{2}
        \bigl( \sum_{i \in G_k^\ast} \boldsymbol{\alpha}^S_i \bigr)
        \mathbf{1}_n
        + 
        \frac{
|G_k^\ast|
        }{2}
        \boldsymbol{\alpha}^S   
		- 
		(\mathbf{X}_{S, \cdot}^\top \mathbf{X}_{S, \cdot}) \mathbf{1}_{G_k^\ast}
		- 
		\mathbf{B}^S   \mathbf{1}_{G_k^\ast}.
        \numberthis \label{W_S_one_G_k_ast}
    \end{align*}
Then the restriction of $\mathbf{W}^S \mathbf{1}_{G_k^\ast}$ to the indices in $G_k^\ast$ is given by:
	\begin{equation} \label{W_s_1_gk_gk}
(\mathbf{W}^S 
		\mathbf{1}_{G_k^\ast})_{G_k^\ast}
	\overset{(i)}{=}
		\lambda^S \mathbf{1}_{|G_k^\ast|} 
		+  
		\frac{1}{2}
		\bigl( \sum_{i \in G_k^\ast} 
		\boldsymbol{\alpha}^S_i
		\bigr) 
		\mathbf{1}_{|G_k^\ast|}
		+ 
		\frac{|G_k^\ast|}{2} \boldsymbol{\alpha}^S_{G_k^\ast}
		- 
		|G_k^\ast|
		\mathbf{X}_{S, G_k^\ast}^\top
		\overline{\mathbf{X}}_{S, k}^\ast
		\overset{(ii)}{=}
        0,
	\end{equation}
    where
    step $(i)$ uses \eqref{empirical_cluster_mean_calculation} and \eqref{B_one_equals_zero},
    which  require (C4),
    and step $(ii)$ uses \eqref{w_one_equals_zero}, which requires (C2) and (C3).
    Rearranging the terms, 
	this computation implies:
	\begin{equation}\label{W_s_1_gk_gk_zero}  
		\boldsymbol{\alpha}^S_{G_k^\ast}
		=
		2\mathbf{X}_{S, G_k^\ast}^\top
		\overline{\mathbf{X}}_{S, k}^\ast
		-
		\biggl\{
		\frac{2 \lambda^S }{|G_k^\ast|}
		+
		\frac{1}{|G_k^\ast|}
		\bigl( \sum_{i \in G_k^\ast} 
		\boldsymbol{\alpha}^S_i
		\bigr) 
		\biggr\}
		\mathbf{1}_{|G_k^\ast|}.
	\end{equation} 
It remains to express $\sum_{i \in G_k^\ast} \boldsymbol{\alpha}^S_i$ on the right-hand side solely in terms of the data and $\lambda^S$. To this end, we compute:
	\begin{align*}
		\mathbf{1}_{G_k^\ast}^\top
		\mathbf{W}^S 
		\mathbf{1}_{G_k^\ast}
		& \overset{(i)}{=}
        \mathbf{1}_{G_k^\ast}^\top
        \biggl\{
\lambda^S
         \mathbf{1}_{G_k^\ast}
		+ 
		\frac{1}{2}
        \bigl( \sum_{i \in G_k^\ast} \boldsymbol{\alpha}^S_i \bigr)
        \mathbf{1}_n
        + 
        \frac{
|G_k^\ast|
        }{2}
        \boldsymbol{\alpha}^S   
		- 
		(\mathbf{X}_{S, \cdot}^\top \mathbf{X}_{S, \cdot}) \mathbf{1}_{G_k^\ast}
		- 
		\mathbf{B}^S   \mathbf{1}_{G_k^\ast}
        \biggr\}
		\\&  \overset{(ii)} {=}
		\lambda^S |G_k^\ast| + |G_k^\ast|  \sum_{i \in G_k^\ast} \boldsymbol{\alpha}^S_i
		-
		|G_k^\ast|^2
		\|
		\overline{\mathbf{X}}_{S, k}^\ast
		\|_2^2 \numberthis \label{W_quadratic_form}
		\\& \overset{(iii)}{=}0,
	\end{align*}
    where
    step $(i)$ uses \eqref{W_S_one_G_k_ast}
    step $(ii)$ uses \eqref{B_diagonal_block_zero}, which require (C4),
    and
step $(iii)$ uses \eqref{Ws_quadform_equal_zero},
which require (C2) and (C3).
Rearranging the terms,
this computation implies:
	\begin{equation}\label{alpha_condition_intermediate_alpha_sum}
		\sum_{i \in G_k^\ast} \boldsymbol{\alpha}^S_i
		=
		|G_k^\ast|
		\|
		\overline{\mathbf{X}}_{S, k}^\ast
		\|_2^2
		-
		\lambda^S.
	\end{equation}
	This with \eqref{W_s_1_gk_gk_zero} implies:
	\begin{equation*}
		\boldsymbol{\alpha}^S_{G_k^\ast} = 
		2\mathbf{X}_{S, G_k^\ast}^\top
		\overline{\mathbf{X}}_{S, k}^\ast
		- 
		\biggl\{
		\frac{\lambda^S}{|G_k^\ast|}
		+ 
		\|
		\overline{\mathbf{X}}_{S, k}^\ast
		\|_2^2
		\biggr\}
		\mathbf{1}_{|G_k^\ast|}
		.
	\end{equation*}
This corresponds precisely to condition \ref{alpha_condition_final}.

    \medskip \noindent \textbf{Necessary condition for \( \mathbf{B}^S \).}	
Our goal is to derive condition \eqref{B_condition_rowsum}.
Fix $1 \leq k \neq l \leq K$.
Similar to \eqref{W_S_one_G_k_ast},
only using the definition of \( \mathbf{W}^S \) given in Lemma \ref{lemma:opt_cond},
we compute:
  \begin{equation*}
        \mathbf{W}^S 
		\mathbf{1}_{G_l^\ast}
		=
		\lambda^S
         \mathbf{1}_{G_l^\ast}
		+ 
		\frac{1}{2}
        \bigl( \sum_{i \in G_l^\ast} \boldsymbol{\alpha}^S_i \bigr)
        \mathbf{1}_{n}
        + 
        \frac{
|G_l^\ast|
        }{2}
        \boldsymbol{\alpha}^S 
		- 
		(\mathbf{X}_{S, \cdot}^\top \mathbf{X}_{S, \cdot}) \mathbf{1}_{G_l^\ast}
		- 
		\mathbf{B}^S   \mathbf{1}_{G_l^\ast}.
    \end{equation*}
Then, the restriction of $\mathbf{W}^S \mathbf{1}_{G_l^\ast}$ to the indices in $G_k^\ast$ is given by:
  \begin{equation} \label{W_S_1_G_l_G_k}
  (\mathbf{W}^S 
		\mathbf{1}_{G_l^\ast})_{G_k^\ast}
		\overset{(i)}{=}
		\frac{1}{2}
        \bigl( \sum_{i \in G_l^\ast} \boldsymbol{\alpha}^S_i \bigr)
        \mathbf{1}_{|G_k^\ast|}
        + 
        \frac{
|G_l^\ast|
        }{2}
        \boldsymbol{\alpha}^S_{G_k^\ast}   
		- 
		|G_l^\ast|
		\mathbf{X}_{S, G_k^\ast}^\top
		\overline{\mathbf{X}}^\ast_{S, l}
		- 		
        \mathbf{B}^S_{G_k^\ast, G_l^\ast} \mathbf{1}_{|G_l^\ast|}
        \overset{(ii)}{=}
        0,
    \end{equation}
    where step $(i)$ uses $(\mathbf{1}_{G_l^\ast})_{G_k^\ast}=0$,
    \eqref{empirical_cluster_mean_calculation}, and \eqref{B_one_off_diagonal},
    and step $(ii)$ uses \eqref{w_one_equals_zero},
    which requires (C2) and (C3).
Rearranging the terms, this computation implies:
	\begingroup
\begin{equation*}
		\mathbf{B}^S_{G_k^\ast, G_l^\ast}
		\mathbf{1}_{ |G_l^\ast| }
		= 
		\frac{1}{2}
		\bigl(
		\sum_{i \in G_l^\ast} 
		\boldsymbol{\alpha}^S_i 
		\bigr) 
		\mathbf{1}_{|G_k^\ast|}
		+ 
		\frac{ |G_l^\ast| }{2} \boldsymbol{\alpha}^S_{G_k^\ast}
		- 
		|G_l^\ast|
		\mathbf{X}_{S, G_k^\ast}^\top
		\overline{\mathbf{X}}^\ast_{S, l}.
	\end{equation*}
\endgroup
    Using \eqref{alpha_condition_intermediate_alpha_sum}  with $k$ replaced by $l$, 
	and  
	\eqref{W_s_1_gk_gk_zero}, 
	which require (C2) and (C3),
this computation implies:
	\begin{align*}
		\mathbf{B}^S_{G_k^\ast, G_l^\ast} \mathbf{1}_{ |G_l^\ast| }
		=&
		\biggl\{
		\frac{ |G_l^\ast| }{2} \|\overline{\mathbf{X}}^\ast_{S, l}\|_2^2
		-
		\frac{|G_k^\ast| + |G_l^\ast|}{2 |G_k^\ast|} \lambda^S
		-
		\frac{ |G_l^\ast| }{2}
		\|\overline{\mathbf{X}}_{S, k}^\ast\|_2^2
		\biggr\} \mathbf{1}_{|G_k^\ast|}
		\\&+
		|G_l^\ast| \mathbf{X}_{S, G_k^\ast}^\top
		\overline{\mathbf{X}}_{S, k}^\ast
		-
		|G_l^\ast|
		\mathbf{X}_{S, G_k^\ast}^\top
		\overline{\mathbf{X}}^\ast_{S, l}.
	\end{align*}
This implies that, for $i \in G_k^\ast$, the $i$th row sum of 
	$\mathbf{B}^S_{G_k^\ast G_l^\ast} $
	is expressed as:
	\begin{align*}
		(
		\mathbf{B}^S_{G_k^\ast, G_l^\ast} \mathbf{1}_{ |G_l^\ast| }
		)_i
		\overset{(i)}{=}&
		-\frac{|G_k^\ast| + |G_l^\ast|}{2 |G_k^\ast|} \lambda^S
		+ \frac{ |G_l^\ast| }{2} 
		\bigl(
		\| \overline{\mathbf{X}}^\ast_{S, l}\|_2^2
		-
		2
		\mathbf{X}_{S,i}^\top
		\overline{\mathbf{X}}^\ast_{S, l}
		+
		\| \mathbf{X}_{S,i}\|_2^2
		\bigr) 
		\\&- \frac{ |G_l^\ast| }{2} 
		\bigl(
		\| \overline{\mathbf{X}}_{S, k}^\ast\|_2^2
		-2
		\mathbf{X}_{S, i}^\top
		\overline{\mathbf{X}}_{S, k}^\ast
		+
		\| \mathbf{X}_{S,i}\|_2^2
		\bigr) 
		,
		\\=&-\frac{|G_k^\ast| + |G_l^\ast|}{2 |G_k^\ast|} \lambda^S
		+ \frac{ |G_l^\ast| }{2} 
		\bigl(
		\| \overline {\mathbf{X}}^\ast_{S,l} - \mathbf{X}_{S,i} \|_2^2 
		- 
		\| \overline{\mathbf{X}}^\ast_{S,k} - \mathbf{X}_{S,i} \|_2^2 
		\bigr),
	\end{align*}
	where step $(i)$ adds and subtracts $\frac{|G_l^\ast|}{2} \| \mathbf{X}_{S,i}\|_2^2$.
This concludes the derivation of condition \eqref{B_condition_rowsum} and completes the proof of Lemma \ref{lemma:opt_necessary}.
\end{proof}
\begin{remark}\label{remark:W_s_1_gk_gk}
If we construct $\boldsymbol{\alpha}^S$ and $\mathbf{B}^S_{G_k^\ast, G_l^\ast}$ according to conditions \eqref{alpha_condition_final} and \eqref{B_condition_rowsum}, then step (ii) in \eqref{W_s_1_gk_gk} and  \eqref{W_S_1_G_l_G_k} hold without requiring conditions (C2) and (C3), as the construction is explicitly designed to satisfy these equalities. This observation is useful in the proof of Lemma \ref{lemma:lambda_s_bound}, presented in Appendix \ref{section:proof:lemma:lambda_s_bound}.
\end{remark}

\subsubsection{Proof of Lemma \ref{lemma:lambda_upper_bound}}\label{section:proof:lemma:lambda_upper_bound}
Let $S \subset [p]$ be a support set, and consider the corresponding primal SDP problem \eqref{optim:SDP_submatrix_S}. We  show that constructing  
$\dot{\boldsymbol{\alpha}}^S(\lambda^S)$, $\dot{\mathbf{B}}^S(\lambda^S)$ and $\dot{\mathbf{W}}^S(\lambda^S)$
as in  Lemma \ref{lemma:dual_construction} guarantees the following:
\begin{enumerate}
    \item Conditions (C3) and (C4) in Lemma \ref{lemma:opt_cond} are satisfied for  any $\lambda^S$,
    \item Conditions (C1) and (C5) are satisfied for $\lambda^S < U^S(\mathbf{X})$.
\end{enumerate}

\begin{proof}
For notational simplicity, we omit the notation $\lambda^S$ from
$\dot{\boldsymbol{\alpha}}^S(\lambda^S)$, $\dot{\mathbf{B}}^S(\lambda^S)$, and $\dot{\mathbf{W}}^S(\lambda^S)$
and denote them
$\dot{\boldsymbol{\alpha}}^S$, $\dot{\mathbf{B}}^S$, and $\dot{\mathbf{W}}^S$, respectively.

\medskip \noindent \textbf{1. Conditions (C3) and (C4).}
Condition (C4) follows directly from the block-diagonal structure of both $\mathbf{Z}^\ast = \sum_{k=1}^K |G_k^\ast|^{-1} \mathbf{1}_{G_k^\ast} \mathbf{1}_{G_k^\ast}^\top$ and $\dot{\mathbf{B}}^S $. 
To verify condition (C3), which requires that $\langle \dot{\mathbf{W}}^S , \mathbf{Z}^\ast \rangle = 0$, we use the definition $\mathbf{Z}^\ast = \sum_{k=1}^K |G_k^\ast|^{-1} \mathbf{1}_{G_k^\ast} \mathbf{1}_{G_k^\ast}^\top$ to expand the inner product as
	\begin{align*}
		\langle \dot{\mathbf{W}}^S , \mathbf{Z}^\ast \rangle
		=
		\frac{1}{|G_k^\ast|}
		\sum_{k=1}^K 
		\mathbf{1}_{G_k^\ast}^\top
		\dot{\mathbf{W}}^S  
		\mathbf{1}_{G_k^\ast}.
	\end{align*}
Thus, it suffices to verify that $\mathbf{1}_{G_k^\ast}^\top \dot{\mathbf{W}}^S  \mathbf{1}_{G_k^\ast} = 0$ for all $k = 1, \ldots, K$.
Since our construction sets 
 $\dot{\mathbf{B}}^S_{G_k^\ast, G_k^\ast}  = \mathbf{0}$,
 we can invoke  \eqref{W_quadratic_form} to state that
 \begin{equation*}
 \mathbf{1}_{G_k^\ast}^\top
		\dot{\mathbf{W}}^S 
		\mathbf{1}_{G_k^\ast}
        =
     \lambda^S |G_k^\ast| + |G_k^\ast|  \sum_{i \in G_k^\ast} 
     \bigl( \dot{\boldsymbol{\alpha}}^S({\lambda^S}) \bigr)_i
		-
		|G_k^\ast|^2
		\|
		\overline{\mathbf{X}}_{S, k}^\ast
		\|_2^2.
 \end{equation*}
 Now using our construction of $\dot{\boldsymbol{\alpha}}^S({\lambda^S})$, recalled below:
         \begin{equation*}
			\dot{\boldsymbol{\alpha}}^S_{G_k^\ast}
			= 
			2
			\mathbf{X}_{S, G_k^\ast}^\top
			\overline{\mathbf{X}}^\ast_{S,k}
			- 
			\frac{\lambda^S}{|G_k^\ast|}
            \mathbf{1}_{|G_k^\ast|}
			- 
			\|
			\overline{\mathbf{X}}^\ast_{S,k}
			\|_2^2
            \mathbf{1}_{|G_k^\ast|}
            ,
            \end{equation*}
we can conclude:
	\begin{equation*}
    \mathbf{1}_{G_k^\ast}^\top
		\dot{\mathbf{W}}^S 
		\mathbf{1}_{G_k^\ast}
	 =
		\lambda^S 
		|G_k^\ast| 
		+
		2|G_k^\ast|^2
		\|
		\overline{\mathbf{X}}_{S, k}^\ast
		\|_2^2
		- 
		 \lambda^S |G_k^\ast|
		-
		|G_k^\ast|^2
		\|
		\overline{\mathbf{X}}_{S, k}^\ast
		\|_2^2
		-
		|G_k^\ast|^2
		\|
		\overline{\mathbf{X}}_{S, k}^\ast
		\|_2^2
 =0.
	\end{equation*}
This concludes the verification of condition (C3).

\medskip \noindent \textbf{2. Conditions (C1) and (C5).}
We now show that  the condition $\lambda^S < U^S(\mathbf{X})$  ensures conditions (C1) and (C5). Since our construction sets the diagonal block $\dot{\mathbf{B}}^S_{G_k^\ast, G_k^\ast} $ as  a zero matrix, it suffices to show that if $\lambda^S < U^S(\mathbf{X})$, then 
	$ \dot{\mathbf{B}}^S_{G_k^\ast, G_l^\ast}  > 0
	$ 
 for all $1 \leq k \neq l \leq K$.
 By our construction  in \eqref{construction:B_off_diagonal}, 
 for $i \in G_k^\ast$ and $j \in G_l^\ast$,
the $(i,j)$th element of $\dot{\mathbf{B}}^S_{G_k^\ast, G_k^\ast} $  is
\begingroup\color{magenta}
\begin{equation*}
\frac{
 r_i^{(k,l,S)} \, r_j^{(l,k,S)} 
}{
\sum_{j \in G_l^\ast}
 r_j^{(l,k,S)}.
}
\end{equation*}
\endgroup
where
\begin{equation*}
r_i^{(k,l,S)}
		=-\frac{|G_k^\ast| + |G_l^\ast|}{2 |G_k^\ast|} \lambda^S
		+ \frac{ |G_l^\ast| }{2} 
		\bigl(
		\| \overline{\mathbf{X}}^\ast_{S,l}- \mathbf{X}_{S, i} \|_2^2 
		- 
		\| \overline{\mathbf{X}}^\ast_{S,k}- \mathbf{X}_{S, i} \|_2^2 
		\bigr),
\end{equation*}
and
\begin{equation*}
r_j^{(l,k,S)}
		=-\frac{|G_l^\ast| + |G_k^\ast|}{2 |G_l^\ast|} \lambda^S
		+ \frac{ |G_k^\ast| }{2} 
		\bigl(
		\| \overline{\mathbf{X}}^\ast_{S,k}- \mathbf{X}_{S, j} \|_2^2 
		- 
		\| \overline{\mathbf{X}}^\ast_{S,l}- \mathbf{X}_{S, j} \|_2^2 
		\bigr).
\end{equation*}
Therefore, all elements of $\dot{\mathbf{B}}^S_{G_k^\ast, G_k^\ast} $ for all $1 \leq k \neq l \leq K$ are strictly positive if,
for all $1 \leq k \neq l \leq K$
and $i \in G_k^\ast$,
the inequality
$r_i^{(k,l,S)}>0$ holds, or  equivalently,
	\begin{equation*}
		\lambda^S < 	\frac{|G_l^\ast| |G_k^\ast|}{|G_k^\ast| + |G_l^\ast|} 
		\bigl(
		\| \overline{\mathbf{X}}_{S,l}^\ast - \mathbf{X}_i \|_2^2 
		- 
		\| \overline{\mathbf{X}}_{S, k}^\ast - \mathbf{X}_i \|_2^2 
		\bigr).
	\end{equation*}
This condition is precisely $\lambda^S < U^S(\mathbf{X})$, completing the proof of Lemma \ref{lemma:lambda_upper_bound}.
\end{proof}

\subsubsection{Proof of Lemma \ref{lemma:lambda_s_bound}}\label{section:proof:lemma:lambda_s_bound}
The proof proceeds in four steps:
\begin{enumerate}
    \item Reduction to uniform control of a quadratic form over a subspace (equation \eqref{W_psd_proof_goal}),
    \item Derivation of $L^S_1(\mathbf{X})$ (equation \eqref{W_psd_intermediate_goal}),
    \item Derivation of $t_{lk}^S(\lambda^S)$ (equation \eqref{vbsv}),
    \item Derivation of $T^S_{1,kl}$, $T^S_{2,kl}$, and  $T^S_{3,kl}$ (equation \ref{derive_T}).
\end{enumerate}

\begin{proof}
Given ${\lambda^S}$, we construct $\dot{\mathbf{B}}({\lambda^S})$, $\dot{\boldsymbol{\alpha}}^S({\lambda^S})$, and $\dot{\mathbf{W}}^S({\lambda^S})$ according to Lemma~\ref{lemma:dual_construction} and  denote them by $\dot{\boldsymbol{\alpha}}^S$, $\dot{\mathbf{B}}^S$, and $\dot{\mathbf{W}}^S$, respectively,
for notational simplicity.

\medskip \noindent \textbf{1. Reduction to uniform control of a quadratic form over a subspace.}
By Lemma \ref{lemma:lambda_upper_bound}, these constructed dual variables satisfy condition (C4) of Lemma~\ref{lemma:opt_cond}. Therefore,
by Remark \ref{remark:W_s_1_gk_gk}, for any $k =1, \ldots, K$, we have
$\bigl(
\dot{ \mathbf{W}}^S
		\mathbf{1}_{G_k^\ast}
        \bigr)_{G_k^\ast}
	=
        \mathbf{0},
$	from \eqref{alpha_condition_final}, and for all $l \neq k$, we have
$
  (
  \dot{\mathbf{W}}^S  
		\mathbf{1}_{G_k^\ast})_{G_l^\ast}
=
       \mathbf{0}$ by  \eqref{B_condition_rowsum}.
    Therefore, we have
    $\dot{\mathbf{W}}^S  
		\mathbf{1}_{G_k^\ast} = \mathbf{0}$ for $k = 1, \ldots, K$. 
This implies that $\mathbf{1}_{G_1^\ast}, \ldots, \mathbf{1}_{G_K^\ast}$ are eigenvectors of $\dot{\mathbf{W}}^S$ corresponding to zero eigenvalues.
	Since any vector in \( \mathbb{R}^n \) can be decomposed into the sum of its projections onto  
 \( \operatorname{span} \left( \mathbf{1}_{G_k^\ast} : k \in [K] \right) \)
 and its orthogonal complement, denoted as $\Gamma_K $,
	to show that \( \dot{\mathbf{W}}^S \succeq 0 \), it suffices to prove that  
\begin{equation}\label{W_psd_proof_goal}
	\mathbf{v}^\top \dot{\mathbf{W}}^S \mathbf{v} \geq 0  
	\quad \text{for all } \mathbf{v} \in \Gamma_K \text{ such that } \|\mathbf{v}\|_2 = 1.  
\end{equation}

\medskip \noindent \textbf{2. Derivation of $L^S_1(\mathbf{X})$.}
By the definition of ${\Gamma_K}$, $\mathbf{v} \in {\Gamma_K}$ implies
	$
	\mathbf{v}^\top \mathbf{1}_{G_1^\ast}
	=
	\mathbf{v}^\top \mathbf{1}_{G_2^\ast}
	=
	\ldots
	=
	\mathbf{v}^\top \mathbf{1}_{G_K^\ast} =0,
	$
	which implies
	\begin{equation}\label{vi_sum_zero}
		\sum_{j \in G_k^\ast}^n v_j = \sum_{i =1}^n v_i = 0
        ~\text{for all}~k=1, \ldots, K.
	\end{equation}
Now we compute the quadratic form $\mathbf{v}^\top \dot{\mathbf{W}}^S \mathbf{v}$
for
$\mathbf{v} \in \Gamma_K$ such that $\|\mathbf{v}\|_2 = 1$:
	\begin{align*}
		\mathbf{v}^\top \dot{\mathbf{W}}^S \mathbf{v}
		& \overset{(i)}{=}
		\lambda^S 
		\mathbf{v}^\top \textbf{I}_n \mathbf{v} 
		+ 
		\frac{1}{2} 
		\mathbf{v}^\top
		\mathbf{1}_n (\boldsymbol{\alpha}^S)^\top
		\mathbf{v}  
		+
		\frac{1}{2}
		\mathbf{v}^\top
		\boldsymbol{\alpha}^S
		\mathbf{1}_n^\top
		\mathbf{v} 
		-
		\mathbf{v}^\top
		\mathbf{X}_{S, \cdot}^\top \mathbf{X}_{S, \cdot} 
		\mathbf{v} 
		-
		\mathbf{v}^\top
		\dot{\mathbf{B}}^S
		\mathbf{v}
		\\&\overset{(ii)}{=} 
		\lambda^S 
		+ 
		( \mathbf{v}^\top \boldsymbol{\alpha}_S )
	 \sum_{i=1}^n v_i 
		- 
		\|		
		\mathbf{X}_{S, \cdot} \mathbf{v}  
        \|_2^2
		- 
		\mathbf{v}^\top \dot{\mathbf{B}}^S \mathbf{v}
		\\&\overset{(ii)}{=} 
		\lambda^S
		- 
		\|		
		\mathbf{X}_{S, \cdot} \mathbf{v}  
        \|_2^2
		- 
		\mathbf{v}^\top \dot{\mathbf{B}}^S \mathbf{v},
	\end{align*}
where
step $(i)$ uses the construction
	$
	\dot{\mathbf{W}}^S = \lambda^S \textbf{I}_n 
    +
    \frac{1}{2}
    \mathbf{1}_n
    (
    \dot{\boldsymbol{\alpha}}^S)^\top 
    + 
    \frac{1}{2}
    \dot{\boldsymbol{\alpha}}^S \mathbf{1}_n^\top - \mathbf{X}_{S, \cdot}^\top \mathbf{X}_{S, \cdot} 
    - 
    \dot{ \mathbf{B}}^S
	$,
    step $(ii)$
uses
	$\mathbf{v}^\top \textbf{I}_n \mathbf{v} = \|\mathbf{v}\|^2=1$,
	and
	step $(ii)$
	uses \eqref{vi_sum_zero}.
Using our model and notation in Definition \ref{def:theorem_1_notation},
	\begin{equation*}
    \mathbf{X}_{S, \cdot} \mathbf{v}
    =
		\sum_{i =1}^n 
		v_i \mathbf{X}_{S,i}  
		= 
		\sum_{k=1}^K
        \biggl(
		\sum_{i \in G_k^*}
		v_i 
		(\mathbf{c}_k)_S
		+
		\sum_{i =1}^n 
		v_i 
        \mathcal{E}_{S,i}
        \biggr)
		\overset{(i)}{=} 	
		\sum_{i=1}^n
		v_i 
		\mathcal{E}_{S,i}
		=
		\mathcal{E}_{S, \cdot} \mathbf{v}
		,
	\end{equation*}
	where step $(i)$ uses  \eqref{vi_sum_zero}.
		Therefore, a sufficient condition for \eqref{W_psd_proof_goal} is:
    \begin{equation}\label{W_psd_intermediate_goal}
  \lambda^S
  \geq
  \underbrace{
  \sup_{\mathbf{v} \in \Gamma_K, \|\mathbf{v}\|_2 = 1}
  \hspace{-1.3em}
        \|		
		\mathcal{E}_{S, \cdot} \mathbf{v} 
        \|_2^2
        }_{L^S_1(\mathbf{X})}
		+
         \sup_{\mathbf{v} \in \Gamma_K, \|\mathbf{v}\|_2 = 1}
         \hspace{-1.3em}
		\mathbf{v}^\top \dot{\mathbf{B}}^S \mathbf{v}.
    \end{equation}
This completes the derivation of $L^S_1(\mathbf{X})$.

\medskip \noindent \textbf{3. Derivation of  $	t_{lk}^S(\lambda^S)$.}
Now we start to derive $
    L_2^S(\mathbf{X}, \lambda^S)$.
Recall from
Lemma \ref{lemma:dual_construction} that 
given $\lambda^S$,
we construct $\dot{\mathbf{B}}$ as a block-diagonal matrix, where the diagonal blocks are zero matrices and
for $1 \leq k \neq l \leq K$, the $(k,l)$th off-diagonal block is defined as:
\begin{equation*}
\dot{\mathbf{B}}^S_{G_k^\ast, G_l^\ast}  =
\frac{1}{\mathbf{1}_{ |G_l^\ast| }^\top \mathbf{r}^{(l,k,S)}}
\, \mathbf{r}^{(k,l,S)} \, (\mathbf{r}^{(l,k,S)})^\top,
\end{equation*}
with $\mathbf{r}^{(k,l,S)} \in \mathbb{R}^{|G_k^\ast|}$ defined componentwise by
\begin{equation*}
r_i^{(k,l,S)}
		=-\frac{|G_k^\ast| + |G_l^\ast|}{2 |G_k^\ast|} \lambda^S
		+ \frac{ |G_l^\ast| }{2} 
		\bigl(
		\| \overline{\mathbf{X}}^\ast_{S,l}- \mathbf{X}_{S, i} \|_2^2 
		- 
		\| \overline{\mathbf{X}}^\ast_{S,k}- \mathbf{X}_{S, i} \|_2^2 
		\bigr).
\end{equation*}
Using this definition, we compute
	$\mathbf{v}^\top \dot{\mathbf{B}}^S \mathbf{v}$ in \eqref{W_psd_intermediate_goal}: 
	\begin{equation}\label{vbsv}
		\mathbf{v}^\top \dot{\mathbf{B}}^S \mathbf{v}
\overset{(i)}{=}  
		\sum_{1 \leq k \neq l \leq K}
        \mathbf{v}_{G_k^\ast}^\top
        \dot{\mathbf{B}}^S_{G_k^\ast, G_l^\ast}
        \mathbf{v}_{G_l^\ast} 
 \overset{(ii)}{=} 
		\sum_{1 \leq k \neq l \leq K} \sum_{i \in G_k^\ast} \sum_{j \in G_l^\ast} 
\frac{
(
\mathbf{v}_{G_k^\ast}^\top \mathbf{r}^{(k,l,S)}
)
(
\mathbf{v}_{G_l^\ast}^\top \mathbf{r}^{(l,k,S)}
)
}{
\underbrace{
    \mathbf{1}_{ |G_l^\ast| }^\top  \mathbf{r}^{(l,k,S)}
    }_{=t_{lk}^S(\lambda^S)}
},
	\end{equation}
	where
    step $(i)$ uses 
 $\dot{\mathbf{B}}^S_{G_k^\ast, G_k^\ast} = \mathbf{0}$ for $k =1, \ldots, K$,
	and
    step $(ii)$ uses 
	our construction of $\dot{\mathbf{B}}^S_{G_k^\ast, G_l^\ast}$.

\medskip \noindent \textbf{4. Derivation of $T^S_{1,kl}$, $T^S_{2,kl}$, and  $T^S_{3,kl}$ .}
From \eqref{vbsv}, we analyze one term in the numerator:
	\begin{align*}
    &\mathbf{v}_{G_k^\ast}^\top \mathbf{r}^{(k,l,S) }
    \\& \quad =
		\sum_{i \in G_k^\ast}
		v_i
		r_i^{(k,l,S)}
    \\& \quad =
		\sum_{i \in G_k^\ast}
		v_i
        \left\{
		-\frac{|G_k^\ast| + |G_l^\ast|}{2 |G_k^\ast|} \lambda^S
		+ \frac{ |G_l^\ast| }{2} 
		\bigl(
		\| \overline{\mathbf{X}}^\ast_{S,l}- \mathbf{X}_{S, i} \|_2^2 
		- 
		\| \overline{\mathbf{X}}^\ast_{S,k}- \mathbf{X}_{S, i} \|_2^2 
		\bigr)
        \right\}
    \\& \quad \stackrel{(i)}{=}
    		\frac{ |G_l^\ast| }{2} 
		\sum_{i \in G_k^\ast}
		v_i
		\bigl(
		\| \overline{\mathbf{X}}^\ast_{S,l}- \mathbf{X}_{S, i} \|_2^2 
		- 
		\| \overline{\mathbf{X}}^\ast_{S,k}- \mathbf{X}_{S, i} \|_2^2 
		\bigr)
    \\& \quad =
    		\frac{ |G_l^\ast| }{2} 
		\sum_{i \in G_k^\ast}
		v_i
		\bigl(
		\| \overline{\mathbf{X}}^\ast_{S,l}\|_2^2  
        +
        \|\mathbf{X}_{S, i} \|_2^2 
        - 2 \langle  \overline{\mathbf{X}}^\ast_{S,l}, \mathbf{X}_{S, i} \rangle
		- 
		\| \overline{\mathbf{X}}^\ast_{S,k} \|_2^2  
        - \| \mathbf{X}_{S, i} \|_2^2 
        + 2 \langle  \overline{\mathbf{X}}^\ast_{S,k}, \mathbf{X}_{S, i} \rangle
		\bigr)
    \\& \quad \overset{(ii)}{=}
   |G_l^\ast|
		\sum_{i \in G_k^\ast}
		v_i
         \langle  
         \overline{\mathbf{X}}^\ast_{S,k}-\overline{\mathbf{X}}^\ast_{S,l},\;
         \mathbf{X}_{S, i} \rangle
   \\& \quad \overset{(iii)}{=}
   |G_l^\ast|
		\sum_{i \in G_k^\ast}
		v_i
         \langle  
         (\mathbf{c}_k^\ast- \mathbf{c}_l^\ast)_{S}
         +
		\overline {\mathcal{E}}_{S,k}^\ast
		-
		\overline {\mathcal{E}}_{S,l}^\ast
         ,\;
         {\mathcal{E}}_{S,i}\rangle
   \\& \quad =
   |G_l^\ast|
		\sum_{i \in G_k^\ast}
		v_i
         \langle  
         (\mathbf{c}_k^\ast- \mathbf{c}_l^\ast)_{S \cap S^\ast}
         ,\;
         {\mathcal{E}}_{S \cap S^\ast,i}\rangle
         +
          |G_l^\ast|
		\sum_{i \in G_k^\ast}
		v_i
         \langle  
		\overline {\mathcal{E}}_{S,k}^\ast
		-
		\overline {\mathcal{E}}_{S,l}^\ast
         ,\;
         {\mathcal{E}}_{S,i}\rangle
   \\& \quad =
  I^S_{kl}(\mathbf{v}, \mathbf{X})
         +
  N_{kl}^S (\mathbf{v}, \mathbf{X})
	\end{align*}
    where step $(i)$ and $(ii)$ use \( \sum_{i \in G_k^\ast} v_i = 0 \) for  \( k =1, \ldots, K \),
    and step $(iii)$ comes from our model and notation in Definition \ref{def:theorem_1_notation}.
	Similarly,   we have
	\begin{equation*}
		\mathbf{v}_{G_l^\ast}^\top \mathbf{r}^{(l,k,S) }
		= I^S_{lk}(\mathbf{v}, \mathbf{X})
         +
  N_{lk}^S (\mathbf{v}, \mathbf{X}).	
	\end{equation*}
    Therefore, we have
\begin{align*}
    \mathbf{v}^\top \dot{\mathbf{B}}^S \mathbf{v} &= 
    \sum_{1 \leq k \neq l \leq K} \sum_{i \in G_k^\ast} \sum_{j \in G_l^\ast}
    \frac{1}{t_{lk}^S(\lambda^S)}
    \bigl(
     I^S_{kl}(\mathbf{v}, \mathbf{X})
         +
  N_{kl}^S (\mathbf{v}, \mathbf{X})
  \bigr)
  \bigl(
  I^S_{lk}(\mathbf{v}, \mathbf{X})
         +
  N_{lk}^S (\mathbf{v}, \mathbf{X})
  \bigr)
  \\&=
    \sum_{1 \leq k \neq l \leq K} \sum_{i \in G_k^\ast} \sum_{j \in G_l^\ast}
    \frac{1}{t_{lk}^S(\lambda^S)}
    \bigl(
    T^S_{1,kl}(\mathbf{v})
    +
    T^S_{2,kl}(\mathbf{v})
    +
    T^S_{3,kl}(\mathbf{v})
  \bigr).
  \numberthis \label{derive_T}
\end{align*}
Plugging this into \eqref{W_psd_intermediate_goal} completes the proof of Lemma \ref{lemma:lambda_s_bound}.

\end{proof}

\subsubsection{Proof of Lemma \ref{lemma:bound_of_upper_bound}}
\label{section:chisq_tail_application}
For notational simplicity, we assume $\sigma^2 = 1$ here. 
The proof proceeds in the following steps:
\begin{enumerate}
    \item Decomposition of $D^S_{kli}(\mathbf{X}) $ into Chi-squares, Gaussian  and inner product of Gaussians,
    \item Bounding  the Gaussian and Chi-squares using the tail bounds,
    \item Bounding the  Gaussian inner product using conditioning, Gaussian tail bounds and separation condition,
    \item Defining the uniform bound event,
    \item Bounding the probability of the uniform bound event.
\end{enumerate}
\begin{proof}We proceed by following the steps outlined above in sequence.

\medskip \noindent \textbf{1. Decomposition of $D^S_{kli}(\mathbf{X}) $ into Chi-squares, Gaussian  and inner product of  Gaussians.}

Fix indices \( 1 \leq k \neq l \leq K \) and choose \( i \in G_k^\ast \).
{Define the leave-one-out average noise by}
\[
\textcolor{blue}{\overline{\mathcal{E}}_{S,k\setminus\{i\}}^\ast := \frac{1}{|G_k^\ast|-1}\sum_{\ell\in G_k^\ast\setminus\{i\}}\mathcal{E}_{S,\ell}.}
\]
Based on our model and notation in Definition \ref{def:theorem_1_notation}, 
we consider the following decomposition of \( D^S_{kli}(\mathbf{X}) \):

\begingroup
\begin{align*}
D^S_{kli}(\mathbf{X}) 
~=~& 
\| \mathbf{X}_{S,i} - \overline{\mathbf{X}}_{S, l}^\ast \|_2^2
-
\| \mathbf{X}_{S,i} - \overline{\mathbf{X}}_{S,k}^\ast \|_2^2
\\
~=~&
\| (\mathbf{c}_k^\ast - \mathbf{c}_l^\ast )_S
+ \mathcal{E}_{S,i}
- \overline{\mathcal{E}}_{S,l}^\ast \|_2^2
-
\| \mathcal{E}_{S,i} - \overline{\mathcal{E}}_{S,k}^\ast \|_2^2
\\
~=~&
\| (\mathbf{c}_k^\ast - \mathbf{c}_l^\ast )_S
+ \mathcal{E}_{S,i}
- \overline{\mathcal{E}}_{S,l}^\ast \|_2^2
-
\left( \frac{|G_k^\ast|-1}{|G_k^\ast|} \right)^2
\| \mathcal{E}_{S,i} - \overline{\mathcal{E}}_{S,k \setminus \{i\}}^\ast \|_2^2
\\
\overset{(i)}{=}~&
\| (\mathbf{c}_k^\ast - \mathbf{c}_l^\ast )_{S \cap S^\ast} \|_2^2
+ \|\mathcal{E}_{S,i}\|_2^2
+ \| \overline{\mathcal{E}}_{S,l}^\ast \|_2^2
+ 2 \left\langle (\mathbf{c}_k^\ast - \mathbf{c}_l^\ast )_S, \mathcal{E}_{S,i} \right\rangle
\\
& - 2 \left\langle (\mathbf{c}_k^\ast - \mathbf{c}_l^\ast )_{S \cap S^\ast}, \overline{\mathcal{E}}_{S \cap S^\ast, l}^\ast \right\rangle
- 2 \left\langle \overline{\mathcal{E}}_{S, l}^\ast, \mathcal{E}_{S,i} \right\rangle
\\
& -
\left( \frac{|G_k^\ast|-1}{|G_k^\ast|} \right)^2
\left(
\| \mathcal{E}_{S,i} \|_2^2
+ \| \overline{\mathcal{E}}_{S,k \setminus \{i\}}^\ast \|_2^2
- 2 \left\langle \mathcal{E}_{S,i}, \overline{\mathcal{E}}_{S,k \setminus \{i\}}^\ast \right\rangle
\right)
\\
~=~&
\| (\mathbf{c}_k^\ast - \mathbf{c}_l^\ast )_{S \cap S^\ast} \|_2^2
+ \| \overline{\mathcal{E}}_{S,l}^\ast \|_2^2
+ \frac{2|G_k^\ast|-1}{|G_k^\ast|^2} \| \mathcal{E}_{S,i} \|_2^2
\\
& - \left( \frac{|G_k^\ast|-1}{|G_k^\ast|} \right)^2 \| \overline{\mathcal{E}}_{S,k \setminus \{i\}}^\ast \|_2^2
- 2 \left\langle (\mathbf{c}_k^\ast - \mathbf{c}_l^\ast )_{S \cap S^\ast}, \overline{\mathcal{E}}_{S \cap S^\ast, l}^\ast \right\rangle
\\
& + 2 \left\langle 
(\mathbf{c}_k^\ast - \mathbf{c}_l^\ast )_S
- \overline{\mathcal{E}}_{S,l}^\ast 
+ \left( \frac{|G_k^\ast|-1}{|G_k^\ast|} \right)^2 \overline{\mathcal{E}}_{S, k \setminus \{i\}}^\ast,
\mathcal{E}_{S,i}
\right\rangle,
\end{align*}
\endgroup

where step $(i)$ uses the sparsity assumption on cluster center difference.

The six terms in this decomposition are classified into the following categories:

    \begin{itemize}
        \item Signal:  $\| (\mathbf{c}^\ast_k - \mathbf{c}^\ast_l)_{S \cap S^\ast} \|_2^2$
        \item Chi-square noise:
        \begin{align*}
             \operatorname{CN}_l^S &:= \|\overline{\mathcal{E}}_{S,l}^\ast\|_2^2,
            \\ \operatorname{CN}_i^S &:= \frac{2|G^\ast_k|-1}{|G^\ast_k|^2}  \|\mathcal{E}_{S,i}\|_2^2,~\text{and}
            \\ \operatorname{CN}_{ki}^S &:= -  \left( \frac{|G^\ast_k| - 1}{|G^\ast_k|} \right)^2 \|\overline{\mathcal{E}}_{S,
			k \setminus \{i\}
		}
		\|_2^2 
        \end{align*}

        \item Gaussian signal-noise interaction:
        \begin{equation*}
     \operatorname{GSN}_{kl}^S := - 2 \langle (\mathbf{c}^\ast_k - \mathbf{c}^\ast_l)_{S \cap S^\ast}, \overline{\mathcal{E}}_{S \cap S^\ast, l}^\ast \rangle     
        \end{equation*}
        
        \item Inner product of  Gaussians:
        \begin{equation*}
        \mathrm{IP}_{kli}^S :=2 \left\langle 
		\bigl(
		\mathbf{c}_k^\ast - \mathbf{c}_l^\ast 
        \bigr)_S
        - \overline{\mathcal{E}}_{S,l}^\ast + \left( \frac{|G^\ast_k| - 1}{|G^\ast_k|} \right)^2 
		\overline{\mathcal{E}}_{S, k \setminus \{i\}}^\ast		
		, \;
        \mathcal{E}_{S,i} 
		\right\rangle,
    \end{equation*}
    \end{itemize}
    In summary, we have:
	\begin{equation}\label{dskli_decomposition}
		D^S_{kli}(\mathbf{X}) 
		~=~ 
		 \| (\mathbf{c}^\ast_k - \mathbf{c}^\ast_l)_{S \cap S^\ast} \|_2^2
		+ \operatorname{CN}_l^S 
		+    \operatorname{CN}_i^S 
        +
        \operatorname{CN}_{ki}^S
        +
        \operatorname{GSN}_{kl}^S
        +
        \mathrm{IP}_{kli}^S.
	\end{equation}
    
\medskip \noindent \textbf{2. Bounding  the Gaussian and Chi-squares using the tail inequalities.}
Chi-square noise and Gaussian signal-noise interaction terms
$(\operatorname{CN}_l^S,
		\operatorname{CN}_i^S, 
        \operatorname{CN}_{ki}^S$, and $
        \operatorname{GSN}_{kl}^S)$
in \eqref{dskli_decomposition} are straightforward to bound from below. Define the following lower bound events:
	\begin{align*}
		\mathcal{B}^{(i,1)}_{kl, S} 
        &:=
        \left\{
        \operatorname{CN}_l^S 
        \geq 
        \frac{|S| - 2\sqrt{|S|\zeta^S}}{|G_l^\ast|} 
        \right\},
        \numberthis \label{dskli_bound_1}
        \\
		\mathcal{B}^{(i,2)}_{kl, S} &:= \left\{  
        \operatorname{CN}_i^S 
        \geq 
        \frac{2|G^\ast_k|-1}{|G^\ast_k|^2} 
        \bigl(
        |S| - 2\sqrt{|S|\zeta^S}
        \bigr)
        \right\}, 
                \numberthis \label{dskli_bound_2}
        \\
		\mathcal{B}^{(i,3)}_{kl, S} 
        &:= 
        \left\{ 
        \operatorname{CN}_{ki}^S
        \geq 
        \frac{1-|G_k^\ast|}{|G_k^\ast|^2} 
        \bigl(
        |S| + 2\sqrt{|S|\zeta^S}  + 2\zeta^S
        \bigr)
        \right\}, 
                \numberthis \label{dskli_bound_3}
        \\
		\mathcal{B}^{(i,4)}_{kl, S} 
        &:= 
        \left\{ 
        \operatorname{GSN}_{kl}^S 
        \geq 
        -2\sqrt{ \frac{2 \zeta^S}{|G_l^\ast|} } \|(\mathbf{c}_k^\ast - \mathbf{c}_l^\ast)_{S \cap S^\ast}\|_2 \right\}, 
                \numberthis \label{dskli_bound_4}
	\end{align*}
	By setting $ \zeta^S = \log(n^2|S|^2  {\binom{p}{|S|}})$,
    Lemma \ref{lemma:chisq_tail} implies that each event has a probability of at least  
	$1-1/(n^2K^2 |S|^2 {\binom{p}{|S|}})$, for any $S \in \mathcal S$.
    
\medskip \noindent \textbf{3. Bounding the  Gaussian inner product using conditioning, Gaussian tail bounds and separation condition.}
Here, we use the separation condition to bound the term. 
If we  condition on \( \overline{\mathcal{E}}_{S,l}^\ast \) and \( \overline{\mathcal{E}}_{S, k \setminus \{i\}}^\ast \),
then  $ -\mathrm{IP}_{kli}^S$ has a univariate Gaussian distribution with mean zero and variance as the following:
	\begin{align*}
		\mathrm{Var}(
		-\mathrm{IP}_{kli}^S
		\; |\;
		\overline{\mathcal{E}}_{S,l}^\ast , \overline{\mathcal{E}}_{S, k \setminus \{i\}}^\ast 
		)&=\left\| (\mathbf{c}_k^\ast - \mathbf{c}_l^\ast)_S - \overline{\mathcal{E}}_{S,l}^\ast 
		+ (1 - |G_k^\ast|^{-1})^2 
        \overline{\mathcal{E}}_{S, k \setminus \{i\}}^\ast
        \right\|_2^2
		\\&  \overset{(i)}{=}
		\left\| (\mathbf{c}_k^\ast - \mathbf{c}_l^\ast)_{S  \cap S^\ast}\right\|_2^2
		+
        \underbrace{
		\left\|
		\overline{\mathcal{E}}_{S,l}^\ast 
		- (1 - |G_k^\ast|^{-1})^2 \overline{\mathcal{E}}_{S, k \setminus \{i\}}^\ast \right\|_2^2
        }_{:=\operatorname{CN}_{kli}^S}
		\\&\quad +
        \underbrace{
		2
		\bigl\langle
		(\mathbf{c}_k^\ast - \mathbf{c}_l^\ast)_{S \cap S^\ast},
        \;
		(1 - |G_k^\ast|^{-1})^2 
        \overline{\mathcal{E}}_{S \cap S^\ast, k \setminus \{i\}}^\ast
		-
		\overline{\mathcal{E}}_{S \cap S^\ast, l}^\ast
		\bigr\rangle
        }_{:=
        \operatorname{GN}_{kli}^S
        }
        ,
	\end{align*}
    where step $(i)$ uses our sparsity assumption on cluster center difference.

Next, we define events that bound this variance from above:
\begin{align*}
		\mathcal{B}^{(i,5)}_{kl, S} 
        &:= 
        \left\{ 
        \operatorname{CN}_{kli}^S
        \leq (|G_l^\ast|^{-1} + |G_k^\ast|^{-1}) (|S| + 2\sqrt{|S|\zeta^S}  + 2\zeta^S) \right\}, \notag \\
		\mathcal{B}^{(i,6)}_{kl, S}
        &:=
        \left\{ 
\operatorname{GN}_{kli}^S 
		\leq 2\sqrt{2(|G_l^\ast|^{-1} + |G_k^\ast|^{-1})\zeta^S} \|(\mathbf{c}_k^\ast - \mathbf{c}_l^\ast)_{S \cap S^\ast}\|_2 \right\}.
	\end{align*}
By setting $ \zeta^S = \log(n^2 |S|^2  {\binom{p}{|S|}})$,
    Lemma \ref{lemma:chisq_tail} implies that each event has a probability of at least  
	$1-1/(n^2 |S|^2 {\binom{p}{|S|}})$, uniformly overall all $S \in \mathcal S$.

	Thus, conditioned on events $\mathcal{B}^{(i,5)}_{kl, S}$ and
	$\mathcal{B}^{(i,6)}_{kl, S}$, the variance can be bouneded as 
	\begin{align*}
		\mathrm{Var}(
		- \mathrm{IP}_{kli}^S
		\; |\ \;
		\overline{\mathcal{E}}_{S,l}^\ast , \overline{\mathcal{E}}_{S, k \setminus \{i\}}^\ast 
		)
		&\leq
		\left\| (\mathbf{c}_k^\ast - \mathbf{c}_l^\ast)_{S  \cap S^\ast}\right\|_2^2
		+ 
		(|G_l^\ast|^{-1} + |G_k^\ast|^{-1}) (|S| + 2\sqrt{|S|\zeta^S}  + 2\zeta^S)
		\\&\quad 	+
		2 \sigma^2\sqrt{2(|G_l^\ast|^{-1} + |G_k^\ast|^{-1})\zeta^S} \|(\mathbf{c}_k^\ast - \mathbf{c}_l^\ast)_{S \cap S^\ast}\|_2.
	\end{align*}
    To further bound this variance, 
    it is convenient to remove the first-order signal term $\|(\mathbf{c}_k^\ast - \mathbf{c}_l^\ast)_{S \cap S^\ast}\|_2$. To this end, 
    we collect some conditions.
    First, recall from Theorem \ref{theorem:separation_condition} that
   \begin{equation*}
        m = 2\min_{1 \leq k \neq l \leq K}
		\bigl(
		|G_l^\ast|^{-1} + |G_k^\ast|^{-1}
		\bigr)^{-1}.
    \end{equation*} 
    From \eqref{def:S-set}, recall that for 
     $S \in \mathcal S$, we have
    \begin{equation}\label{separation_condition_recall}
\min_{1 \leq k \neq l \leq K} \| (\mathbf{c}_l^\ast - \mathbf{c}_k^\ast )_{S \cap S^\ast} \|_2^2
    \gtrsim 
    \sigma^2
		\left(\log n +
		\frac{|S| \log p}{m}
		+
		\sqrt{\frac{|S| \log p}{m}}
		\right),
    \end{equation}
    and  $|S| \leq \sqrt{p}$.
    Under these conditions, by treating $K$ as a costant,
    we have 
    \begin{equation*}
        \zeta^S = \log(n^2 |S|^2  {\binom{p}{|S|}}) \lesssim  |S| \log p +  \log n,
    \end{equation*}
  which implies
    \begin{equation*}
		2\sqrt{2(|G_l^\ast|^{-1} + |G_k^\ast|^{-1})
        \zeta^S
        } \left\| (\mathbf{c}_k^\ast - \mathbf{c}_l^\ast)_{S  \cap S^\ast}\right\|_2
        \leq
        2\sqrt{
       \frac{\zeta^S}{m}
        } \left\| (\mathbf{c}_k^\ast - \mathbf{c}_l^\ast)_{S  \cap S^\ast}\right\|_2
        \lesssim
        	\left\| (\mathbf{c}_k^\ast - \mathbf{c}_l^\ast)_{S  \cap S^\ast}\right\|_2^2.
	\end{equation*}
	Therefore, we can further bound the variance as
	\begin{align*}
		\mathrm{Var}( -\mathrm{IP}_{kli}^S)
		&\lesssim
		\left\| (\mathbf{c}_k^\ast - \mathbf{c}_l^\ast)_{S  \cap S^\ast}\right\|_2^2
		+
		(|G_l^\ast|^{-1} + |G_k^\ast|^{-1}) (|S| + 2\sqrt{|S|\zeta^S}  + 2\zeta^S).
	\end{align*}
	Therefore, by the Gaussian tail inequality, we have 
	\begin{align*}
        \mathrm{CP}_{kli}^S &:=
		\mathbb{P}
		\left(
		-\mathrm{IP}_{kli}^S >
		\left\| (\mathbf{c}_k^\ast - \mathbf{c}_l^\ast )_{S  \cap S^\ast}\right\|^2_2 
		\; | \;
		\overline{\mathcal{E}}_{S,l}^\ast , \overline{\mathcal{E}}_{S, k \setminus \{i\}}^\ast  
		\right)
		\\& \leq 
		\exp
		\left(
		\frac{
			-\left\| (\mathbf{c}_k^\ast - \mathbf{c}_l^\ast)_{S  \cap S^\ast} \right\|_2^4
		}{2 	\mathrm{Var}(
			-\mathrm{IP}_{kli}^S
			\; | \;
			\overline{\mathcal{E}}_{S,l}^\ast , \overline{\mathcal{E}}_{S, k \setminus \{i\}}^\ast 
			)}
		\right)
		\\&  \lesssim
		\exp
		\left(
		\frac{
			-\left\| (\mathbf{c}_k^\ast - \mathbf{c}_l^\ast)_{S  \cap S^\ast} \right\|_2^4
		}{
			\left\| (\mathbf{c}_k^\ast - \mathbf{c}_l^\ast)_{S  \cap S^\ast}\right\|_2^2
			+
			2(|G_l^\ast|^{-1} + |G_k^\ast|^{-1}) (|S| + 2\sqrt{|S|\zeta^S}  + 2\zeta^S)}
		\right),
                        \numberthis \label{dskli_bound_5}
	\end{align*}
    where the last step uses the variance bound.
    We want to bound the last term by $C/n^2$, with $C$ a constant.
	To this end, it suffices to show
	\begin{align*}
		\frac{
			\left\| (\mathbf{c}_k^\ast - \mathbf{c}_l^\ast)_{S  \cap S^\ast} \right\|_2^4
		}{
			\left\| (\mathbf{c}_k^\ast - \mathbf{c}_l^\ast)_{S  \cap S^\ast}\right\|^2
			+
			2(|G_l^\ast|^{-1} + |G_k^\ast|^{-1}) (|S| + 2\sqrt{|S|\zeta^S}  + 2\zeta^S)}
	 \gtrsim \log n,
	\end{align*}
	which is indeed satisfied under our separation condition in \eqref{separation_condition_recall}.
    
\medskip \noindent \textbf{4. Defining the uniform bound event.}
We collect the bounds in 
         \eqref{dskli_bound_1},
         \eqref{dskli_bound_2},
         \eqref{dskli_bound_3},
         \eqref{dskli_bound_4}, and
         \eqref{dskli_bound_5}.
         From \eqref{dskli_decomposition}, recalled below,  
         \begin{equation*}
		D^S_{kli}(\mathbf{X}) 
		~=~ 
		 \| (\mathbf{c}^\ast_k - \mathbf{c}^\ast_l)_{S \cap S^\ast} \|_2^2
		+ \operatorname{CN}_l^S 
		+    \operatorname{CN}_i^S 
        +
        \operatorname{CN}_{ki}^S
        +
        \operatorname{GSN}_{kl}^S
        +
        \mathrm{IP}_{kli}^S.
	\end{equation*}
    we can say
 \begin{align*}
		D^S_{kli}(\mathbf{X}) 
		~\gtrsim &~
		\| (\mathbf{c}_l^\ast - \mathbf{c}_k^\ast )_{S \cap S^\ast}\|_2^2
		+
        \frac{|S| - 2\sqrt{|S|\zeta^S}}{|G_l^\ast|} 
		 +
		\frac{2|G^\ast_k|-1}{|G^\ast_k|^2} 
        \bigl(
        |S| - 2\sqrt{|S|\zeta^S}
        \bigr)
		\\&~+~
		        \frac{1-|G_k^\ast|}{|G_k^\ast|^2} 
        \bigl(
        |S| + 2\sqrt{|S|\zeta^S}  + 2\zeta^S
        \bigr)
        -
        2\sqrt{ \frac{2 \zeta^S}{|G_l^\ast|} } \|(\mathbf{c}_k^\ast - \mathbf{c}_l^\ast)_{S \cap S^\ast}\|_2
		\\& ~ +~
		\left\| (\mathbf{c}_k^\ast - \mathbf{c}_l^\ast)_{S  \cap S^\ast}\right\|_2^2,
        \\ ~\gtrsim&~ 
        \| (\mathbf{c}_l^\ast - \mathbf{c}_k^\ast )_{S \cap S^\ast}\|_2^2 
         -
        \sqrt{ \frac{ \zeta^S}{|G_l^\ast|} } \|(\mathbf{c}_k^\ast - \mathbf{c}_l^\ast)_{S \cap S^\ast}\|_2
        \\&~-~\biggl( \frac{1}{|G_k^\ast|} + \frac{1}{|G_l^\ast|} \biggr)|S|
        -
         \biggl( \frac{1}{|G_k^\ast|} + \frac{1}{|G_l^\ast|} \biggr)\sqrt{|S| \zeta^S}
          -
         \frac{1}{|G_k^\ast|}   \zeta^S
        \\ ~=&~ 
        \| (\mathbf{c}_l^\ast - \mathbf{c}_k^\ast )_{S \cap S^\ast}\|_2^2 
         ~-~\biggl( \frac{1}{|G_k^\ast|} + \frac{1}{|G_l^\ast|} \biggr)|S|
         -
        R^S_{kl},
	\end{align*} 
    where 
    \begin{equation*}
         R^S_{kl} =  \sqrt{ \frac{ \zeta^S}{|G_l^\ast|} } \| (\mathbf{c}_l^\ast - \mathbf{c}_k^\ast )_{S \cap S^\ast}\|_2 
          +
          \biggl( \frac{1}{|G_k^\ast|} + \frac{1}{|G_l^\ast|} \biggr)\sqrt{|S| \zeta^S}
          +
         \frac{1}{|G_k^\ast|}   \zeta^S.
    \end{equation*}
 As a result, we define an event
	\begingroup
\begin{align*}
		\mathcal{A}^{   S   }_{kli} 
		:= \Bigg\{ 
		D^S_{kli}(\mathbf{X})
		\gtrsim 
         \| (\mathbf{c}_k^\ast - \mathbf{c}_l^\ast)_{S \cap S^\ast} \|_2^2
        +
        \biggl( \frac{1}{|G_k^\ast|} + \frac{1}{|G_l^\ast|} \biggr)
        |S|  
		 - R^S_{kl} 
		\Bigg\},
	\end{align*}
\endgroup
	with the index  $(k, l, i)$  ranging $1 \leq k \neq l \leq K$
	and all  $i \in G^*_k$, and $S$ ranging over $\mathcal S$.
The proof is complete once we show that the intersection event
    	\begin{equation*}
		\bigcap_{1 \leq k \neq l \leq K} 
        \bigcap_{ i \in G_k^\ast}
        \bigcap_{S \in \mathcal S}
    \mathcal{A}^{   S   }_{kli} 
	\end{equation*}
holds with high probability.

    \medskip \noindent \textbf{5. Bounding the probability of the uniform bound event.}
	To prove the lemma, it suffices to show that 
	\begin{equation*}
		\mathbb{P} \left( 
		\bigcup_{1 \leq k \neq l \leq K} 
        \bigcup_{ i \in G_k^\ast}
        \bigcup_{S \in \mathcal S}
        { \bigl( \mathcal{A}^{   S   }_{kli} \bigr)}^C
		\right) \lesssim \frac{1}{n}.
	\end{equation*}
    To show this, first denote
    	\begin{equation*}
		\mathcal{B}^{(i)}_{kl, S} := \bigcap_{\ell=1}^6 \mathcal{B}^{(i, \ell)}_{kl, S}.
	\end{equation*}
Then, we have
	\begin{align*}
		\mathbb{P} \biggl( 
		\bigcup_{1 \leq k \neq l \leq K} &
        \bigcup_{i \in G_k^\ast} 
        \bigcup_{S \in \mathcal S}
        {  (\mathcal{A}^{   S}_{kli})^C}
		\biggr) 
       \\ &\leq
		\sum_{1 \leq k \neq l \leq K}
		\sum_{i \in G_k^\ast}
		\sum_{S \in \mathcal{S}}
		\mathbb{P} { \bigl( (\mathcal{A}^{   S}_{kli})^C \bigr)}
        \\& =
        		\sum_{1 \leq k \neq l \leq K}
		\sum_{i \in G_k^\ast}
		\sum_{S \in \mathcal{S}}
        \mathbb{E}
        \bigl[
		\mathbb{P}  \bigl( (\mathcal{A}^{   S}_{kli})^C 
        \; | \;
        \overline{\mathcal{E}}_{S,l}^\ast, \mathcal{E}_{S,i}, \overline{\mathcal{E}}_{S,
			k \setminus \{i\}}
        \bigr)
        \bigr]
                \\& \leq
        		\sum_{1 \leq k \neq l \leq K}
		\sum_{i \in G_k^\ast}
		\sum_{S \in \mathcal{S}}
        \mathbb{E}
        \bigl[
		\mathbb{P}  \bigl( (\mathcal{A}^{   S}_{kli})^C 
        \; | \;
        \overline{\mathcal{E}}_{S,l}^\ast, \mathcal{E}_{S,i}, \overline{\mathcal{E}}_{S,
			k \setminus \{i\}}
        \bigr) \mathbbm{1}(\mathcal{B}^{(i)}_{kl,S})
        + 
        \mathbbm{1}(\mathcal{B}^{(i)}_{kl,S})^C
        \bigr]
		\\& \overset{(i)}{\leq}
        \left(
		\sum_{1 \leq k \neq l \leq K}
		\sum_{i \in G_k^\ast}
		\sum_{q=1}^{\lfloor \sqrt{p} \rfloor}
		\sum_{S \in \mathcal{S}, |S|=q}
        \mathbb{E}
        \bigl[
		\mathrm{CP}_{kli}^S
        \mathbbm{1}(
		\mathcal{B}^{(i)}_{kl,S} )
        \bigr]
        \right)
		+
		\frac{\pi^2}{n}
		\\& \overset{(ii)}{\lesssim}
		\frac{K^2\pi^2}{6n}
		+
		\frac{K^2\pi^2}{n},
	\end{align*}
	where 
    step $(i)$ uses 
      $
	\sum_{q=1}^\infty
    (1 / q^2)
    = \pi^2 / 6
	$, and
    step $(ii)$ uses \eqref{dskli_bound_5}.
	Treating $K$ as a constant, we conclude that
    \begin{equation*}
        D^S_{kli}(\mathbf{X})
		\gtrsim 
         \| (\mathbf{c}_k^\ast - \mathbf{c}_l^\ast)_{S \cap S^\ast} \|_2^2
        +
        \biggl( \frac{1}{|G_k^\ast|} + \frac{1}{|G_l^\ast|} \biggr)
        |S|  
		 - R^S_{kl}.
    \end{equation*}
    uniformly over $S \in \mathcal{S}$, with probability at least $1-C/n$, where $C>0$ is a constant.
This completes the proof of Lemma \ref{lemma:bound_of_upper_bound}.

\end{proof}
\subsubsection{Proof of Lemma \ref{lemma:L1}}\label{section:proof:lemma:L1}

\begin{proof}

Given a support set $S \subset [p]$ with $|S| \leq \sqrt{p}$,
we have
\begin{align*}
	L_1^S(\mathbf{X}) 
    =  
 \sup_{\mathbf{v} \in \Gamma_K, \|\mathbf{v}\|_2 = 1}
  \hspace{-1.3em}
        \|		
		\mathcal{E}_{S, \cdot} \mathbf{v} 
        \|_2^2
	\leq
	\sup_{\mathbf{v} \in \mathbb{R}^n, \|\mathbf{v}\|_2 = 1}
	        \|		
		\mathcal{E}_{S, \cdot} \mathbf{v} 
        \|_2^2
	=
	\|\mathcal{E}_{S, \cdot} \|_{\mathrm{op}}^2.
\end{align*}
Here,
$\mathcal{E}_{S, \cdot}$ is a $|S| \times n$ random matrix with i.i.d. Gaussian entries distributed as \( \mathcal{N}(0, \sigma^2) \). 
Therefore,
	by Lemma \ref{lemma:gaussian_matrix_opnorm},
    if we define an event
\begingroup
\begin{equation*}
\mathcal{C}_L^S
:=
\bigl\{
		L_1^S(\mathbf{X})  
		\leq
		\sigma^2(n+ |S| + \xi^S)
		\bigr\},
\end{equation*}
\endgroup
where $ \xi^S = \log(n  |S|^2  {\binom{p}{|S|}})$,
then we have
$
\mathbb{P}(\mathcal{C}_L^S) \geq 1-1/(n |S|^2 {\binom{p}{|S|}}).
$
 Now we show that the event $\mathcal{C}_L^S$ holds uniformly across all $S \subset [p]$ such that $|S| \leq \sqrt{p}$, with probability at least $1-C_7/n$, where $C_7>0$ is a constant.
    \begin{equation*}
		\mathbb{P}
        \biggl(
        \bigcup_{S \subset [p], |S| \leq \sqrt{p}}
        (\mathcal{C}_L^{S})^C
        \biggr)
		\leq
		\sum_{q = 1}^{\lfloor \sqrt{p} \rfloor} 
		\sum_{|S| = q}
        \mathbb{P}((\mathcal{C}_L^{S})^C)
			\leq
		\sum_{q = 1}^{\lfloor \sqrt{p} \rfloor} 
		\sum_{|S| = q}
		\frac{1}{n |S|^2  {\binom{p}{|S|}}}
			 =
		\sum_{q = 1}^{\lfloor \sqrt{p} \rfloor} 
		\frac{1}{n q^2}
			 \leq
		\frac{\pi^2}{6n },
	\end{equation*}
where the last
	step  uses
	$
	\sum_{q=1}^\infty = \pi^2/6.
	$
Since $|S| \leq \sqrt{p}$, $|S| \log p$ dominates $\log|S|$. Therefore, we conclude that
uniformly across $S \in [p]$ such that $|S| \leq \sqrt{p}$,
with probability at least $1 - C_7/n$, it holds that
\begin{equation*}
    L_1^S(\mathbf{X})  
		\lesssim
		\sigma^2(n + \log n + |S| \log p).
\end{equation*}
	This completes the proof of Lemma \ref{lemma:L1}.
\end{proof}

\subsubsection{Proof of Lemma \ref{lemma:lambda_between}}\label{section:proof:lemma:lambda_between}
The proof proceeds in the following steps:
\begin{enumerate}
    \item The first inequality 
    $L_1^S(\mathbf{X}).
\lesssim
\dot{\lambda}^S$.
    \item  The second inequality 
    $\dot{\lambda}^S 
\lesssim
U^S(\mathbf{X})$.
\end{enumerate}

\begin{proof} For simplicity, we assume $\sigma^2=1$ here.

\medskip \noindent \textbf{1. The first inequality.} We aim to show
    $L_1^S(\mathbf{X}) 
\lesssim
\dot{\lambda}^S$.
 From \eqref{def:S-set}, recall that for 
     $S \in \mathcal S$, we have
    \begin{equation}\label{separation_condition_recall_2}
    \Delta_{S \cap S^\ast}^2=
\min_{1 \leq k \neq l \leq K} \| (\mathbf{c}_l^\ast - \mathbf{c}_k^\ast )_{S \cap S^\ast} \|_2^2
    \gtrsim 
    \sigma^2
		\left(\log n +
		\frac{|S| \log p}{m}
		+
		\sqrt{\frac{|S| \log p}{m}}
		\right),
    \end{equation}
    where 
    \begin{equation*}
         2\min_{1 \leq k \neq l \leq K}
\left\{
		\biggl(
		\frac{1}{ |G_k^\ast| }
		+
		\frac{1}{ |G_l^\ast| }
		\biggr)^{-1}
        \right\}.
    \end{equation*}
    Therefore,
    with probability at least $1 - C_7/n$ uniformly across all $S \in \mathcal S$, where $C_7>0$ is a constant from Lemma \ref{lemma:L1}, it holds that
    \begin{align*}
    L_1^S(\mathbf{X})  
		&\lesssim
		(n + \log n + |S| \log p)
        \\&\overset{(i)}{\lesssim}
          |S|
         +
         \frac{1}{4}
         \bigl(
         m \log n + |S| \log p + \sqrt{m |S| \log p}
         \bigr)
         \\&\lesssim
        |S| + \frac{m}{4} \Delta_{S \cap S^\ast }^2
         \\&= \dot{\lambda}^S,
\end{align*}
where step $(i)$ uses Lemma \ref{lemma:L1}.
This completes the derivation of the first inequality.

\medskip \noindent \textbf{2. The second inequality.} We aim to show
    $\dot{\lambda}^S 
\lesssim
U^S(\mathbf{X})$.
 Recall from 
\eqref{U_s_X} that
\begin{equation*}
		U^S(\mathbf{X})
		=
		\min_{1 \leq k \neq l \leq K}
		\biggl\{
		\biggl(
		\frac{1}{ |G_k^\ast| }
		+
		\frac{1}{ |G_l^\ast| }
		\biggr)^{-1}
		\min_{i \in G_k^\ast}
		D^S_{kli}(\mathbf{X}) 
		\biggr\}.
\end{equation*}
By Lemma \ref{lemma:bound_of_upper_bound}, with probability at least
$1 - C_6/n$ uniformly across $S \in \mathcal{S}$ and 
uniformly over $(k,l,i)$ where $1 \leq k \neq l \leq K$ and $i \in G_k^\ast$, we have
   \begin{equation*}
   \biggl(
		\frac{1}{ |G_k^\ast| }
		+
		\frac{1}{ |G_l^\ast| }
		\biggr)^{-1}
        \hspace{-0.45em}
        D^S_{kli}(\mathbf{X})
		\gtrsim 
        \underbrace{
        \biggl(
		\frac{1}{ |G_k^\ast| }
		+
		\frac{1}{ |G_l^\ast| }
		\biggr)^{-1}  \hspace{-0.45em}
         \| (\mathbf{c}_k^\ast - \mathbf{c}_l^\ast)_{S \cap S^\ast} \|_2^2
        +
        |S|  }_{\text{Part A}}
          \underbrace{
		 - 
         \biggl(
		\frac{1}{ |G_k^\ast| }
		+
		\frac{1}{ |G_l^\ast| }
		\biggr)^{-1}  \hspace{-0.45em}
         R^S_{kl}}_{\text{Part B}}
         , 
    \end{equation*}
where
 \begin{equation*}
         R^S_{kl} =  \sqrt{ \frac{ \xi^S}{|G_l^\ast|} } \| (\mathbf{c}_l^\ast - \mathbf{c}_k^\ast )_{S \cap S^\ast}\|_2 
          +
          \biggl( \frac{1}{|G_k^\ast|} + \frac{1}{|G_l^\ast|} \biggr)\sqrt{|S| \xi^S}
          +
         \frac{1}{|G_k^\ast|}   \xi^S,
    \end{equation*} 
    and $ \xi^S = \log(n^2K^2 |S|^2  {\binom{p}{|S|}})$.
 Under the separation condition for $\mathcal{S}$, recalled in \eqref{separation_condition_recall_2}, and noting that $|S| \leq \sqrt{p}$ for all $S \in \mathcal{S}$, Part A dominates Part B in the inequality above. Therefore, we have
\begin{align*}
	\min_{1 \leq k \neq l \leq K}
    \left\{
     \biggl(
		\frac{1}{ |G_k^\ast| }
		+
		\frac{1}{ |G_l^\ast| }
		\biggr)^{-1} \hspace{-0.45em}
        D^S_{kli}(\mathbf{X})
        \right\}
        &\gtrsim \min_{1 \leq k \neq l \leq K}
                \biggl(
		\frac{1}{ |G_k^\ast| }
		+
		\frac{1}{ |G_l^\ast| }
		\biggr)^{-1} \hspace{-0.45em}
         \| (\mathbf{c}_k^\ast - \mathbf{c}_l^\ast)_{S \cap S^\ast} \|_2^2
        +
        |S|
               \\&\gtrsim
        |S| + \frac{m}{4} \Delta_{S \cap S^\ast }^2
         \\&= \dot{\lambda}^S.
\end{align*}
This concludes the derivation of the second inequality and thereby completes the proof.
\end{proof}

\subsubsection{Proof of Lemma \ref{lemma:boun(D^t)^Csl}}\label{section:proof:lemma:boun(D^t)^Csl}
\begin{proof}
For notational simplicity, let $\sigma^2 = 1$.
	Fix \( S \in \mathcal{S} \) and \( (l, k) \) with \( 1 \leq l \neq k \leq K \). Condition on the high-probability event in Lemma \ref{lemma:bound_of_upper_bound}. Then we have 
	\begin{align*}
		t^S_{lk}(\dot{\lambda}^S)
		&\overset{(i)}{=}
		\sum_{j \in G_l^\ast}
	\biggl\{
	\frac{ |G_k^\ast| }{2} 
	D^S_{lkj}(\mathbf{X})
	-\frac{|G_k^\ast|+ |G_l^\ast|}{2|G_l^\ast|} \dot{\lambda}^S
	\biggr\} 
 \\&\overset{(ii)}{=}
		\sum_{j \in G_l^\ast}
	\biggl\{
	\frac{ |G_k^\ast| }{2} 
	D^S_{lkj}(\mathbf{X})
	-\frac{|G_k^\ast|+ |G_l^\ast|}{2|G_l^\ast|} 
     \bigl( |S| + \frac{m}{4} \Delta_{S \cap S^\ast }^2 \bigr)
	\biggr\} 
 \\&\overset{(iii)}{\gtrsim}
		\sum_{j \in G_l^\ast}
	\biggl\{
	\frac{ |G_k^\ast| }{2} 
    \bigl(
	|S| + 	
		\|(\mathbf{c}_l^\ast - \mathbf{c}_k^\ast)_{S \cap S^\ast}\|_2^2
        \bigr)
	-\frac{|G_k^\ast|+ |G_l^\ast|}{2|G_l^\ast|} 
     \bigl( |S| + \frac{m}{4} \Delta_{S \cap S^\ast }^2 \bigr)
	\biggr\} 
\\&\overset{(iii)}{\geq} 
\sum_{j \in G_l^\ast}
\biggl(
    \frac{|G_k^\ast|}{2} 
		 |S| + 	
		\frac{|G_k^\ast|}{2} 
		\|(\mathbf{c}_k^\ast - \mathbf{c}_l^\ast)_{S \cap S^\ast}\|_2^2
		\\& \hspace{4.5em}
        -
		\frac{|G_k^\ast| + |G_l^\ast|}{2|G_l^\ast|}  |S|
        -
		\frac{|G_k^\ast| + |G_l^\ast|}{2|G_l^\ast|}
		\frac{m}{4}  \| (\mathbf{c}_l^\ast - \mathbf{c}_k^\ast )_{S \cap S^\ast}\|_2^2 		
\biggr)
        \\
		& \overset{(iv)}{\geq}
		\sum_{j \in G_l^\ast}
        \biggl\{
        \biggl(
		\frac{|G_k^\ast|}{2} 
        -
		\frac{|G_k^\ast| + |G_l^\ast|}{2|G_l^\ast|}
        \biggr)
        |S|  	
		+
		\frac{|G_k^\ast|}{4}
		\| (\mathbf{c}_l^\ast - \mathbf{c}_k^\ast )_{S \cap S^\ast}\|_2^2
        \biggr\}
		\\
		&=
		\frac{|G_l^\ast| |G_k^\ast|}{4} \|\mathbf{c}_k^\ast - \mathbf{c}_l^\ast\|_2^2
		+\frac{|G_l^\ast| |G_k^\ast|}{2} |S|
		-
		\frac{|G_k^\ast| + |G_l^\ast|}{2}  |S|.
		\\
		&\overset{(v)}{\gtrsim} |G_l^\ast| |G_k^\ast| 
        \bigl(
        \|\mathbf{c}_k^\ast - \mathbf{c}_l^\ast\|_2^2 + |S| \bigr)
	\end{align*}
	where 
    step \( (i) \) follows from the definition of $t^S_{lk}(\lambda^S)$ in \eqref{def:tkls},
    step $(ii)$ follows from the definition of $\textcolor{blue}{\dot{\lambda}^S}$ in Lemma \ref{lemma:lambda_between},
    step $(iii)$ uses
$-\Delta_{S \cap S^\ast }^2 > - \|(\mathbf{c}_k^\ast - \mathbf{c}_l^\ast)_{S \cap S^\ast}\|_2^2$,
step $(iv)$ uses
   \begin{equation*}
        -m \geq -2
\left\{
		\frac{ |G_k^\ast| \cdot |G_l^\ast| }{
|G_k^\ast|+|G_l^\ast|
        }
        \right\},
    \end{equation*} 
and step $(v)$ uses $|G_k^\ast|\geq2$ and $|G_l^\ast| \geq2$.
This completes the proof of Lemma \ref{lemma:boun(D^t)^Csl}.
\end{proof}

\subsubsection{Proof of Lemma \ref{lemma:T1Sv}}\label{section:proof:lemma:T1Sv}
the proof proceeds in the following steps.
    \begin{enumerate}
    \item Individual bound by Chi-square tail bound,
        \item Uniform bound.
    \end{enumerate}
\begin{proof}
	For notational simplicity, let $\sigma^2 = 1$.
    
\medskip \noindent \textbf{1. Individual bound by Chi-square tail bound.}
    Fix $(k, l)$ such that $1 \leq k \neq l \leq K$ and
	 $S \in \mathcal{S}$.
Then we have
    \begin{align*}
	 	\sup_{\mathbf{v} \in \Gamma_K, \|\mathbf{v}\|_2=1}
I^S_{kl}(\mathbf{v}, \mathbf{X}) 
&=      
\sup_{\mathbf{v} \in \Gamma_K, \|\mathbf{v}\|_2=1}
|G_l^\ast|
	\sum_{i \in G_k^\ast} v_i 
	\langle
	(\mathbf{c}_k^\ast
	-
	\mathbf{c}_l^\ast)_{S \cap S^\ast},
\;
	\mathcal{E}_{S \cap S^\ast,i}
	\rangle   
\\& \leq
\sup_{\mathbf{v} \in \Gamma_K, \|\mathbf{v}\|_2=1}
    |G_l^\ast|
    \biggl|
	\sum_{i \in G_k^\ast} v_i 
	\langle
	(\mathbf{c}_k^\ast
	-
	\mathbf{c}_l^\ast)_{S \cap S^\ast},
\;
	\mathcal{E}_{S \cap S^\ast,i}
	\rangle
    \biggr|
        \\& \overset{(i)}{\leq}
\sup_{\mathbf{v} \in \Gamma_K, \|\mathbf{v}\|_2=1}
        |G_l^\ast|
    \biggl( \sum_{i \in G_k^\ast} v_i^2 \biggr)^{1/2}
	\biggl( \sum_{i \in G_k^\ast} 
	\bigl \langle 
	(\mathbf{c}_k^\ast - \mathbf{c}_l^\ast)_{S \cap S^\ast}, \mathcal{E}_{S \cap S^\ast,i}
	\bigr \rangle^2
	\biggr)^{1/2}
        \\& \leq
        |G_l^\ast|
	\biggl( \sum_{i \in G_k^\ast} 
	\bigl \langle 
	(\mathbf{c}_k^\ast - \mathbf{c}_l^\ast)_{S \cap S^\ast}, \mathcal{E}_{S \cap S^\ast,i}
	\bigr \rangle^2
	\biggr)^{1/2},
    \end{align*}
where step $(i)$ uses   the Cauchy-Schwarz inequality.

    Now we derive an inequality for $ \sum_{i \in G_k^\ast} 
	\bigl \langle 
	(\mathbf{c}_k^\ast - \mathbf{c}_l^\ast)_{S \cap S^\ast}, \mathcal{E}_{S \cap S^\ast,i}
	\bigr \rangle^2$.
	Since
	\[
	\frac{1}{\|
		(\mathbf{c}_k^\ast - \mathbf{c}_l^\ast)_{S \cap S^\ast}
		\|_2}
	\bigl \langle 
	(\mathbf{c}_k^\ast - \mathbf{c}_l^\ast)_{S \cap S^\ast}, \mathcal{E}_{S \cap S^\ast,i}
	\bigr \rangle
	\overset{\text{i.i.d.}}{\sim} \mathcal{N}(0, 1), \quad i = 1, \ldots, |G_k^\ast|,
	\]
	the square of this quantity follows a chi-square distribution with \( |G_k^\ast|\) degrees of freedom, provided that \( S \) is nonempty. Applying the chi-square tail inequality (Lemma \ref{lemma:chisq_tail}), if we define
\begingroup
\begin{equation*}
\mathcal{C}_{Ikl}^S
:=
\biggl\{
		\sup_{\mathbf{v} \in \Gamma_K, \|\mathbf{v}\|_2=1}
I^S_{kl}(\mathbf{v}, \mathbf{X})  
		\leq
\|
		(\mathbf{c}_k^\ast - \mathbf{c}_l^\ast)_{S \cap S^\ast}
		\|_2
		\bigl(
        |G_k^\ast| + 2\sqrt{ |G_k^\ast| \zeta_{I}^S} + 2 \zeta_{I}^S
        \bigr)^{1/2}
        \biggr\},
\end{equation*}
\endgroup
where
 $ \zeta_{I}^S = \log(nK^2 |S|^2  {\binom{p}{|S|}})$,
 then we have
$
\mathbb{P}
\bigl(
\mathcal{C}_{Ikl}^S
\bigr) 
\geq 1-1/(n^2K^2 |S|^2 {\binom{p}{|S|}})
$.

\medskip \noindent \textbf{2. Uniform bound.}
 Now we show that the event $\mathcal{C}_{Ikl}^S$ holds uniformly across all $S \subset \mathcal S$ and $1 \leq k \neq l \leq K$, with probability at least $1-C_9/n$, where $C_9>0$ is a constant.
    \begin{equation*}
		\mathbb{P}
        \bigl(
        \bigcup_{1 \leq k \neq l \leq K}
                \bigcup_{S \in  \mathcal S}
        (\mathcal{C}_{Ikl}^{S})^C
        \bigr)
		\leq
        \hspace{-.4em}
        \sum_{1 \leq k \neq l \leq K}
		\sum_{q = 1}^{\lfloor \sqrt{p} \rfloor} 
		\sum_{|S| = q}
        \mathbb{P}((\mathcal{C}_{Ikl}^{S})^C)
			 \overset{(i)}{\leq }
		\sum_{q = 1}^{\lfloor \sqrt{p} \rfloor} 
		\sum_{|S| = q}
		\frac{1}{n |S|^2  {\binom{p}{|S|}}}
			 =
		\sum_{q = 1}^{\lfloor \sqrt{p} \rfloor} 
		\frac{1}{n q^2}
			 \leq
		\frac{\pi^2}{6n },
	\end{equation*}
where the last
	step  uses
	$
	\sum_{q=1}^\infty = \pi^2/6.
	$
Since $|S| \leq \sqrt{p}$, $|S| \log p$ dominates $\log|S|$. Therefore, we conclude that
uniformly across $S \in \mathcal S$ and $(k,l)$ such that $1 \leq k \neq l \leq K$,
with probability at least $1 - C_9/n$, it holds that
\begin{align*}
 &\sup_{\mathbf{v} \in \Gamma_K, \|\mathbf{v}\|_2=1}
I^S_{kl}(\mathbf{v}, \mathbf{X})   
		\\& \quad \quad \lesssim
        |G_l^\ast|
        \|
		(\mathbf{c}_k^\ast - \mathbf{c}_l^\ast)_{S \cap S^\ast}
		\|_2
                \biggl(
                |G_k^\ast|
                +
                \sqrt{
                |G_k^\ast| \log n
                }
                +
                \sqrt{
                |G_k^\ast| |S| \log p
                }+
                \log n + |S| \log p
        \biggr)^{1/2}.
\end{align*}
	This completes the proof of Lemma \ref{lemma:T1Sv}.
\end{proof}

\subsubsection{Proof of Lemma \ref{lemma:T1_dominance}}\label{section:proof:lemma:T1_dominance}
\begin{proof}
For notational simplicity, 
let $\sigma^2 = 1$ and
for $k = 1, \ldots, K$, denote
\begin{equation*}
    \mathrm{IB}_k :=
    \biggl(
                |G_k^\ast|
                +
                \sqrt{
                |G_k^\ast| \log n
                }
                +
                \sqrt{
                |G_k^\ast| |S| \log p
                }+
                \log n + |S| \log p
        \biggr)^{1/2}.
\end{equation*}
Then we have
\begin{align*}
 \sup_{\mathbf{v} \in \Gamma_K, \| \mathbf{v} \|_2  =1}
\sum_{1 \leq k \neq l \leq K} 
		\frac{T^S_{1,kl}(\mathbf{v})}{t^S_{lk}(\dot{\lambda}^S)} 
&\overset{(i)}{\lesssim}
 \sup_{\mathbf{v} \in \Gamma_K, \| \mathbf{v} \|_2  =1}
\sum_{1 \leq k \neq l \leq K} 
		\frac{I^S_{kl}(\mathbf{v}, \mathbf{X})
    \times 
    I^S_{lk}(\mathbf{v}, \mathbf{X})}{
        |G_l^\ast| \cdot |G_k^\ast|
        \cdot
        \|
        (\mathbf{c}^\ast_k -\mathbf{c}^\ast_l)_{S \cap S^\ast}
        \|_2^2 + |S|
        }
\\&{\leq}
 \sup_{\mathbf{v} \in \Gamma_K, \| \mathbf{v} \|_2  =1}
\sum_{1 \leq k \neq l \leq K} 
		\frac{I^S_{kl}(\mathbf{v}, \mathbf{X})
    \times 
    I^S_{lk}(\mathbf{v}, \mathbf{X})}{
        |G_l^\ast| \cdot |G_k^\ast|
        \cdot
        \|
        (\mathbf{c}^\ast_k -\mathbf{c}^\ast_l)_{S \cap S^\ast}
        \|_2^2 
        } 
\\&\overset{(ii)}{\lesssim}
\sum_{1 \leq k \neq l \leq K} 
\mathrm{IB_k^S}
\times \mathrm{IB_l^S}
\\&\leq
\sum_{k=1}^K 
\mathrm{IB_k^S}
\;
\sum_{l=1}^K
 \mathrm{IB_l^S}
 \\& \overset{(iii)}{\lesssim}
	n + \sqrt{ n \log n} + 
    \sqrt{n |S| \log p}
    +
    \log n + |S| \log p
\\& 
\overset{(iv)}{\lesssim}
	|S|	+
	m
	\log n + 
	\sqrt{m
		|S| \log p
	}
+
	|S| \log p
	\\& \overset{(v)}{\lesssim}
 |S|+	
	\frac{m}{4} \Delta_{S \cap S^\ast}^2
    =
\dot{\lambda}^S,
\end{align*}
where step $(i)$ uses Lemma \ref{lemma:boun(D^t)^Csl}
and
definition
    $T^S_{1, k l}(\mathbf{v})=
	I^S_{kl}(\mathbf{v}, \mathbf{X})
    \times 
    I^S_{lk}(\mathbf{v}, \mathbf{X})$
presented in \eqref{def:T1},
step $(ii)$ uses Lemma
\ref{lemma:T1Sv}, 
 step $(iii)$ uses the identity
$\sum_{k=1}^K |G_k^\ast| = n$
and 
inequality $\sum_{k=1}^K \sqrt{|G_k^\ast|} \leq \sqrt{K \sum_{k=1}^K |G_k^\ast|} \lesssim \sqrt{n}$,
step $(iv)$ uses the condition 
$
		m \gtrsim
n / \log n
$ of Theorem \ref{theorem:separation_condition},
and step $(v)$ uses the separation condition of $\mathcal S$, presented in \eqref{def:S-set}.
This completes the proof of Lemma \ref{lemma:T1_dominance}.
\end{proof}

\subsubsection{Proof of Lemma \ref{lemma:upperbound_G1ksv}}\label{section:proof:lemma:upperbound_G1ksv}
With appropriate algebraic manipulation, the problem reduces to an application of the concentration inequality for an unbounded empirical process developed by \citet{adamczak_tail_2008} (Lemma \ref{lemma:unbounded_empirical_process}). This requires estimating the expectation, variance, and sub-exponential norm of the empirical process. Accordingly, the proof proceeds in the following steps.
    \begin{enumerate}
    \item Reformulation  into a quadratic form,
        \item Bounding the operator norm dependent terms,
        \item Bounding the expectation of the supremum,
        \item Individual bound using Hanson-wright inequality,
        \item Uniform bound.
    \end{enumerate}
\begin{proof} For notational simplicity, we set $\sigma^2 = 1$ and assume, without loss of generality, that $G_k^\ast = {1, \dots, |G_k^\ast|}$ throughout  the proof.

\medskip \noindent \textbf{1. Transformation into a quadratic form.}
Fix $(k,l)$ such that $1 \leq k \neq l \leq K$ and $S \subset [p]$ such that $|S| \leq \sqrt{p}$.
We can express \( {G^S_{1, kl}}(\mathbf{v}) \) as a scaled quadratic form:
	\begin{align*}
		G^S_{1, kl}(\mathbf{v})
		&= 
\frac{ |G_l^\ast| }{|G_k^\ast|} 
		\sum_{
       \{(i, j) \in G_k^\ast \}
        }
        \hspace{-1em}
        v_i 
		\left\langle
		\mathcal{E}_{S,i},
		\mathcal{E}_{S,j}
		\right\rangle
        \numberthis
        \label{uprocess_assymmetric}
		\\&= %
\frac{ |G_l^\ast| }{|G_k^\ast|} 
		\sum_{\{(i, j) \in G_k^\ast\}} 
        \hspace{-1em}
        \frac{1}{2}(v_i + v_j) \left\langle
		\mathcal{E}_{S,i},
		\mathcal{E}_{S,j}
		\right\rangle 
	\\ &= 
    \frac{ |G_l^\ast| }{|G_k^\ast|} 
    \underbrace{
        \mathrm{vec}(\mathcal{E}_{S,G_k^\ast})^\top 
		( \mathbf{A}^{\mathbf{v}} \otimes \mathbf{I}_{|S|} )\;
		\mathrm{vec}(\mathcal{E}_{S,G_k^\ast})
        }_{\mathrm{QF}_{kl}^S(
\mathbf{A}^{\mathbf{v}} \otimes \mathbf{I}_{|S|}, \mathbf{X}
        )}
        ,
	\end{align*}
 Here,
	 the $(i,j)$th element of the matrix
    $
    \mathbf{A}^{\mathbf{v}}
    \in
    \mathbb{R}^{|G_k^\ast| \times |G_k^\ast|}
    $
    is defined as $A^{\mathbf{v}}_{i,j} := (v_i + v_j)/2$, and
	$\mathrm{vec}(\mathcal{E}_{S,G_k^\ast})$ denotes the $|S|\cdot|G_k^\ast|$-dimensional vector obtained by stacking the columns of $\mathcal{E}_{S,G_k^\ast}$ in order, with the first column at the top and the last column at the bottom,
    and $\mathbf{A}^{\mathbf{v}} \otimes \mathbf{I}_{|S|}$ denotes the tensor product:
\begin{equation*}
\mathbf{A}^{\mathbf{v}} \otimes \mathbf{I}_{|S|} =
\begin{bmatrix}
A^{\mathbf{v}}_{1,1} \mathbf{I}_{|S|} & \cdots & A^{\mathbf{v}}_{1,|G_k^\ast|} \mathbf{I}_{|S|} \\
\vdots & \ddots & \vdots \\
A^{\mathbf{v}}_{|G_k^\ast|,1} \mathbf{I}_{|S|} & \cdots & A^{\mathbf{v}}_{|G_k^\ast|,|G_k^\ast|} \mathbf{I}_{|S|}
\end{bmatrix}.
\end{equation*}
The quadratic form 
     $\mathrm{QF}_{kl}^S(
\mathbf{A}^{\mathbf{v}} \otimes \mathbf{I}_{|S|}, \mathbf{X}
        )$
 is centered because
\begin{align*}
\mathbb{E}
\bigl[
    \mathrm{QF}_{kl}^S(
\mathbf{A}^{\mathbf{v}} \otimes \mathbf{I}_{|S|}, \mathbf{X}
        )
        \bigr]
        =
        \mathbb{E}
        \sum_{
       \{(i, j) \in G_k^\ast \}
        }
        \hspace{-1em}
        v_i 
        \mathbb{E}
\bigl[
		\left\langle
		\mathcal{E}_{S,i},
		\mathcal{E}_{S,j}
		\right\rangle
        \bigr]
        =
        \sum_{
       i \in G_k^\ast 
        }
        v_i 
        \mathbb{E}
\bigl[
\|
		\mathcal{E}_{S,i}
        \|_2^2
        \bigr]
                \overset{(i)}{=}
                 |S|
        \sum_{
       i \in G_k^\ast 
        }
        v_i   
= 0,
\end{align*}   
where step $(i)$ uses $\sum_{i =1}^n v_i = 0$, shown in \eqref{vi_sum_zero},
and
         indexed by 
        	\[
	\mathcal{M}^S_{kl} = \{ \mathbf{A}^{\mathbf{v}} \otimes \mathbf{I}_{|S|} \mid \mathbf{v} \in \Gamma_K, \|\mathbf{v}\|_2 = 1 \}.
	\]  
To derive the uniform bound, we apply the uniform Hanson-Wright inequality (Lemma \ref{lemma:uniform_hanson}) to obtain the concentration inequality for 
\begin{equation*}
    \sup_{\mathbf{M} \in \mathcal{M}^S_{kl}} \mathrm{QF}_{kl}^S(
\mathbf{M}, \mathbf{X}
        ) 
        -
        \mathbb{E}
        \biggl[
         \sup_{\mathbf{M} \in \mathcal{M}^S_{kl}} \mathrm{QF}_{kl}^S(
\mathbf{M}, \mathbf{X}
        )
        \biggr].
\end{equation*}%
To apply Lemma \ref{lemma:uniform_hanson}, 
 we need to bound the following two quantities:
    \begin{align*}
 \sup_{\mathbf{M} \in {\mathcal{M}^S_{kl}}} \|\mathbf{M}\|_{\mathrm{op}}
 \quad\text{and}\quad
   \|\mathrm{vec}(\mathcal{E}_{S,G_k^\ast})\|_{{\mathcal{M}^S_{kl}}} &:=
	\mathbb{E} 
    \bigl[
    \sup_{\mathbf{M} \in {\mathcal{M}^S_{kl}}} \|
	(\mathbf{M} + \mathbf{M}^\top)
    \mathrm{vec}(\mathcal{E}_{S,G_k^\ast})\|_2 
    \bigr],
    \end{align*}
where bounding both terms critically depends on $\|\mathbf{A}^\mathbf{v}\|_{\mathrm{op}}$,
    and the expectation of supremum
    \begin{equation*}
        \mathbb{E}
        \biggl[
         \sup_{\mathbf{M} \in \mathcal{M}^S_{kl}} \mathrm{QF}_{kl}^S(
\mathbf{M}, \mathbf{X}
        )
        \biggr].
    \end{equation*}
    We present the bounds in order.
    
\medskip \noindent \textbf{2. Bounding the operator norm dependent terms.}
{Let $\mathbb{V}_k:=\{\mathbf{v}_{G_k^\ast}:\mathbf{v}\in\Gamma_K,\|\mathbf{v}\|_2=1\}$.}
	We start by bounding $\sup_{\mathbf{M} \in {\mathcal{M}^S_{kl}}} \|\mathbf{M}\|_{\mathrm{op}}$.
	First note that for $\mathbf{M} \in {\mathcal{M}^S_{kl}}$, its operator norm can be simplified into:
\begin{equation*}
   \|\mathbf{M}\|_{\mathrm{op}}=\|\mathbf{A}^{\mathbf{v}} \otimes \mathbf{I}_{|S|}\|_{\text{op}} = \|\mathbf{A}^{\mathbf{v}}\|_{\text{op}} \|\mathbf{I}_{|S|}\|_{\text{op}} = \|\mathbf{A}^{\mathbf{v}}\|_{\text{op}}. 
\end{equation*}
Therefore it suffices to bound the supremum of 	$\|\mathbf{A}^{\mathbf{v}}\|_{\text{op}}$:
	\begin{align*}
		\|\mathbf{A}^{\mathbf{v}}\|_{\text{op}} 
        &= 
        \max_{
        \|\mathbf{u}\|_2=1,
        \mathbf{u} \in \mathbb{R}^{|G_k^\ast|}
        } \mathbf{u}^T \mathbf{A}^{\mathbf{v}} \mathbf{u} 
		\\&= 
        \max_{
        \|\mathbf{u}\|_2=1,
        \mathbf{u} \in \mathbb{R}^{|G_k^\ast|}
        } 
\sum_{\{(i, j) \in G_k^\ast\}} 
\hspace{-1em}
        u_i u_j 
        \frac{v_i + v_j}{2}
		\\&=
		\max_{
        \|\mathbf{u}\|_2=1,
        \mathbf{u} \in \mathbb{R}^{|G_k^\ast|}
        } 
		\bigl(
		\sum_{i=1}^{|G_k^\ast|} u_i v_i 
		\bigr) 
		\bigl(
		\sum_{j=1}^{|G_k^\ast|} u_j 
		\bigr)
		\\&\stackrel{(i)}{\leq} 
        \max_{
        \|\mathbf{u}\|_2=1,
        \mathbf{u} \in \mathbb{R}^{|G_k^\ast|}
        } \bigl(\sum_{i=1}^{|G_k^\ast|} u_i^2 \bigr)^{1/2} \bigl(\sum_{i=1}^{|G_k^\ast|} v_i^2 \bigr)^{1/2} \bigl(\sum_{j=1}^{|G_k^\ast|} u_j^2 \bigr)^{1/2} \sqrt{|G_k^\ast|}
		\\&
        \overset{(ii)}{\leq}  \sqrt{|G_k^\ast|},
	\end{align*}
	where step $(i)$ uses Cauchy-Schwarz inequality,
	and
	step  $(ii)$ uses $\| \mathbf{v} \|_2 \leq 1$ for $\mathbf{v} \in \mathbb{V}_k$.
	Since the last term do not depend on $\mathbf{v}$, we have
	\begin{equation}\label{sup_A_op}
		\sup_{\mathbf{M} \in {\mathcal{M}^S_{kl}}} \|\mathbf{M}\|_{\mathrm{op}} \leq  \sqrt{|G_k^\ast|}.
	\end{equation}
	 Next, we bound $\|\mathrm{vec}(\mathcal{E}_{S,G_k^\ast})\|^2_{{\mathcal{M}^S_{kl}}}$:
	\begin{align*}
		\|\mathrm{vec}(\mathcal{E}_{S,G_k^\ast})\|_{{\mathcal{M}^S_{kl}}}^2
		&= 
		\bigl(
		\mathbb{E} 
		\bigl[ 
		\sup_{\mathbf{M} \in {\mathcal{M}^S_{kl}}} \|(\mathbf{M} + \mathbf{M}^\top) \; \mathrm{vec}(\mathcal{E}_{S,G_k^\ast})\|_2 
		\bigr]
		\bigr)^2
		\\&\leq  
		\bigl(
		\mathbb{E} 
		\bigl[ \sup_{\mathbf{M} \in {\mathcal{M}^S_{kl}}} \|(\mathbf{M} + \mathbf{M}^\top)\|_{\mathrm{op}} \| \; \mathrm{vec}(\mathcal{E}_{S,G_k^\ast}) \|_2 
		\bigr]
		\bigr)^2
		\\&
		\stackrel{(i)}{\leq}
		\bigl(
		\mathbb{E} 
		\bigl[ 2 |G_k^\ast|^{1/2}  \| \mathrm{vec}(\mathcal{E}_{S,G_k^\ast}) \|_2 \bigr]
		\bigr)^2
		\\& \stackrel{(ii)}{\leq}
		4 |G_k^\ast| \;
		\mathbb{E} \bigl[ \| \mathrm{vec}(\mathcal{E}_{S,G_k^\ast})\|_2^2
        \bigr]
\\& =
		4 |G_k^\ast| 
        \sum_{i \in G_k^\ast}
		\mathbb{E} \bigl[ \|\mathcal{E}_{S,i}\|_2^2
        \bigr]
		\\& =
		4 |G_k^\ast|^2  |S|,
	\end{align*}
	where step $(i)$ uses \eqref{sup_A_op}, 
	step $(ii)$ applies Jensen's inequality to a convex function $x \mapsto x^2$.

  \medskip \noindent \textbf{3. Bounding the expectation of the supremum.}
It is more convenient to start from \eqref{uprocess_assymmetric} and decompose the quadratic form $\mathrm{QF}_{kl}^S(\mathbf{A}^{\mathbf{v}}, \mathbf{X})$ into the following two terms:
\begin{equation*}
  \mathrm{QF}_{kl}^S(
\mathbf{A}^{\mathbf{v}}, \mathbf{X})
= \hspace{-1em}
		\sum_{ \{(i, j) \in G_k^\ast : i \neq j\} } 
        \hspace{-1em}
        v_j \bigl\langle
		\mathcal{E}_{S,i},
		\mathcal{E}_{S,j}
		\bigr\rangle
        +
		\sum_{i \in G_k^*} v_i 
		\| \mathcal{E}_{S,i} \|_2^2.
\end{equation*}
For these two terms, the index set ${\mathcal{M}^S_{kl}}$, which indexes the quadratic form, is more suitably represented by
	\[
	\mathbb{V}_k := \left\{ \mathbf{v}_{G_k^\ast} : \mathbf{v} \in \Gamma_K, \|\mathbf{v}\|_2 = 1 \right\}.
	\]  
We provide the bound for each of these two terms, starting with the first one.
	First note that given $\mathbf{v} \in \mathbb{V}_k$, we have
	\begin{align*}
\sup_{\mathbf{v} \in \mathbb{V}_k}
		\sum_{ \{(i, j) \in G_k^\ast : i \neq j\} }  
        \hspace{-1em}
        v_j \bigl\langle
		\mathcal{E}_{S,i},
		\mathcal{E}_{S,j}
		\bigr\rangle
		&=
		\sup_{\mathbf{v} \in \mathbb{V}_k}
		\bigl|
		\sum_{ \{(i, j) \in G_k^\ast : i \neq j\} }  
        \hspace{-1em}
        v_j \bigl\langle
		\mathcal{E}_{S,i},
		\mathcal{E}_{S,j}
		\bigr\rangle
		\bigr|
		\\
		&= 
		\sup_{\mathbf{v} \in \mathbb{V}_k}
		\bigl| \sum_{j \in G_k^\ast} v_j \bigl\langle 
        \mathcal{E}_{S,j}, 
        \sum_{i  \in G_k^\ast, i \neq j} \mathcal{E}_{S,i} \bigr\rangle \bigr|
		\\
		&\stackrel{(i)}{\leq} 
		\sup_{\mathbf{v} \in \mathbb{V}_k}
		\bigl( \sum_{j \in G_k^\ast} v_j^2 \bigr)^{1/2} 
		\bigl( \sum_{j \in G_k^\ast}  
		\bigl\langle 
        \mathcal{E}_{S,j}, 
        \sum_{i \in G_k^\ast, i \neq j} \mathcal{E}_{S,i}
		\bigr\rangle ^2 \bigr)^{1/2}
		\\&\stackrel{(ii)}{\leq} 
        \bigl( \sum_{j \in G_k^\ast}
		\bigl\langle 
        \mathcal{E}_{S,j}, 
        \sum_{i \in G_k^\ast, i \neq j} \mathcal{E}_{S,i}
		\bigr\rangle ^2 \bigr)^{1/2},
	\end{align*}
	where
	step $(i)$ uses Cauchy-Schwarz inequality
	and
	step $(ii)$ uses $\| \mathbf{v} \|_2 \leq 1$ for $\mathbf{v} \in \mathbb{V}_k$.  
	Using this bound, we can upper bound the expectation as follows:
	\begin{align*}
		\mathbb{E} 
                \biggl[
        \sup_{\mathbf{v} \in \mathbb{V}_k}
		\sum_{ \{(i, j) \in G_k^\ast : i \neq j\} }  
        \hspace{-1em}
        v_j \bigl\langle
		\mathcal{E}_{S,i},
		\mathcal{E}_{S,j}
		\bigr\rangle
        \biggr]
		&\leq
		\mathbb{E}
        \biggl[
        \bigl( \sum_{j \in G_k^\ast}
			\bigl\langle 
        \mathcal{E}_{S,j}, 
        \sum_{i \in G_k^\ast, i \neq j} \mathcal{E}_{S,i}
		\bigr\rangle ^2 \bigr)^{1/2}
    \biggr]
		\\&
		\stackrel{(i)}{\leq} \biggl( \sum_{j \in G_k^\ast} \mathbb{E} \bigl[ \bigl\langle 
		(\boldsymbol{\varepsilon}_j)_S, \sum_{i \neq j, i \in [n_k]} \mathcal{E}_{S,i} 
		\bigr\rangle^2 \bigr] \biggr)^{1/2}
		\\&
		\stackrel{(ii)}{=} \bigl( \sum_{j \in G_k^\ast} \mathbb{E}
		\| \hskip -0.7em
		\sum_{i \in G_k^\ast, i \neq j}
        \hskip -0.7em
		\mathcal{E}_{S,i}
		\|_2^2\;
		\bigr)^{1/2}
		\\&
		\stackrel{(iii)}{\leq} 
        \bigl(
        |G_k^\ast| (|G_k^\ast|-1)|S| \bigr)^{1/2}
		\\&
		\leq   |G_k^\ast| \sqrt{|S|},
        \numberthis \label{esupvee}
	\end{align*}
	where step $(i)$ uses Jensen's inequality to a concave function $x \mapsto x^{1/2}$,
	step $(ii)$ the independence and zero-mean assumption,
	and step $(iii)$ uses isotropic Gaussian assumption.

      Next, we bound the expectation of the supremum of $\sum_{i \in G_k^*} v_i 
		\| \mathcal{E}_{S,i} \|_2^2$.
	\begin{align*}
		\mathbb{E} \sup_{\mathbf{v} \in \mathbb{V}_k} 
		\bigl|
		\sum_{i \in G_k^*} v_i 
		\| \mathcal{E}_{S,i} \|_2^2 
		\bigr|
		&\overset{(i)}{=} \mathbb{E} \sup_{\mathbf{v} \in \mathbb{V}_k} 
		\bigl|
		\sum_{i \in G_k^*} v_i 
		\bigr(
		\| \mathcal{E}_{S,i} \|_2^2 
		-
		|S|
		\bigl)
		\bigr|
		\\
		&\stackrel{(ii)}{\leq}
		\mathbb{E} \sup_{\mathbf{v} \in \mathbb{V}_k} 
		\bigl( 
		\sum_{i \in G_k^*} v_i^2 
		\bigr)^{1/2} 
		\biggl\{  \sum_{i \in G_k^*} \left( \|  \mathcal{E}_{S,i} \|_2^2  
		-
		 |S|
		\right)^2 \biggr\}^{1/2}
		\\& \stackrel{(iii)}{\leq}
		\mathbb{E} 
		\biggl\{  \sum_{i \in G_k^*} \left( \|  \mathcal{E}_{S,i} \|_2^2  
		-
		 |S|
		\right)^2 \biggr\}^{1/2}
		\\&
		\stackrel{(iv)}{\leq}
		\biggl\{ \sum_{i \in G_k^*} \mathbb{E} \left( \|\mathcal{E}_{S,i}\|_2^2 -  |S| \right)^2 \biggr\}^{1/2} 
		\\&=
		\bigl( \sum_{i \in G_k^*}
		\mathrm{Var}  \|\mathcal{E}_{S,i}\|_2^2  \bigr)^{1/2} 
		\\&\overset{(v)}{=}
		\sqrt{
			2 |G_k^\ast|
			|S| },
            \numberthis \label{esvesq}
	\end{align*}
where step $(i)$ uses $\sum_{i \in G_k^\ast} v_i = 0$, which follows from the fact that ${\Gamma_K}$ is orthogonal to $\mathbbm{1}_{G_k^\ast}$,
 step $(ii)$ uses     Cauchy-Schwarz inequality,
step $(iii)$     uses  $\|\mathbf{v}\|_2 \leq 1$, 
step $(iv)$ 	uses Jensen's inequality to a concave function $x \mapsto x^{1/2}$,
    and step $(v)$ uses our Gaussian noise assumption and the variance formula for Chi-square random variables.

Combining \eqref{esupvee} and \eqref{esvesq}, we conclude that
\begin{equation*}
    \mathbb{E}
        \biggl[
         \sup_{\mathbf{M} \in \mathcal{M}^S_{kl}} \mathrm{QF}_{kl}^S(
\mathbf{M}, \mathbf{X}
        )
        \biggr] 
        \lesssim 
         |G_k^\ast| \sqrt{|S|}.
\end{equation*}

    \medskip \noindent \textbf{4.  Individual bound using Hanson-wright inequality.}
	Applying Lemma \ref{lemma:uniform_hanson} with the bounds established in the previous steps and using the formulation \eqref{uprocess_assymmetric} of $G^S_{1, kl}(\mathbf{v})$, we define an event
	\begin{equation*}
		\mathcal{C}_{N, kl}^S :=
		\biggl\{
        \sup_{ \mathbf{v} \in \mathbb{V}_k }
                  \frac{|G_l^\ast|}{|G_k^\ast|}
\sum_{
       \{(i, j) \in G_k^\ast \}
        }
        \hspace{-1em}
        v_i 
		\left\langle
		\mathcal{E}_{S,i},
		\mathcal{E}_{S,j}
		\right\rangle
		\lesssim
          \frac{|G_l^\ast|}{|G_k^\ast|}
          \underbrace{
			 |G_k^\ast| \sqrt{|S|}
                }_{   
                \hspace{-10em}
                \mathbb{E}
        \bigl[
         \sup_{\mathbf{M} \in \mathcal{M}^S_{kl}} \mathrm{QF}_{kl}^S(
\mathbf{M}, \mathbf{X}
        )
        \bigr] \lesssim}
                ~+~
                \frac{|G_l^\ast|}{|G_k^\ast|}
		\zeta^S_{N, kl}
		\biggr\},
	\end{equation*}
  where 	$\zeta^S_{N, kl} :=   |G_k^\ast|  \sqrt{
		|S| (  \log n + 2\log |S| + \log {{\binom{p}{|S|}}})
	}$, then we have
	\begin{align*}
		\mathbb{P}\biggl(
		(\mathcal{C}_{N, kl}^{S})^C
		\biggr) 
		 &\lesssim 
          \exp 
          \biggl[ 
          - \min 
          \biggl( 
          \underbrace{
          \frac{(\zeta^S_{N, kl})^2}{|G_k^\ast|^2 |S|}
          }_{
         \hspace{-7em}
        \|\mathrm{vec}(\mathcal{E}_{S,G_k^\ast})\|_{{\mathcal{M}^S_{kl}}}^2 \lesssim 
         }
          , \quad
          \underbrace{
		\frac{\zeta^S_{N, kl}}{
       \sqrt{ |G_k^\ast|}
        } }_{
        \hspace{-3em}
\sup_{\mathbf{M} \in {\mathcal{M}^S_{kl}}} \|\mathbf{M}\|_{\mathrm{op}}
\lesssim 
        }
        \biggr) 
        \biggr]
		\\& \lesssim \frac{1}{n |S|^2  {\binom{p}{|S|}}}.
	\end{align*}
The proof is complete once we show that the intersection event
\begin{equation*}
        \bigcap_{S \in  [p], |S| \leq \sqrt{p}}
        \bigcap_{1 \leq k \neq l \leq K}
        \hspace{-1em}
        \mathcal{C}_{N, kl}^{S}
    \end{equation*}
    holds with high probability.
    
\medskip \noindent \textbf{5. Uniform bound.}
 Now we show that the event $\mathcal{C}_{N, kl}^S$ holds uniformly across all $S \in [p]$ such that $|S| \leq \sqrt{p}$ and $1 \leq k \neq l \leq K$, with probability at least $1-C_{11}/n$, where $C_{11}>0$ is a constant.
  \begin{align*}
		\mathbb{P}
        \bigl(
        \hspace{-1em}
        \bigcup_{S \in  [p], |S| \leq \sqrt{p}}
        \bigcup_{1 \leq k \neq l \leq K}
        \hspace{-1em}
        (\mathcal{C}_{N, kl}^{S})^C
        \bigr)
		&\leq
        \hspace{-.4em}
        \sum_{1 \leq k \neq l \leq K}
		\sum_{q = 1}^{\lfloor \sqrt{p} \rfloor} 
		\sum_{|S| = q}
        \mathbb{P}((\mathcal{C}_{N, kl}^{S})^C)
			\\ &\lesssim
		\sum_{q = 1}^{\lfloor \sqrt{p} \rfloor} 
		\sum_{|S| = q}
		\frac{K^2}{n |S|^2  {\binom{p}{|S|}}}
			\\& =
		\sum_{q = 1}^{\lfloor \sqrt{p} \rfloor} 
		\frac{K^2}{n q^2}
			\\& \leq
		\frac{K^2 \pi^2}{6n },
	\end{align*}
where the last
	step  uses
	$
	\sum_{q=1}^\infty = \pi^2/6.
	$
Since $|S| \leq \sqrt{p}$, $|S| \log p$ dominates $\log|S|$. Also, we treat $K$ as a constant. Therefore, we conclude we conclude that
uniformly across $S \in \mathcal S$ and $(k,l)$ such that $1 \leq k \neq l \leq K$,
with probability at least $1 - C_{11}/n$, it holds that
\begin{equation*}
\sup_{\mathbf{v} \in \Gamma_K, \|\mathbf{v} \|_2 = 1}
G^S_{1, kl}(\mathbf{v})
\lesssim
|G_l^\ast|
\sqrt{
|S| (  \log n
+ 
|S| \log p
)}
\end{equation*}
	This completes the proof of Lemma \ref{lemma:upperbound_G1ksv}.
\end{proof}

\subsubsection{Proof of Lemma \ref{lemma:ub_g2klsv}}\label{section:proof:lemma:ub_g2klsv}
The proof proceeds in the following steps.
\begin{enumerate}
    \item Conditional analysis using a concentration inequality for the supremum of a Gaussian process,
    \item Individual bound by Chi-square tail bound,
    \item Uniform bound
\end{enumerate}
\begin{proof}	For notational simplicity, let $\sigma^2 = 1$.

\medskip \noindent \textbf{1.  Conditional analysis.}
Since the sum of inner products of Gaussian random vectors is challenging to analyze directly, we first condition on \( \overline{\mathcal{E}}_{S,l}^\ast \) and then apply the concentration inequality for the supremum of a Gaussian process (Lemma \ref{lemma:suprema_concentration}).
Fix $(k,l)$ such that $1 \leq k \neq l \leq K$ and $S \subset [p]$ such that $|S| \leq \sqrt{p}$.
	Conditioned on \( \overline{\mathcal{E}}_{S,l}^\ast \), \( G^S_{2, kl}(\mathbf{v}) \) can be viewed as a canonical Gaussian process \( 
    \langle G, \mathbf{w} \rangle 
    \), where \( G \sim \mathcal{N}(0, \mathbf{I}_{|G_k^\ast|}) \) and \( \mathbf{w} \in \mathbb{W}_k \), with  
	\[
	\mathbb{W}_k := \left\{ \| \overline{\mathcal{E}}_{S,l}^\ast\|_2 \mathbf{v}_{\upharpoonright G_k^\ast} : \mathbf{v} \in \Gamma_K, \|\mathbf{v}\|_2 = 1 \right\},
	\]  
	where \( \mathbf{v}_{\upharpoonright G_k^*} \) is the restriction of \( \mathbf{v} \in \Gamma_K \) onto \( G_k^\ast \), and  the canonical metric on \( \mathbb{W}_k \) is the \( \ell_2 \)-metric.

    To apply Lemma \ref{lemma:suprema_concentration}, we have to bound
    \begin{align*}
\mathrm{EC}_{kl}^S(\overline{\mathcal{E}}_{S,l}^\ast) 
&:= 
        \mathbb{E}\bigl[
\sup_{\mathbf{v} \in \Gamma_K, \|\mathbf{v}\|_2 = 1}
        \sum_{i \in G_k^*}  v_i 
		\bigl \langle
\overline{\mathcal{E}}_{S,l}^\ast, \mathcal{E}_{S,i}
		\bigr \rangle \;| \;\overline{\mathcal{E}}_{S,l}^\ast 
        \bigr]
        =
   \mathbb{E}\bigl[
\sup_{\mathbf{w} \in \mathbb{W}_k}
      \langle G, \mathbf{w} \rangle \;| \;\overline{\mathcal{E}}_{S,l}^\ast 
        \bigr]~\text{, and}
\\
\mathrm{SVC}_{kl}^S(\overline{\mathcal{E}}_{S,l}^\ast ) &:= 
\sup_{\mathbf{v} \in {\Gamma_K}, \|\mathbf{v}\|_2 = 1}
\mathrm{Var}
\bigl(
        \sum_{i \in G_k^*}  v_i 
		\bigl \langle
\overline{\mathcal{E}}_{S,l}^\ast, \mathcal{E}_{S,i}
		\bigr \rangle \;| \;\overline{\mathcal{E}}_{S,l}^\ast 
        \bigr)
=
\sup_{\mathbf{w} \in \mathbb{W}_k} \mathrm{Var}
\bigl(
\langle G, \mathbf{w} \rangle \;| \;
\overline{\mathcal{E}}_{S,l}^\ast \bigr).
    \end{align*}
    We bound $ \mathrm{EC}_{kl}^S(\overline{\mathcal{E}}_{S,l}^\ast)$ using Dudley's integral inequality (Lemma \ref{lemma:dudley}):
\begin{align*}
\mathrm{EC}_{kl}^S(\overline{\mathcal{E}}_{S,l}^\ast) 
		\lesssim \int_0^\infty \sqrt{\log N(\mathbb{W}_k, \epsilon)} \, d\epsilon
		&\overset{(i)}{\lesssim}\|\overline{\mathcal{E}}_{S,l}^\ast\|_2 \int_0^\infty \sqrt{|G_k^\ast| \log(1 / \varepsilon)} \, d\varepsilon
		&=  \|\overline{\mathcal{E}}_{S,l}^\ast\|_2
        \sqrt{|G_k^\ast|},
	\end{align*} 
    where step $(i)$ uses the fact that the $\varepsilon$-covering entropy of the unit sphere in $\mathbb{R}^{|G_k^\ast|}$ is at most $ |G_k^\ast| \log(1/\varepsilon)$, up to a constant factor,  for any $\varepsilon \in (0, 1)$.

	Next, we bound \( \mathrm{SVC}_{kl}^S(\overline{\mathcal{E}}_{S,l}^\ast ) \):  
	\[
\mathrm{SVC}_{kl}^S(\overline{\mathcal{E}}_{S,l}^\ast ) 
= 
\sup_{\mathbf{w} \in \mathbb{W}_k} \mathrm{Var}
\bigl(
\langle G, \mathbf{w} \rangle \;| \;
\overline{\mathcal{E}}_{S,l}^\ast \bigr) 
	= 
    \sup_{\mathbf{w} \in \mathbb{W}_k} 
\|\overline{\mathcal{E}}_{S,l}^\ast \|_2^2 
    \sum_{i \in G_k^\ast} v_i^2 
	\leq
\|\overline{\mathcal{E}}_{S,l}^\ast\|_2^2,
	\]  
	where the last inequality follows from \( \|\mathbf{v}\|_2 = 1 \) for \( \mathbf{v} \in \Gamma_K \).  
Let us denote $ \xi^S = \log(n  |S|^2  {\binom{p}{|S|}})$.
Define the following conditional probability
      as follows:
	\begin{equation*}
		\mathrm{CP}_{N,kl}^S:=
        \mathbb{P}
		\biggl(
		\sup_{\mathbf{v} \in \Gamma_K, \| \mathbf{v} \|_2 = 1}
		|G_{2, kl}^S(\mathbf{v})|
		\gtrsim
           |G_l^\ast| 
        \underbrace{
      \|\overline{\mathcal{E}}_{S,l}^\ast\|_2 \sqrt{|G_k^\ast|  
		}
        }_{
       \gtrsim  \mathrm{EC}_{kl}^S(\overline{\mathcal{E}}_{S,l}^\ast)   }
        +
           |G_l^\ast|
        \underbrace{
        \|\overline{\mathcal{E}}_{S,l}^\ast\|_2
        }_{
\gtrsim
       \sqrt{ \mathrm{SVC}_{kl}^S(\overline{\mathcal{E}}_{S,l}^\ast })  
        }
		\sqrt{\xi^S }
        \; | \; \overline{\mathcal{E}}_{S,l}^\ast
		\biggr),
	\end{equation*}
Then by Lemma \ref{lemma:suprema_concentration},
we have \begin{equation}\label{cpnkls}
\mathrm{CP}_{N,kl}^S
\leq
\frac{1}{n |S|^2 {\binom{p}{|S|}}}.
\end{equation}

\medskip \noindent \textbf{2. Chi-square tail bound}
    Now we use the Chi-square tail inequality to further bound 
    	$ \|\overline{\mathcal{E}}_{S,l}^\ast\|_2$ and $ \|\overline{\mathcal{E}}_{S,l}^\ast\|_2^2$.
        If we define the following event:
	$$
	\mathcal{B}_{N,kl}^S
	:= 
	\left\{\|\overline{\mathcal{E}}_{S,l}^\ast\|_2^2 
    \leq 
    \frac{1}{|G_l^\ast|} \left( |S| + 2\sqrt{
		|S| 
		 \xi^S 
	} + 2 \xi^S \right) \right\}.
	$$
	By Lemma \ref{lemma:chisq_tail} , we have
	$$
	\mathbb{P}
	\bigl(
	 (\mathcal{B}_{N,kl}^S)^C
	\bigr)
	\leq
\frac{1}{n |S|^2 {\binom{p}{|S|}}}.
	$$
Combining the bounds in $\mathrm{CP}_{N,kl}^S$ and $\mathcal{B}_{N,kl}^S$, we define an event:
	\begin{equation*}
		\mathcal{A}_{N,kl}^S:=
		\biggl\{
		\sup_{\mathbf{v} \in \Gamma_K, \| \mathbf{v} \|_2 = 1}
		|G_{2, kl}^S(\mathbf{v})|
		\lesssim
       \sqrt{|G_l^\ast|}
      \sqrt{ |S| + 2\sqrt{
		|S| 
		 \xi^S 
	} + 2 \xi^S }
     ( \sqrt{|G_k^\ast|  +\sqrt{\xi^S })
		}	
		\biggr\}.
	\end{equation*}
The proof is complete once we show that the intersection event
	\begin{equation*}
		\bigcap_{1 \leq k \neq l \leq K} 
        \bigcap_{S \in [p], |S| \leq \sqrt{p}}
    \mathcal{A}^{S}_{N,kl}
        \end{equation*}
        holds with high probability.
        
    \medskip \noindent \textbf{3. Uniform bound.}
	We have
	\begin{align*}
		\mathbb{P} \biggl( 
		\bigcup_{1 \leq k \neq l \leq K} &
        \bigcup_{S \in [p], |S| \leq \sqrt{p}}
        {  (\mathcal{A}^{S}_{N,kl})^C}
		\biggr) 
        \\&\leq
		\sum_{1 \leq k \neq l \leq K}
		\sum_{q=1}^{\lfloor \sqrt{p} \rfloor}
        \sum_{|S| = q}
		\mathbb{P} { \bigl( (\mathcal{A}^{S}_{N,kl})^C \bigr)}
        \\& =
        		\sum_{1 \leq k \neq l \leq K}
		\sum_{q=1}^{\lfloor \sqrt{p} \rfloor}
        \sum_{|S| = q}
        \mathbb{E}
        \bigl[
		\mathbb{P}  \bigl( (\mathcal{A}^{S}_{N,kl})^C 
        \; | \;
        \overline{\mathcal{E}}_{S,l}^\ast
        \bigr)
        \bigr]
                \\& \leq
        		\sum_{1 \leq k \neq l \leq K}
			\sum_{q=1}^{\lfloor \sqrt{p} \rfloor}
        \sum_{|S| = q}
        \mathbb{E}
        \bigl[
		\mathbb{P}  \bigl( (\mathcal{A}^{S}_{N,kl})^C 
        \; | \;
        \overline{\mathcal{E}}_{S,l}^\ast
        \bigr) \mathbbm{1}(\mathcal{B}^{S}_{N, kl})
        + 
        \mathbbm{1}
        \bigl( (\mathcal{B}^{S}_{N, kl})^C
        \bigr)
        \bigr]
		\\& \overset{(i)}{\leq}
        \left(
		\sum_{1 \leq k \neq l \leq K}
		\sum_{q=1}^{\lfloor \sqrt{p} \rfloor}
		\sum_{ |S|=q}
        \mathbb{E}
        \bigl[
		\mathrm{CP}_{N,kl}^S \;
        \mathbbm{1}(
		\mathcal{B}^{S}_{N, kl} )
        \bigr]
        \right)
		+
		\frac{K^2 \pi^2}{6n}
		\\& \overset{(ii)}{\lesssim}
		\frac{K^2\pi^2}{6n}
		+
		\frac{K^2\pi^2}{6n},
	\end{align*}
    where step $(i)$ uses  $
	\sum_{q=1}^\infty
    (1/q^2)
    = \pi^2/6
	$, and step $(ii)$ also uses it and additionally uses \eqref{cpnkls}.
	
	Treating $K$ as a constant, we conclude that uniformly across $S \in \mathcal S$ and $(k,l)$ such that $1 \leq k \neq l \leq K$,
with probability at least $1 - C_{12}/n$, where $C_{12}>0$ is a constant, it holds that
	\begin{equation*}
		\sup_{\mathbf{v} \in \Gamma_K, \| \mathbf{v} \|_2 = 1}
		|G_{2, kl}^S(\mathbf{v})|
		\lesssim
       \sqrt{|G_l^\ast|}
      \sqrt{ |S| + 2\sqrt{
		|S| 
		 \xi^S 
	} + 2 \xi^S }
      \sqrt{|G_k^\ast|  +\sqrt{\xi^S }
		}.	
	\end{equation*}
	This completes the proof of Lemma \ref{lemma:ub_g2klsv}.
\end{proof}

\subsubsection{Proof of Lemma \ref{lemma:T2_dominance}}\label{section:proof:lemma:dominance_T2}
Leveraging the decomposition of $T_{2, kl}^S$ into $G_{1, kl}^S$ and $G_{2, kl}^S$ given in Appendix \ref{section:proof:NkliS}, the proof proceeds in the following intermediate steps:
\begin{enumerate}
    \item Problem decomposition
    \item Inequality for the $G_1 \times G_1$ term
    \item Inequality for the $G_2 \times G_2$ term
    \item Conclusion
\end{enumerate}

\begin{proof}
For notational simplicity, let $\sigma^2 = 1$.

\medskip \noindent \textbf{1. Problem decomposition.}
Recall that  we defined
$\Delta_{S \cap S^\ast}^2 = 
\min_{1 \leq k \neq l \leq K} \| (\mathbf{c}_l^\ast - \mathbf{c}_k^\ast )_{S \cap S^\ast} \|_2^2$ in Theorem \ref{theorem:separation_condition}, and
$ \dot{\lambda}^S = |S| + m \Delta_{S \cap S^\ast }^2/4$ in Lemma \ref{lemma:lambda_between}.
\begin{align*}
  &\sup_{\mathbf{v} \in \Gamma_K, \| \mathbf{v} \|_2  =1}
\sum_{1 \leq k \neq l \leq K} 
		\frac{T^S_{2,kl}(\mathbf{v})}{t^S_{lk}(\dot{\lambda}^S)}  
   \\& \overset{(i)} {=}
   \sum_{1 \leq k \neq l \leq K} 
   \sup_{\mathbf{v} \in \Gamma_K, \| \mathbf{v} \|_2  =1}
		\frac{
        N^S_{kl}(\mathbf{v}, \mathbf{X})
\;
    N^S_{lk}(\mathbf{v}, \mathbf{X})
        }{
        t^S_{lk}(\dot{\lambda}^S)
        }   
  \\&  \overset{(ii)} {\lesssim}
  \sum_{1 \leq k \neq l \leq K} 
   \sup_{\mathbf{v} \in \Gamma_K, \| \mathbf{v} \|_2  =1}
\frac{N^S_{kl}(\mathbf{v}, \mathbf{X}) \;N^S_{lk}(\mathbf{v}, \mathbf{X})}{
 |G_l^\ast|  |G_k^\ast|
 \bigl(
         \|
        (\mathbf{c}^\ast_k -\mathbf{c}^\ast_l)_{S \cap S^\ast}
        \|_2^2 + |S| \bigr)
}
  \\&  =
  \sum_{1 \leq k \neq l \leq K} 
   \sup_{\mathbf{v} \in \Gamma_K, \| \mathbf{v} \|_2  =1}
\frac{
\bigl( 
G_{1,kl}^S(\mathbf{v}) + G_{2,kl}^S(\mathbf{v})
\bigr)
\bigl(
G_{1,lk}^S(\mathbf{v}) + G_{2,lk}^S(\mathbf{v})
\bigr)
}{
 |G_l^\ast|  |G_k^\ast|
 \bigl(
         \|
        (\mathbf{c}^\ast_k -\mathbf{c}^\ast_l)_{S \cap S^\ast}
        \|_2^2 + |S| \bigr)
}
\end{align*}
where step $(i)$ uses the definition of $  T^S_{2,kl}(\mathbf{v})$ stated in \eqref{def:T2}
and step $(ii)$ uses Lemma \ref{lemma:boun(D^t)^Csl}.
Expanding the numerator yields four terms. We establish the inequality for two of them, the $G_1 \times G_1$ and $G_2 \times G_2$ terms, as the cross terms can be easily handled by leveraging the results from these two cases. Before proceeding, we recall from \eqref{def:S-set} that $\mathcal{S}$ is defined as
\begin{equation*}
      \mathcal S   
      =
      \left\{S: S\subset [p], 
    |S|\leq \sqrt p, 	\Delta_{S \cap S^\ast}^2 \gtrsim
		\left(\log n +
        \frac{|S| \log p}{m}
		+
		\sqrt{
        \frac{|S| \log p}{m}
        }
		\right)\right\}.
\end{equation*}

\medskip \noindent \textbf{2. Inequality for the $G_1 \times G_1$ term.}
  By Lemma \ref{lemma:upperbound_G1ksv},  uniformly across $S \in \mathcal S$ and $(k,l)$ such that $1 \leq k \neq l \leq K$,
with probability at least $1 - C_{11}/n$, it holds that
\begin{align*}
  \sum_{1 \leq k \neq l \leq K} &
   \sup_{\mathbf{v} \in \Gamma_K, \| \mathbf{v} \|_2  =1}
\frac{
G_{1,kl}^S(\mathbf{v})
G_{1,lk}^S(\mathbf{v}) 
}{
 |G_l^\ast|  |G_k^\ast|
 \bigl(
         \|
        (\mathbf{c}^\ast_k -\mathbf{c}^\ast_l)_{S \cap S^\ast}
        \|_2^2 + |S| \bigr)
}
\\
& \overset{(i)}{\lesssim}
  \sum_{1 \leq k \neq l \leq K}
\frac{
		|S| (  \log n + |S| \log p)
}{
 \bigl(
         \|
        (\mathbf{c}^\ast_k -\mathbf{c}^\ast_l)_{S \cap S^\ast}
        \|_2^2 + |S| \bigr)
}
\\
& \lesssim
\frac{
		|S|  \log n + |S|^2 \log p
}{
 \bigl(
        \Delta_{S \cap S^\ast }^2 + |S| \bigr)
}.
\end{align*}
Therefore it suffices to show that
\begin{equation*}
		|S|  \log n + |S|^2 \log p
\lesssim 
\bigl(
|S| + m \Delta_{S \cap S^\ast }^2
\bigr)
\bigl(
        \Delta_{S \cap S^\ast }^2 + |S| \bigr),
\end{equation*}
which is true
since $|S|\Delta_{S \cap S^\ast }^2 \gtrsim |S| \log n$ and $m|S| \Delta_{S \cap S^\ast }^2 \gtrsim |S|^2 \log p$ under the separtion condition of $\mathcal{S}$.

\medskip \noindent \textbf{3. Inequality for the $G_2 \times G_2$ term.}
 By Lemma \ref{lemma:ub_g2klsv},  uniformly across $S \in \mathcal S$ and $(k,l)$ such that $1 \leq k \neq l \leq K$,
with probability at least $1 - C_{12}/n$, it holds that
\begin{align*}
  \sum_{1 \leq k \neq l \leq K} &
   \sup_{\mathbf{v} \in \Gamma_K, \| \mathbf{v} \|_2  =1}
\frac{
G_{2,kl}^S(\mathbf{v})
G_{2,lk}^S(\mathbf{v}) 
}{
 |G_l^\ast|  |G_k^\ast|
 \bigl(
         \|
        (\mathbf{c}^\ast_k -\mathbf{c}^\ast_l)_{S \cap S^\ast}
        \|_2^2 + |S| \bigr)
}
\\
& \overset{(i)}{\lesssim}
  \sum_{1 \leq k \neq l \leq K}
\frac{
	 \sqrt{|G_l^\ast|}
      \sqrt{ |S| + 2\sqrt{
		|S| 
		 \xi^S 
	} + 2 \xi^S }
      \sqrt{|G_k^\ast|  +\sqrt{\xi^S }
		}
     \sqrt{|G_k^\ast|}
      \sqrt{ |S| + 2\sqrt{
		|S| 
		 \xi^S 
	} + 2 \xi^S }
      \sqrt{|G_l^\ast|  +\sqrt{\xi^S }
		}
}{
 |G_l^\ast|  |G_k^\ast| 
        \bigl(  \|
        (\mathbf{c}^\ast_k -\mathbf{c}^\ast_l)_{S \cap S^\ast}
        \|_2^2 + |S| \bigr)
}
\\
& =
  \sum_{1 \leq k \neq l \leq K}
\frac{
  \bigl( |S| + 2\sqrt{
		|S| 
		 \xi^S 
	} + 2 \xi^S \bigr)
      \sqrt{
      |G_k^\ast|  |G_l^\ast|
      +
       \sqrt{\xi^S }
      \bigl(
      |G_k^\ast| +
      |G_l^\ast|
      \bigr)
      +
     \xi^S 
		}
}{
\sqrt{ |G_l^\ast|  |G_k^\ast|}
\bigl(
         \|
        (\mathbf{c}^\ast_k -\mathbf{c}^\ast_l)_{S \cap S^\ast}
        \|_2^2 + |S| \bigr)
}
\\
& =
  \sum_{1 \leq k \neq l \leq K}
\frac{
  \bigl( |S| + 2\sqrt{
		|S| 
		 \xi^S 
	} + 2 \xi^S \bigr)
    \bigl(
     \sqrt{
      |G_k^\ast|  |G_l^\ast|
      }
      +
      \sqrt{
      \sqrt{\xi^S }
      \bigl(
      |G_k^\ast| +
      |G_l^\ast|
      \bigr)
      }
      +
        \sqrt{
     \xi^S}
    \bigr)
		}
{
\sqrt{ |G_l^\ast|  |G_k^\ast|}
\bigl(
         \|
        (\mathbf{c}^\ast_k -\mathbf{c}^\ast_l)_{S \cap S^\ast}
        \|_2^2 + |S| \bigr)
},
\end{align*}
where
$ \xi^S = \log(n  |S|^2  {\binom{p}{|S|}})$.
The last term can be decomposed into three parts, and we establish an inequality for each one in turn, starting with the first term:
\begin{align*}
      \sum_{1 \leq k \neq l \leq K}&
\frac{
  \bigl( |S| + 2\sqrt{
		|S| 
		 \xi^S 
	} + 2 \xi^S \bigr)
    \bigl(
     \sqrt{
      |G_k^\ast|  |G_l^\ast|
      }
    \bigr)
		}
{
\sqrt{ |G_l^\ast|  |G_k^\ast|}
\bigl(
         \|
        (\mathbf{c}^\ast_k -\mathbf{c}^\ast_l)_{S \cap S^\ast}
        \|_2^2 + |S| \bigr)
}
\\&=
      \sum_{1 \leq k \neq l \leq K}
\frac{
  \bigl( |S| + 2\sqrt{
		|S| 
		 \xi^S 
	} + 2 \xi^S \bigr)
		}
{
\bigl(
         \|
        (\mathbf{c}^\ast_k -\mathbf{c}^\ast_l)_{S \cap S^\ast}
        \|_2^2 + |S| \bigr)
}
\\& \lesssim
\frac{
  \bigl( |S| + 
  |S| \sqrt{\log p } +  \sqrt{|S| \log n}
 + 
\log n + |S| \log p 
    \bigr)
		}
{
         \Delta_{S \cap S^\ast}^2
 + |S| 
}
\\& \lesssim
|S| + m \Delta_{S \cap S^\ast}^2,
\end{align*}
where the last inequality holds due to the sepration condition of $\mathcal{S}$.

Now, for the second term, 
we first recall from Theorem \ref{theorem:separation_condition} that
$m = 2\min_{1 \leq k \neq l \leq K}
\left\{
		\frac{|G_k^\ast| |G_l^\ast|}{ |G_k^\ast| + |G_l^\ast|}
        \right\}$.
Then we have:
\begin{align*}
      \sum_{1 \leq k \neq l \leq K}&
\frac{
  \bigl( |S| + 2\sqrt{
		|S| 
		 \xi^S 
	} + 2 \xi^S \bigr)
    \bigl(
      \sqrt{
      \sqrt{\xi^S }
      \bigl(
      |G_k^\ast| +
      |G_l^\ast|
      \bigr)
      }
    \bigr)
		}
{
\sqrt{ |G_l^\ast|  |G_k^\ast|}
\bigl(
         \|
        (\mathbf{c}^\ast_k -\mathbf{c}^\ast_l)_{S \cap S^\ast}
        \|_2^2 + |S| \bigr)
},
\\& \lesssim
\frac{
  \bigl( |S| + 2\sqrt{
		|S| 
		 \xi^S 
	} + 2 \xi^S \bigr)
      (\xi^S )^{1/4}
		}
{\sqrt{m}
\bigl(
        \Delta_{S \cap S^\ast}^2 + |S| 
        \bigr)
}
\\&\lesssim 
|S| + m \Delta_{S \cap S^\ast}^2,
\end{align*}
where the last inequality holds because under the separation condition of $\mathcal{S}$, we have
$ \bigl( |S| + 2\sqrt{
		|S| 
		 \xi^S 
	} + 2 \xi^S \bigr) \lesssim |S| + m \Delta_{S \cap S^\ast}^2$ and  $(\xi^S )^{1/4} \lesssim \sqrt{m}
\bigl(
        \Delta_{S \cap S^\ast}^2 + |S| 
        \bigr)$.

Finally, for the third term, we have
\begin{align*}
      \sum_{1 \leq k \neq l \leq K}&
\frac{
  \bigl( |S| + 2\sqrt{
		|S| 
		 \xi^S 
	} + 2 \xi^S \bigr)
        \sqrt{
     \xi^S}
		}
{
\sqrt{ |G_l^\ast|  |G_k^\ast|}
\bigl(
         \|
        (\mathbf{c}^\ast_k -\mathbf{c}^\ast_l)_{S \cap S^\ast}
        \|_2^2 + |S| \bigr)
}
\\& \lesssim 
      \sum_{1 \leq k \neq l \leq K}
\frac{
  \bigl( |S| + 2\sqrt{
		|S| 
		 \xi^S 
	} + 2 \xi^S \bigr)
        \sqrt{
     \xi^S}
		}
{
\sqrt{m}
\bigl(
         \|
        (\mathbf{c}^\ast_k -\mathbf{c}^\ast_l)_{S \cap S^\ast}
        \|_2^2 + |S| \bigr)
}
\\&\lesssim 
|S| + m \Delta_{S \cap S^\ast}^2,
\end{align*}
where the last inequality holds because under the separation condition of $\mathcal{S}$, we have
$ \bigl( |S| + 2\sqrt{
		|S| 
		 \xi^S 
	} + 2 \xi^S \bigr) \lesssim |S| + m \Delta_{S \cap S^\ast}^2$ and  $(\xi^S )^{1/2} \lesssim \sqrt{m}
\bigl(
        \Delta_{S \cap S^\ast}^2 + |S| 
        \bigr)$.

        \medskip \noindent \textbf{4. Conclusion.}
The inequalities for the cross terms can be verified in a straightforward manner. Combining all the inequalities established above, we conclude that, uniformly over $S \in \mathcal{S}$ and all $(k, l)$ with $1 \leq k \neq l \leq K$, the following holds with probability at least $1 - C_{13}/n$, where $C_{13} > 0$ is a constant, it holds that:
        \begin{equation*}
            \sup_{\mathbf{v} \in \Gamma_K, \| \mathbf{v} \|_2  =1}
\sum_{1 \leq k \neq l \leq K} 
		\frac{T^S_{2,kl}(\mathbf{v})}{t^S_{lk}(\dot{\lambda}^S)} .
        \end{equation*}
     This completes the    proof of Lemma \ref{lemma:T2_dominance}.
\end{proof}

\subsection{Proof of the Minimax Lower Bound}\label{section:proof:theorem:minimax_lower_bound}
The proof proceeds in the following intermediate steps:
\begin{enumerate}
\item Reduction to symmetric two-cluster problem \eqref{reduction:smaller_set},
\item Reduction to average case risk \eqref{reduction:average},
\item Reduction to  Bayes risk \eqref{def:bayes_rate},
\item Bounding the Bayes risk \eqref{bound_bayes_risk}.
\end{enumerate}
\begin{proof} We proceed by following the steps outlined above in sequence.
\paragraph*{1.  Reduction to symmetric two-cluster problem.}
We denote the separation by $\rho > 0$ and leave its value unspecified for now.
First, we define a reduced class of distributions consisting of two symmetric, sparse cluster centers.
\begin{equation*}
    \tilde{\Psi}(s,  \rho, 2) :=
    \tilde{\mathcal{M}}(s,  \rho, 2) \times \mathcal{H}(2)
\end{equation*}
where
\begin{align*}
	\tilde{\mathcal{M}}(s,  \rho, 2) := \bigl\{ 	
	( 
	\mathbf{c}_1,  \mathbf{c}_2
	):&
	\mathbf{c}_1 = \mathbf{c},
        \mathbf{c}_2 = -\mathbf{c},
        \mathbf{c} \in \mathbb{R}^p,
	\| \mathbf{c}_1  - \mathbf{c}_2\|_2^2 \geq \rho^2,
    \\
& \mathrm{supp}(\mathbf{c}) =  S^\ast \subset [p], |S^\ast| \leq s
	\bigr\},
\end{align*}
Then we have (by augmenting the matrices in $\mathcal{H}(2)$ with zero columns to match the  dimensions):
\begin{equation*}
	\tilde{\Psi}(s,  \rho, 2) \subset \Psi (s,  \rho , K).
\end{equation*}
This inclusion immediately implies that  
\begin{equation}\label{reduction:smaller_set}
	\inf_{\hat{\mathbf{H}}}
	\sup_{\boldsymbol{\psi} \in 	\Psi (s,  \rho , K)}
	\mathbb{P}_{\boldsymbol{\psi}} (\hat{\mathbf{H}}(\mathbf{X}) \neq \mathbf{H})
	\geq
	\inf_{\hat{\mathbf{H}}}
	\sup_{\boldsymbol{\psi} \in 	\tilde{\Psi}(s,  \rho, 2)}
	\mathbb{P}_{\boldsymbol{\psi}} (\hat{\mathbf{H}}(\mathbf{X}) \neq \mathbf{H}),
\end{equation}
since the supremum over a smaller set is less than or equal to that over a larger set.

\paragraph*{2. Reduction to average case risk.}
Next, we lower bound the worst-case risk in \eqref{reduction:smaller_set} by the  average risk.
\begin{align*}
	\inf_{\hat{\mathbf{H}}}
	\sup_{\boldsymbol{\psi} \in 	\tilde{\Psi}(s,  \rho, 2)}
	\mathbb{P}_{\boldsymbol{\psi}} (\hat{\mathbf{H}}(\mathbf{X}) \neq \mathbf{H})
    &=
    \inf_{\hat{\mathbf{H}}}
	\sup_{
    (\mathbf{c}_1,  \mathbf{c}_2) \in 
    \tilde{\mathcal{M}}(s,  \rho, 2)}
  \sup_{\mathbf{H} \in  \mathcal{H}(2)}
	\mathbb{P}_{\boldsymbol{\psi}} (\hat{\mathbf{H}}(\mathbf{X}) \neq \mathbf{H})
	\\& \geq
	\inf_{\hat{\mathbf{H}}}
	\sup_{
    (\mathbf{c}_1,  \mathbf{c}_2) \in 
    \tilde{\mathcal{M}}(s,  \rho, 2)}
  \mathbb{E}_{\mathbf{H}(\mathbf{X}) \sim   \pi}
	\mathbb{P}_{\boldsymbol{\psi}} (\hat{\mathbf{H}} \neq \mathbf{H}),
    \numberthis
    \label{reduction:average}
\end{align*}
where \( \pi \) is  a uniform prior on \( \mathcal{H}(2) \), formally given by,
for $ i=1, \ldots, n$,
\[
H_{i,1} \overset{\text{i.i.d.}}{\sim} \mathrm{Ber}(1/2), \quad H_{i2} = 1 - H_{i1}.
\]
\paragraph*{3. Reduction to Bayes risk.}
The risk \eqref{reduction:average} is equivalent to the minimax risk under zero-one loss, where the data is generated from a uniform hierarchical Gaussian mixture model defined as follows:\begin{align*}
H_{i,1} &\sim \mathrm{Ber}(1/2), \\
\mathbf{X}_i \mid H_{i,1} &\sim \mathcal{N}\left( (-1)^{H_{i,1}} \mathbf{c} \;, \mathbf{I}_p \right).
\end{align*}
In this setting, the algorithm that minimizes
$
    \mathbb{E}_{\mathbf{H} \sim \pi}  \mathbb{P}_{\boldsymbol{\psi}}\bigl( \hat{\mathbf{H}}(\mathbf{X}) \neq \mathbf{H} \bigr)
$
is the Bayes classifier $\mathbf{H}^{\mathbf{c}}$, 
which assumes knowledge of the true cluster centers. It operates by thresholding the Gaussian likelihood ratio, and under symmetric cluster centers with isotropic covariance, this yields the  decision rule
$
H^{\mathbf{c}}_{i,1} = \mathbbm{1} \left( \mathbf{c}^\top \mathbf{X}_i > 0 \right)$ for $i=1, \ldots, n$.
Its error probability for one observation, known as the Bayes rate, is explicitly given by  \citep{cai_high_2019}:
\begin{equation}\label{def:bayes_rate}
\mathcal{R}^\ast_{\mathbf{c}_1, \mathbf{c}_2}:=
    \Phi \biggl(- 
    \frac{\|
    \mathbf{c}_1 - \mathbf{c}_2
    \|_2}{2\sigma}
     \biggr),
\end{equation}
where $\Phi$ denotes the cumulative distribution function of standard normal distribution.
Therefore we can bound \eqref{reduction:average} further from below:
\begin{align*}
    \inf_{\hat{\mathbf{H}}}
	\sup_{
    (\mathbf{c}_1,  \mathbf{c}_2) \in 
    \tilde{\mathcal{M}}(s,  \rho, 2)}
  \mathbb{E}_{\mathbf{H}(\mathbf{X}) \sim   \pi}
	\mathbb{P}_{\boldsymbol{\psi}} (\hat{\mathbf{H}} \neq \mathbf{H})
    &~\overset{(i)}{\geq}
    	\sup_{
    (\mathbf{c}_1,  \mathbf{c}_2) \in 
    \tilde{\mathcal{M}}(s,  \rho, 2)}
   \inf_{\hat{\mathbf{H}}} \;
  \mathbb{E}_{\mathbf{H}(\mathbf{X}) \sim   \pi}
	\mathbb{P}_{\boldsymbol{\psi}} (\hat{\mathbf{H}} \neq \mathbf{H}) 
    \\
     &~\overset{(ii)}{=}
    	\sup_{
    (\mathbf{c}_1,  \mathbf{c}_2) \in 
    \tilde{\mathcal{M}}(s,  \rho, 2)}
  1-(1 - \mathcal{R}^\ast_{\mathbf{c}_1, \mathbf{c}_2})^n
  \\&~\overset{(iii)}{=}
    1-\bigl(1-\Phi( - \rho/\sigma )\bigr)^n,
    \numberthis \label{reduction:bayes}
\end{align*}
where step $(i)$ applies the max–min inequality (see, for example, Section 5.4.1 of Boyd and Vandenberghe, 2004),
step $(ii)$ follows from the optimality of Bayes classifier, and
step $(iii)$ uses \eqref{def:bayes_rate}.
\paragraph*{4. Bounding the Bayes risk.}
Let $C_4>0$ be a constant.
If we set $ \rho^2 = C_4 \sigma^2 \log n $, we have
\begin{equation*}
\Phi( - \rho/\sigma )=
	\Phi \bigl(- \sqrt{C_4\log n} \bigr)
	=
	1 - \mathbb{P}(Z > - \sqrt{C_4\log n})
	\overset{(i)}{\geq}
	1 - \exp(-\frac{C_4}{2}\log n)
	=
	1 - n^{-C_4/2},
\end{equation*}
where \( Z \) denotes a standard normal random variable and step $(i)$ uses Gaussian tail bound.  
Leveraging this, for any \( n \geq 2 \), we further bound \eqref{reduction:bayes} as:
\begin{equation*}
	1-\bigl\{1 - \Phi \bigl(- \sqrt{C_4\log n} \bigr) \bigr\}^n
	\geq
	1-n^{-C_4n/2}
	> C_5
    \numberthis \label{bound_bayes_risk}
    ,
\end{equation*}
where $C_5>0$ depends on $C_4$.
Collecting the reductions \eqref{reduction:smaller_set}, \eqref{reduction:average}, \eqref{def:bayes_rate}, and the bound  \eqref{bound_bayes_risk}, we conclude that
\begin{equation*}
	\inf_{\hat{\mathbf{H}}}
	\sup_{\boldsymbol{\psi} \in 	\Psi (s,   \sigma^2 \log n , K)}
	\mathbb{P}_{\boldsymbol{\psi}} (\hat{\mathbf{H}}(\mathbf{X}) \neq \mathbf{H})
	>C_5.
\end{equation*}
This completes the proof of the lower bound part of Theorem \ref{theorem:separation_condition}.
\end{proof}
\begin{remark}
The reduction in step $(i)$ of \eqref{reduction:bayes} is not tight in some regime. A sharper analysis can be obtained by decomposing the average risk
$ \mathbb{E}_{\mathbf{H}(\mathbf{X}) \sim \pi} \mathbb{P}_{\boldsymbol{\psi}}(\hat{\mathbf{H}} \neq \mathbf{H}) $
into the Bayes risk $\mathcal{R}^\ast_{\mathbf{c}_1, \mathbf{c}_2}$ and the excess risk
$ 
\mathbb{E}_{\mathbf{H}(\mathbf{X}) \sim \pi} 
\mathbb{P}_{\boldsymbol{\psi}}(\hat{\mathbf{H}} \neq \mathbf{H}) - \mathcal{R}^\ast_{\mathbf{c}_1, \mathbf{c}_2}$.
\citet{cai_chime_2019} shows that the excess risk is of order $(s \log p) / n$ in a setting slightly different from ours. Rather than modifying their analysis, we assume a standard high-dimensional sparse regime $s \log p / n = o(1)$, so that the Bayes risk term $\log n$ dominates. In this regime, this lower bound matches the separation bound established in first claim of Theorem \ref{theorem:separation_condition}.
\end{remark}

\subsection{Technical Lemmas and Definitions}\label{section:technical_lemma}
This section presents technical lemmas and definitions that are used throughout the main proofs, beginning with the tail bounds for the chi-square distribution.
\begin{lemma}[Lemma 1 of \citealt{laurent_adaptive_2000}]\label{lemma:chisq_tail}
If $Z$ has a chi-square distribution with $p$ degrees of freedom, then for all $t > 0$, we have
	\begin{equation*}
		\mathbb{P} (Z \geq p + 2\sqrt{pt} + 2t) \leq e^{-t},
        \quad \text{and} \quad
		 \mathbb{P} (Z \leq p - 2\sqrt{pt} )  \leq e^{-t}.
	\end{equation*}
\end{lemma}
Before presenting the next two lemmas, we formally define the sub-exponential norm.
\begin{definition}\label{def:sub_exponential_norm}
Let $\psi_1(x) := e^x - 1$. The sub-exponential norm of a random variable $X$, denoted by $\|X\|_{\psi_1}$, is defined as
\[
\|X\|_{\psi_1} := \inf\left\{ \lambda > 0 : \mathbb{E}\left[\psi_1\left(\frac{|X|}{\lambda}\right)\right] \leq 1 \right\}.
\]
\end{definition}
Using this definition, Lemma \ref{lemma:subexp_norm} provides an upper bound on the subexponential norm of a chi-square random variable.
\begin{lemma}[Lemma VIII.2 of \citealt{chen_cutoff_2021}]\label{lemma:subexp_norm}
	Let $\mathbf{X}_1, \mathbf{X}_2 \stackrel{i.i.d.}{\sim}\mathcal{N}(0, I_q)$. Then there exists a  constant $C_5>0$ such that
	\[
	\|\|\mathbf{X}_1\|_2^2 - q\|_{\psi_1} + \|\langle \mathbf{X}_1, \mathbf{X}_2 \rangle\|_{\psi_1} \leq C_5 \sqrt{q}.
	\]
\end{lemma}
Next, Lemma \ref{lemma:subexp_maximal} provides a maximal inequality for sub-exponential norm.
\begin{lemma}[Lemma 2.2.2 of \citealt{vandervaartWeakConvergenceEmpirical2023}]\label{lemma:subexp_maximal}
  For  random variables $X_1, \ldots, X_m$,  there exists a constant $C_{7}$ such that
\[
\left\| \max_{1 \le i \le m} X_i \right\|_{\psi_1} \le C_{7} \, \log(m+1) \max_{1 \le i \le m} \|X_i\|_{\psi_1}.
\]
\end{lemma}
Next, Lemma~\ref{lemma:gaussian_matrix_opnorm} provides an upper bound on the $\ell_2$ operator norm of an $m \times n$ random matrix with i.i.d. Gaussian entries.
\begin{lemma}[Corollary 5.35 of \citealt{vershynin_introduction_2012}]\label{lemma:gaussian_matrix_opnorm}
	Let $\mathbf{A} \in \mathbb{R}^{m \times n}$ have i.i.d. $\mathcal{N}(0, \sigma^2)$ entries. Then, for any $t > 0$, we have
	\begin{equation*}
		\mathbb{P} \bigl(
		\| \mathbf{A}\|_{\mathrm{op}} \geq  \sigma (\sqrt{m} + \sqrt{n} +\sqrt{2t})
		\bigr)
		\leq
		e^{-t}.	
	\end{equation*}
\end{lemma}
Next, Lemma \ref{lemma:unbounded_empirical_process} provides a concentration inequality for empirical processes indexed by an unbounded function class.
\begin{lemma}[Theorem 4 of \citealt{adamczak_tail_2008}]\label{lemma:unbounded_empirical_process}
	Let $X_1, \dots, X_n$ be independent random variables taking values in a measurable space $(S, \mathcal{B})$, and let $\mathcal{F}$ be a countable class of measurable functions $f: S \to \mathbb{R}$.  Assume that for all $f \in \mathcal{F}$ and all $i = 1, \ldots, n$, it holds that $\mathbb{E}[f(X_i)] = 0$ and
    \[
\bigl\| \sup_{f \in \mathcal{F}} |f(X_i)| \bigr\|_{\psi_1} < \infty,
\]
    Define
	\[
	Z := \sup_{f \in \mathcal{F}} 
	\bigl| \sum_{i=1}^n f(X_i) 
	\bigr|,~
	\gamma^2 := \sup_{f \in \mathcal{F}} \sum_{i=1}^n \mathrm{Var} [f(X_i)],~
    \text{and}~M_{\psi_1} := \| \max_{i \in [n]} \sup_{f \in \mathcal{F}} |f(X_i)| \|_{\psi_1}
	\]
Then there exists a constant $C_7 > 0$ such that for all $t > 0$, we have 
	\[
	\mathbb{P} \left( Z \geq 2 \mathbb{E}[Z] + t \right)
	\leq \exp \left( -\frac{t^2}{3\gamma^2} \right) + 3 \exp \left( - \frac{t}{C_7 M_{\psi_1}}  \right).
	\]
\end{lemma}

Next, Lemma \ref{lemma:suprema_concentration} presents a  concentration inequality for the suprema of the Gaussian process.
\begin{lemma}[Theorem 5.8 of \citealt{boucheron_concentration_2016}]\label{lemma:suprema_concentration}
Let $(X_t)_{t \in \mathcal{T}}$ be an almost surely continuous, centered Gaussian process indexed by a totally bounded set $\mathcal{T}$. Define
$$
\gamma^2 := \sup_{t \in \mathcal{T}} \operatorname{Var}(X_t^2), \quad \text{and} \quad Z := \sup_{t \in \mathcal{T}} X_t.
$$
Then, we have $\operatorname{Var}(Z) \leq \gamma^2$. Moreover, for any $t > 0$,
$$
\mathbb{P}(Z - \mathbb{E}[Z] \geq t) \leq \exp\left(-\frac{t^2}{2\gamma^2}\right).
$$
\end{lemma}
The next lemma, Dudley's integral inequality, bounds the expected supremum of a Gaussian process.
\begin{lemma}[Theorem 8.1.3 of \citealt{vershynin_highdimensional_2018}]\label{lemma:dudley}
Let $(T, d)$ be a metric space, and let $N(T, d, \varepsilon)$ denote the $\varepsilon$-covering number of $T$ with respect to $d$.
Let $(X_t)_{t \in T}$ be a mean-zero random process on a metric space with sub-Gaussian increments. Then
\[
\mathbb{E} \sup_{t \in T} X_t \leq C_8 K_{\psi_2} \int_0^\infty \sqrt{\log N(T, d, \varepsilon)} \, d\varepsilon,
\]
where $C_8$ is an absolute constant, and $K_{\psi_2}$ is the sub-Gaussian parameter.
\end{lemma}

Finally, Lemma \ref{lemma:uniform_hanson} presents
the uniform Hanson-Wright inequality for Gaussian quadratic forms.
\begin{lemma}[Lemma VIII.4 of \citealt{chen_cutoff_2021}]\label{lemma:uniform_hanson}
	Let $\mathbf{X} \sim \mathcal{N}(0, I_{q})$ and $\mathcal{A}$ be a bounded class of $q \times q$ matrices.
     Define the random variable
	\[
	Z = \sup_{\mathbf{A} \in \mathcal{A}}
	\biggl(
	\mathbf{X}^\top \mathbf{A}
	\mathbf{X}
	-
	\mathbb{E}[	
	\mathbf{X}^\top \mathbf{A}
	\mathbf{X}] 
	\biggr),
	\]
which corresponds to the supremum of a centered random  process.
Then, there exists a constant $C_9 > 0$ such that for any $t > 0$,
	\[
	\mathbb{P}(|Z - \mathbb{E}[Z]| \geq t) 
	\leq 
	2 \exp\left[-C_9 \min\left(\frac{t^2}{
		\|\mathbf{X}\|_{\mathcal{A}}^2},
	\frac{t}{\sup_{\mathbf{A} \in \mathcal{A}} \|\mathbf{A}\|_{\mathrm{op}}} \right)\right],
	\]
	where $
	\|\mathbf{X}\|_{\mathcal{A}}
	:=
	\mathbb{E}\left[\sup_{ \mathbf{A} \in \mathcal{A}} \|(\mathbf{A} + \mathbf{A}^T)\mathbf{X}\|_2 \right]$.
\end{lemma}

 \subsection{List of Notations}\label{section:list_of_notations}
 For the reader's convenience, we list the   notations used throughout the proof of Theorem \ref{theorem:separation_condition}.  To facilitate the reading of the proof, these notations are organized into three thematic categories: variables associated with the semidefinite programming (SDP) relaxation and its dual formulation, variables defining the data matrix and true cluster model, and the bounding and reformulation statistics constructed for the proof.
\subsubsection{SDP and Dual Variables}

\begin{itemize}
    \item  $\mathbf{Q}^S$: dual variable corresponding to the PSD constraint of the SDP associated with the variable set $S$.
    \item $\lambda^S$: dual variable corresponding to the trace constraint in the SDP associated with the variable set $S$.
    \item $\boldsymbol{\alpha}^S$: dual variable corresponding to the symmetry constraint in the SDP associated with the variable set $S$.
    \item $\dot{\boldsymbol{\alpha}}^S(\lambda^S)$: for a given value of $\lambda^S$, the unique value of $\boldsymbol{\alpha}^S$ determined by condition~\eqref{alpha_condition_final}.
    \item $\mathbf{B}^S$: dual variable corresponding to the entrywise nonnegativity constraint in the SDP associated with the variable set $S$.
    \item $\dot{\mathbf{B}}^S(\lambda^S)$: for a given value of $\lambda^S$, the value of $\mathbf{B}^S$ implied by condition~\eqref{B_condition_rowsum}.
    \item $\mathbf{W}^S := \lambda^S \mathbf{I}_n - \mathbf{B}^S + \frac{1}{2} \mathbf{1}_n (\boldsymbol{\alpha}^S)^\top + \frac{1}{2} \boldsymbol{\alpha}^S \mathbf{1}_n^\top - \mathbf{X}_{S,\cdot}^\top \mathbf{X}_{S,\cdot} $: a function of the dual variables, introduced to simplify notation.
    \item $\dot{\mathbf{W}}^S(\lambda^S)$: the constructed function of the parameterized dual variables based on $\lambda^S$.
    \item $\mathcal{L}(\mathbf{Z}, \mathbf{Q}^S, \lambda^S, \boldsymbol{\alpha}^S, \mathbf{B}^S)$: the Lagrangian function associated with the primal SDP problem.
    \item $\mathbf{r}^{(k,l,S)}$: the unique vector determining the row-wise sum of the $(k, l)$ off-diagonal block of the dual certificate $\dot{\mathbf{B}}^S$.
    \item $t_{lk}^S(\lambda^S)$: the sum of all elements in the $(l,k)$th off-diagonal block of the constructed dual matrix $\dot{\mathbf{B}}^S (\lambda^S)$.
\end{itemize}

\subsubsection{Data and True Cluster Model}
\begin{itemize}
    \item $\mathbf{Z}^\ast$: a symmetric block-diagonal real matrix corresponding to the true cluster partition $G^\ast$.
    \item $\mathbf{M}^\ast$: the matrix containing the true cluster means column-wise.
    \item $\mathcal{E}$: the matrix collecting the random noise vectors.
    \item $\mathbf{X}_{S, i}$: the $i$th observation restricted to the indices corresponding to the variable set $S$.
    \item $\mathcal{E}_{S, i}$: the noise vector of the $i$th observation restricted to the variable set $S$.
    \item $\overline{\mathbf{X}}^\ast_{S,k}$: the average of the observations in true cluster $k$, restricted to the indices corresponding to the variable set $S$.
    \item $\overline{\mathcal{E}}^\ast_{S, k}$: the  average of the noise of true cluster $k$, restricted to the indices corresponding to the variable set $S$.
    \item $\mathcal{S}$: the collection of support sets satisfying the required separation and set size conditions.
    \item $\Delta_{S \cap S^\ast}^2$: the minimum squared Euclidean distance between true cluster centers restricted to $S \cap S^\ast$.
    \item $m$: the minimum harmonic-mean-like quantity of the true cluster sizes, defined as $\min_{k \neq l} 2|G_l^\ast||G_k^\ast| / (|G_l^\ast| + |G_k^\ast|)$.
\end{itemize}

\subsubsection{Bounding and Reformulation Statistics}
\begin{itemize}
    \item $D^S_{kli}(\mathbf{X})$: the difference between the squared Euclidean distances from observation $\mathbf{X}_{S, i}$ to the true cluster centers $l$ and $k$.
    \item $U^S(\mathbf{X})$: a data-dependent upper bound for the dual variable $\lambda^S$.
    \item $\Gamma_K$: the linear subspace of $\mathbb{R}^n$ defined as the orthogonal complement to the span of the cluster indicator vectors ($\mathbf{1}_{G_k^\ast}$).
    \item $L_1^S(\mathbf{X})$: the supremum of the noise quadratic form process over the subspace $\Gamma_K$.
    \item $I^S_{kl}(\mathbf{v}, \mathbf{X})$: the fundamental component representing signal--noise interactions.
    \item $N^S_{kl}(\mathbf{v}, \mathbf{X})$: the fundamental component representing noise--noise interactions.
    \item $T^S_{1,kl}(\mathbf{v}), T^S_{2,kl}(\mathbf{v}), T^S_{3,kl}(\mathbf{v})$: composite random quantities representing combinations of signal-noise and noise-noise interactions.
    \item $L_2^S(\mathbf{X}, \lambda^S)$: a supremum quantity defining the lower bound condition for $\lambda^S$.
    \item $\dot{\lambda}^S$: a specific chosen value for the dual variable $\lambda^S$, defined as $\sigma^2 |S| + m \Delta_{S \cap S^\ast }^2/4$.
\end{itemize}
\section{Details for Robust MMD Statistic }\label{section:details_MMD}

\subsection{Bounded Transformation}\label{section:bounded_transformation}
The reward step uses
robust MMD test (Algorithm~\ref{alg:robust_dc}) to generate stochastic rewards.
theory of the robust MMD test \citep{schrabRobustKernelHypothesis2025} requires the observations to lie within a bounded range. 
We therefore introduce an appropriate transformation, denoted by $\texttt{BoundedTransform}$ in Algorithm~\ref{alg:robust_dc}, to ensure that the boundedness condition is satisfied.

\medskip
\noindent
\textbf{Transformation under Known Covariance Setting.}
Let $\Phi$ denote the standard normal distribution function.  We use the fixed coordinatewise transformation
\begin{equation}
\label{eq:probit_reward_transform}
X_{ij}^{\mathrm R}
:=\Phi\!\left(\frac{\widetilde X_{ij}}{s_j}\right),
\qquad
s_j:=\{(\boldsymbol\Omega^\ast)_{jj}\}^{1/2}.
\end{equation}
Thus $X_{ij}^{\mathrm R}\in(0,1)$ for every possible observation. Therefore the reward step proceeds as follows: compute
$\mathbf X^{\mathrm R}$ once before TVS starts, hold it fixed across all
iterations, and permute only the group assignments in
Algorithm~\ref{alg:robust_dc}.  The transformation is used only for the
reward: the clustering block continues to use the original innovated matrix
$\widetilde{\mathbf X}=\boldsymbol\Omega^\ast\mathbf X$.

\medskip
\noindent
\textbf{Transformation under Unknown Covariance Setting.}
We retain the diagonal entries of the
block precision estimates already computed by
Algorithm~\ref{alg:ISEE_subroutine}, and use the plug-in scale
\(
\widehat s_j^t
:=\{\max(\widehat\Omega_{jj}^t,\eta)\}^{1/2}
\)
and
\(
X_{ij}^{\mathrm R,t}
:=\Phi(\widehat{\widetilde X}_{ij}^t/\widehat s_j^t)
\), where a small numerical floor (for example,   $\eta=10^{-8}$) is applied to prevent division
by zero. The reward step proceeds as follows: After each ISEE update, form this bounded matrix once and hold it
fixed throughout all coordinate tests and permutations at iteration $t$.
This requires no additional covariance fit.  It is the practical extension
of \eqref{eq:probit_reward_transform}; the finite-sample statement below
concerns the fixed known-covariance map.  Thus implementation adds only one
vectorized normal-CDF pass per newly formed innovated matrix and leaves the
permutation routine unchanged.

\subsection{Connection to Linear MMD}
Let us denote
\(
\widetilde c_{kj}^\ast
:=(\boldsymbol\Omega^\ast\mathbf c_k^\ast)_j
\).
Then we have
\(
\widetilde X_{ij}\mid i\in G_k^\ast
\sim N(\widetilde c_{kj}^\ast,s_j^2)
\).
Using Gaussian convolution identity
\begin{equation*}
  Z\sim N(\mu,1)
\quad\Longrightarrow\quad
\mathbb E[\Phi(Z)]
=
\Phi\left(\frac{\mu}{\sqrt{2}}\right),    
\end{equation*}
we have
\begin{equation}\label{gaussian_convolution}
\mathbb E(X_{ij}^{\mathrm R}\mid i\in G_k^\ast)
=\Phi\!\left(\frac{\widetilde c_{kj}^\ast}{\sqrt2\,s_j}\right).
\end{equation}
Let $P_j^{\mathrm R}$ and $Q_j^{\mathrm R}$ denote the two population laws
after bounded transform \eqref{eq:probit_reward_transform}.
Leveraging \eqref{gaussian_convolution}, for the linear kernel $k(v,v')=vv'$, the population squared MMD of the
bounded reward distributions is
\begin{equation}\label{eq:reward_gap}
  \operatorname{MMD}^2_k(P_j^{\mathrm R},Q_j^{\mathrm R})
  =\left\{
  \mathbb E_{P_j^{\mathrm R}}X^{\mathrm R}
  -\mathbb E_{Q_j^{\mathrm R}}X^{\mathrm R}
  \right\}^2
  =
  \left|
\Phi\!\left(\frac{\widetilde c_{1j}^\ast}{\sqrt2\,s_j}\right)
-\Phi\!\left(\frac{\widetilde c_{2j}^\ast}{\sqrt2\,s_j}\right)
\right|^2
:=
(d_j^{\mathrm R})^2.
\end{equation}
For the  symmetric model \eqref{simulation_setting_symmetric_gaussian} used in the numerical study in Section \ref{section:simulations},
we have $\widetilde c_{2j}^\ast=-\widetilde c_{1j}^\ast$, which implies
\begingroup
\begin{equation*}
d_j^{\mathrm R}
=2\Phi\!\left(
\frac{|\widetilde c_{1j}^\ast-\widetilde c_{2j}^\ast|}
{2\sqrt2\,s_j}
\right)-1
\asymp
\min\!\left\{
\frac{|\widetilde c_{1j}^\ast-\widetilde c_{2j}^\ast|}{s_j},1
\right\}.
\end{equation*}
\endgroup
Hence in this case. the transformed gap $(d_j^{\mathrm R})^2$ is proportional to the standardized innovated mean gap
for weak signals and remains bounded away from zero for strong signals.

As the empirical counterpart of \eqref{eq:reward_gap}, for $\mathbf X_j^{\mathrm R}=(X_{1j}^{\mathrm R},\ldots,
X_{nj}^{\mathrm R})^\top$, the corresponding V-statistic is defined as
\begin{align*}
\widehat{\operatorname{MMD}}_k^2(G^\ast,\mathbf X_j^{\mathrm R})
&:=\frac{1}{|G_1^\ast|^2}\sum_{\ell,m\in G_1^\ast}X_{\ell j}^{\mathrm R}X_{mj}^{\mathrm R}
 +\frac{1}{|G_2^\ast|^2}\sum_{\ell,m\in G_2^\ast}X_{\ell j}^{\mathrm R}X_{mj}^{\mathrm R}\\
&\quad-\frac{2}{|G_1^\ast||G_2^\ast|}
\sum_{\ell\in G_1^\ast}\sum_{m\in G_2^\ast}X_{\ell j}^{\mathrm R}X_{mj}^{\mathrm R}
\\&=\Lambda(G^\ast,\mathbf X_j^{\mathrm R})^2.
\end{align*}
Therefore, the cluster mean difference statistic in \eqref{eq:def-Tj} can be viewed as a linear MMD statistic, allowing us to directly leverage the existing theory for MMD tests to analyze our reward generators. We note that $\texttt{rMMD}$ and Algorithm \ref{alg:robust_dc} refers to permutation test based on statistic $\widehat{\operatorname{MMD}}_k(G^\ast,\mathbf X_j^{\mathrm R})$.

\subsection{Power Guarantee}\label{section:MMD_theory}
Given observations transformed via mappings in Appendix \ref{section:bounded_transformation},
for a fixed partition $G=(G_1,G_2)$, changing one bounded  observation 
changes $\Lambda$ by at most
\begin{equation}
\label{eq:reward_sensitivity}
\Gamma_j(G)
\leq\frac{1}{\min(|G_1|,|G_2|)}.
\end{equation}
The map in \eqref{eq:probit_reward_transform} is fixed and bounded for every
raw input.
In contrast, empirical min--max scaling can change all normalized
observations when a single raw observation changes and therefore does not
justify the sensitivity in \eqref{eq:reward_sensitivity}.

Leveraging this sensitivity, Theorem~\ref{theorem:mmd} specializes the finite-sample two-sample testing error guarantees established in Theorem~1 of \citet{schrabRobustKernelHypothesis2025}.
\begin{theorem}[Finite-sample calibration of Algorithm~\ref{alg:robust_dc}]
\label{theorem:mmd}
Fix desired type-I and type-II errors $a,b\in(0,1)$, let
$n_{\min}:=\min(n_1,n_2)$, and suppose $n_1 \asymp n_2$.  Let $r$ be an upper
bound on the total number of arbitrarily corrupted observations in the pooled
two-sample data, where $0\leq r<n_{\min}$.  
If the number of permutations
$B$ satisfies
\(
B>3a^{-2}\{\log(8/b)+a(1-a)\},
\)
then, under $P_j^{\mathrm R}=Q_j^{\mathrm R}$, the robust permutation test has marginal type-I
error at most $a$ uniformly over every corruption affecting at most $r$
pooled observations.  Moreover, there is a constant $C_{\mathrm R}>0$ such that, if
\[
d_j^{\mathrm R}\geq C_{\mathrm R}\max\!\left\{
\sqrt{\frac{\max\{\log(e/a),\log(e/b)\}}{n_{\min}}},
\frac{r}{n_{\min}}
\right\},
\]
its marginal power satisfies
\begingroup
\begin{equation*}
\mathbb P_{P_j^{\mathrm R},Q_j^{\mathrm R}}
\{\text{reject }H_0\mid\text{at most $r$ pooled observations are corrupted}\}
\geq1-b.
\end{equation*}
\endgroup
Algorithm~\ref{alg:robust_dc} uses the test level $a=\tau_\epsilon-\alpha$; the desired type-II bound $b$ is specified separately when applying this theorem.  If one optionally imposes the symmetric power target $1-b=\tau_\epsilon+\alpha$ (which additionally requires $\alpha<1-\tau_\epsilon$), then $b=1-\tau_\epsilon-\alpha$.
\end{theorem}
Originally, Theorem~1 of \citet{schrabRobustKernelHypothesis2025} requires the kernel to be characteristic. This condition is necessary for MMD-based two-sample testing because it ensures that
\[
\operatorname{MMD}(P,Q)=0 \quad \Longleftrightarrow \quad P=Q.
\]
Although the linear kernel is not characteristic in general, under our Gaussian mixture model assumption \eqref{def:gaussian_mixture_model} with known isotropic covariance, equality of the linear MMD is nevertheless equivalent to equality of the reward distributions.

The budget $r$ appears in both the type I and II error conditions and rejection threshold.
   In fact, $0\leq\Lambda\leq1$ and the empirical permutation
quantile is nonnegative, so the correction $2r/n_{\min}$ makes rejection
impossible when $r\geq n_{\min}/2$.  Accordingly, the   calibration
requires $r_n<n_{\min}/2$;
the fraction $r_n/n_{\min}$ need not vanish, but
useful power requires it to remain sufficiently smaller than the relevant
separation.
The lower bound on $B$ need not contain $r$: $B$ controls Monte Carlo
quantile error, whereas corruption is paid for through the enlarged rejection threshold
and the power requirement.

\section{Proofs of Theoretical Results for TVS}\label{section:proof:TVS_theory}

\subsection{Target Setting for Feature Subset Recovery}\label{section:target_for_feature_recovery}
Recall from Theorem~\ref{theorem:separation_condition} that
$\mathbf{Z}^\ast=\sum_{k=1}^K|G_k^\ast|^{-1}
\mathbf{1}_{G_k^\ast}\mathbf{1}_{G_k^\ast}^\top$ is the block-diagonal
representation of the true partition $G^\ast=(G_1^\ast,G_2^\ast)$, and that
$\mathcal S$ defined in 
\eqref{def:S-set}
is the family of feature sets for which exact recovery is guaranteed to hold with high probability.  Our analysis is conditioned on the following \say{clean} uniform event defined by Theorem \ref{theorem:separation_condition} and $\mathcal{S}$:
\begin{equation*}
\mathcal{E}_{\mathrm{SDP}}
~:=~
\bigl\{\,
\widehat{\mathbf{Z}}(S)=\mathbf{Z}^\ast,\ \forall S\in\mathcal{S}
\,\bigr\}.
\end{equation*}
By Theorem \ref{theorem:separation_condition}, we have
   $\mathbb{P}(\mathcal{E}_{\mathrm{SDP}}) \ge 1- {C_2}K/n$,
and that for every possibly data-dependent selected set $\widehat S$:
\begingroup
\begin{equation*}
\mathbb P\{\widehat{\mathbf Z}(\widehat S)\neq\mathbf Z^\ast\}
\leq
\frac{C_2K}{n}+\mathbb P(\widehat S\notin\mathcal S).
\end{equation*}
\endgroup  
Therefore, to establish exact clustering, we do not need to recover the precise set $S^\ast$; it suffices to show that $\widehat{S} \in \mathcal{S}$.

\subsection{Proof of Theorem \ref{theorem:posterior_surplus}}\label{section:proof:theorem:posterior_surplus}
\subsubsection{Proof of Oracle Target Convergence}\label{section:proof_oracle_target_convergence}
We first show that, under Conditions~\ref{condition:core_margin}
and~\ref{condition:oracle_screening},
the oracle reward target
$S_\epsilon^\ast$ satisfies
\begin{equation*}
\mathbb P_{\mathbf X}(S_\epsilon^\ast\in\mathcal S)\longrightarrow1.
\end{equation*}
\begin{proof}
For notational simplicity,
put $\tau:=\tau_\epsilon$.
For each feature index $j \in[p]$, define
\begin{equation*}
A_j:=\mathbbm 1\{\theta_j^{\mathrm{orc}}>\tau\}.
\end{equation*}
Then we can leverage this notation to express the oracle reward target
$S_\epsilon^\ast$ as:
\begin{equation*}
S_\epsilon^\ast=\{j:A_j=1\}.
\end{equation*}
Recall that in 
\eqref{feature_wise_clustering_signal},
for $K=2$,
we denoted the feature-wise clustering signal as
\begin{equation*}
w_j :=
\left\{\bigl[\boldsymbol\Omega^\ast
(\mathbf c_1^\ast-\mathbf c_2^\ast)\bigr]_j\right\}^2,
\end{equation*}
so that we can write
$\Delta_A^2=\sum_{j\in A}w_j$ for every subset $A\subseteq S^\ast$.
Let us denote $L_p=\lfloor\sqrt p\rfloor$ and
$s_\ast:=|S^\ast|$.

We have shown in \eqref{eq:marginal_target_bridge} that,
for every $j\notin S^\ast$, by Markov's inequality,
\[
\mathbb P_{\mathbf X}(A_j=1)
\leq
\frac{\bar\theta_j^{\mathrm{orc}}}{\tau}.
\]
Thus, if $N_\epsilon:=|S_\epsilon^\ast\setminus S^\ast|$, then
by \eqref{descrp_upper_bounds}, we have
\[
\mathbb E_{\mathbf X}(N_\epsilon)
\leq
\frac{1}{\tau}
\sum_{j\notin S^\ast}\bar\theta_j^{\mathrm{orc}}
=o(L_p),
\]
where the last relation uses Condition \ref{condition:oracle_screening} and
$\tau\geq\tau_0$.  Hence we have
\begin{equation*}
N_\epsilon=o_{\mathbb P_{\mathbf X}}(L_p) 
\end{equation*}
by Markov's inequality.

For $j\in D$, the event $A_j=0$ implies
$1-\theta_j^{\mathrm{orc}}\geq1-\tau$. We have shown in \eqref{eq:marginal_target_bridge} that
\[
\mathbb P_{\mathbf X}(A_j=0)
\leq
\frac{1-\bar\theta_j^{\mathrm{orc}}}{1-\tau}.
\]
Let
$H_\epsilon:=\sum_{j\in D}w_j(1-A_j)$ be the core signal omitted by
the oracle cutoff. Then by
Condition \ref{condition:oracle_screening}, we have
\[
\mathbb E_{\mathbf X}(H_\epsilon)
\leq
\frac{1}{1-\tau}
\sum_{j\in D}w_j(1-\bar\theta_j^{\mathrm{orc}})
=o(\Delta_D^2).
\]
Therefore, we have
\begin{equation*}
    H_\epsilon=o_{\mathbb P_{\mathbf X}}(\Delta_D^2).
\end{equation*}
 Recall from 
 \eqref{T_l} that
 for an integer $L \geq 0$, we denoted the SDP signal threshold defined in \eqref{def:S-set} for a selected set of size $L$ as
\begin{equation*}
T_L
:=
C_{\mathcal S}\sigma^{-2}
\left\{
\log n+\frac{L\log p}{m}
+\sqrt{\frac{L\log p}{m}}
\right\}.
\end{equation*}
  Because
$N_\epsilon$ and $H_\epsilon$ are functions of the data, both of
$N_\epsilon=o_{\mathbb P_{\mathbf X}}(L_p) $
and
$H_\epsilon=o_{\mathbb P_{\mathbf X}}(\Delta_D^2)$
also
hold under the joint law used for the adaptive part*.  In particular, the event
\[
\mathcal E_{\mathrm{orc},n}
:=\left\{N_\epsilon\leq\kappa L_p,\quad
H_\epsilon\leq\frac{\gamma}{1+\gamma}\Delta_D^2\right\}
\]
has probability tending to one.  On this event, the fixed margins in
\eqref{eq:clean_core_margin} give
\[
|S_\epsilon^\ast|
\le
s_\ast + |S_\epsilon^\ast\setminus S^\ast|
= s_\ast+N_\epsilon
\leq L_p,
\qquad
\Delta_{S_\epsilon^\ast\cap S^\ast}^2
\geq\Delta_D^2-H_\epsilon
\geq\frac{\Delta_D^2}{1+\gamma}
\geq T_{L_p}
\geq T_{|S_\epsilon^\ast|},
\]
where the final inequality uses the monotonicity of $T_L$.  Thus
$S_\epsilon^\ast\in\mathcal S$ with probability tending to one,
which proves 
$\mathbb P_{\mathbf X}(S_\epsilon^\ast\in\mathcal S)\longrightarrow1$
.
\end{proof}
\subsubsection{Proof of TVS Algorithm Convergence}
Next, we show that
under Conditions~\ref{condition:core_margin}, \ref{condition:oracle_screening} and
\ref{condition:posterior_learning}, we have
\begin{equation*}
    \mathbb P\!\left(
\widehat S^{t_n}\in\mathcal S,
\ \widehat G^{t_n}=G^\ast
\right)\longrightarrow1.
\end{equation*}

\begin{proof}
For notational simplicity, let
us denote $t=t_n$.
Conditioning on $\mathcal H_t$, define
\[
R_t:=|\widehat S^t\setminus(S_\epsilon^\ast\cup S^\ast)|,
\qquad
Q_t:=\sum_{j\in D\cap S_\epsilon^\ast}
w_j\mathbbm 1\{j\notin\widehat S^t\}.
\]
By definitions in \eqref{eq:posterior_target_errors}, we have
\begin{equation*}
\mathbb E(R_t\mid\mathcal H_t)=\operatorname{Out}_t,
\qquad
\mathbb E(Q_t\mid\mathcal H_t)=\operatorname{Miss}_t.
\end{equation*}
Recall that
$L_p = \lfloor\sqrt p\rfloor$.
For every fixed $\eta>0$, conditional Markov's inequality yields
\[
\mathbb P(R_t>\eta L_p)
\leq
\mathbb E\!\left[1\wedge
\frac{\operatorname{Out}_t}{\eta L_p}\right]
\longrightarrow0,
\]
because of
the condition
$\frac{\operatorname{Out}_{t_n}}{L_p}\xrightarrow{\mathbb P}0$
in Condition \ref{condition:posterior_learning}.
Using the same argument with
$\frac{\operatorname{Miss}_{t_n}}{\Delta_D^2}\xrightarrow{\mathbb P}0$ results in
\begin{equation*}
\mathbb P(Q_t>\eta\Delta_D^2)
\leq
\mathbb E\!\left[1\wedge
\frac{\operatorname{Miss}_t}{\eta\Delta_D^2}\right]
\longrightarrow0.
\end{equation*}
Combining these with the results in Appendix \ref{section:proof_oracle_target_convergence}, we have
\begin{equation*}
N_\epsilon+R_t=o_{\mathbb P}(L_p)
\quad
\text{and}
\quad
H_\epsilon+Q_t=o_{\mathbb P}(\Delta_D^2),
\end{equation*}
where we recall that
$o_{\mathbb P}$ is under the joint distribution of the data and all algorithmic randomness contained in $\mathcal H_{t_n}$.

Therefore the event
\[
\mathcal E_{\mathrm{ad},n}
:=\left\{N_\epsilon+R_t\leq\kappa L_p,\quad
H_\epsilon+Q_t\leq\frac{\gamma}{1+\gamma}\Delta_D^2\right\}
\]
has probability tending to one.

On the event $\mathcal{E}_{\mathrm{ad},n}$, we first bound the size of $\widehat{S}^t$. By decomposing $\widehat{S}^t$, we observe that the definition of $R_t$ does not penalize selected coordinates in $S^\ast \setminus S_\epsilon^\ast$, as these are already absorbed by $s_\ast$. Combining this with Condition~\ref{condition:core_margin} gives
\begin{equation*}
    |\widehat{S}^t|
    \leq |S^\ast| + |S_\epsilon^\ast \setminus S^\ast| + |\widehat{S}^t \setminus (S_\epsilon^\ast \cup S^\ast)|
    = s_\ast + N_\epsilon + R_t
    \leq (1-\kappa)L_p + \kappa L_p
    = L_p.
\end{equation*}
Furthermore, because $D \subseteq S^\ast$, the core signal missed by $\widehat{S}^t$ is bounded by the sum of the signal missed by the oracle ($H_\epsilon$) and the signal missed by TVS ($Q_t$). Hence, the retained signal satisfies
\begin{equation*}
    \Delta_{\widehat{S}^t \cap S^\ast}^2
    \geq \Delta_{\widehat{S}^t \cap D}^2
    \geq \Delta_D^2 - H_\epsilon - Q_t
    \geq \Delta_D^2 - \frac{\gamma}{1+\gamma}\Delta_D^2
    = \frac{\Delta_D^2}{1+\gamma}
    \geq T_{L_p}
    \geq T_{|\widehat{S}^t|},
\end{equation*}
where the last step uses the monotonicity of the threshold $T_L$. Consequently, we conclude that $\widehat{S}^t \in \mathcal{S}$ on the event $\mathcal{E}_{\mathrm{ad},n}$.

We have established the two conditions of $\mathcal{S}$:
$|\widehat{S}^t| \le L_p$ and
$\Delta_{\widehat{S}^t \cap S^\ast}^2
 \geq T_{|\widehat{S}^t|}$. These imply
 $\mathbb P(\widehat S^{t_n}\in\mathcal S)\to1$.

 On $\mathcal E_{\mathrm{ad},n}\cap\mathcal E_{\mathrm{SDP}}$, the selected
set belongs to $\mathcal S$ and the subset SDP satisfies
$\widehat{\mathbf Z}(\widehat S^{t_n})=\mathbf Z^\ast$; its rounding step
therefore returns $G^\ast$ up to the   permutation of the two cluster
labels.  Consequently, Theorem~\ref{theorem:separation_condition} and $K=2$
give the explicit joint failure bound
\[
\mathbb P\!\left\{\widehat S^{t_n}\notin\mathcal S
\ \text{or}\ \widehat G^{t_n}\neq G^\ast\right\}
\leq \mathbb P(\mathcal E_{\mathrm{ad},n}^{\mathsf c})
+\frac{2C_2}{n}\longrightarrow0,
\]
which is \eqref{eq:clean_exact_recovery}.
\end{proof}

\subsection{Proof of Corollary~\ref{corollary:rmmd_oracle_screening}}
\label{section:proof:corollary:rmmd_oracle_screening}
\begin{proof}
The proof uses notations from Appendix \ref{section:details_MMD}.
Condition~\ref{condition:core_margin} gives
$\tau_\epsilon\geq\tau_0$.  Because $a_n=o(p^{-1/2})$ and hence $a_n\to0$,
$\alpha_n=\tau_\epsilon-a_n\in(0,\tau_\epsilon)$ for all sufficiently large
$n$, so the stated rMMD test level is well defined.

For $j\notin S^\ast$, the definition of $S^\ast$ gives
$\widetilde c_{1j}^\ast=\widetilde c_{2j}^\ast$.  Under the Gaussian model in
  Condition~\ref{condition:core_margin}, the two innovated
coordinate distributions therefore coincide, and applying the same fixed
probit map gives $P_j^{\mathrm R}=Q_j^{\mathrm R}$.
By the definition of permutation test with level $a_n$, we have
\begin{equation*}
\bar\theta_j^{\mathrm{orc}}
=
\mathbb P_{\mathbf X,U}\{Y_j(G^\ast)=1\}
\leq a_n,
\qquad j\notin S^\ast.
\end{equation*}
By  Theorem~\ref{theorem:mmd}, for every $j\in D$, the lower bound condition in
\eqref{eq:detectable_core} 
implies that the power is bounded below by
$1-b_n$, which means
\[
\bar\theta_j^{\mathrm{orc}}
=
\mathbb P_{\mathbf X,U}\{Y_j(G^\ast)=1\}
\geq
1-b_n,
\qquad j\in D.
\]
Since $L_p=\lfloor\sqrt p\rfloor\asymp\sqrt p$, these coordinatewise bounds
imply
\[
\sum_{j\notin S^\ast}\bar\theta_j^{\mathrm{orc}}
\leq p a_n=o(L_p),
\qquad
\sum_{j\in D}w_j(1-\bar\theta_j^{\mathrm{orc}})
\leq b_n\sum_{j\in D}w_j
=b_n\Delta_D^2=o(\Delta_D^2).
\]
These are precisely the two requirements in  
Condition~\ref{condition:oracle_screening}.
\end{proof}

\subsection{Proof of Corollary~\ref{corollary:posterior_certificate}}\label{section:proof:corollary:posterior_certificate}
\begin{proof}
We first record a conditional Beta-tail bound.  Let
$Z\sim\operatorname{Beta}(a,b)$, put $q:=a+b$ and $\mu:=a/q$, and define
\[
\operatorname{kl}(u,x)
:=u\log\!\left(\frac{u}{x}\right)
+(1-u)\log\!\left(\frac{1-u}{1-x}\right).
\]
For $x>\mu$, use the representation
$Z=G_a/(G_a+G_b)$, where
$G_a\sim\operatorname{Gamma}(a,1)$ and
$G_b\sim\operatorname{Gamma}(b,1)$ are independent.  Chernoff's inequality
with $\lambda=(x-\mu)/\{x(1-x)\}$ gives
\begin{align*}
\mathbb P(Z\geq x)
&=\mathbb P\{(1-x)G_a-xG_b\geq0\}\\
&\leq
\{1-\lambda(1-x)\}^{-a}
\{1+\lambda x\}^{-b}\\
&=\exp\{-q\operatorname{kl}(\mu,x)\}
\leq\exp\{-2q(x-\mu)^2\},
\end{align*}
where the last inequality is Pinsker's inequality for Bernoulli
distributions.  The analogous lower-tail calculation gives, for $x<\mu$,
\begingroup
\begin{equation*}
\mathbb P(Z\leq x)
\leq\exp\{-2q(\mu-x)^2\}.
\end{equation*}
\endgroup
By the nature of Thompson sampling, these bounds remain valid conditional on $\mathcal{H}_{t_n}$ (Definition~\ref{def:sigma_field}), even though the posterior parameters are random.

Let $\mathcal A_n$ be the event in the corollary, and define
\(
\pi_{j,t_n}
:=\mathbb P(\vartheta_j^{t_n}>\tau_\epsilon\mid\mathcal H_{t_n})
\).
On $\mathcal A_n$,  for $j\in\mathcal O_n$, using 
\begin{equation*}
    \mu_{j,t_n}\leq\tau_\epsilon-\delta
\quad
\text{and}
\quad
\nu_{j,t_n}\geq c_{\rm out}\log p,
\end{equation*}
we can derive
\begin{equation*}
\pi_{j,t_n}
\leq
\exp(
-2 \nu_{j, t_n}(\tau_\epsilon- \mu_{j,t})^2
)
\leq
\exp
\bigl(
-(2 c_{\rm out}\delta^2) \log p 
\bigr)
=
p^{-2c_{\rm out}\delta^2}
\end{equation*}
On $\mathcal A_n$,  for $j\in\mathcal C_n$, using 
\begin{equation*}
 \mu_{j,t_n}\geq\tau_\epsilon+\delta
\quad
\text{and}
\quad
\nu_{j,t_n}\geq q_n,
\end{equation*}
we can derive
\begin{equation*}
  1-\pi_{j,t_n}\leq e^{-2q_n\delta^2}.
\end{equation*}
By conditional linearity of expectation,
\[
\operatorname{Out}_{t_n}
=\sum_{j\in\mathcal O_n}\pi_{j,t_n},
\qquad
\operatorname{Miss}_{t_n}
=\sum_{j\in\mathcal C_n}w_j(1-\pi_{j,t_n}).
\]
Because $L_p=\lfloor\sqrt p\rfloor$ and $p/L_p\leq2\sqrt p$ for all
sufficiently large $p$, on $\mathcal A_n$,
\[
\frac{\operatorname{Out}_{t_n}}{L_p}
\leq2p^{1/2-2c_{\rm out}\delta^2}=o(1),
\]
where the \say{sufficiently large} requirement in the corollary may be taken as
$c_{\rm out}>1/(4\delta^2)$.  Similarly,
\[
\frac{\operatorname{Miss}_{t_n}}{\Delta_D^2}
\leq e^{-2q_n\delta^2}
\frac{\sum_{j\in\mathcal C_n}w_j}{\Delta_D^2}
\leq e^{-2q_n\delta^2}=o(1).
\]
Since $\mathbb P(\mathcal A_n)\to1$, these two bounds hold in probability
and verify  Condition~\ref{condition:posterior_learning}.  The
conclusion follows from 
Theorem~\ref{theorem:posterior_surplus}.
\end{proof}

\section{Details for Numerical Study}\label{section:simulation_details}
This section provides additional details on the numerical studies, omitted from Section \ref{section:simulation}.
All simulations are conducted on a Linux machine equipped with an AMD EPYC 7542 processor (32 cores, 2.90 GHz) and 16 GB of RAM. Table \ref{tab:synthetic_summary} presents detailed parameter settings for our synthetic data study in Section \ref{section:simulation}.

\begin{itemize}
    \item Appendix~\ref{section:sdp_solver_detail} details the implementation of the SDP solver.
    \item Appendix~\ref{section:hyperparameter} specifies the hyperparameters used across all simulations and empirical experiments.
    \item Appendix~\ref{section:baselinemethod} reviews the three baseline methods introduced in Section~\ref{section:method_unknown_cov}.
    \item  Appendix \ref{section:greedy_stopping_criteria} introduces stopping criterion for our TVS and greedy algorithm.
    \item Appendix~\ref{section:autoencoder} outlines the text preprocessing pipeline used for the real data application.
\end{itemize}

\subsection{Details on Initial Clustering and SDP K-means Implementation}\label{section:sdp_solver_detail}
We use SDP K-means \citep{chen_cutoff_2021} for the initial clustering step, implemented as \verb|sdp_kmeans.R| in our code repository.
The SDP is solved using a custom first-order ADMM solver \citep{boyd2011distributed} with adaptive penalty updates and an efficient simplex projection algorithm proposed by \cite{duchi2008efficient}.
To improve convergence stability and reduce sensitivity to the initial choice of the ADMM penalty parameter $\rho$, we implement a residual balancing scheme that uses hyperparamters $\mu$ and $\tau$.
In this scheme, $\mu$ serves as the imbalance threshold, ensuring the primal and dual residuals remain within the same order of magnitude. The parameter $\tau$ acts as a scaling factor  that dictates the magnitude of adjustments to the penalty parameter $\rho$. This adaptive mechanism enables the solver to dynamically balance the trade-off between objective optimization and constraint satisfaction. For our implementation, we employ the following hyperparameter configuration: $\rho = 1.0$, $\mu = 10.0$, and $\tau = 2.0$, with a convergence tolerance of $10^{-3}$ and a maximum of $2,000$ iterations.

 In the final step of the SDP-relaxed $K$-means, we extract the leading eigenvector (an $n$-dimensional vector) from the output matrix $\mathbf{Z}$ and apply the \texttt{kmeans} function from the  R \texttt{stats} package.
 
\subsection{Hyperparameter Settings}\label{section:hyperparameter}
This section specifies the hyperparamter settings used in synthetic and real data study in the main text.

\subsubsection{Synthetic Data Study}
For the TVS algorithm, the hyperparameters are defined as follows:
\begin{itemize}
    \item Maximum number of iterations: $T = 300$,
    \item Significance threshold: $\tau_{\epsilon} - \alpha = 0.5$,
    \item Total-corruption budget for rMMD: $r = 1$,
    \item Beta prior parameters: $a_j^0 = b_j^0 = 1$ for all $j = 1, \ldots, p$,
    \item Regularization parameter $\epsilon$: selected via grid search over $\{0.3, 0.4, 0.5, 0.6, 0.7, 0.8, 0.9\}$ based on shilloutte index.
\end{itemize}
For the greedy algorithm introduced in 
Section \ref{section:method_unknown_cov} 
and Appendix \ref{section:method_known_cov},
the hyperparameters are specified as follows:
\begin{itemize}
    \item Target false discovery rate (FDR): $q = 0.4$,
    \item Maximum number of iterations: $T = 100$,
    \item Stopping criterion: the algorithm terminates if the adjusted Rand index (ARI) between clusterings from consecutive iterations equals 1 for 10 successive iterations,
    \item FDR control method: the Benjamini--Hochberg (BH) procedure
\end{itemize}

\subsubsection{Real Data Study}
For the TVS algorithm, the hyperparameters are defined as follows:
\begin{itemize}
    \item Maximum number of iterations: $T = 2000$,
    \item Significance threshold: $\tau_{\epsilon} - \alpha = 0.3$,
    \item Total-corruption budget for rMMD: $r = 5$,
    \item Beta prior parameters: $a_j^0 = b_j^0 = 1$ for all $j = 1, \ldots, p$,
    \item Regularization parameter $\epsilon$: selected via grid search over $\{0.3, 0.4, 0.5\}$ based on shilloutte index.
\end{itemize}
For the greedy algorithm introduced in 
Section \ref{section:method_unknown_cov} 
and Appendix \ref{section:method_known_cov}, the hyperparameters are specified as follows:
\begin{itemize}
    \item Target false discovery rate (FDR): $q = 0.4$,
    \item Maximum number of iterations: $T = 100$,
    \item Stopping criterion: the algorithm terminates if the adjusted Rand index (ARI) between clusterings from consecutive iterations equals 1 for 10 successive iterations,
    \item FDR control method:
    \begin{itemize}
        \item Without ISEE: the Benjamini--Hochberg (BH) procedure,
        \item With ISEE: the Significance Analysis of Microarrays (SAM) framework \citep{tusherSignificanceAnalysisMicroarrays2001}.
    \end{itemize}
\end{itemize}

\subsection{Baseline Methods}\label{section:baselinemethod}
\begin{itemize}
    \item Appendix~\ref{section:baselinemethod_existing} provides details on existing sparse clustering methods.
    \item Appendix~\ref{section:method_known_cov} presents the permutation construction, FDR control, and pseudocode for the greedy version of our algorithm.
    \item Appendix \ref{section:isee_detail} provides details for ISEE subroutine.
\end{itemize}
\subsubsection{Sparsity-aware clustering methods used for benchmarking}\label{section:baselinemethod_existing} 
This section describes the baseline sparsity-aware methods used for comparison with our algorithms in the numerical studies presented in Section~\ref{section:simulation}.

\medskip
\noindent
\textbf{IFPCA.}
Influential Feature PCA (IFPCA; \citealp{jinInfluentialFeaturesPca2016}) is a two step method. 
In the feature Choose Step, features with the highest Kolmogorov-Smirnov (KS) scores are selected, with the selection threshold determined in a data-driven manner using the Higher Criticism principle~\citep{donoho_higher_2004}.  
In the clustering step, spectral clustering is applied: the top \( K - 1 \) left singular vectors are extracted from the normalized data matrix after feature selection and used as inputs to the classical \( K \)-means algorithm to estimate cluster labels. For simulations, we use the authors' \href{https://www.stat.cmu.edu/~jiashun/Research/software/HCClustering/}{MATLAB implementation}.

\medskip
\noindent
\textbf{SAS.}
Sparse Alternate Similarity  clustering (SAS;  \citealp{arias-castroSimpleApproachSparse2017}) is an iterative method that uses the fact that the within-cluster dissimilarity of $K$-means, measured by the squared $\ell_2$ distances, can be decomposed coordinate-wise. The algorithm alternates between clustering and support estimation: given a support set, it applies $K$-means to estimate the clusters; given the cluster assignments, it selects the top $\hat{|S^\ast|}$ coordinates with the largest average within-cluster dissimilarity. The tuning parameter $\hat{|S^\ast|}$, which denotes the number of signal features, is selected using  the following permutation-based approach: For each $\hat{|S^\ast|} = 1, \ldots, p$,  run the SAS algorithm to obtain corresponding cluster and support estimates. For each run,  compare the within-cluster dissimilarity of the selected features under the estimated cluster assignment to that under a randomly permuted cluster assignment. The discrepancy between the two is quantified using the gap statistic \citep{tibshiraniEstimatingNumberClusters2001}. Then choose the value of $\hat{|S^\ast|}$ that yields the largest gap statistic.
	We use the authors'  \href{https://github.com/victorpu/SAS_Hill_Climb}{R implementation}.
    
\medskip
\noindent
\textbf{SKM.}~
Sparse $K$-means (SKM; \citealp{witten_framework_2010} )
is an iterative method which exploits the fact that the within-cluster dissimilarity in $K$-means, measured by squared \( \ell_2 \) distances, can be decomposed coordinate-wise. The method assigns a feature weight \( Y_j \) to each variable \( j = 1, \ldots, p \), and constructs a feature-weighted version of the $K$-means objective. It then optimizes this objective over both feature weights and cluster assignments, subject to a combined lasso and ridge penalty. The proposed algorithm alternates between maximizing the objective with respect to the cluster assignments (given fixed feature weights) and with respect to the feature weights (given fixed cluster assignments).
For simulations, we use the authors' \href{https://cran.r-project.org/web/packages/sparcl/index.html}{R package \texttt{sparcl}}.

\medskip
\noindent
\textbf{CHIME.} CHIME~\citep{cai_chime_2019} is an  Etype iterative method, where the E-step updates the cluster assignments based on current Gaussian mixture model parameters, and the step updates the model parameters with cluster labels fixed. 
Notably, CHIME avoids explicit covariance estimation by employing an \( \ell_1 \)-regularized estimation of the sparse discriminative direction \( \boldsymbol{\beta}^\ast = \boldsymbol{\Omega}^\ast(\mathbf{c}_1^\ast - \mathbf{c}_2^\ast) \).  
After a suitable number of iterations, the estimated \( \boldsymbol{\beta}^\ast \) is plugged into the Fisher discriminant rule to produce the final cluster assignments.
 For the simulations, we translated the official \href{https://github.com/drjingma/gmm}{MATLAB implementation} into R. This algorithm requires careful initialization of the model parameters. The initialization procedure suggested in the original paper (Hardt-Price) is exponential-time and is not practical in high-dimensional settings. Therefore, in our simulations, we initialize the model parameters using the estimated clusters obtained via SKM.

\medskip
\noindent
\textbf{SCVX.}   Sparse Convex Clustering (SCVX; \citealp{wangSparseConvexClustering2018}) extends convex clustering to high-dimensional scenarios by addressing the performance degradation caused by uninformative features. The method simultaneously clusters observations and conducts feature selection by formulating convex clustering as a regularization problem, applying an adaptive group-lasso penalty term to the cluster centers. To optimally balance the trade-off between cluster fitting and sparsity, SCVX employs a tuning criterion based on clustering stability. Theoretically, the estimator guarantees a finite sample error bound and establishes variable selection consistency. Code and further details are provided in the authors' online supplementary material.

\subsubsection{Memoryless Greedy Benchmark}\label{section:method_known_cov}
Under the known isotropic covariance assumption, similar to TVS algorithm,
our greedy algorithm maintains the same alternating structure between the selection block and the clustering block. While the clustering block is identical to our TVS algorithm, the selection block differs: it replaces posterior memory and Thompson sampling with a deterministic current-reward rule. Given the previous partition $\widehat G^{t-1}$, the selection block proceeds in two steps:
\begin{itemize}
\item \textbf{Reward step.} Each variable $j$ receives a binary reward $Y_j^t$ based on the permutation significance of its clustering score, with false discovery rate (FDR) control.
\item \textbf{Choose step.} Every variable with $Y_j^t=1$ is selected, so $\widehat S^t=\{j:Y_j^t=1\}$.
\end{itemize}
Figure \ref{fig:alg_schematic_dual} schematically compares TVS algorithm and greedy algorithm.
We do not employ the \textup{\texttt{rMMD}} test (Algorithm \ref{alg:robust_dc}) for the greedy algorithm, because its conservativeness yields excessively small variable sets. As shown in Theorem \ref{theorem:separation_condition}, this under-selection degrades clustering performance in the subsequent iteration. In contrast, TVS decouples variable selection from the immediate reward. Although the \textup{\texttt{rMMD}} test produces conservative rewards, the accumulated evidence in the Beta posteriors of the TVS algorithm acts as a buffer. This prevents a strict reward from disproportionately restricting the next Choose Step, thereby enabling the moderate over-selection necessary for successful clustering of TVS algorithm.

The permutation test for the greedy algorithm 
 is based on the  transformed cluster mean statistic defined in \eqref{eq:def-Tj}, which was also used for TVS algorithm.
To evaluate $\Lambda(\widehat G^{t-1},\widetilde{\mathbf X}_j)$, let $\pi_b$ be a uniformly sampled permutation of $[n]$ and construct a permuted partition that preserves the estimated cluster sizes:
\begin{equation*}
\widehat G_1^{t-1,(b)}
:=
\{\pi_b(1),\ldots,\pi_b(|\widehat G_1^{t-1}|)\},
\qquad
\widehat G_2^{t-1,(b)}
:=
\{\pi_b(|\widehat G_1^{t-1}|+1),\ldots,\pi_b(n)\}.
\end{equation*}
Repeating this construction $B$ times gives the empirical null statistics
$\{\Lambda(\widehat G^{t-1,(b)},\widetilde{\mathbf X}_j)\}_{b=1}^B$ and hence an empirical $p$-value for each feature.   We then apply FDR control across features.  When the covariance is known, we use the Benjamini--Hochberg procedure \citep{benjaminiControllingFalseDiscovery1995}.
When covariance is unknown, since the tests can be highly correlated across features, so we use Significance Analysis of Microarrays \citep{tusherSignificanceAnalysisMicroarrays2001}, which estimates the FDR directly from permutations.

\begin{algorithm}[H]
\caption{Greedy selection block at iteration $t$ (reward + choose)}
\label{alg:known_cov}
\begin{algorithmic}[1]
\Require Transformed data $\widetilde{\mathbf X}$, partition $\widehat G^{t-1}$, permutation count $B$, FDR target $q$, FDR procedure $Q$
\State $\lambda_j^{(0)}\gets\Lambda(\widehat G^{t-1},\widetilde{\mathbf X}_j)$ for $j=1,\ldots,p$ \Comment{Observed statistics from \eqref{eq:def-Tj}}
\For{$b=1,\ldots,B$}
    \State Sample $\pi_b$ and construct $\widehat G^{t-1,(b)}$ preserving the cluster sizes
    \State $\lambda_j^{(b)}\gets\Lambda(\widehat G^{t-1,(b)},\widetilde{\mathbf X}_j)$ for $j=1,\ldots,p$ \Comment{Permuted statistics from \eqref{eq:def-Tj}}
\EndFor
\State $(Y_1^t,\ldots,Y_p^t)\gets Q\bigl(\{\lambda_j^{(b)}\}_{j=1,b=0}^{p,B};q\bigr)$ \Comment{Reward step}
\State \Return $\widehat S^t\gets\{j\in[p]:Y_j^t=1\}$ \Comment{Choose step}
\end{algorithmic}
\end{algorithm}

\begin{figure}

\begin{tikzpicture}[
    node distance=0.6cm and 3.5cm,
    auto,
    block/.style={
        rectangle, 
        draw=black, 
        align=center, 
        minimum width=5.5cm, 
        minimum height=1.4cm,
        inner sep=8pt,
        font=\linespread{0.85}\small\rmfamily\selectfont
    },
    line/.style={
        draw, 
        -latex, 
        thick,
        font=\small\rmfamily
    },
    gridtable/.style={
        matrix of nodes,
        nodes={
            draw=black, 
            thin,
            anchor=center, 
            align=center,
            text width=3.5em, 
            text height=1.2em, 
            text depth=0.5em,
            inner sep=0pt,
            font=\linespread{0.85}\small\rmfamily\selectfont
        },
        column 1/.style={nodes={fill=black!5, text width=2.5em}},
        row 1/.style={nodes={fill=black!15, font=\linespread{0.85}\small\bfseries\rmfamily\selectfont}},
        row sep=-\pgflinewidth,
        column sep=-\pgflinewidth,
        ampersand replacement=\&
    }
]

\node[font=\large\bfseries\rmfamily] (title1) at (0, 5.5) {Greedy Selection};
\node[font=\large\bfseries\rmfamily] (title2) at (7.2, 5.5) {Thompson Sampling};

\node [block, fill=white, minimum height=0.9cm, inner sep=3pt, below=0.5cm of title1] (init_greedy)  {
Clustering on all features\\
to get initial $\hat{G}^0$, set $\hat{S}^0 = [p]$
};

\node [block, fill=white, minimum height=0.9cm, inner sep=3pt, below=0.5cm of title2] (init_ts)   {
Clustering on all features\\
to obtain initial $\hat{G}^0$
\\
Sample $\vartheta_j^0 \sim \operatorname{Beta}(a_j^0,b_j^0)$\\
$\hat{S}^0 = \{ j : \vartheta_j^0 > \tau_{\epsilon} \}$
};

\coordinate (MergeStart) at ($(init_greedy.south)!0.5!(init_ts.south)$);

\node [block, below=1.5cm of MergeStart, fill=white, minimum width=12.7cm, inner sep=3pt] (trans_shared) {
    \textbf{Transformation step} \\
    $\widetilde{\mathbf{X}} = \boldsymbol{\Omega}^\ast \mathbf{X}$ ($\boldsymbol{\Omega}^\ast$  known) \quad \textbf{OR} \quad
    $\widetilde{\mathbf{X}} = \texttt{ISEE}(\mathbf{X}, \hat{G}^{t-1})$
    ($\boldsymbol{\Omega}^\ast$  unknown; Algorithm \ref{alg:ISEE_subroutine})
};

\node [block, below=0.6cm of {trans_shared.south -| init_greedy}, inner sep=3pt] (reward1) {
    \textbf{Reward  (Algorithm \ref{alg:known_cov})} \\
    Determine $Y_{j}^t \in \{0, 1\},~j=1,\ldots,p$ \\
    via permutation test with FDR control
};

\node [block, below=0.6cm of {trans_shared.south -| init_ts}, inner sep=3pt] (reward2) {
    \textbf{Reward  (Algorithm \ref{alg:tvs_step})} \\
    Determine $Y_{j}^t \in \{0, 1\},~j \in \hat{S}^{t-1}$ via\\
     robust permutation test (Algorithm \ref{alg:robust_dc})
};


\node [block, below=0.5cm of reward2, fill=black!10, inner sep=3pt] (update2) {
    \textbf{Update  (Algorithm \ref{alg:tvs_step})} \\
    $a_{j}^t \gets a_{j}^{t-1} + Y_{j}^{t-1},~j \in \hat{S}^{t-1}$ \\
    $b_{j}^t \gets b_{j}^{t-1} + ( 1 - Y_{j}^{t-1} ),~j \in \hat{S}^{t-1}$
};

\matrix (tbl2) [gridtable, below=-\pgflinewidth of update2] {
    $j$ \& $a_j^t$ \& $b_j^t$ \\
    1 \& $a_1^t$ \& $b_1^t$ \\
    $\vdots$ \& $\vdots$ \& $\vdots$ \\
    $p$ \& $a_p^t$ \& $b_p^t$ \\
};

\node [block, below=0.4cm of tbl2, fill=black!10, inner sep=3pt] (choose2) {
    \textbf{Choose  (Algorithm \ref{alg:tvs_step})} \\
    Sample $\vartheta_{j}^t \sim \operatorname{Beta}
    (
    a_{j}^t, b_{j}^t
     )$ \\
$\hat{S}^t = \bigl\{
j   : \vartheta_{j}^t > 
\tau_{\epsilon}
\bigr\}$
};

\node (ghost_tbl1) [
    draw=black!40, dashed, 
    rectangle,
    minimum width=5.5cm, 
    minimum height = 2cm,
    fit={($(reward1.south) + (0,-0.5)$) ($(reward1.south |- tbl2.south)+ (0,0.25)$)},
    inner sep=0pt,
    label={[black!60, font=\small\rmfamily\itshape, anchor=north, yshift=-10pt]north:No State Storage}
] {};

\node [block, at=(ghost_tbl1 |- choose2),  minimum height=1cm, inner sep=3pt] (choose1) {
    \textbf{ Choose  (Algorithm \ref{alg:known_cov})} \\
    $\hat{S}^t = \{j : Y_{j}^t = 1\}$
};


\coordinate (MergeStart2) at ($(choose1.south)!0.5!(choose2.south)$);

\node [block, below=1cm of MergeStart2, fill=white, minimum width=10.7cm, inner sep=3pt] (cluster_shared) {
    $\hat{G}^t = \texttt{SDPclust}(\widetilde{\mathbf{X}}_{\hat{S}^t, \cdot}, \boldsymbol{\Sigma}^\ast_{\hat{S}^t, \hat{S}^t})$
    ($\boldsymbol{\Sigma}^\ast$  known)
    ~\textbf{OR}\\
    $\hat{G}^t = \texttt{SDPclust}(\hat{\widetilde{\mathbf{X}}}_{\hat{S}^t, \cdot}, \hat{\boldsymbol{\Sigma}}_{\hat{S}^t, \hat{S}^t})$
    ($\boldsymbol{\Sigma}^\ast$  unknown)
};

\node [block, below=0.7cm of cluster_shared, fill=white, minimum width=8.7cm, minimum height=0.7cm, inner sep=3pt] (out_shared) {
 Final Clustering Output $\hat{G}^T$
};

\coordinate (LogicTop) at ($(update2.north) + (0, 0.8cm)$);
\coordinate (LogicBottom) at ($(choose2.south) - (0, 0.4cm)$);

\begin{scope}[on background layer]
    \node (greedy_box) [
        draw=black, thin, fill=white,
        minimum width=6cm,
        fit={(LogicTop -| choose1.west) (LogicBottom -| choose1.east)},
        label={[font=\small\bfseries\rmfamily, black, anchor=north, yshift=-5pt]north:Greedy deterministic selection}
    ] {};

    \node (memory_box) [
        draw=black, thin, fill=white,
        minimum width=6.cm,
        fit={(LogicTop -| update2.west) (LogicBottom -| update2.east)},
        label={[font=\small\bfseries\rmfamily, black, align=center, anchor=north, yshift=-5pt]north:Thompson sampling selection}
    ] {};

    \coordinate (AlignLeft) at ($(greedy_box.west) - (0.3cm, 0)$);
    \coordinate (AlignRight) at ($(memory_box.east) + (0.3cm, 0)$);

    \coordinate (SelectionTop) at ($(trans_shared.north) + (0, 0.2cm)$);
    \coordinate (SelectionBottom) at (greedy_box.south);
    
    \coordinate (ClusterTop) at ($(cluster_shared.north) + (0, 0.2cm)$);
    \coordinate (ClusterBottom) at (cluster_shared.south);

    \node (selection_block) [
        draw=black!80, thick, dotted, fill=black!2, 
        inner sep=15pt,        
        fit={(AlignLeft |- SelectionTop) (AlignRight |- SelectionBottom)}, 
        label={[font=\bfseries\rmfamily, black, anchor=north, yshift=-3pt]north:Selection Block}
    ] {};
    
    \node (clustering_block) [
        draw=black!80, thick, dotted, fill=black!5, 
        inner sep=15pt,        
        fit={(AlignLeft |- ClusterTop) (AlignRight |- ClusterBottom)}, 
        label={[font=\bfseries\rmfamily, black, anchor=north, yshift=-3pt]north:Clustering Block (Algorithm \ref{alg:sdp_subroutine})}
    ] {};
\end{scope}


\draw [line] (init_greedy.south) -- (trans_shared.north -| init_greedy);
\draw [line] (init_ts.south) --  (trans_shared.north -| init_ts);

\draw [line] (trans_shared.south -| init_greedy) -- (reward1.north -| init_greedy);
\draw [line] (trans_shared.south -| init_ts) -- (reward2.north -| init_ts);
\path [line, -] (reward1.south) --  (ghost_tbl1);
\path [line] (ghost_tbl1) -- (choose1);
\path [line] (reward2) -- (update2);
\path [line] (tbl2) -- (choose2);

\draw [line] (choose1.south) -- ++(0, -0.4) -| (cluster_shared.north -| choose1);
\draw [line] (choose2.south) -- ++(0, -0.4) -| (cluster_shared.north -| choose2);

\draw [line] (cluster_shared.west) -- ++(-1.9, 0) |- (trans_shared.west);
\draw [line] (cluster_shared.east) -- ++(1.9, 0) |- (trans_shared.east);

\draw [line] (cluster_shared) -- (out_shared);

\end{tikzpicture}
    \caption{Schematic comparison of the block coordinate optimization algorithms. For Thompson sampling algorithm $\tau_{\epsilon} = \log(1/\epsilon)/\log((\epsilon+1)/\epsilon)$.
    }
    \label{fig:alg_schematic_dual}
\end{figure}

\subsubsection{Details for ISEE Subroutine}\label{section:isee_detail}
This section provides 
the omitted details for the ISEE subroutine introduced in Section \ref{section:method_unknown_cov} of the main text. We proceed in the following steps:
\begin{enumerate}
    \item  Regression relationship on the data sub-vectors,
    \item Key identities on transformed cluster center and noise,
    \item High-dimensional regression method used for ISEE.
\end{enumerate}
Without loss of generality, we focus on the case where \( i \in G_1^\ast \).  
The derivations build on those in \citet{fan_innovated_2016}, with the key distinction that they only consider  the special case \( \mathbf{c}_1^\ast = \boldsymbol{0} \).

We start with regression relationship on the data sub-vectors.
Using the conditional distribution property of multivariate Gaussians and the Schur complement, we derive the following regression relationship between sub-vectors of $\mathbf{X}_i \sim \mathcal{N}(\mathbf{c}_1^\ast, \boldsymbol{\Sigma}^\ast)$ for \( i \in G_1^\ast \), as presented in equation~\eqref{eq:regression} in Section \ref{section:method_unknown_cov} of the main text.
Given a small index  set \( A \subset [p] \) of size 2 or 3,
we recall the following relationship:
\begin{equation*}
\hspace{-1.7em}
\underbrace{
\mathbf{X}_{A,i}
}_{
\hspace{1.1em}
\text{response} 
\in \mathbb{R}^{|A|}} 
\hspace{-0.7em}
= 
\hspace{0.7em}
\underbrace{(\mathbf{c}_1^\ast)_{A} + (\boldsymbol{\Omega}^\ast_{A,A})^{-1} \boldsymbol{\Omega}^\ast_{A,A^c} (\mathbf{c}_1^\ast)_{A^c}}_{:=(\boldsymbol{\alpha}_1)_A \in \mathbb{R}^{|A|}~\text{(intercept)}} 
~-~
\underbrace{
(\boldsymbol{\Omega}^\ast_{A,A})^{-1} \boldsymbol{\Omega}^\ast_{A,A^c}}_{
\hspace{-2.6em}
\text{slope} \in \mathbb{R}^{|A| \times (p - |A|)}
} 
\hspace{-1.6em}
\underbrace{
\mathbf{X}_{A^c, i}
}_{
\hspace{1.1em}
\text{predictor} \in \mathbb{R}^{p - |A|}
} 
\hspace{-1.3em}
+ 
\hspace{-1.53em}
\underbrace{ \mathbf{E}_{A,i}}_{
\hspace{2.1em}
\text{residual} \in \mathbb{R}^{|A|}
}\hspace{-2.1em},
\end{equation*}
where \( \mathbf{E}_{A, i} \sim \mathcal{N}(0, (\boldsymbol{\Omega}^\ast_{A,A})^{-1}) \).
Let us partition \( \mathbf{X}_i \), \( \mathbf{c}_1^\ast \), and \( \boldsymbol{\Omega}^\ast \) according to :
\[
\mathbf{X}_i = 
\begin{bmatrix}
\mathbf{X}_{A,i} \\
\mathbf{X}_{A^c,i}
\end{bmatrix}, \quad
\mathbf{c}_1^\ast = 
\begin{bmatrix}
(\mathbf{c}_1^\ast)_A \\
(\mathbf{c}_1^\ast)_{A^c}
\end{bmatrix}, \quad
\boldsymbol{\Sigma}^\ast = 
\begin{bmatrix}
\boldsymbol{\Sigma}^\ast_{A,A} & \boldsymbol{\Sigma}^\ast_{A,A^c} \\
\boldsymbol{\Sigma}^\ast_{A^c,A} & \boldsymbol{\Sigma}^\ast_{A^c,A^c}
\end{bmatrix}.
\]
Then the conditional distribution \( \mathbf{X}_{A,i}  \mid \mathbf{X}_{A^c,i} \) follows $
\mathcal{N}
(
\bar{\mathbf{c}},
\bar{\boldsymbol{\Sigma}}
)
$, where
\begin{equation}\label{def:mu_bar}
\bar{\mathbf{c}}
:=
(
\mathbf{c}_1^\ast)_A + \boldsymbol{\Sigma}^\ast_{A,A^c} (\boldsymbol{\Sigma}^\ast_{A^cA^c})^{-1} 
\bigl(
\tilde{\mathbf{X}}_{A^c,i}  - (\mathbf{c}_1^\ast)_{A^c} 
\bigr),
\end{equation}
and
\begin{equation}\label{def:Sigma_bar}
\bar{\boldsymbol{\Sigma}} :=
\boldsymbol{\Sigma}^{\ast}_{A,A}
- \boldsymbol{\Sigma}^{\ast}_{A,A^c}
\left( \boldsymbol{\Sigma}^{\ast}_{A^c,A^c} \right)^{-1}
\boldsymbol{\Sigma}^{\ast}_{A^c,A}.
\end{equation}
Next, to apply the Schur complement, we first introduce the following simplfied notation for the matrix partition:
\begin{equation}\label{Sigma_partition}
    \boldsymbol{\Sigma}^\ast = 
\begin{bmatrix}
\boldsymbol{\Sigma}^\ast_{A,A} & \boldsymbol{\Sigma}^\ast_{A,A^c} \\
\boldsymbol{\Sigma}^\ast_{A^c,A} & \boldsymbol{\Sigma}^\ast_{A^c,A^c}
\end{bmatrix}
=
\begin{bmatrix}
\mathbf{A} & \mathbf{B} \\
\mathbf{C} & \mathbf{D}
\end{bmatrix}, 
\end{equation}
Then by the Schur complement, we have:
\begin{equation}\label{shur}
\boldsymbol{\Omega}^\ast = (\boldsymbol{\Sigma}^\ast)^{-1}
=
\begin{bmatrix}
\boldsymbol{\Omega}^\ast_{A,A} & \boldsymbol{\Omega}^\ast_{A,A^c} \\
\boldsymbol{\Omega}^\ast_{A^c,A} & \boldsymbol{\Omega}^\ast_{A^c,A^c}
\end{bmatrix}
=
\begin{bmatrix}
\mathbf{W} & -\mathbf{W} \mathbf{B} \mathbf{D}^{-1} \\
- \mathbf{D}^{-1} \mathbf{C} \mathbf{W} & \mathbf{D}^{-1} + \mathbf{D}^{-1} \mathbf{C} \mathbf{W} \mathbf{B} \mathbf{D}^{-1}
\end{bmatrix}, 
\end{equation}
where 
\begin{equation}\label{def:double_u}
    \mathbf{W} := (\mathbf{A} - \mathbf{B} \mathbf{D}^{-1} \mathbf{C})^{-1}.
\end{equation}
Leveraging this representation of $\boldsymbol{\Omega}^\ast$, we can derive:
\[
- \boldsymbol{\Sigma}_{A,A^c}^\ast (\boldsymbol{\Sigma}_{A^c,A^c}^\ast)^{-1}
\overset{(i)}{=} 
- \mathbf{B} \mathbf{D}^{-1}
\overset{(ii)}{=} 
\mathbf{W}^{-1} (-\mathbf{W} \mathbf{B} \mathbf{D}^{-1})
\overset{(iii)}{=} 
(\boldsymbol{\Omega}_{A,A}^\ast)^{-1} \boldsymbol{\Omega}_{A,A^c}^\ast,
\]
where 
step $(i)$ uses \eqref{Sigma_partition},
step $(ii)$ uses \eqref{def:double_u}, and
step $(iii)$ uses
\eqref{shur}.
This lets us rewrite $\bar{\mathbf{c}}$  in \eqref{def:mu_bar} as follows:
\begin{align*}
\bar{\mathbf{c}}
&=
(
\mathbf{c}_1^\ast)_A + \boldsymbol{\Sigma}^\ast_{A,A^c} (\boldsymbol{\Sigma}^\ast_{A^cA^c})^{-1} 
\bigl(
\tilde{\mathbf{X}}_{A^c,i}  - (\mathbf{c}_1^\ast)_{A^c} 
\bigr)
\\
&=
(
\mathbf{c}_1^\ast)_A - (\boldsymbol{\Omega}_{AA}^\ast)^{-1} \boldsymbol{\Omega}_{AA^c}^\ast
\bigl(
\tilde{\mathbf{X}}_{A^c,i}  - (\mathbf{c}_1^\ast)_{A^c} 
\bigr)
\\
&=
\underbrace{
(
\mathbf{c}_1^\ast)_A 
+
(\boldsymbol{\Omega}_{A,A}^\ast)^{-1} \boldsymbol{\Omega}_{A,A^c}^\ast
(\mathbf{c}_1^\ast)_{A^c} 
}_{ (\boldsymbol{\alpha}_1)_A }
- (\boldsymbol{\Omega}_{A,A}^\ast)^{-1} \boldsymbol{\Omega}_{A,A^c}^\ast
\tilde{\mathbf{X}}_{A^c,i} 
,
\end{align*}
Likewise, we can also simplify $\bar{\boldsymbol{\Sigma}} $ in \eqref{def:Sigma_bar} as follows:
\begin{equation*}
\bar{\boldsymbol{\Sigma}} 
=
\boldsymbol{\Sigma}^{\ast}_{A,A}
- \boldsymbol{\Sigma}^{\ast}_{A,A^c}
\left( \boldsymbol{\Sigma}^{\ast}_{A^c,A^c} \right)^{-1}
\boldsymbol{\Sigma}^{\ast}_{A^c,A}
\overset{(i)}{=}
\mathbf{A} - \mathbf{B} \mathbf{D}^{-1} \mathbf{C}
\overset{(ii)}{=}
\mathbf{W}^{-1} = 
(\boldsymbol{\Omega}^\ast_{A, A})^{-1},
\end{equation*}
where
step $(i)$ uses \eqref{Sigma_partition},
step $(ii)$ uses \eqref{def:double_u}, and
step $(iii)$ uses
\eqref{shur}.
Substituting these expressions  into the conditional distribution
$ \mathbf{X}_{A,i}  \mid \mathbf{X}_{A^c,i} \sim
\mathcal{N}
(
\bar{\mathbf{c}},
\bar{\boldsymbol{\Sigma}}
)
$,
we obtain the regression relationship \eqref{eq:regression}
in Section \ref{section:method_unknown_cov} of the main text.

Next, we turn to the two key identities presented  
in equation \eqref{ISEE_estimation_key_identity} in Section \ref{section:method_unknown_cov} of the main text
:
\begin{equation*}
(\tilde{\mathbf{c}}_1^\ast)_{A}
	= \boldsymbol{\Omega}^\ast_{A,A} (\boldsymbol{\alpha}_1)_{A}
    \quad
    \text{and}
    \quad
         (\tilde{\mathcal{E}}_i)_{A} = \boldsymbol{\Omega}_{A,A } \mathbf{E}_{A, i},
\end{equation*}
which play a central role in the ISEE subroutine.

We start with the first identity.
By pre-multiplying  $\boldsymbol{\Omega}_{A,A}^\ast$ to 
$ (\boldsymbol{\alpha}_1)_A$, we have:
    \begin{equation*}
    \boldsymbol{\Omega}_{A,A}^\ast
 (\boldsymbol{\alpha}_1)_A =
\boldsymbol{\Omega}_{A,A}^\ast
(\mathbf{c}_1^\ast)_A 
+
\boldsymbol{\Omega}_{A,A^c}^\ast
(\mathbf{c}_1^\ast)_{A^c} .
    \end{equation*}    
Therefore, we can express $(\tilde{\mathbf{c}}_1^\ast)_{A}$ as follows:
\begin{equation*}
		(\tilde{\mathbf{c}}_1^\ast)_{A}
		=
		\boldsymbol{\Omega}^\ast_{A,\cdot}\;
		\mathbf{c}_1^\ast 
		= 
		\boldsymbol{\Omega}^\ast_{A,A}
		(\mathbf{c}_1^\ast)_{A}
		+
		\boldsymbol{\Omega}^\ast_{A,A^c}
		(\mathbf{c}_1^\ast)_{A^c}
		=  
		\boldsymbol{\Omega}^\ast_{A,A} (\boldsymbol{\alpha}_1)_{A}.
\end{equation*}
This completes the derivation of the first identity.

Next, we derive the second identity:
	\begin{align*}
		\tilde{\mathcal{E}}_{A,i}
		&=
		\boldsymbol{\Omega}^\ast_{A,\cdot }\;
		\mathbf{X}_{i}	
		-
		\boldsymbol{\Omega}^\ast_{A,\cdot }\;
		\mathbf{c}^\ast_{1}	
		\\&=
		\boldsymbol{\Omega}^\ast_{A,A }
		\mathbf{X}_{A,i}	
		+
		\boldsymbol{\Omega}^\ast_{A,A^c }
		\mathbf{X}_{A^c, i}
		-
		\boldsymbol{\Omega}^\ast_{A,A }
		(\mathbf{c}_{1}^\ast)_{A}
		+
		\boldsymbol{\Omega}^\ast_{A,A^c }
		(\mathbf{c}_{1}^\ast)_{A^c}
		\\&=
		\boldsymbol{\Omega}^\ast_{A,A }
		\biggl(
		\mathbf{X}_{A,i}	
		+
		(\boldsymbol{\Omega}^\ast_{A,A })^{-1}
		\boldsymbol{\Omega}^\ast_{A,A^c }
		\mathbf{X}_{A^c, i}
		-
		(\mathbf{c}^\ast_{1})_{A}
		+
		(\boldsymbol{\Omega}^\ast_{A,A })^{-1}
		\boldsymbol{\Omega}^\ast_{A,A^c }
		(\mathbf{c}^\ast_{1})_{A^c}
		\biggr)
		\\&=
		\boldsymbol{\Omega}^\ast_{A,A }\mathbf{E}_{A,i}.
	\end{align*}
	where the last equality uses the regression relationship \eqref{eq:regression} in Section \ref{section:method_unknown_cov} of the main text.
	This completes the derivation of the second identity.

Finally, we provide information on high-dimensional regression method used for ISEE.
We use the lasso \citep{tibshiraniRegressionShrinkageSelection1996} for high-dimensional regression between nodes in the ISEE subroutine (Algorithm~\ref{alg:ISEE_subroutine}). At each iteration, the regularization parameter is tuned by computing the lasso path and selecting the value that minimizes the Akaike Information Criterion (AIC). This procedure is implemented using functions from the R package \texttt{glmnet} \citep{glmnet}.

\subsection{Stopping Criteria}\label{section:greedy_stopping_criteria}
For the TVS algorithm, we use a fixed $T$ iterations.
In contrast, the greedy algorithm does not involve accumulated learning and we empirically observe that it can enter short cycles, repeatedly visiting a limited set of states. To address this, we introduce an early stopping criterion based on the adjusted Rand index (ARI; \citealp{hubert1985comparing}).

The ARI measures the similarity between two clustering assignments by comparing all pairs of observations and evaluating whether they are assigned to the same or different clusters in both partitions. It corrects for agreement expected by chance by normalizing the Rand Index with respect to its null expectation under random labeling. As a result, the ARI takes the value 1 for identical clusterings, 0 for random agreement, and can be negative when agreement is worse than chance, making it a standardized measure of clustering concordance across settings with different numbers of clusters.

For the greedy algorithm, we monitor the ARI between clusterings from consecutive iterations and terminate the procedure when the ARI equals 1 for 10 consecutive iterations. We compute the ARI using the \texttt{adjustedRandIndex} function from the R package \texttt{mclust}.

\subsection{Text Preprocessing Details}\label{section:autoencoder}
We use the \texttt{tm} package in R. 
\begin{enumerate}
    \item All text were converted to lowercase. 
    \item Punctuation marks and numeric characters  were systematically excised. 
  \item Structural language tokens are discarded by applying the standard English stop word dictionary built into the \texttt{tm} package. This dictionary, derived from the standard Snowball stemmer list, contains exactly 174 high-frequency grammatical tokens, including pronouns (e.g., ``we'', ``itself'', ``they''), articles (``a'', ``the''), conjunctions (``and'', ``because''), and auxiliary verbs (``is'', ``would'', ``having'').
\item Excess whitespace was stripped.

\end{enumerate}

\section{Additional Numerical Results}\label{section:additional_simulation}
This section supplements Section \ref{section:simulation} by presenting  additional simulation results.
\begin{itemize}
    \item Section \ref{section:simul_known_noniso} evaluates our greedy algorithm under known non-isotropic covariance setting.
    \item Section \ref{section:ISEE_variants}
    demonstrates the benefits of ISEE bypassing the full covariance structure estimation.
    \item Section \ref{section:assumption_robustness:same_cov} demonstrates the robustness of our greedy algorithm to homogeneous covariance assumption.
    \item Section \ref{section:assumption_robustness:sparsity} shows the robustness of our greedy algorithm to sparsity assumptions.
    \item Section \ref{section:solver} shows the scalability of our algorithms to large sample sizes, via approximate SDP solvers.
   \item Section \ref{section:JL} compares our sparsity-based algorithms to subspace clustering approaches. 
\end{itemize}

\subsection{Evaluation of  Greedy Algorithm under  Known Non-isotropic Covariance}\label{section:simul_known_noniso}
As discussed in Appendix \ref{section:method_known_cov}, the subset SDP problem \eqref{def:subset_sdp} 
captures the principal sparse clustering signal encoded in $\boldsymbol{\Omega}^\ast(\mathbf{c}_1^\ast - \mathbf{c}_2^\ast)$.
To numerically validate this claim, we evaluate the greedy algorithm from Appendix \ref{section:method_known_cov} under a non-isotropic covariance setting, where the true covariance matrix is provided to the procedure. Specifically, we implement a block coordinate optimization algorithm that uses Algorithm~\ref{alg:known_cov} for the Choose Step and Algorithm~\ref{alg:sdp_subroutine} for the clustering step, omitting the ISEE subroutine (Algorithm~\ref{alg:ISEE_subroutine}).
The synthetic data are generated according to \eqref{simulation_setting_symmetric_gaussian}, with parameter settings (including the covariance structure) matching those in line 1 of Table~\ref{tab:appendix_sim_summary}.

\begin{figure} 
	\center
	\includegraphics[width=0.7\linewidth]{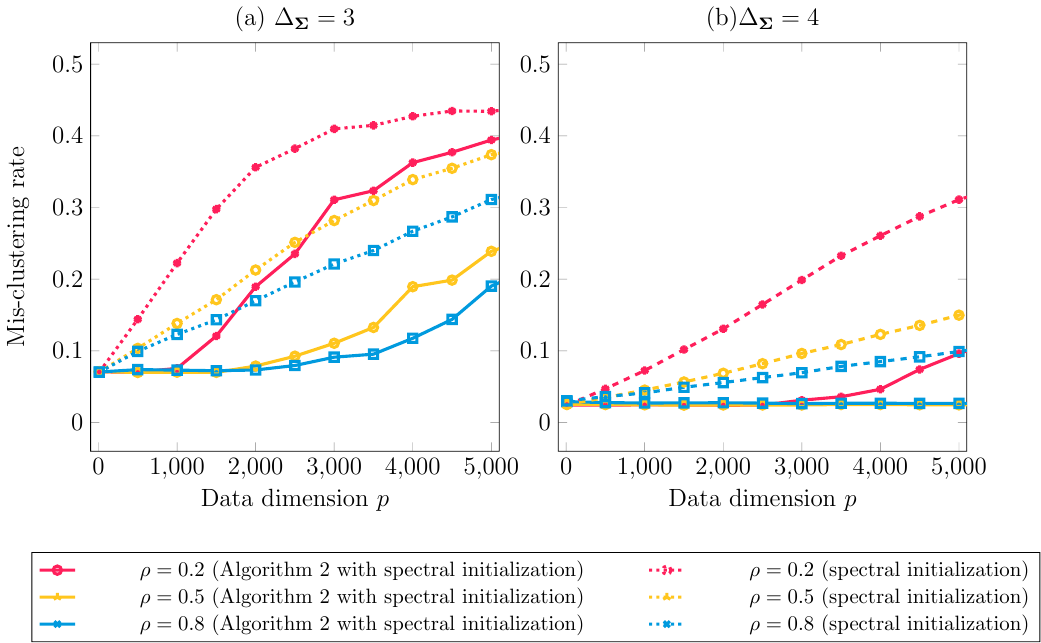}
	\caption{Validation of the subset SDP: under a known non-isotropic (AR(1)) covariance, we solve \eqref{def:subset_sdp}
via a greedy block coordinate method (greedy selection + SDP clustering, without ISEE), with data generated as in \eqref{simulation_setting_symmetric_gaussian} ans Table \ref{tab:appendix_sim_summary}. The $y$-axis reports misclustering rates.
}\label{fig:known_nonidentity_cov}
\end{figure}

The results in Figure~\ref{fig:known_nonidentity_cov} reveal two main findings. First, the method adapts effectively to the sparsity of $\boldsymbol{\Omega}^\ast(\mathbf{c}_1^\ast - \mathbf{c}_2^\ast)$: when the total clustering signal is fixed (in particular, when $\Delta_{S^\ast} = 4$), performance remains stable as ambient dimension increases with the clustering signal fixed. Second, when the covariance matrix is known, stronger correlations facilitate clustering. As $\rho$ increases, the mis-clustering rate deteriorates more slowly with increasing sparsity, again with the total clustering signal held fixed. This behavior is expected, since knowledge of a highly correlated structure is effectively equivalent to operating in a lower-dimensional space.

\begin{table}
\caption{Summary of Parameter Settings for Model \eqref{simulation_setting_symmetric_gaussian}  used in Appendices \ref{section:simul_known_noniso} and \ref{section:ISEE_variants}, 
with $i,j = 1, \ldots, p$.
}
\label{tab:appendix_sim_summary}
\vspace{1em}
\centering
\begin{tabular}{llllll}
\toprule
 Section
& Covariance structure
& $p$
& $n$ 
& $\Delta_{S^\ast}$
\\
\midrule
 \ref{section:simul_known_noniso}
& \makecell[l]{known, $\Sigma^\ast_{i,j} = \rho^{|i-j|}$,\\ $\rho \in \{0.2, 0.5, 0.8\}$}
& 50-5000   
& 200
& $3, 4$
\\
 \ref{section:ISEE_variants}
& \makecell[l]{unknown, $\Omega^*_{i,i+1} = \Omega^*_{i+1,i} = \rho$,\\ $\Omega^*_{i,i} = 1$, $\Omega^*_{i,j} = 0$~o.w., 
\\ 
$\rho = 0.45$ }
& 400
& 500
& $3, 4$
\\
\bottomrule
\end{tabular}
\end{table}

\subsection{Demonstrating the Benefits of ISEE Bypassing Full Covariance Structure Estimation}\label{section:ISEE_variants}
We argued in Section \ref{section:method_unknown_cov} that a key strength of the greedy algorithm with ISEE is its ability to avoid full estimation of the precision matrix. In this section, we provide numerical evidence supporting this claim by comparing the greedy ISEE-based algorithm with  a specially designed variant.
At each iteration \( t \), the variant algorithm estimates the full precision matrix using the graphical lasso \citep{friedman_sparse_2008}, and then uses this estimate to compute the required quantities, namely \( \tilde{\mathbf{X}} \) and \( \boldsymbol{\Omega}^\ast(\mathbf{c}_1^\ast - \mathbf{c}_2^\ast) \). The submatrix \( \boldsymbol{\Sigma}_{\hat{S}^t, \hat{S}^t} \) is estimated in the same manner as in the greedy algorithm with ISEE. These estimates are subsequently used in the clustering block (Algorithm \ref{alg:sdp_subroutine}) and selection block (Algorithm \ref{alg:known_cov}) of our greedy algorithm.
We use synthetic data generated with   the setting specified in line 2 of Table \ref{tab:appendix_sim_summary}.
The results in Table~\ref{tab:graphical_lasso} show that the graphical lasso--based greedy algorithm performs substantially worse than the ISEE--based approach, indicating that avoiding full estimation of the precision matrix leads to improved performance. 
 
\begin{table}[htpb]
\centering
\caption{Mis-clustering rate of graphical-lasso based and ISEE-based greedy algorithms,  with data generated as in \eqref{simulation_setting_symmetric_gaussian} ans Table \ref{tab:appendix_sim_summary}.}\label{tab:graphical_lasso}
\begin{tabular}{ccccc}
\hline
& \multicolumn{2}{c}{$\Delta_{S^\ast} = 3$} & \multicolumn{2}{c}{$\Delta_{S^\ast} = 4$} \\
\cline{2-3} \cline{4-5}
Ambient Dimension ($p$) & Graphical Lasso & ISEE & Graphical Lasso & ISEE \\
\hline
50 & 0.28 & 0.19 & 0.06 & 0.04 \\
100 & 0.31 & 0.22 & 0.07 & 0.05 \\
150 & 0.33 & 0.23 & 0.10 & 0.07 \\
200 & 0.33 & 0.26 & 0.14 & 0.08 \\
250 & 0.37 & 0.27 & 0.18 & 0.08 \\
300 & 0.39 & 0.29 & 0.22 & 0.11 \\
350 & 0.40 & 0.30 & 0.24 & 0.10 \\
400 & 0.41 & 0.31 & 0.25 & 0.11 \\
\hline
\end{tabular}
\end{table}

\subsection{Robustness to the Assumption of Covariance Homogeneity}\label{section:assumption_robustness:same_cov}
When cluster covariances differ, our theoretical results no longer apply and the optimal (Bayes) clustering boundary becomes nonlinear (quadratic for Gaussian distributions). Blindly applying our algorithm under covariance heterogeneity presents a common model misspecification challenge, as our algorithm learns a linear boundary. 
Under mild heterogeneity, we expect our greedy algorithm to retain strong performance. To verify this, we conduct a simulation in which the data distribution remains Gaussian but with two different covariance matrices, $(1+\delta)\mathbf{I}_p$ and $(1-\delta)\mathbf{I}_p$. The results,  reported in Table \ref{table:robustness:same_cov:iso},
show an interesting phenomenon of higher clustering accuracy for larger $\delta$. A plausible explanation is that larger $\delta$ corresponds to larger difference in clusters and hence an easier clustering problem. Theoretical understanding of the impact of such heterogeneity is an interesting future research direction.
\begin{table}[h!]
\centering
\caption{Comparison of clustering accuracy for the greedy algorithm under heterogeneous and homogeneous covariances. All other settings follow line 1 of Table \ref{tab:synthetic_summary} with $\Delta_{S^\ast} = 4$.}\label{table:robustness:same_cov:iso} \vspace{1em}
\begin{tabular}{lccccc}
\toprule
$\delta$  & $p=1000$ & $p=2000$ & $p=3000$ & $p=4000$ & $p=5000$ \\
\midrule
0 (homogeneous covariance)    & 0.97 & 0.93 & 0.86 & 0.74 & 0.68 \\
0.05 (homogeneous covariance) & 0.98 & 0.92 & 0.84 & 0.76 & 0.74 \\
0.1 (heterogeneous covariance)  & 0.98 & 0.92 & 0.91 & 0.87 & 0.88 \\
\bottomrule
\end{tabular}
\end{table}

\subsection{Assumption of  Sparsity}\label{section:assumption_robustness:sparsity}
Rather than the cluster centers themselves being exactly sparse,
we assume that their difference $
( \mathbf{c}^\ast_1 - \mathbf{c}^\ast_2 )
$ is exactly sparse.
When this difference is approximately sparse, it implies that all entries have clustering power, just that some have weak signals.
Our separation bound in Theorem \ref{theorem:separation_condition} is derived for the difference $\mathbf{c}^\ast_1 - \mathbf{c}^\ast_2$, and it does not rule out the approximately sparse scenario where $\|\mathbf{c}^\ast_1 - \mathbf{c}^\ast_2\|_0 \approx s$. Indeed, our result ensures that for any subset $S$ containing enough signals such that the cumulative signal strength satisfies the inequality in \eqref{def:S-set}, exact recovery of cluster IDs can be achieved.

In practice, having weak signals can cause challenges in feature selection. However, since our goal is clustering and feature selection is conducted only to reduce noise, including a few weak signals does not necessarily make clustering much worse.
We conduct additional experiments to support this claim.
First, in the known covariance setting of line 1 of Table \ref{tab:synthetic_summary}, we add 
$\delta \in \{0.02, 0.05\}$
to 
$\lfloor p/10 \rfloor$ noise entries of
$\mathbf{c}^\ast_1$.
The result presented in Table \ref{table:robustness:sparsity:mean} shows that our greedy algorithm retains strong performance under approximate sparsity of the cluster centers.
Note that the slight increase in clustering accuracy stems from the increased cumulative signal strength from these weak signals.
\begin{table} 
\centering
\caption{Comparison of clustering accuracy for our greedy algorithm under exact  and approximate sparsity of the cluster centers. All other settings follow line 1 of Table \ref{tab:synthetic_summary} with $\Delta_{S^\ast} = 4$.}
\label{table:robustness:sparsity:mean}
\vspace{1em}
\begin{tabular}{cccccc}
\toprule
$\delta$ & $p=1000$ & $p=2000$ & $p=3000$ & $p=4000$ & $p=5000$ \\
\midrule
0 (exact sparsity)   & 0.97 & 0.93 & 0.86 & 0.74 & 0.68 \\
0.02 (approximate sparsity) & 0.98 & 0.95 & 0.81 & 0.70 & 0.63 \\
0.05 (approximate sparsity) & 0.97 & 0.96 & 0.85 & 0.75 & 0.68 \\
\bottomrule
\end{tabular}
\end{table}

Under the unknown covariance setting, our greedy algorithm  depends on ISEE, which depends on the exact sparsity of the precision matrix. We demonstrate numerically that our greedy algorithm with ISEE remains robust under approximate sparsity of the precision matrix. Starting from an exactly sparse precision matrix $\boldsymbol{\Omega}^\ast$, we introduce approximate sparsity in two ways. First, we add $\delta \in \{0.02, 0.05\}$ to 10\% of the zero entries of $\boldsymbol{\Omega}^\ast$. Second, we add $0.01$ to all zero entries of $\boldsymbol{\Omega}^\ast$.
The result presented in Table \ref{table:robustness:sparsity:cov} shows that our greedy algorithm with ISEE retains strong performance under approximate sparsity of the precision matrix.
\begin{table}
\centering
\caption{Comparison of clustering accuracy for our greedy algorithm under  exact  and approximate sparsity of the precision matrix. All other settings follow line 2 of Table \ref{tab:synthetic_summary} with $\Delta_{S^\ast} =  4$.}
\label{table:robustness:sparsity:cov}
\vspace{1em}
\begin{tabular}{lcccc}
\toprule
$\delta$ & $p=100$ & $p=200$ & $p=300$ & $p=400$ \\
\midrule
0 (exact sparsity)             & 0.94 & 0.95 & 0.93 & 0.92 \\
0.02 (added to 10\% of zero entries)     & 0.95 & 0.96 & 0.93 & 0.94 \\
0.05 (added to 10\% of zero entries)     & 0.96 & 0.91 & 0.91 & 0.88 \\
0.001 (added to all zero entries)     & 0.96 & 0.95 & 0.95 & 0.93 \\
\bottomrule
\end{tabular}
\end{table}

\subsection{Scalability to Large Samples and Alternative Solvers }\label{section:solver}
Our focus is the high-dimensional, low-sample-size regime, but we also address scalability: for large $n$, we recommend approximate SDP solvers, and our greedy algorithm shows successful scaling up to $n=100{,}000$.

The computational cost of the clustering block of our algorithms (both greedy and TVS)
is largely determined by the sample size $n$, since the decision variable in the SDP  is an $n \times n$ matrix. 
The  selection block prunes irrelevant variables,
so $p$ primarily affects only the initial clustering stage. 
However, the ADMbased solver   does not scale well beyond $n > 1000$ 
(see Table~5 of \citealp{sun_sdpnal_2020} for details). 

For moderately large sample sizes (up to $n=10{,}000$), instead of solving the full SDP, we recommend solving its nonnegative low-rank restriction using a nonconvex Burer–Monteiro factorization \citep{zhuang_statistically_2023}. 
\citet{zhuang_statistically_2023} shows that
this approach retains the same statistical clustering guarantees as the full SDP problem. Under the same setting as line 1 of Table \ref{tab:synthetic_summary}, we numerically compare the clustering accuracy of our greedy algorithm, where the clustering step is performed either by solving the original SDP formulation   or its low-rank restriction. The results in Table \ref{tab:scalability:accuracy} show that the low-rank SDP formulation achieves comparable clustering accuracy in the small-sample setting. For $p=5000$, the average end-to-end runtime was $121.33$\,s for the original SDP-based iterative algorithm and $26.66$\,s for the low-rank SDP-based version (Intel Core i7-1270P CPU, 16\,GB RAM).

\begin{table}[b]
\centering
\caption{Clustering accuracy of our greedy algorithm with  different clustering step forumlations.
Original SDP solved by ADMbased solver; 
low-rank restriction solved via Burer–Monteiro factorization. 
Other settings follow line~1 of Table~\ref{tab:synthetic_summary} with $\Delta_{S^\ast}=4$.
}
\label{tab:scalability:accuracy}
\vspace{1em}
\begin{tabular}{lccccc}
\toprule
Clustering step formulation  & $p=1000$ & $p=2000$ & $p=3000$ & $p=4000$ & $p=5000$ \\
\midrule
Original SDP  & 0.97 & 0.93 & 0.86 & 0.74 & 0.68 \\
Low-rank restriction of SDP    & 0.97 & 0.93 & 0.83 & 0.74 & 0.66 \\
\bottomrule
\end{tabular}
\end{table}
We further verify numerically that the low-rank restriction–based algorithm efficiently handles sample sizes up to $n=10{,}000$, achieving reliable clustering accuracy (Table~\ref{tab:scalability:large}). For $n=10{,}000$, the average end-to-end runtime of the iterative algorithm was about seven minutes (Intel Core i7-1270P CPU, 16\,GB RAM).

\begin{table}
\centering
\caption{Clustering accuracy of our greedy algorithm with clustering step formulated as low-rank restriction of SDP, under varying $n$ with fixed ambient dimension $p=10000$. 
Low-rank restriction solved via Burer–Monteiro factorization. 
Other settings follow line~1 of Table~\ref{tab:synthetic_summary} with $\Delta_{S^\ast}=3$.}
\label{tab:scalability:large}
\vspace{1em}
\begin{tabular}{lcccc}
\toprule
Clustering step formulation   & $n=7000$ & $n=8000$ & $n=9000$ & $n=10000$ \\
\midrule
Low-rank restriction of SDP & 0.92 & 0.92 & 0.92 & 0.93 \\
\bottomrule
\end{tabular}
\end{table}
To further demonstrate scalability to extremely large sample sizes (e.g., $n=100{,}000$), 
we incorporate the sketch-and-lift technique of \citet{zhuang_sketchandlift_2022}. 
This method randomly subsamples observations, solves the smaller clustering problem, and then propagates the clustering results. 
\citet{zhuang_sketchandlift_2022} show that this approach enjoys a similar statistical guarantees as the inner algorithm applied to the sub-problems. 
In our implementation, we solve the low-rank restriction of the SDP proposed by \citet{zhuang_statistically_2023} for the sub-problems of sample size $n^{3/7}$. 
Table~\ref{tab:scalability:sketchlift} reports strong clustering accuracy and favorable end-to-end timing of this approach,
for sample sizes up to $n=100{,}000$ (Intel Core i7-1270P CPU, 16\,GB RAM).
The runtime increases only modestly with $n$, 
since feature selection and the sketch-and-lift technique keep the per-iteration cost manageable. 
Moreover, the algorithm tends to stop earlier for larger sample sizes, further reducing the overall runtime. 
\begin{table}
\centering
\caption{Scalability of  our greedy algorithm with known covariance, with clustering step formulated as sketch-and-lift with low-rank restriction of SDP, for fixed ambient dimension $p=10,000$ and varying sample sizes. Other settings follow line~1 of Table~\ref{tab:synthetic_summary} with $\Delta_{S^\ast}=3$.}
\label{tab:scalability:sketchlift}
\vspace{1em}
\begin{tabular}{lccc}
\toprule
& $n=60,000$ & $n=80,000$ & $n=100,000$ \\
\midrule
Clustering accuracy              & 0.957 & 0.962 & 0.969 \\
End-to-end timing (s)            & 196.80 & 277.58 & 345.03 \\
Iterations until stop & 11.26 & 10.54 & 10.34 \\
\bottomrule
\end{tabular}
\end{table}

\subsection{Comparison with Subspace Clustering Approach}\label{section:JL}
While both subspace clustering and sparse clustering involve dimensionality reduction, they differ fundamentally. Subspace clustering achieves dimensionality reduction via linear combinations of features, whereas feature selection explicitly identifies and prunes individual variables. Beyond this conceptual distinction, we empirically compared our method against a subspace clustering approach with theoretical guarantees: the Johnson-Lindenstrauss (JL) transforbased $K$-means (JL $K$-means; \citealp{boutsidisRandomProjectionsKmeans2010}). After applying their JL transform, we ran the SDP-relaxed $K$-means \citep{chen_cutoff_2021} on the dimension-reduced data. This JL transform is theoretically guaranteed to preserve the $K$-means objective up to a constant. Under the setting of line~1 of Table~\ref{tab:synthetic_summary}), which has 10 signal features, we provided the JL algorithm with intrinsic dimension parameters $t=10$ and $t=20$. The results, displayed in Table~\ref{table:JL}, show that our greedy algorithm outperforms   JL $K$-means  in terms of clustering accuracy.

\begin{table}
\centering
\caption{Comparison of clustering accuracy of our greedy algorithm with JL $K$-means with different intrinsic dimension parameter $t$.  Other settings follow line~1 of Table~\ref{tab:synthetic_summary} with $\Delta_{S^\ast}=4$.}\label{table:JL}
\vspace{1em}
\begin{tabular}{c c c c c}
\toprule
 & $p=2000$ & $p=3000$ & $p=4000$ & $p=5000$ \\
\midrule
JL $K$-means ($t=10$) & 0.53 & 0.54 & 0.53 & 0.52 \\
JL $K$-means ($t=20$)  & 0.53 & 0.53 & 0.52 & 0.54 \\
Our Greedy Algorithm   & $\mathbf{0.93}$ & 
$\mathbf{0.86}$ & 
$\mathbf{0.74}$ & 
$\mathbf{0.68}$ \\
\bottomrule
\end{tabular}
\end{table}

\vskip 0.2in
\bibliography{sample}

\end{document}